\documentclass[1p]{elsarticle}

\usepackage{graphicx}
\usepackage{stackengine}
\usepackage{amsmath}
\usepackage{amssymb}

\newcommand{\cN}{\mathcal{N}}
\newcommand{\qso}{Q_{s,0}}
\newcommand{\lqcd}{\Lambda_{\mathrm{QCD}}}
\newcommand{\gev}{\ \textrm{GeV}}

\newcommand{\ben}{\begin{eqnarray}}
\newcommand{\een}{\end{eqnarray}}

\newcommand{\bef}{\begin{figure}[htpb]\centering}
\newcommand{\eef}{\end{figure}}

\def\beq{\begin{equation}}
\def\eeq{\end{equation}}
\def\bea{\begin{eqnarray}}
\def\eea{\end{eqnarray}}
\def\eq#1{{Eq.~(\ref{#1})}}

\newcommand{\state}[4]{{^{#1}\hspace{-0.1mm}#2_{#3}^{[#4]}}}

\newcommand\CScSa{\state{3}{\rm S}{1}{1}}

\newcommand\COaSz{\state{1}{\rm S}{0}{8}}

\newcommand\COcSa{\state{3}{\rm S}{1}{8}}
\newcommand\COcPz{\state{3}{\rm P}{0}{8}}

\newcommand\COcPj{\state{3}{\rm P}{J}{8}}
\newcommand{\vt}[1]{{{\boldsymbol #1}_T}}
\newcommand{\vtn}[2]{{{\boldsymbol #1}_{#2T}}}

\newcommand{\vp}{{\vt{p}}}
\newcommand{\vk}{{\vt{k}}}
\newcommand{\vka}{{\vtn{k}{1}}}

\newcommand{\vtp}[1]{{{\boldsymbol #1}'_T}}

\newcommand{\vkp}{{\vtp{k}}}

\newcommand{\jpsi}{{J/\psi}}

\newcommand\mo{{\mathcal O}}
\newcommand{\LDME}[2]{\langle\mo^{#1}(#2)\rangle}
\newcommand\mops{\LDME{\jpsi}{\CScSa}}
\newcommand\mopa{\LDME{\jpsi}{\COaSz}}
\newcommand\mopb{\LDME{\jpsi}{\COcSa}}
\newcommand\mopc{\LDME{\jpsi}{\COcPz}}

\begin{document} 

\begin{frontmatter}
  
  \title{Predictions for Cold Nuclear Matter Effects in
    $p+$Pb Collisions at $\sqrt{s_{_{NN}}} = 8.16$ TeV}

\author{Javier L. Albacete} 
\address{CAFPE and Departamento de F\'{\i}sica Te\'orica y del Cosmos, 
Universidad de Granada E-18071 Campus de Fuentenueva, Granada, Spain}

\author{Fran\c{c}ois Arleo}
\address{Laboratoire Leprince-Ringuet, \'Ecole Polytechnique,
CNRS/IN2P3, Universit\'e Paris-Saclay, 91128, Palaiseau, France}


\author{Gergely G. Barnaf\"oldi} 
\address{Wigner Research Centre for Physics of the Hungarian Academy of 
Sciences, 29-33 Konkoly-Thege Mikl\'os Str, H-1121 Budapest, Hungary}

\author{G\'abor B\'ir\'o}
\address{Wigner Research Centre for Physics of the Hungarian Academy of 
Sciences, 29-33 Konkoly-Thege Mikl\'os Str, H-1121 Budapest, Hungary}
\address{E\"otv\"os Lor\'and University, 1/A P\'azm\'any P. S\'et\'any, 
H-1117, Budapest, Hungary} 


\author{David d'Enterria}
\address{CERN, EP Department, 1211 Geneva, Switzerland}


\author{Bertrand Duclou\'e}
\address{
Department of Physics, University of Jyv\"askyl\"a,
 P.O. Box 35, 40014 University of Jyv\"askyl\"a, Finland
}
\address{
Helsinki Institute of Physics, P.O. Box 64, 00014 University of Helsinki,
Finland
}

\author{Kari J. Eskola}
\address{University of 
Jyv\"{a}skyl\"{a}, Department of Physics, P.O. Box 35, FI-40014 University of 
Jyv\"{a}skyl\"{a}, Finland}
\address{Helsinki Institute of Physics, P.O. Box 64, FIN-00014 University 
of Helsinki, Finland}

\author{Elena G. Ferreiro}
\address{Departamento de F{\'\i}sica de Part{\'\i}culas, Universidade de
  Santiago de Compostela, 15782 Santiago de Compostela, Spain}

\author{Miklos Gyulassy} 
\address{Wigner Research Centre for Physics of the Hungarian Academy of Sciences, 29-33 Konkoly-Thege Mikl\'os Str, H-1121 Budapest, Hungary}
\address{Pupin Lab MS-5202, Department of Physics, Columbia University, New York, NY 10027, USA}
\address{Nuclear Science Division, MS 70R0319, Lawrence Berkeley National Laboratory, Berkeley, California 94720 USA}
\address{Institute of Particle Physics, Central China Normal University, Wuhan 430079, China}

\author{Szilvester Mikl\'os Harangoz\'o}
\address{Wigner Research Centre for Physics of the Hungarian Academy of 
Sciences, 29-33 Konkoly-Thege Mikl\'os Str, H-1121 Budapest, Hungary}
\address{E\"otv\"os Lor\'and University, 1/A P\'azm\'any P. S\'et\'any, 
H-1117, Budapest, Hungary}


\author{Ilkka Helenius}
\address{Institute for Theoretical Physics, T\"{u}bingen University, Auf der 
Morgenstelle 14, 72076 T\"{u}bingen, Germany}

\author{Zhong-Bo Kang}  
\address{Department of Physics and Astronomy, University of California, Los 
Angeles, CA 90095, USA}
\address{Mani L. Bhaumik Institute for Theoretical Physics, University of 
California, Los Angeles, CA 90095, USA}
\address{Theoretical Division, MS B283, Los Alamos National Laboratory, 
Los Alamos, NM 87545, USA}

\author{Piotr Kotko}
\address{Department of Physics, Penn State University,
University Park, PA 16803, USA}

\author{Sergey A. Kulagin}
\address{Institute for Nuclear Research of the Russian Academy of Sciences,
  Moscow 117312, Russia}

\author{Krzysztof Kutak}
\address{The H.\ Niewodnicza\'nski Institute of Nuclear Physics PAN, 
Radzikowskiego 152, 31-342 Krak\'ow, Poland}

\author{Jean Philippe Lansberg}
\address{IPNO, Universit\'e Paris-Sud, CNRS/IN2P3, Universit\'e Paris-Saclay,
91406 Orsay Cedex, France}

\author{Tuomas Lappi}
\address{
Department of Physics, University of Jyv\"askyl\"a,
P.O. Box 35, 40014 University of Jyv\"askyl\"a, Finland
}
\address{
Helsinki Institute of Physics, P.O. Box 64, 00014 University of Helsinki,
Finland
}

\author{P\'eter L\'evai}
\address{Wigner Research Centre for Physics of the Hungarian Academy of 
Sciences, 29-33 Konkoly-Thege Mikl\'os Str, H-1121 Budapest, Hungary}

\author{Zi-Wei Lin}
\address{Department of Physics, East
  Carolina University, Greenville, NC 27858, USA} 
\address{Key Laboratory of Quarks and Lepton Physics (MOE) and 
Institute of Particle Physics,
Central China Normal University, Wuhan 430079, China}

\author{Guoyang Ma} 
\address{Institute of Particle Physics, Central China Normal University, 
Wuhan 430079, China}

\author{Yan-Qing Ma}
\address{School of Physics and State Key Laboratory of Nuclear Physics and 
Technology, Peking University, Beijing 100871, China}
\address{Center for High Energy Physics, Peking University, Beijing 100871, 
China}

\author{Heikki M\"antysaari}
\address{
Physics Department, Brookhaven National Laboratory, Upton, NY 11973, USA
}

\author{Hannu Paukkunen}
\address{University of 
Jyv\"{a}skyl\"{a}, Department of Physics, P.O. Box 35, FI-40014 University of 
Jyv\"{a}skyl\"{a}, Finland}
\address{Helsinki Institute of Physics, P.O. Box 64, FIN-00014 University 
of Helsinki, Finland}

\author{G\'abor Papp}
\address{E\"otv\"os Lor\'and University, 1/A P\'azm\'any P. S\'et\'any, 
H-1117, Budapest, Hungary} 


\author{Roberto Petti}
\address{Department of Physics and Astronomy, University of South Carolina, 
Columbia SC 29208, USA}


\author{Amir H. Rezaeian}
\address{Departamento de F\'\i sica, Universidad T\'ecnica
Federico Santa Mar\'\i a, Avda. Espa\~na 1680,
Casilla 110-V, Valpara\'iso, Chile}
\address{Centro Cient\'\i fico Tecnol\'ogico de Valpara\'\i so (CCTVal),
  Universidad T\'ecnica
Federico Santa Mar\'\i a, Casilla 110-V, Valpara\'\i so, Chile }

\author{Peng Ru}
\address{School of Physics $\&$ Optoelectronic Technology, Dalian
University of Technology, Dalian, 116024 China}
\address{Institute of Particle Physics,
Central China Normal University, Wuhan, 430079 China}

\author{Sebastian Sapeta}
\address{The H.\ Niewodnicza\'nski Institute of Nuclear Physics 
PAN, Radzikowskiego 152, 31-342 Krak\'ow, Poland}

\author{Bj\"orn Schenke}
\address{Physics Department, Brookhaven National Laboratory, Upton, New York 
  11973-5000, USA}

\author{S\"oren Schlichting}
\address{Department of Physics, University of Washington, Seattle, WA
98195-1560, USA}

\author{Hua-Sheng Shao}
\address{Sorbonne Universit\'es, UPMC Univ. Paris 06, UMR 7589, LPTHE, F-75005
  Paris, France}
\address{CNRS, UMR 7589, LPTHE, F-75005 Paris, France}


\author{Prithwish Tribedy}
\address{Physics Department, Brookhaven National Laboratory, Upton, New York 
11973-5000, USA}

\author{Raju Venugopalan}
\address{Physics Department, Brookhaven National Laboratory, Upton, New York 
11973-5000, USA}

\author{Ivan Vitev}  
\address{Theoretical Division, MS B283, Los Alamos National Laboratory, 
Los Alamos, NM 87545, USA}

\author{Ramona Vogt\fnref{email}}
\address{Physics Division, Lawrence Livermore National Laboratory, 
Livermore, CA 94551, USA}
\address{Physics Department, University of California at Davis, 
Davis, CA 95616, USA}\fntext[email]{vogt@physics.ucdavis.edu}

\author{Enke Wang}
\address{Key Laboratory of Quark $\&$ Lepton Physics (MOE) and 
Institute of Particle Physics, Central China Normal University, Wuhan 430079, 
China}

\author{Xin-Nian Wang}
\address{Nuclear Science Division, MS 70R0319, Lawrence Berkeley National 
Laboratory, Berkeley, California 94720 USA}
\address{Institute of Particle Physics, Central China Normal University, 
Wuhan 430079, China}

\author{Hongxi Xing}
\address{Department of Physics and Astronomy, Northwestern University, 
Evanston, IL 60208, USA}
\address{High Energy Physics Division, Argonne National Laboratory, Argonne,
IL 60439, USA}

\author{Rong Xu}
\address{Key Laboratory of Quark and Lepton Physics (MOE) and Institute of 
Particle Physics, Central China Normal University, Wuhan 430079, China}

\author{Ben-Wei Zhang}
\address{Key Laboratory of Quark $\&$ Lepton Physics (MOE) and 
Institute of Particle Physics, Central China Normal University, Wuhan 430079, 
China}

\author{Hong-Fei Zhang}
\address{School of Science, Chongqing University of Posts and 
Telecommunications, Chongqing 400065, China}

\author{Wei-Ning Zhang}
\address{ School of Physics $\&$ Optoelectronic
Technology, Dalian University of
Technology, Dalian, 116024 China}
\address{Physics Department, Harbin
Institute of Technology, Harbin 150006, China}


\begin{abstract}
  Predictions for cold nuclear matter effects on
  charged hadrons, identified light hadrons, quarkonium and
  heavy flavor hadrons, Drell-Yan dileptons, jets, photons, gauge bosons and top
  quark pairs produced in $p+$Pb collisions at $\sqrt{s_{_{NN}}} = 8.16$ TeV are
  compiled and, where possible, compared to each other.  Predictions of the
  normalized ratios of $p+$Pb to $p+p$ cross sections are also presented for
  most of the observables, providing new insights into the expected role of
  cold nuclear matter effects.  In particular, the role of nuclear parton
  distribution functions on particle production can now be probed over a wider
  range of phase space than ever before.
\end{abstract}

\begin{keyword}
  \texttt{perturbative QCD, hard and electromagnetic probes,
    cold nuclear matter, charged hadron production}
\end{keyword}

\end{frontmatter}
  
\section{Introduction}
\label{sec:Intro}

This paper compiles cold matter
predictions for the 8.16 TeV $p+$Pb run at the LHC that
occurred in November 2016.  While it appears after the completion of the
run, the predictions were all gathered before any data appeared.  However, the
preliminary data that have become available after the run, namely $J/\psi$
rapidity and $p_T$ dependence in the forward and backward rapidity regions from
ALICE \cite{Enrico} and LHCb \cite{Aaij:2017cqq} are included
for comparison in the appropriate sections.

This work follows the format of the predictions for $p+$Pb run at 
$\sqrt{s_{_{NN}}} = 5.02$~TeV \cite{Albacete:2013ei}.  Section~\ref{sec:models}
describes the models that specifically address unidentified light charged
hadron production.
These include saturation approaches, Monte Carlo
event generators, and perturbative 
QCD-based calculations.  Section~\ref{sec:charged_hadrons} 
compares predictions obtained
from models described in Sec.~\ref{sec:models} with each other.
The next several sections present predictions for specific
observables including
quarkonium and heavy flavor hadrons (Sec.~\ref{sec:Onia_HF}),
Drell-Yan lepton pairs (Sec.~\ref{sec:DY}),
jets (Sec.~\ref{sec:jets}),
direct photons (Sec.~\ref{sec:photons}),
massive gauge bosons (Sec.~\ref{sec:WZ_prod}), and
top quarks (Sec.~\ref{sec:topQ}).  Some of the calculations were made at
$\sqrt{s_{_{NN}}} = 8$~TeV
while others were made at the actual center-of-mass energy of 8.16~TeV.  The 2\%
difference in $\sqrt{s_{_{NN}}}$, does not have a significant effect on most
calculated observables and especially not on ratios such as the nuclear
modification factor $R_{p{\rm Pb}}$.  The energies at which the calculations are
carried out are noted throughout.

Note that, as in the 2013 $p+$Pb run at 5.02 TeV, the proton direction is
defined to be toward forward rapidity, similar to a fixed-target configuration
where the nucleus is the target.  This is assumed to be the case, even though
data are taken in two different experimental configurations, one with the
proton moving toward forward rapidity and one where the beam directions are
reversed.  As before, changing the beam direction is necessary for
the forward detectors of ALICE and LHCb to be able to cover the full phase
space.

There are advantages with the 8.16~TeV run that were missing from the run
at 5.02 TeV.  The Run 2 $p+$Pb 8.16~TeV luminosity was nearly a factor of
five higher than the 2013 Run 1 5.02~TeV $p+$Pb luminosity
so that the rates for hard processes should be considerably higher than in the
earlier $p+$Pb run.  In addition, while there was not a 5~TeV $p+p$ run for
a baseline comparison at the time of
the initial $p+$Pb run, $p+p$ data was taken at 8~TeV
in Run 1.  Therefore, when constructing the nuclear suppression factor
$R_{p{\rm Pb}}$, there is no need to rely on interpolations between runs at
different energies because a more direct comparison can be made.
In addition, the 8~TeV
$p+p$ comparison data was taken during a long LHC proton run rather than a short
heavy-ion run, as was the case for the $p+p$ comparison data at 2.76~TeV used
to extrapolate the $p+p$ baseline at 5.02~TeV.
All these factors combine to make it more likely that the data can better
discriminate between approaches and constrain models.

It is noteworthy that the LHC Run 2 has also included a
short $p+p$ run at 5.02~TeV, the same energy as the earlier $p+$Pb run and
also the same as the Pb+Pb Run 2 energy.
Thus it is possible to return to the previously
released $R_{p {\rm Pb}}$ results to form a measurement-based ratio rather than
employing an extrapolated $p+p$ denominator, allowing some clarification of
previous controversial results, see Ref.~\cite{Albacete:2016veq}.   In addition,
for the first time at the LHC, data from $p+p$, $p+$Pb and Pb+Pb collisions
are now available at the same energy and thus can be compared on the same level.

One physics outcome from the 5~TeV $p+$Pb run is the new set of nuclear
parton distribution functions by Eskola and collaborators, EPPS16
\cite{Eskola:2016oht}.
This set is the first to include the LHC data, specifically
that of $W^\pm$ and $Z^0$ production from
CMS \cite{Khachatryan:2015hha,Khachatryan:2015pzs}
and ATLAS \cite{Aad:2015gta} as well as the dijet
data from CMS \cite{Chatrchyan:2014hqa}.  
One advantage of these results is that they are all
forward-backward asymmetry data and do not rely on a $p+p$ baseline at the same
energy.  They also added, for the first time for the Eskola {\it et al}
sets, the neutrino deep-inelastic scattering data from CHORUS
\cite{Onengut:2005kv}.
Incorporating the LHC and neutrino data into the analysis allowed more detailed
flavor separation for the quark sets.  In particular, the LHC data allowed
them to increase the fit range in
momentum fraction, $x$, and factorization scale, $Q^2$,
to regions heretofore
unavailable.  Unfortunately, even with the dijet data from CMS, the gluon
distribution in the nucleus, particularly at low $x$ and moderate
$Q^2$, is still not well constrained.
These sets were not yet available at the time most of the predictions for
this paper were collected.  Therefore there are no calculations with these
sets presented here except for the top quark predictions in
Sec.~\ref{sec:topQ}.  However, it is worth noting that the central EPPS16 set
gives results
quite similar to those calculated with EPS09 NLO.  The largest change, for
gluon-dominated processes, is the increase in the uncertainty band due to the
increased number of parameters required for flavor separation and the relaxing
of some previous constraints.  See Ref.~\cite{Eskola:2016oht}
for details and comparison to the 5.02 TeV $p+$Pb data included in the global
analysis.

One might expect further global analyses of the nuclear parton densities after
the 8.16 TeV data become available.  At a given $p_T$, the $x$ value probed in
a hard scattering process is a factor of 0.62 smaller at 8.16~TeV than 5.02~TeV.
In addition, the higher energy allows a somewhat broader reach in rapidity so
that some processes, such as $Z^0$ production at LHCb, see the discussion in
Ref.~\cite{Albacete:2016veq}, measured near the edge of
phase space, can expect higher statistics and perhaps high enough significance
to be included in future global fits.  Similarly, the $p_T$ reach of most
processes is increased.

A further physics outcome of the 5.02~TeV $p+$Pb run, particularly in the most
central collisions, along with high multiplicity $p+p$ data, showed a
smooth transition between these high particle density $p+p$ and $p+$Pb
collisions and Pb+Pb colliisons for some observables \cite{Loizides:2016tew}.
Applications of hydrodynamics to these small systems with large pressure
anisotropies have been remarkably
successful \cite{Dusling:2015gta}
despite the short distances and system evolution
times.  It has been argued that hydrodynamics is applicable as long as
hydrodynamic modes dominate the evolution, independent of the system anisotropy
\cite{Romatschke:2016hle}.
However, the short lifetimes implied for small systems make the non-equilibrium
stage of hydrodynamics more important, as was shown in the case of photon
production \cite{Berges:2017eom}.
Furthermore, flow observables are sensitive to the substructure of the proton
projectile \cite{Mantysaari:2016ybx} in $p+$Pb collisions, as demonstrated
in Ref.~\cite{Mantysaari:2017cni}.

However, other approaches can mimic the signatures of hydrodynamics.
As has been shown previously, collective behavior can arise
from models such as $\mathtt{AMPT}$ \cite{Albacete:2016veq} where only a few
collisions are required to produce an anisotropy.  Correlations can also arise
in the saturation picture because initial-state correlations can be carried
into the final-state, including long range correlations in rapidity
\cite{Dusling:2015gta}.

Observables to distinguish between the different approaches
have been suggested, including mass ordering of the anisotropy coefficients,
multi-particle cumulants, odd flow harmonics, and jet quenching
\cite{Schenke:2017bog}.   Measurements with different
collision systems, such as the $p+$Au, d+Au and $^3$He+Au studies at RHIC
\cite{Adare:2014keg,Adare:2015ctn} and modeled in {\it e.g.}
Refs.~\cite{Schenke:2014gaa,Romatschke:2015gxa} are also important.  For more
details and further references, see Ref.~\cite{Schenke:2017bog}.
This interesting topic will not be further covered here
since the focus is on cold nuclear matter effects in these collisions.


\section{Inclusive charged hadron production models (J. Albacete, G. G. Barnaf\"oldi,
  G. B\'ir\'o, A. Dumitru,
  M. Gyulassy, Sz. M. Harangoz\'o, T. Lappi, Z.-B. Kang,
  P. L\'evai, Z. Lin, G. Ma,
  H. M\"antysaari, G. Papp, A. Rezaeian, B. Schenke, S. Schlichting, P. Tribedy,
  R. Venugopalan, I. Vitev, X.-N.
  Wang, H. Xing, B.-W. Zhang)}
\label{sec:models}

Here the models employed for inclusive
charged hadron production are described.  They
include saturation models, event generators, and perturbative QCD, assuming
collinear factorization.  

\subsection{Saturation models}
\label{sec:saturation}

Three saturation models are discussed here: the running-coupling
Balitsky-Kovchegov (rcBK) hybrid approach, the
impact-parameter Color Glass Condensate (bCGC) approach, and the
IP-Glasma model.

\subsubsection{rcBK (J. Albacete, A. Dumitru, T. Lappi, H. M\"antysaari) and bCGC (A. Rezaeian)}
\label{sec:sinc}

The discussion in this section is based on that of Lappi and M\"antysaari in
Ref.~\cite{Lappi:2013zma} using the rcBK hybrid approach with a color glass
condensate (CGC)
initial condition for the nucleus and collinear factorization for
the proton in the forward direction.  They have provided the transverse
momentum
dependence of the nuclear suppression factor for charged hadrons,
$R_{p {\rm Pb}}(p_T)$ at mid and forward rapidity.  Albacete and Dumitru
provided the charged hadron multiplicity distribution in the lab and center of
mass frames based on the work reported in Ref.~\cite{Albacete:2010sy} and
also shown in the compilation of predictions and results for 5 TeV in
Refs.~\cite{Albacete:2013ei,Albacete:2016veq}.  Rezaeian provided the charged
hadron multiplicity distribution based on the bCGC saturation model in
the center of mass frame and the
transverse momentum dependence of the nuclear suppression factor based on the
rcBK saturation approach at
midrapidity \cite{Rezaeian:2012ye}.
The details of the calculations can be found in
Ref.~\cite{Rezaeian:2012ye}.

\paragraph{Input from HERA}

In the rcBK
approach, particle production is calculated consistently with the HERA
deep inelastic scattering data in the CGC framework, as
discussed in more detail in Ref.~\cite{Lappi:2013zma}. First, the proton
structure function is calculated in terms of the virtual photon-proton
cross section
\begin{equation}
  \sigma_{T,L}^{\gamma^*p}(x,Q^2) = \sigma_0 \sum_f \int d z \int d^2 b_T
  |\Psi_{T,L}^{\gamma^* \to q\overline q}|^2 \cN(r_T, x) \, \, ,
\end{equation}
where $\cN(r_T,x)$ is the dipole-proton scattering amplitude, $r_T$ is the
transverse size of the dipole, and the proton
transverse area, $\sigma_0/2$, is obtained by assuming a factorizable impact
parameter profile, $\int d^2 b_T \to \sigma_0/2$. The virtual photon splitting
function, $\Psi_{T,L}$, describes the $\gamma^* \to q\overline q$ splitting for
transverse ($T$) and longitudinal ($L$) photons. Only light quark flavors
($q=\{u,d,s\}$) are considered here.

The QCD dynamics are included in $\cN(r_T,x)$.
The Bjorken-$x$ evolution of the amplitude
is given by the rcBK equation. The
initial condition for Balitsky-Kovchegov evolution is parameterized as
\begin{equation}
\label{eq:aamqs-n}
\cN(r_T, x=0.01) = 1 - \exp \left[ -\frac{r_T^2 \qso^2}{4} \ln
  \left(\frac{1}{|r_T| \lqcd}+e_c \cdot e\right)\right] \, \, .
\end{equation}
The initial saturation scale at $x=0.01$ is parameterized by $\qso^2$. Instead
of introducing an anomalous dimension, $\gamma$, in the dipole amplitude, in
the calculations of Lappi and M\"antysaari, the
infrared cutoff of the McLerran-Venugopalan (MV)
model is modified by introducing an additional fit
parameter, $e_c$, which also affects the saturation scale at the initial
condition. An advantage of this parameterization \cite{Lappi:2013zma}
over the AAMQS
fit by Albacete {\it et al.}~\cite{Albacete:2010sy} is that, in the ``MV$^e$''
parameterization used here, the dipole amplitude in momentum space (and thus
the unintegrated gluon distribution) is positive definite.

The parameters $\sigma_0$, $\qso$ and $e_c$ are obtained by fitting the
combined HERA proton structure function data~\cite{Aaron:2009aa}.
When solving the rcBK equation,
the strong coupling constant is parameterized as
\begin{equation}
  \alpha_s(r_T) = \frac{12\pi}{(33 - 2N_f) \log \left(\frac{4C^2}{r_T^2\lqcd^2}
    \right)} \, \, ,
\end{equation}
where $C^2$ is also a fit parameter.  The last free
parameter, $C^2$, is the scale at which the strong coupling constant
$\alpha_s$ is evaluated in coordinate space.  The best fit values are
$\qso=0.06\gev^2$, $e_c=18.9$, $\sigma_0/2=16.36$~mb and $C^2=7.2$,
corresponding to the saturation scale $Q_s^2=0.238\gev^2$ at initial
momentum fraction $x=0.01$.

[Note that the prediction by Albacete and Dumitru uses the AAMQS fit
  with initial condition
  \begin{equation}
\label{eq:aamqs-AD}
\cN(r_T, x=0.01) = 1 - \exp \left[ -\frac{(r_T^2 \qso^2)^\gamma}{4} \ln
  \left(\frac{1}{|r_T| \lqcd}+ e\right)\right] \, \, .
\end{equation}
  They used $\qso = 0.20\gev^2$, $\gamma =1$ and $\lqcd = 0.241$~GeV in
  their calculations for this work.]

The dipole amplitude for nuclei is obtained by requiring that, in the dilute
limit, the dipole-nucleus cross section is $A$ times the dipole-proton cross
section, and that, for large dipoles, $\cN\to 1$. The dipole-nucleus
scattering amplitude is then
\begin{eqnarray}
  \cN^A(r_T, b_T, x=0.01) & = & 1 -  \\
  & & \mbox{} \exp\left[ -A T_A(b_T) \frac{\sigma_0}{2}
    \frac{r_T^2 \qso^2}{4}  \ln \left(\frac{1}{|r_T|\lqcd}+e_c
    \cdot e\right) \right] \, \, . \nonumber
\end{eqnarray}
The nuclear thickness function, $T_A$, is obtained from the Woods-Saxon
distribution. No additional nuclear parameters are introduced because
$\sigma_0$, $\qso$ and $e_c$ are obtained from a fit to DIS data. The
dipole-nucleus amplitude is evolved to smaller values of $x$ independently
for each impact parameter using the rcBK equation.

\paragraph{Single inclusive cross section}

At midrapidity, both the proton and the nucleus are probed at small $x$ and
the invariant gluon yield is obtained from the $k_T$ factorization
result~\cite{Gribov:1984tu,Kovchegov:2001sc}
\begin{equation}
  \frac{d N(b_T)}{d y d^2 k_T} = \frac{\sigma_0/2}{(2\pi)^2}
  \frac{C_F}{2\pi^2 k_T^2 \alpha_s} \int \frac{d^2 q_T }{(2\pi)^2} q_T^2
  S^p(q_T)  (k_T-q_T)^2  S^A(k_T-q_T) \, \, ,
\label{eq:sinc}
\end{equation}
where $ S^p(k_T) = \int d^2 r_T e^{i k_T \cdot r_T} \tilde \cN(r_T)$ and the dipole
amplitude is evaluated in the adjoint representation, $\tilde \cN =2\cN-\cN^2$.
The $x$ dependence of $S$ is left implicit. The amplitude
$S^A$ in Eq.~(\ref{eq:sinc}) is obtained from the
Fourier transformation of the dipole-nucleus amplitude $\cN^A$.

Proton-proton scattering is described by replacing $S^A$ by $S^p$ and, instead
of $\sigma_0/2$, the geometric area multiplying the expression becomes
$(\sigma_0/2)^2/\sigma_{\rm in}$, see Ref.~\cite{Lappi:2013zma}.
The inelastic
proton-proton cross section is taken to be $\sigma_{\rm in}=75$~mb.

For particle production at forward rapidity, the proton becomes dilute and can
be described with parton distribution functions obtained from collinear
factorization. The invariant quark or gluon scattering yield in proton-nucleus
collisions is then 
\begin{equation}
  \frac{d N^{q/g+A \to q/g + X}(b_T)}{d y d^2 k_T} = \frac{1}{2\pi}
  xg(x, \mu^2) S^A(k_T) \, \, .
  \label{eq:scyield}
\end{equation}
The dipole-nucleus amplitude in the definition of $S^A$ is evaluated in the
fundamental representation for $u$, $d$, and $s$ quarks and in the adjoint
representation for gluons. In proton-proton collisions, the result in
Eq.~(\ref{eq:scyield}) is
multiplied by $(\sigma_0/2)/\sigma_{\rm in}$ with $S^A$ replaced by
$S^p$~\cite{Lappi:2013zma}. 

To calculate the results at the hadron level, the parton level yields are
convoluted with the leading order
DSS~\cite{deFlorian:2007aj} fragmentation functions and
the integral over impact parameter is calculated within the optical Glauber
model. Note that $R_{pA}\to 1$ at high $|p_T|$ in both the $k_T$-factorization
and hybrid formalisms.

\subsubsection{IP-Glasma (B. Schenke, S. Schlichting, P. Tribedy and
  R. Venugopalan)}

Several interesting observations in small collision systems ($p+p$ and $p+$Pb)
have been made in the high multiplicity events which populate the tails of the
respective multiplicity distributions. A first principles explanation of the
origin of such events can be obtained in the framework of the CGC approach
where high-multiplicity events are attributed to initial-state
fluctuations that lead to rare configurations of the parton distribution
in the colliding hadrons and nuclei. Detailed properties of the shape of the
underlying multiplicity distribution are determined by the mechanism of
correlated multiparticle production from the Glasma gluon fields, generated
shortly after the collision of high energy hadrons and nuclei. Based on
perturbative calculations in this framework, it was shown that multiparticle
production leads to a negative binomial distribution with its mean and width
related to the saturation scales of the colliding hadrons and
nuclei~\cite{Gelis:2009wh}. Beyond the perturbative approach, recent progress
in understanding the origin and features of high-multiplicity events has been
based on the development of the IP-Glasma model~\cite{Schenke:2012wb}.
Multiparticle production in the IP-Glasma model is computed nonperturbatively
from the numerical solution of classical Yang-Mills equations. By including
different sources of initial-state fluctuations, an accurate description of the
experimental multiplicity distribution can be obtained in this framework for a
wide range of collision systems~\cite{Schenke:2013dpa}. 

The IP-Glasma model includes different sources of initial state fluctuations
such as collision geometry, the position of nucleons in the nucleus, intrinsic
fluctuations in the saturation scale and the distribution of color charge
density in the nucleons~\cite{Schenke:2013dpa,McLerran:2015qxa}. In particular,
the sub-nucleonic color charge fluctuations in the IP-Glasma model are
constrained by the saturation scale $Q_s$ extracted from the HERA data
employing the IP-Sat dipole model~\cite{Kowalski:2003hm,Rezaeian:2012ji}. For a
detailed discussion on the implementation of the IP-Glasma model, see
Refs.~\cite{Schenke:2012wb,Schenke:2013dpa,Schenke:2015aqa,Schenke:2016lrs}.

\subsection{Event generators}
\label{sec:generators}

Predictions are reported for two
event generators, $\mathtt{HIJING++}$ and $\mathtt{AMPT}$.

The first, $\mathtt{HIJING++}$, is a
new version of the well known HIJING generator by Gyulassy and Wang
\cite{HIJING}.  This version is still in development so some that of the first
results calculated with $\mathtt{HIJING++}$ are presented here.
Predictions are given for the charged hadron multiplicity
distribution in the center of mass frame, charged hadron transverse momentum
distributions, and the nuclear suppression factor as a function of transverse
momentum at midrapidity for charged hadrons and identified pions, kaons and
protons as well as quarkonium and heavy flavor hadrons.

The second, $\mathtt{AMPT}$,  has been updated since the predictions shown in
Refs.~\cite{Albacete:2013ei,Albacete:2016veq}.
The updates are discussed here
and the differences between the calculations of the results at 5 TeV are
shown.  Predictions are given for the charged hadron multiplicity distribution,
both non-diffractive and as a function of centrality, the transverse momentum
spectrum at midrapidity,
and the elliptic
flow moments $v_2$, $v_3$ and $v_4$ as a function of transverse momentum.

\subsubsection{$\mathtt{HIJING++}$, (G. G. Barnaf\"oldi, G. B\'ir\'o,
  M. Gyulassy, Sz. M. Harangoz\'o, P. L\'evai, G. Ma, G. Papp, X.-N.
  Wang, B.-W. Zhang)}
\label{sec:HIJING++}

Collaborators from Budapest, Wuhan and Berkeley have developed
a new version of the $\mathtt{HIJING}$~\cite{HIJING}
(Heavy Ion Jet INteraction Generator)
Monte Carlo model first developed by Gyulassy and Wang:
$\mathtt{HIJING++}$~\cite{HP2017}.  $\mathtt{HIJING}$
employs minijets in proton-proton
($p+p$), proton-nucleus ($p+A$) and nucleus-nucleus ($A+A$) reactions over a
wide range of center-of-msas energies,
from 5 GeV to a few TeV. The original program
was written in FORTRAN since it was based on the FORTRAN version of
$\mathtt{PYTHIA}$, $\mathtt{PYTHIA5}$~\cite{PYTHIAv5}, as well as the
$\mathtt{FRITIOF}$~\cite{FRITIOF} and $\mathtt{ARIADNE}$~\cite{ARIADNE}
packages along with the parton distribution function package
in the CERN library, 
$\mathtt{PDFLIB}$~\cite{CERNLIB}.  Today, $\mathtt{HIJING}$
is still the most widely used particle event generator for high-energy
heavy-ion collisions both for testing models and for experimental simulations.

The features of the latest FORTRAN version of $\mathtt{HIJING}$,
version 2.552~\cite{HIJING2} with nuclear shadowing~\cite{HIJINGsh},
were embedded in the new $\mathtt{HIJING++}$.
Because new, novel computational techniques require a
shift to more modular programming,
the new version of $\mathtt{HIJING++}$ (version 3.1) was written as
a genuinely modular C++ Monte Carlo event generator, including the most
recent C++ public packages utilized by $\mathtt{HIJING++}$, (e.g.
$\mathtt{PYTHIA8}$~\cite{PYTHIA} and the parton distribution library
$\mathtt{LHAPDF6}$~\cite{LHAPDF6}).

Since $\mathtt{HIJING++}$ is based on $\mathtt{PYTHIA8}$ with the Monash 2013
tune \cite{Monash2013} while $\mathtt{HIJING}$ is based on the FORTRAN version
of $\mathtt{PYTHIA}$ with the Perugia0 \cite{Perugia} tune, one might expect
to see some differences between the two results at the $p+p$ level.  Due to the
different tunes employed, it is likely that the two results will not completely
agree at this level.

It is noteworthy that $\mathtt{HIJING++}$ is suitable
for further parallelization, providing faster and more efficient use of
new parallel architectures.
The $\mathtt{HIJING++}$ development is now at the stage where `preliminary'
predictions are possible.  Such preliminary redictions are presented in this
work, including for light charged particles, $J/\psi$ and heavy flavor
hadrons.

\subsubsection{$\mathtt{AMPT}$ (Z. W. Lin)}
\label{sec:AMPT_ch}

The string melting version of A Multi-Phase Transport 
model, $\mathtt{AMPT}$
\cite{ampt}, is employed to calculate the yields, $p_T$ spectra, and flow
coefficients of charged hadrons produced in $p+$Pb collisions 
at $\sqrt{s_{_{NN}}}=8$ TeV.
$\mathtt{AMPT}$ \cite{ampt,Lin:2004en} is a comprehensive transport
model that 
includes fluctuating initial conditions, parton elastic scatterings,
hadronization through the Lund string fragmentation or quark
coalescence, and hadronic interactions.
The string-melting version of the $\mathtt{AMPT}$ model ($\mathtt{AMPT-SM}$) 
\cite{ampt,Lin:2004en,Lin:2014tya} converts traditional hadronic
strings in the initial state to partonic matter when the energy
density in the overlap volume of the collisions is expected to be
higher than that for the QCD phase transition. It then uses a 
quark coalescence model to describe the bulk hadronization of the
resultant partonic matter to hadron matter. 

The string melting $\mathtt{AMPT}$ version 2.26t7 \cite{ampt}
uses the same parameters
as in earlier studies of Pb+Pb collisions at LHC energies
\cite{Lin:2014tya,Ma:2016fve}.
In particular, Lund string fragmentation is used to generate the initial
hadrons before string melting.  The
parameters
$a=0.30$ and $b=0.15$ GeV$^{-2}$ are used for the Lund symmetric
splitting function.  In addition, the strong coupling constant is fixed at
$\alpha_s=0.33$.  A parton scattering cross section of 3 mb is employed. 
Finally, an upper limit of 0.40 is imposed on the relative production of
strange to non-strange quarks in Lund string fragmentation. 
This set of values has been shown \cite{Lin:2014tya} to reasonably
reproduce the yields, $dN/dy$, $p_T$ spectra, and elliptic flow, $v_2$, of
low $p_T$ pions and kaons in central and mid-central Pb+Pb collisions
at $\sqrt{s_{_{NN}}} = 2.76$~TeV. 

While the parameters listed above have not been tuned to the available 5
TeV $p+$Pb data, it is interesting to note that the parameters employed in
the previous $\mathtt{AMPT}$ string melting version, 2.26t1, could reproduce
the charged
particle yields, $dN/d\eta$, and the elliptic flow coefficients in 5 TeV $p+$Pb
collisions \cite{Albacete:2013ei,Albacete:2016veq}.  However, the charged
hadron $p_T$ spectra were too soft \cite{Albacete:2016veq}. 

\subsection{Perturbative QCD, Collinear Factorization}
\label{sec:pQCD}

Here two perturbative QCD calculations assuming collinear factorization,
described in more detail in Ref.~\cite{Albacete:2013ei}. are
briefly described.  Both include isospin effects, the difference from the proton
results due to the neutron excess in heavy nuclei, transverse momentum
broadening, and nuclear shadowing.  However, there are some differences 
between the calculations.  

The leading order
calculations by Vitev {\it et al.}\
include cold nuclear matter energy loss, not included in the
$\mathtt{kTpQCD}$ calculations by
Barnaf\"oldi {\it et al.}.  Also, shadowing is treated
differently in the two calculations.  Vitev assumes higher-twist dynamical
shadowing, a shift of the target momentum fraction to higher $x$, resulting
in a suppression of the parton density in the nucleus.
The next-to-leading order calculations of Barnaf\"oldi {\it et al.}
employ data-driven nuclear modifications as a ratio of the parton densities in
the nucleus to those in the nucleon such as EPS09.

Vitev {\it et al.}\
provide the nuclear suppression factor as a function of transverse
momentum at $y=0$ and $y = 4$ for charged hadrons, photons, jets, and heavy
flavor mesons.  Barnaf\"oldi {\it et al.} provide calculations of the
transverse momentum distributions and nuclear suppression factor as a function
of transverse momentum.

Note that the NLO result by Eskola and collaborators on the charged hadron
nuclear suppression factor as a function of transverse momentum, also
included, is presented
where that result is discussed in the next section but is not described in
detail here. That calculation includes isospin and the
EPS09 NLO parameterization
of the nuclear parton densities.

\subsubsection{Cold Nuclear Matter in pQCD (I. Vitev, Z.-B. Kang and H. Xing)}
\label{sec:Vitev}

Vitev and collaborators have performed phenomenological calculations including
various cold nuclear matter effects on the production of energetic final
states in $p+$Pb collisions. The ingredients of the calculations, discussed
in detail below, include isospin effects, the Cronin effect, cold nuclear
matter energy loss and dynamical shadowing.  The model was described in more
detail in Sec.~2.5 of Ref.~\cite{Albacete:2013ei}.

A factorized perturbative QCD approach was used to present predictions for 
single inclusive particle production in proton-lead collisions, particularly 
for prompt photon and charged hadron production, heavy flavor production, and 
inclusive jet production. 

{\it Isospin effects} The isospin effect can be easily accounted for on average 
in the parton distribution functions for a nucleus 
with atomic mass $A$ and proton number $Z$
\cite{Gyulassy:2003mc,Vitev:2002pf}  via 
\begin{eqnarray}
  f_{a/A}(x) = \frac{Z}{A} f_{a/p}(x)+ \left(1-\frac{Z}{A}\right)f_{a/n}(x) \,\, ,
  \label{Ivan:isospin}
\end{eqnarray}
where $f_{a/p}(x)$ and $f_{a/n}(x)$ are the parton distribution functions (PDFs)
of a proton and a neutron, 
respectively. The isospin effect plays a role in observables that are flavor
sensitive, for example photon or inclusive hadron production. Conversely, as 
will be discussed later, processes dominated by gluons in the initial state,
such as jets and heavy flavor, are not significantly affected by  
isospin.  Note that energy loss and dynamical shadowing are applied to the
proton and neutron PDFs as described in Eqs.~(\ref{Ivan:eloss})
and (\ref{Ivan:HTshad}) below.

{\it Cronin effect} Theoretical approaches to the Cronin effect are based on 
multiple parton scattering. Recently, calculations have been performed at 
backward rapidity based upon a higher-twist approach~\cite{Kang:2014hha}.  
Traditionally, multiple scatterings have been resummed~\cite{Qiu:2003pm} and 
shown to affect particle production cross sections and back-to-back 
correlations.  As a practical implementation, if the parton distribution 
function $f_{b/A}(x_b, k_{b,T}^2)$ has a normalized Gaussian form, random 
elastic scattering induces further $k_T$-broadening in the 
nucleus~\cite{Vitev:2004gn},
\begin{eqnarray}
  \langle k_{b, T}^2\rangle_{pA} = \langle k_{b, T}^2\rangle_{pp} + \left\langle
  \frac{2\mu^2 L}{\lambda_{q,g}}\right\rangle \xi \, \, ,
\label{Ivan:Cronin}
\end{eqnarray}
where $k_{b,T}$ is the transverse component of the parton in the target 
nucleus,
$\xi=\ln(1+\delta p_T^2)$.  The values $\delta = 0.14$ GeV$^{-2}$, 
$\mu^2=0.12$ GeV$^2$, and $\lambda_g= C_F/C_A \lambda_q$ = 1 fm
\cite{Vitev:2008vk} are chosen. 
These parameter choices can reasonably describe the RHIC
data~\cite{Vitev:2004gn}.  
The Cronin effect is implemented in all calculations in this approach.
To explore the effect of a reduced Cronin enhancement, 50\% longer scattering
lengths, $\lambda_q,\; \lambda_g$, are also tested.  The most recent RHIC
results suggest that the Cronin peak is broader and the maximum value of
$R_{pA}$ is at a slightly higher $p_T$ than the model suggests.  While better
fits to existing data can be pursued in the future, it is important to examine
the possible effect of initial-state multiple scattering on the production of
hard probes at 8.16~TeV at the LHC.

{\it Cold nuclear matter initial-state energy loss} When a parton from the
proton undergoes 
multiple scattering in the nucleus before the hard collision, it can lose 
energy due to medium-induced gluon bremsstrahlung. This effect can be easily 
implemented through a shift in the momentum fraction in the projectile proton
PDFs, 
\begin{eqnarray}
f_{q/p}(x_a) \to  f_{q/p}\left(\frac{x_a}{1-\epsilon_{\rm eff}}\right),
\qquad
f_{g/p}(x_a) \to f_{g/p}\left(\frac{x_a}{1-\epsilon_{\rm eff}}\right),
\label{Ivan:eloss}
\end{eqnarray}
where $x_a$ is the parton momentum fraction of the proton projectile.  The 
energy loss considered in these calculations is the high-energy limit of the
Bertsch-Gunion approach~\cite{Vitev:2007ve}.  Multiple gluon emission,   
$\Delta E =  \sum_i \Delta E_i$, reduces the effect of the mean energy loss.
This is implemented through the relation
$\epsilon_{\rm eff} = 0.7 \, (\Delta E/ E)$.  The mean energy loss 
depends on the momentum transfer per interaction, $\mu$, between the parton and 
the medium and the gluon mean-free path, $\lambda_g$. These parameters, 
constrained by Drell-Yan data \cite{Neufeld:2010dz}, were and found to be
$\mu=0.35$~GeV and $\lambda_g=1$~fm. Incidentally, these values of $\mu$ and
$\lambda_g$ also describe the Cronin effect given in
Eq.~(\ref{Ivan:Cronin}) above.   Enhanced and reduced levels of energy loss
were also considered, see Ref.~\cite{Kang:2015mta}. Larger CNM
energy loss is disfavored, especially by minimum bias jet data. 

{\it Dynamical shadowing}  Final-sate coherent scattering of the struck partons
leads to higher-twist shadowing in the observed cross section
\cite{Qiu:2004qk}. This effect is included through a modification of the
momentum fraction of the target nuclear PDFs,
\begin{eqnarray}
x_b\to x_b \left(1+C_d \frac{\xi^2(A^{1/3}-1)}{-\hat t}\right),
\label{Ivan:HTshad}
\end{eqnarray}
where $x_b$ is the parton momentum fraction in the target nucleus and 
$C_d = C_F$ or $C_A$ for final-state parton $d=q$ or $g$ in the
$2 \rightarrow 2$ partonic scattering $ab\rightarrow cd$.  Here 
$\xi^2$ is a characteristic energy scale of the multiple scattering with 
$\xi^2_q = C_F/C_A \xi^2_g = 0.12$ GeV$^2$.  Resummed coherent power 
corrections are only relevant at low $p_T$.  

\subsubsection{$\mathtt{kTpQCD}$
  (G. G. Barnaf\"oldi, G. B\'ir\'o, Sz. M. Harangoz\'o,
  P. L\'evai, G. Papp)}
\label{kTpQCD}

The NLO $\mathtt{kTpQCD\_v2.0}$ code is based on a phenomenologically enhanced,
perturbative
QCD improved parton model~\cite{YZ02,pgNLO} described in some detail in Sec.~2.6
of Ref.~\cite{Albacete:2013ei}. The model includes a
phenomenologically-generalized parton distribution function in order to
handle nonperturbative effects at relatively low $x$ and small $p_T$. Similar to
$\mathtt{HIJING}$~\cite{HIJING},
multiple scattering in the nucleus is described by the
broadening of the initial intrinsic transverse momenta of the incoming
particles, $\langle k_T^2\rangle$. The broadening appears as a phenomenological
parameter in the calculations and mimics nonperturbative effects. The value
of the intrinsic $k_T$ can be determined from data obtained over a wide energy
range of nucleon-nucleon (predominantly $p+p$) collisions.  It was found to be
$\langle k_T^2 \rangle = 2.5$ GeV$^2$.

In this model, the factorization and renormalization scales are fixed by the
momentum of the intermediate jet,
$Q=Q_R=\kappa  p_q$ with $p_q=p_T/z_c$.  The fragmentation scale is connected
to the final momentum of the hadron, $Q_F=\kappa  p_T$. In all cases, the factor
$\kappa$ multiplying the momentum scale is set to
$2/3$.  The baseline proton parton distribution functions used in the
calculations, assuming collinear factorization, is the MRST central gluon
set, MRST-cg~\cite{MRST01}.  The KKP fragmentation functions \cite{KKP}
are employed for the hadronization process. Both MRST-cg and KKP are applicable
starting from a relatively low squared momentum transfer 
$Q^2\approx 1.25$~GeV$^2$. Thus these calculations are applicable down to
$p_T \geq 2$~GeV.

As in Refs.~\cite{Albacete:2013ei,Adeola:2009,Barnafoldi:2011px}, the
initial-state nuclear effects included in proton-nucleus or nucleus-nucleus
collisions are multiple scattering and shadowing.  Intrinsic
transverse momentum broadening via semihard collisions is related to
multiple scattering in this approach.
For typical large nuclei there are three to four semihard
collisions.  The average broadening per collision in the nucleus is
$C_{\rm sat} = 0.35$~GeV$^2$, independent of $A$.  The only initial-state
energy dependence arises through the average transverse momentum in $p+p$,
$\langle k_T^2 \rangle_{pp}$,
so that the same broadening due to multiple scattering
applies for collisions from SPS to LHC
energies. The model gives a Cronin peak~\cite{Cron75,Antr79} in the
intermediate $p_T$ range, $3 \leq p_T \leq 9$~GeV.

Nuclear shadowing is introduced by modifying the PDFs in the nuclear
environment via a parameterization such as those in
Refs.~\cite{HIJINGsh,Eskola:2009uj}.
Shadowing and isospin effects were previously
taken into account on average using a scale-independent parameterization of the
shadowing function, $S_{a/A}(x)$, adopted from Ref.~\cite{HIJINGsh}.

In the
present work, the results shown are obtained with the
$\mathtt{HIJING}$~\cite{HIJINGsh}
and EPS09 NLO~\cite{Eskola:2009uj} shadowing
parameterizations. Because EPS09 exhibits strong
gluon antishadowing, replicating the Cronin effect -- albeit in the wrong
position and with slower $x$-scaling -- without multiple scattering
to avoid double counting the Cronin effect, the strength of
the transverse momentum broadening due to multiple scattering
is reduced when this set is used.  No reduction in multiple scattering
is required for the $\mathtt{HIJING}$ shadowing
parameterization because it does not include antishadowing.

\section{Charged particle results}
\label{sec:charged_hadrons}

Here the results for charged particle production, calculated using the
approaches described in the previous section, are presented.

\subsection{Multiplicity distribution ($\mathtt{HIJING++}$, $\mathtt{AMPT}$,
  rcBK, bCGC and IP-Glasma)}
\label{sec:dnchdeta}

Results for the charged hadron multiplicity distribution
from $\mathtt{HIJING++}$, $\mathtt{AMPT}$, and the rcBK calculations by
Albacete and Dumitru
are shown here.  A calculation of the probability for inelastic parton-parton
interactions as a function of the charged hadron multiplicity from the IP-Glasma
approach is also shown.

The $\mathtt{AMPT}$ result is calculated in the laboratory
frame.  The rcBK and calculation is given in
both frames.  The event generator results are given over all phase space while
the rcBK and $\mathtt{HIJING++}$ calculations are given for $|\eta|\leq 2.5$.
The results are separated into two different panels, one for
each reference frame.

\begin{figure}[htb]
\centerline{\includegraphics[height=3in]{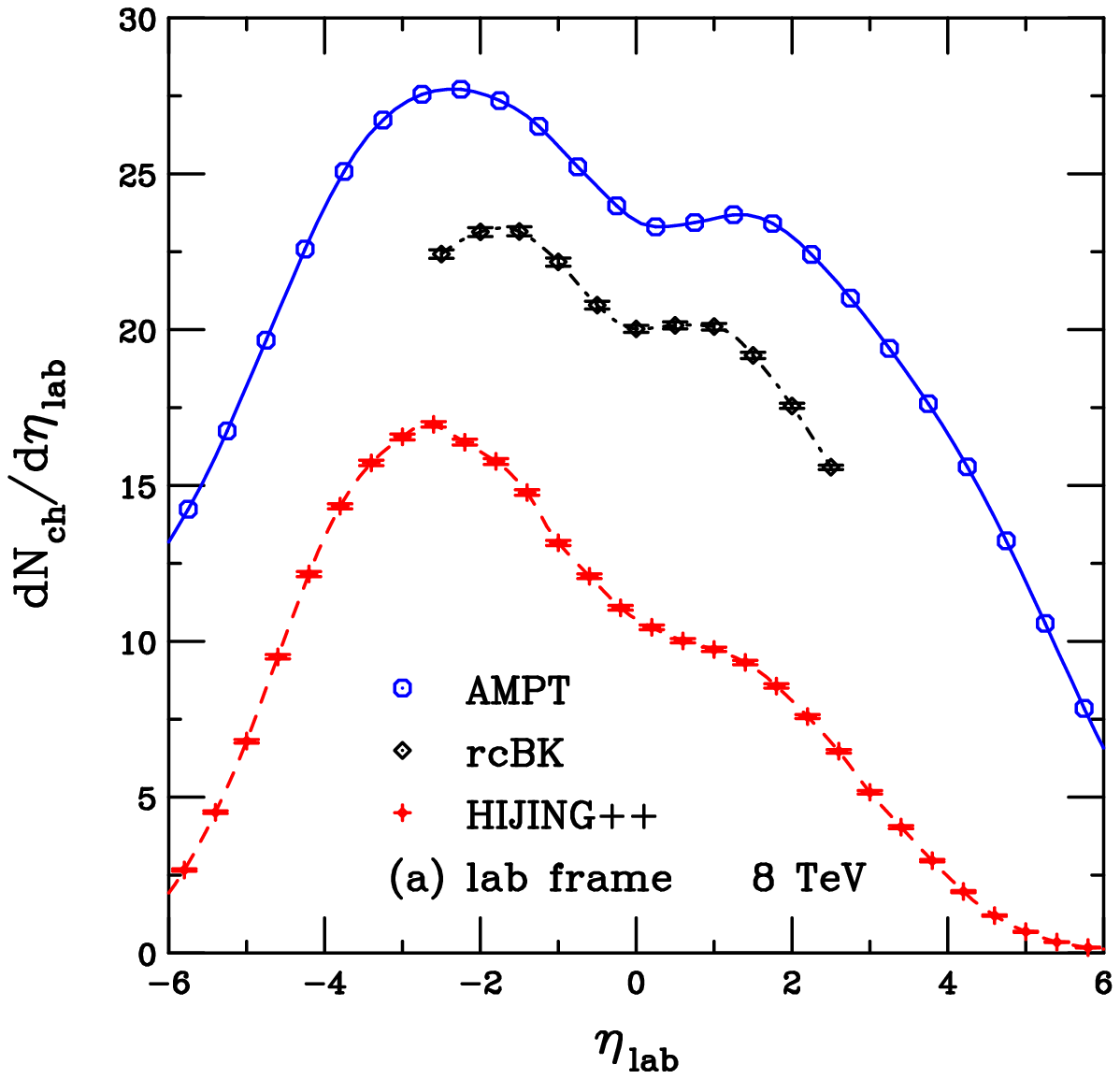}}
\centerline{\includegraphics[height=3in]{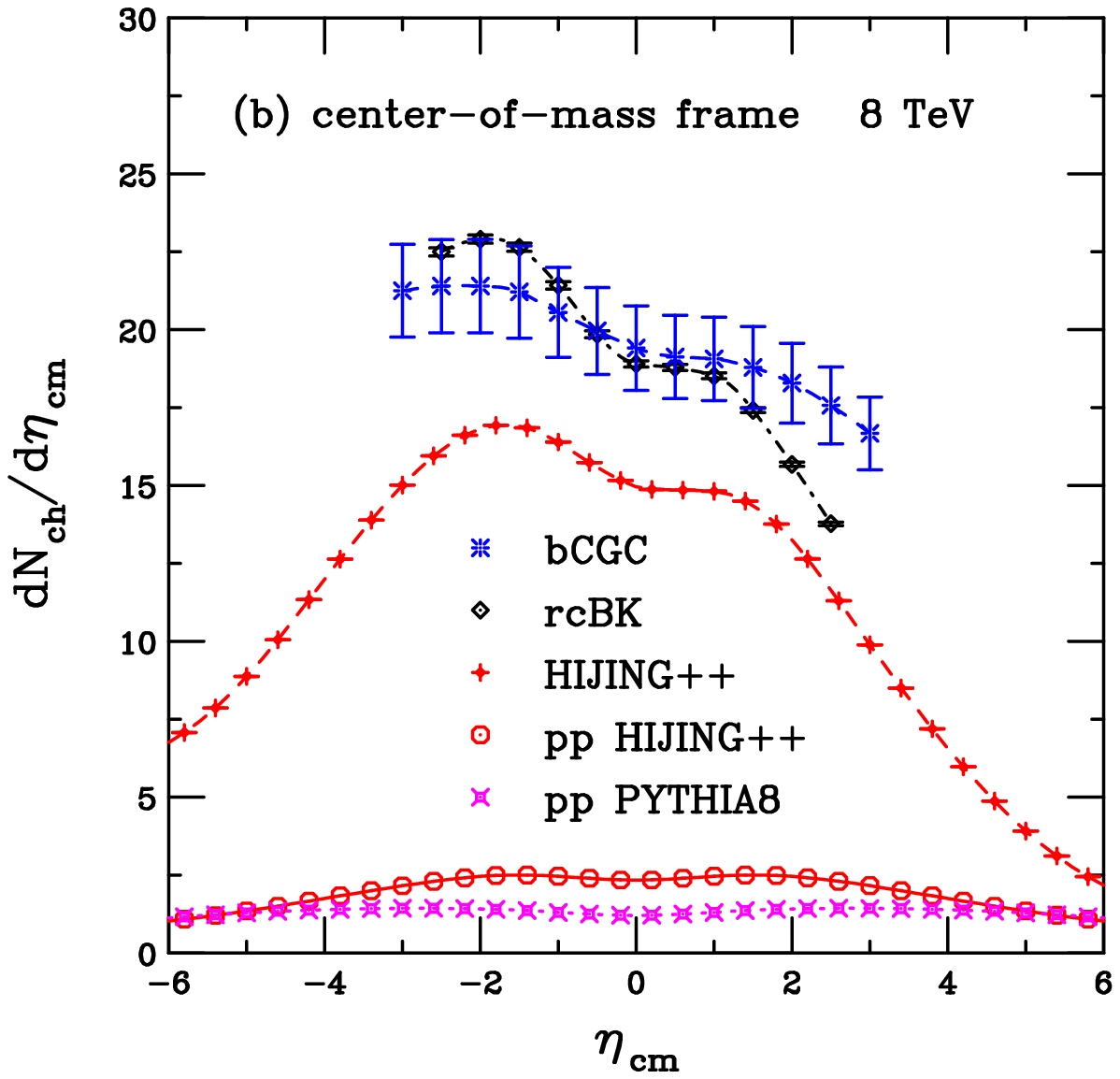}}
\caption[]{(Color online) The charged particle multiplicity distribution
  $dN_{\rm ch}/d\eta$ at $\sqrt{s_{NN}} = 8$~TeV from $\mathtt{AMPT}$
  (solid blue), rcBK from Albacete and Dumitru (black dot dashed),
  bCGC from Rezaeian (blue dashed),
  and $\mathtt{HIJING++}$ (dashed red) in the laboratory (a) and
  center-of-mass (b) frames.  The $\mathtt{AMPT}$ result is
  in the lab frame while the $\mathtt{HIJING++}$ calculation and the rcBK
  result from Albacete and Dumitru
  are given in both frames.} 
\label{dndeta_mb}
\end{figure}

In Ref.~\cite{Albacete:2013ei}, it was explained that the bCGC calculations
of Rezaeian \cite{Rezaeian:2011ia} and the rcBK calculations of
Albacete {\it et al.}
\cite{Albacete:2012xq} depended on the minijet
mass which, in turn, affects the transformation between rapidity, for identified
particles, and pseudorapidity, for unidentified charged particles.
In Ref.~\cite{Albacete:2016veq}, Albacete and Dumitru demonstrated that
$dN_{\rm ch}/d\eta$
depends strongly on the $y \rightarrow \eta$ transformation.
The rcBK calculation depends on the Jacobian of this transformation which is not
uniquely defined in the CGC framework. It is necessary
to assume a fixed minijet mass, related to the pre-hadronization/fragmentation
stage.  In Ref.~\cite{Albacete:2013ei}, they assumed the same 
transformation for $p+p$ and $p+$Pb collisions.
A Jacobian with the hadron momentum modified by
$\Delta P(\eta) = 0.04 \eta [(N_{\rm part}^{\rm proj} + N_{\rm part}^{\rm targ})/2 -1]$
gave very good agreement with the ALICE 5.02 TeV charged hadron multiplicity
distribution \cite{Albacete:2016veq}.
The results were unchanged in the proton direction but modified in the
direction of the lead beam.  The difference shows the sensitivity of this
result to the mean mass and $p_T$ of the unidentified final-state hadrons.
The results with the modified hadron momentum, as in
Ref.~\cite{Albacete:2016veq}, are given in Fig.~\ref{dndeta_mb}.

The results for the charged-particle pseudorapidity 
density in non-single diffractive $p+$Pb collisions calculated by
Rezaeian are given in the center of mass frame.  The boost from the $\eta=0$
laboratory frame to the center of mass frame was accomplished by adding a
rapidity shift of $\Delta y=-0.465$.  The details of calculation can be found in
Ref.~\cite{Rezaeian:2012ye}. The results are based on $k_T$-factorization
\cite{Kovchegov:2001sc}
and the bCGC saturation model
\cite{Rezaeian:2011ia,Levin:2010dw,Levin:2010br,Levin:2010zy,Levin:2011hr,Rezaeian:2013woa}.

The free parameters of the bCGC model were determined by a fit
to the small-$x$ HERA data, including experimental data from diffractive
vector meson production \cite{Rezaeian:2013tka,Armesto:2014sma}.  In the
$k_T$-factorization approach, one
needs to rewrite the rapidity distribution in terms of pseudorapidity using
the Jacobian of rapidity-pseudorapidity transformation
\cite{Rezaeian:2011ia,Rezaeian:2012ye,Levin:2010dw,Levin:2010br,Levin:2010zy,Levin:2011hr,Rezaeian:2013woa}.
As described previously, the Jacobian depends on the minijet
mass $m_{\rm jet}$.  The shape of $dN_{\rm ch}/d\eta$ strongly depends on both
$m_{\rm jet}$ and the Jacobian \cite{Rezaeian:2012ye}. The main theoretical
uncertainties in the bCGC approach come from fitting both the
$K$-factor and the minijet mass to RHIC data \cite{Back:2003hx,Arsene:2004cn}
in minimum-bias collisions.  The RHIC data alone are not enough to uniquely fix
the value of $m_{\rm jet}$.  It was found that $m_{\rm jet}\approx 5$~MeV
gives the best description of RHIC and also describes the ALICE data within a
$7\%$ uncertainty \cite{Rezaeian:2012ye}. The value of $m_{\rm jet}$
is similar to current quark mass.

In the lab frame, the $\mathtt{AMPT-SM}$ result is about 15\% higher than the
rcBK calculation at $\eta_{\rm lab} \sim 0$.
The two shapes are very similar in the
forward, proton, direction but at backward $\eta$ the lead peak is narrower in
the rcBK calculation.  The $\mathtt{HIJING++}$ result is nearly a factor of two
lower than the rcBK calculation at $\eta_{\rm lab} \sim 0$ but the
$\mathtt{HIJING++}$ peak is shifted slightly backward relative to the
$\mathtt{AMPT}$ and rcBK results.

On the other hand, in the center of mass frame, the rcBK calculation is
$\sim 27$\% higher than the $\mathtt{HIJING++}$ $p+$Pb calculation.  There are
also significant differences in the shapes.
In Fig.~\ref{dndeta_mb}, $m_{\rm jet}= 5$ MeV was used to calculate
$dN_{\rm ch}/d\eta$ at 8 TeV in the bCGC approach, as
also assumed for the 5.02 TeV calculations. The band
on the bCGC calculation shows the theoretical uncertainty of $7\%$ due to the
variation of $m_{\rm jet}$ around its central value while still remaining
consistent with the RHIC and ALICE data, see Ref.~\cite{Rezaeian:2012ye}.

The rcBK calculation gives more
enhancement in the lead direction than the $\mathtt{HIJING++}$ calculation.
The bCGC result by Rezaeian is similar in magnitude to the rcBK calculation
by Albacete and Dumitru but flatter in shape with a smaller enhancement in the
lead direction and a milder decrease with $\eta$ in the proton direction.

Also shown are the $p+p$ results at the same energy obtained using
$\mathtt{HIJING++}$ and $\mathtt{PYTHIA8}$
\cite{PYTHIA} with the Monash 2013 tune \cite{Monash2013}.  Here the center
of mass and laboratory frames coincide.  The $p+p$ results are shown in
the center of mass frame in Fig.~\ref{dndeta_mb}(b).
The difference between the two generators on the $p+p$ level arises from the
different tunes and minijet production, which acts up to the minijet cutoff.
The minijet contribution enhances the spectra, especially at midrapidity. 

The IP-Glasma model is now employed to compare multiplicity distributions in
$p+p$ collisions at 7 TeV and $p+$Pb collisions at 5.02 TeV to experimental
measurements and predict the multiplicity distribution for $p+$Pb collisions
at 8 TeV.  In the calculations, approximately 30K IP-Glasma events are generated
for each collision system by uniformly sampling the impact parameter $b$ in the
range from $b_{\rm min}=0$ to $b_{\rm max}$ ($b_{\rm max} = 2.5$~fm for $p+p$
and 10 fm for $p+$Pb) and computing the interaction probability,
$P_{\rm Int}(b)$, for each event, 
\begin{equation}
  P^{\rm event}_{\rm Int}(b) = 1-\exp\left(-T(b) \sigma_{NN}(\sqrt{s_{_{NN}}})\right)
  \,\, ,
\end{equation}
where $\sigma_{NN}(\sqrt{s_{_{NN}}})$ is the nucleon-nucleon interaction cross section
for each center of mass energy, $\sqrt{s_{_{NN}}}$, and $T(b)$ is the collisional
overlap area computed on the basis of individual nucleon-nucleon collisions,
defined as
\begin{equation}
  T(b) = \sum\limits_{i=1}^{A_1}\sum \limits_{j=1}^{A_2}\int d^2b_T T_i(b_T)
  T_j(b_T) \, \, .
\end{equation}
Here, $T_{i,j} (b_T)$ denote the nucleon thickness functions, parameterized as
\begin{equation}
T_{i,j}(b_T)=\frac{1}{2\pi B_{G}} \exp\left({-b_T^{2}\over 2B_{G}}\right)\, \, ,
\label{eq:IPsat-imp-par}
\end{equation}
with the characteristic size scale $B_G=4$~GeV$^{-2}$ extracted from fits to
the diffractive HERA data~\cite{Kowalski:2003hm,Rezaeian:2012ji}.

Event-by-event multiparticle production is computed nonperturbatively from
classical-statistical real-time lattice simulations on $512 \times 512$
lattices with a spacing of 0.02 fm~\cite{Schenke:2015aqa}. Based on the
solutions of the classical Yang-Mills equations, the single inclusive gluon
spectrum, $dN_g/dyd^2{k_T}$, is extracted from correlation functions of the
gauge fields after the collision~\cite{Lappi:2003bi,Krasnitz:2001qu}. By
integrating the gluon spectrum over the range of transverse
momenta $0.25 < k_T < 18$~GeV, the overall gluon multiplicity $N_g$ is obtained
for each event. Based on the CGC+Lund event generator~\cite{Schenke:2016lrs},
matching the IP-Glasma model of multiparticle production to the Lund string
fragmentation model implemented in $\mathtt{PYTHIA}$,
Ref.~\cite{Schenke:2016lrs} demonstrated that including fragmentation effects
does not significantly affect the shape of the multiplicity distribution.
Specifically, the multiplicity distribution
$P(N_{\rm ch}/\langle N_{\rm ch}\rangle)$ is well approximated by
$P(N_g/\langle N_g\rangle)$, such that an estimate of the charged particle
multiplicity distribution can be obtained directly by assuming $N_{\rm ch}$ is
proportional to $N_g$.

\begin{figure}[htb]
  \centerline{\includegraphics[width=0.65\textwidth]{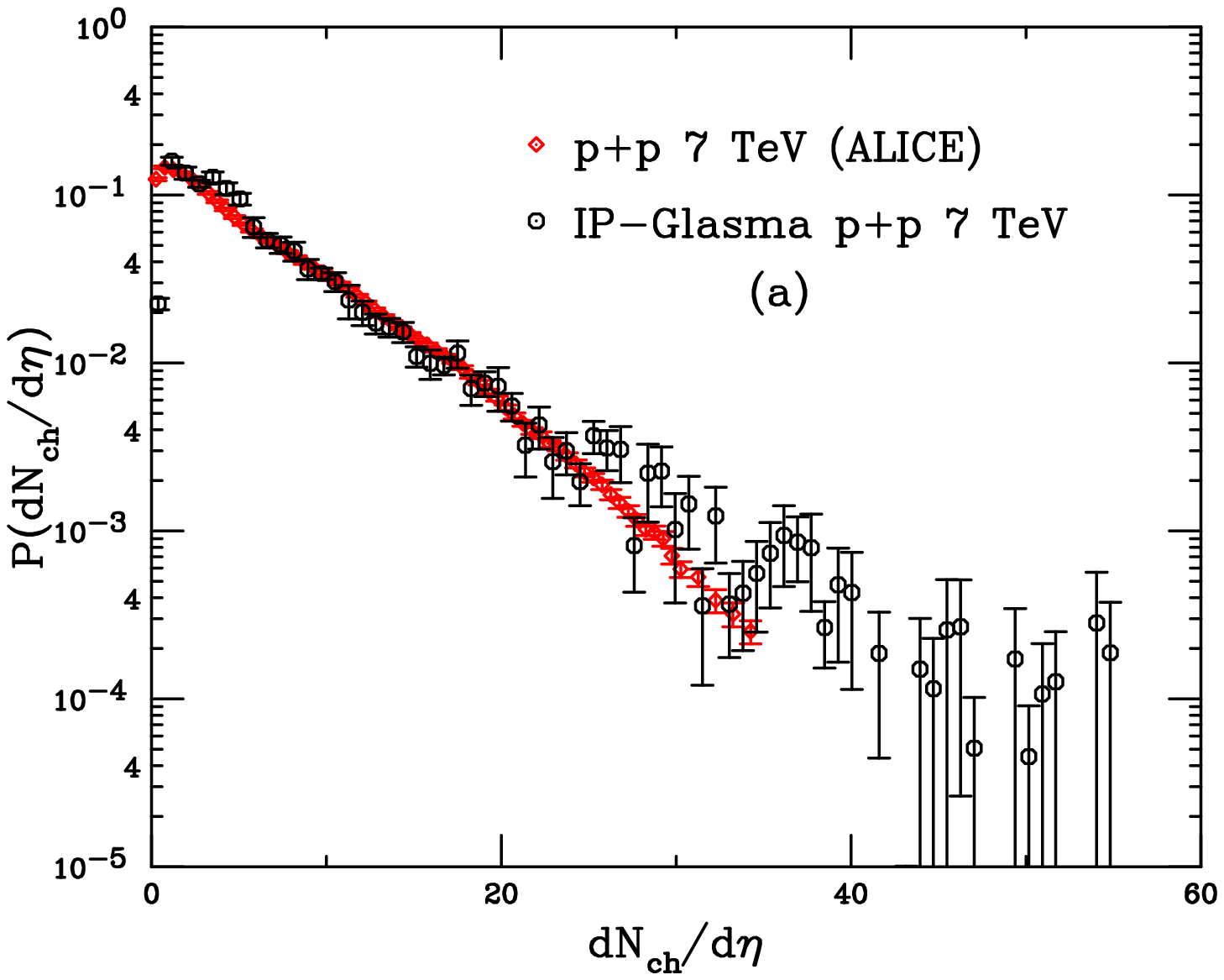}}
  \centerline{\includegraphics[width=0.65\textwidth]{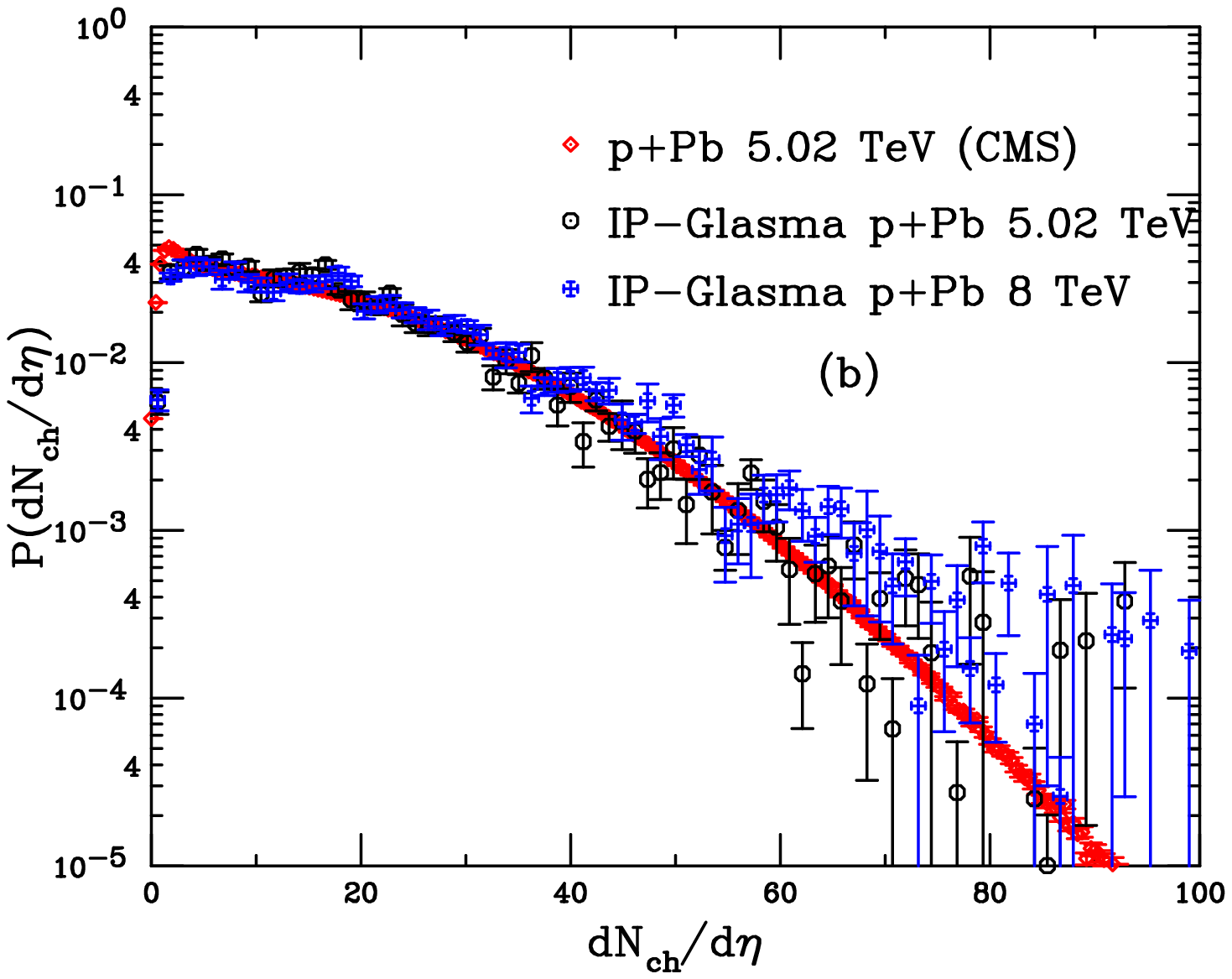}}
\caption[]{(Color online) The IP-Glasma prediction for the charged hadron
  multiplicity distribution in $p+p$ collisions at 7 TeV (a) and $p+$Pb
  collisions at 5.02 and 8~TeV (b) are
  shown.  The ALICE data for $|\eta| < 1$ in 7 TeV $p+p$ collisions
  \protect\cite{Aamodt:2010pp} and the CMS data for $|\eta|<2.4$ in 5.02 TeV
  $p+$Pb collisions \protect\cite{CMS:2012qk} are shown in red.  The
  corresponding IP-Glasma calculations are shown in black while the 8 TeV
  $p+$Pb predictions are given in blue.
  }
\label{IP-Glasma}
\end{figure}

The results for the multiplicity distributions of charged hadrons with the
IP-Glasma model are shown in Fig.~\ref{IP-Glasma}. The multiplicity
distributions are compared to the available data from
ALICE~\cite{Aamodt:2010pp} in $p+p$ collisions at 7 TeV and from
CMS~\cite{CMS:2012qk} in $p+$Pb
collisions at 5.02 TeV. On the same plot, the IP-Glasma
predictions for 8 TeV $p+$Pb collisions are also shown.  The most important
prediction is that no significant change in the multiplicity distribution is
expected between $p+$Pb collisions at 5.02 TeV and 8 TeV.

\subsubsection{Centrality Dependence of $dN_{\rm ch}/d\eta$}
\label{sec:AMPT_cent}


Results for the centrality dependence of $dN_{\rm ch}/d\eta$ in the laboratory
frame calculated with $\mathtt{AMPT}$ are shown here.

Table~\ref{table1} gives information on the different 
centrality classes of $p+$Pb events at 8~TeV
from the $\mathtt{AMPT-SM}$ model, 
including the average, minimum and maximum impact parameter values, 
the average number of participant nucleons in the Pb nucleus per event,
$N_{\rm part}^{\rm Pb}$, and the average number of inelastic participant
nucleons in the Pb nucleus per event,
$N_{\rm part-in}^{\rm Pb}$.  The results are given for the ATLAS centrality
criteria where the average transverse energy per event in the lead-going
direction, $\langle E_T \rangle$ in $-4.9 < \eta_{\rm lab} < -3.1$
\cite{Aad:2015zza},  are also shown.
Diffractive events are excluded.  Thus the results
are non-diffractive events. 
The difference in rapidity of the proton beam in the lab frame and in the
center-of-mass frame is $\delta y \simeq 0.465$.

\begin{table}
  \caption{Centrality classes of 8 and 5~TeV $p+$Pb events from the string
    melting
  version of $\mathtt{AMPT}$, with centrality determined from average transverse
  energy, $\langle E_T \rangle$, in the lead-going direction,
  $-4.9 < \eta_{\rm lab} < -3.1$ \cite{Aad:2015zza}. 
``All'' refers to all simulated non-diffractive events.}
\begin{tabular}{cccccccc}
\hline
Centrality & $\left <b \right >$(fm) & $b_{\rm min}$(fm) & $b_{\rm
  max}$(fm) & $N_{\rm part}^{\rm Pb}$   & $N_{\rm part-in}^{\rm Pb}$ &
$\langle E_T \rangle$ (GeV)\\
\hline \hline
\multicolumn{7}{c}{$\sqrt{s_{_{NN}}} = 8$~TeV} \\ \hline 
All  & 5.72 & 0.0 & 13.2 & 8.64 & 6.11 & 47.9 \\
0-1\% & 2.93 & 0.1 & 6.7 & 19.84 & 15.42 & 158.4 \\
1-5\% & 3.20 & 0.0 & 8.0 & 17.85 & 13.54 & 125.4 \\
5-10\% & 3.47 & 0.0 & 8.2 & 16.38 & 12.20 & 105.3 \\
10-20\% & 3.79 & 0.0 & 8.7 & 14.84 & 10.82 & 87.9 \\
20-30\% & 4.22 & 0.0 & 9.9 & 13.03 & 9.26 & 71.8 \\
30-40\% & 4.70 & 0.0 & 11.1 & 11.21 & 7.80 & 58.7 \\
40-60\% & 5.63 & 0.1 & 12.1 & 8.22 & 5.55 & 41.4 \\
60-90\% & 7.24 & 0.1 & 13.2 & 3.88 & 2.58 & 18.1 \\
90-100\% & 8.17 & 2.2 & 13.2 & 1.95 & 1.28 & 5.2\\ \hline \hline
\multicolumn{7}{c}{$\sqrt{s_{_{NN}}} = 5$~TeV} \\ \hline 
All  & 5.62 & 0.0 & 13.2 & 8.01 & 5.76 & 35.7 \\
0-1\% & 2.84 & 0.1 & 6.5 & 18.73 & 14.92 & 118.1 \\
1-5\% & 3.13 & 0.0 & 7.1 & 16.74 & 13.00 & 92.9 \\
5-10\% & 3.39 & 0.1 & 7.7 & 15.17 & 11.56 & 77.6 \\
10-20\% & 3.73 & 0.0 & 8.6 & 13.67 & 10.15 & 64.7 \\
20-30\% & 4.18 & 0.0 & 9.0 & 11.92 & 8.62 & 52.7 \\
30-40\% & 4.66 & 0.1 & 10.1 & 10.24 & 7.22 & 43.0 \\
40-60\% & 5.55 & 0.0 & 12.5 & 7.57 & 5.20 & 30.6 \\
60-90\% & 7.12 & 0.1 & 13.2 & 3.68 & 2.48 & 14.3 \\
90-100\% & 7.98 & 0.3 & 13.2 & 1.96 & 1.27 & 4.6\\
\hline
\end{tabular}\\[2pt]
\label{table1} 
\end{table}

\begin{figure}[htb]
\centerline{\includegraphics[width=0.495\textwidth]{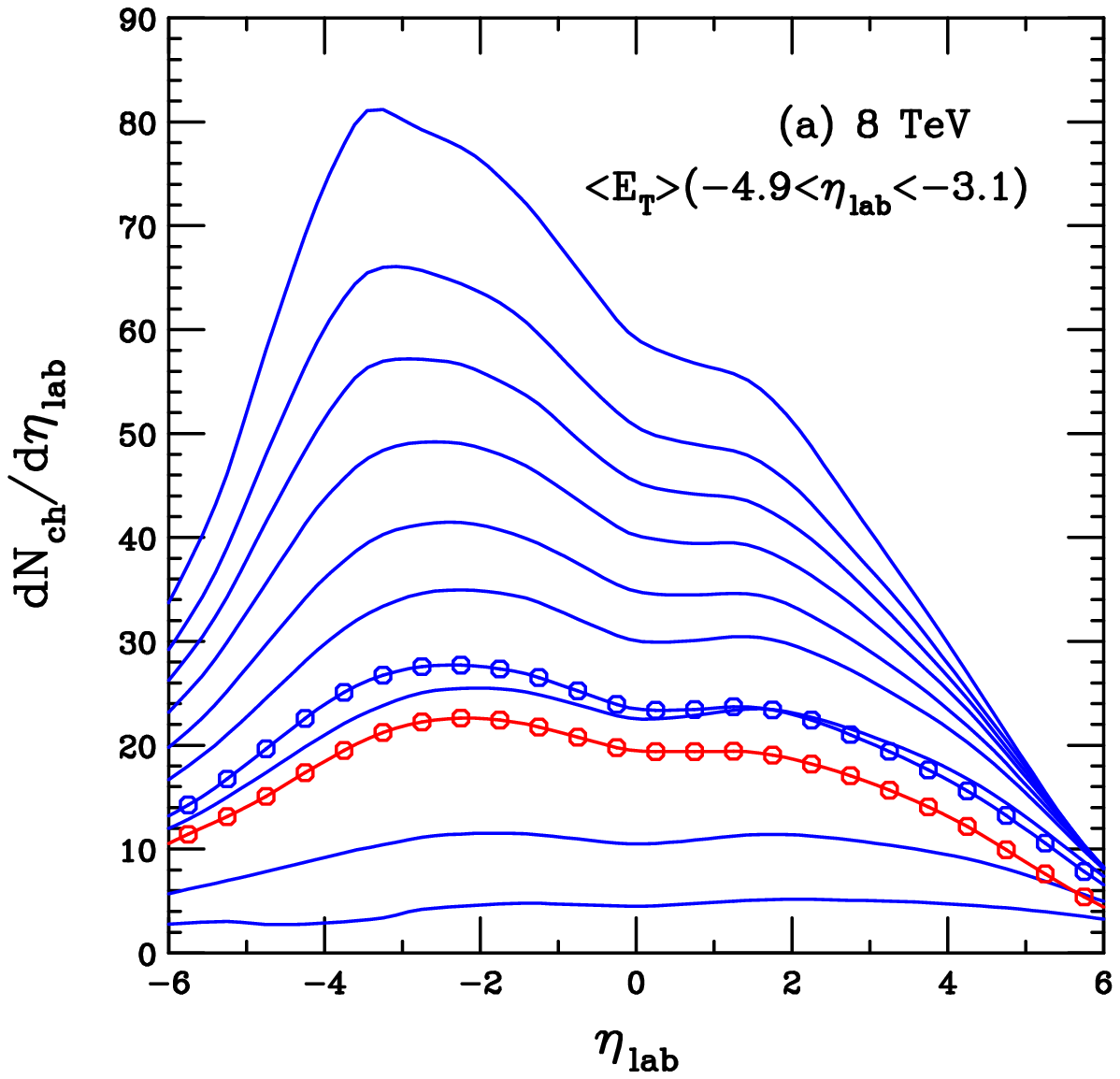}}
\centerline{\includegraphics[width=0.495\textwidth]{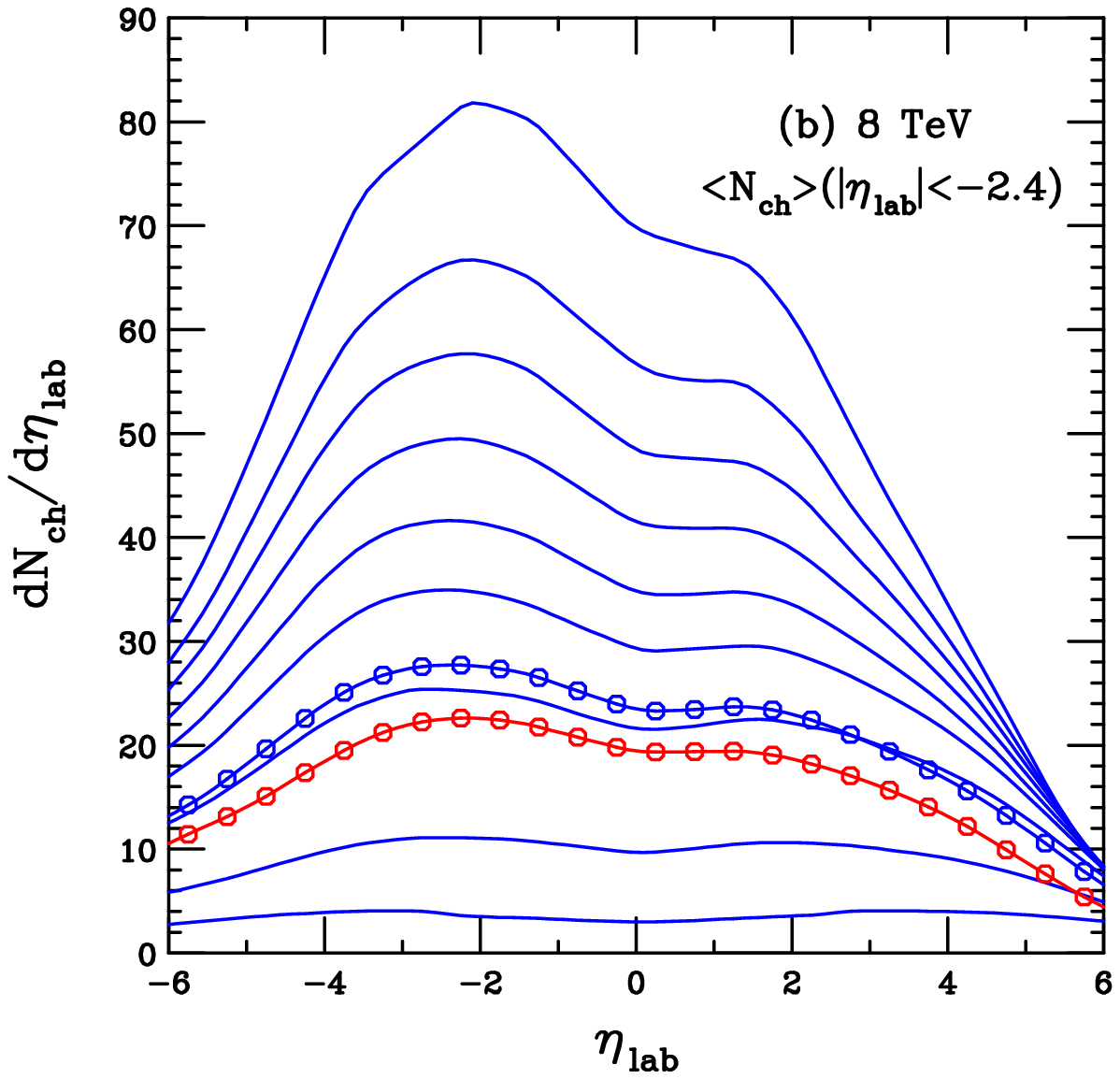}}
\caption[]{(Color online)
The charged hadron multiplicity, $dN_{\rm ch}/d\eta$, in 8 TeV $p+$Pb 
collisions at
different centralities calculated with $\mathtt{AMPT-SM}$ are shown.
The result for all non-diffractive events at 5 TeV (red points
and line) is
also shown for comparison.  The blue lines, from top to bottom are for
centralities of (0-1)\%, (1-5)\%, (5-10)\%, (10-20)\%, (20-30)\%, (30-40)\%,
(40-60)\%, (60-90)\% and (90-100)\%.  The blue points and line shows the
8 TeV non-diffractive multiplicity.  The 8~TeV non-diffractive multiplicity
is very similar to the calculation in the (40-60)\% centrality bin.
The results in (a) are for the ATLAS centrality definition,
$\langle E_T \rangle(-4.9 < \eta_{\rm lab} < -3.1)$
\protect\cite{Aad:2015zza}, while the results in (b)
use the centrality definition based on charged particle multiplicity in the
central region $\langle N_{\rm ch} \rangle(|\eta_{\rm lab}|<2.4)$.}
\label{dndeta8}
\end{figure}

Figure~\ref{dndeta8}(a) shows the results for $dN_{\rm ch}/d\eta$
at the tabulated centralities in the laboratory frame
for non-diffractive events in $p+$Pb collisions at 8 TeV. 
The result for all non-diffractive events at 5 TeV, also calculated with the
same version of $\mathtt{AMPT}$, version 2.26t7, is given by the red line
for comparison.  The overall
increase of multiplicity at 8 TeV is clearly visible.

Because the flow coefficients are shown for the CMS centrality criteria,
based on the number of charged hadrons at central rapidity,
$\langle N_{\rm ch} \rangle (|\eta_{\rm lab}|<2.4)$, the charged particle
pseudorapidity distributions are also shown at 8~TeV for this centrality
definition
in Fig.~\ref{dndeta8}(b).  Note that the distributions based on the central
rapidity criteria are shifted forward for the most central bins,
(0-1)\%, (1-5)\% and
(5-10)\% in particular, as well as for the most peripheral bin, (90-100)\%.
However, the distributions for the two different centrality definitions in the
mid-central and mid-peripheral centrality bins match rather well.  The
centrality classes for this critera are given in Table~\ref{tableCMS}.  Although
the distributions are clearly shifted, the average number of participants
changes no more than 5\% in the most central bins while the difference in the
semi-central bins is even smaller.

\begin{table}
\caption{Centrality classes of 8 TeV $p+$Pb events from the string melting
version of AMPT.  The centrality is determined from the number of
 charged hadrons within $|\eta_{\rm lab}|<2.4$. 
``All'' refers to all simulated non-diffractive events.}
\begin{tabular}{cccccccc}
\hline
Centrality & $\left <b \right >$(fm) & $b_{\rm min}$(fm) & $b_{\rm
  max}$(fm) & $N_{\rm part}^{Pb}$   & $N_{\rm part-in}^{Pb}$ &
$\left <N_{\rm ch}(|\eta_{\rm lab}|<2.4) \right >$\\
\hline
All  & 5.72 & 0.0 & 13.2 & 8.64 & 6.11 & 118.4 \\
0-1\% & 3.24 & 0.0 & 7.1 & 18.95 & 14.70 & 343.6 \\
1-5\% & 3.45 & 0.0 & 8.1 & 17.20 & 13.01 & 280.5 \\
5-10\% & 3.64 & 0.1 & 9.5 & 15.91 & 11.82 & 242.0 \\
10-20\% & 3.90 & 0.0 & 9.0 & 14.53 & 10.59 & 207.8 \\
20-30\% & 4.26 & 0.0 & 9.7 & 12.96 & 9.24 & 175.5 \\
30-40\% & 4.66 & 0.0 & 11.6 & 11.36 & 7.93 & 148.2 \\
40-60\% & 5.50 & 0.0 & 12.2 & 8.58 & 5.82 & 109.9 \\
60-90\% & 7.18 & 0.1 & 13.2 & 3.98 & 2.64 & 49.9 \\
90-100\% & 8.29 & 1.9 & 13.2 & 1.74 & 1.16 & 15.8\\
\hline
\end{tabular}\\[2pt]
\label{tableCMS} 
\end{table}

\begin{figure}[htb]
\centerline{\includegraphics[width=0.495\textwidth]{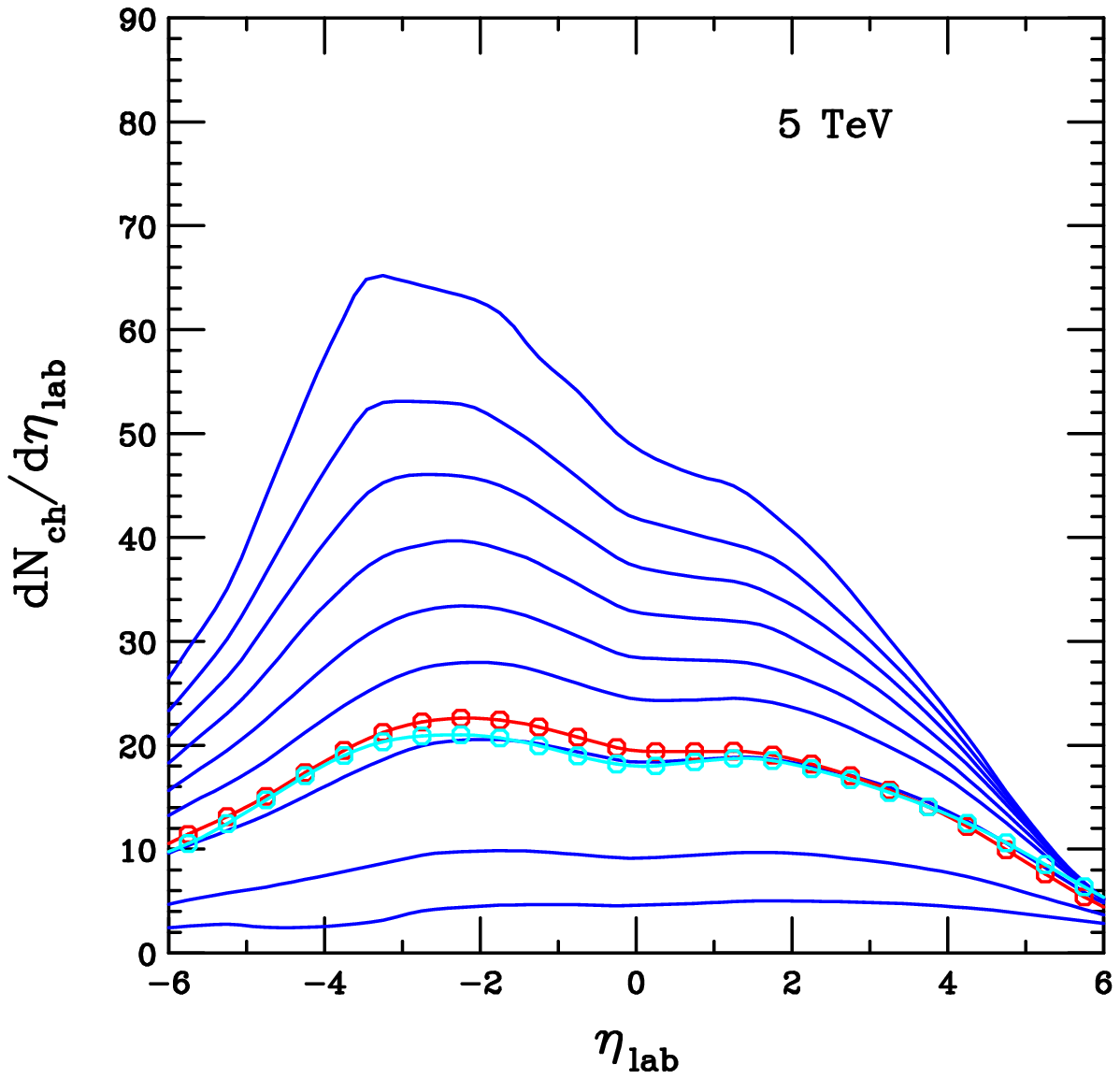}}
\caption[]{(Color online) 
  The charged hadron multiplicity distribution, $dN_{\rm ch}/d\eta$, in 5 TeV
  $p+$Pb collisions at different centralities from $\mathtt{AMPT-SM}$ are shown
  in the laboratory frame.  The previous prediction for minimum-bias events at 5
TeV (cyan points and line)
is shown for comparison. The blue lines, from top to bottom are for
centralities of (0-1)\%, (1-5)\%, (5-10)\%, (10-20)\%, (20-30)\%, (30-40)\%,
(40-60)\%, (60-90)\% and (90-100)\%.  The red points and line shows the
previous $\mathtt{AMPT}$ result for the 5 TeV non-diffractive multiplicity.
The red and cyan curves are somewhat different in shape but similar in
magnitude. The results in are for the ATLAS centrality definition,
$\langle E_T \rangle(-4.9 < \eta_{\rm lab} < -3.1)$ \cite{Aad:2015zza}.} 
\label{dndeta5}
\end{figure}

The results at 5 TeV for different centralities of non-diffractive
events with the current version of $\mathtt{AMPT-SM}$ are shown in
Fig.~\ref{dndeta5}. 
For comparison, the previous prediction for
minimum-bias events at 5 TeV, obtained with $\mathtt{AMPT-SM}$
version 2.26t1 \cite{Albacete:2013ei}, is also shown.  The same centrality
definition as in Fig.~\ref{dndeta8}(a), based on the ATLAS criteria, is used
here.  It is clear that
the distribution $dN_{\rm ch}/d\eta$ of non-diffractive
events is somewhat higher.  Note that the 5 TeV result presented in
Ref.~\cite{Albacete:2013ei}  
was for minimum-bias collisions, including diffractive events.
In addition, different values of the Lund fragmentation parameters,
strong coupling constant, and parton cross section were used. 

\begin{figure}[htb]
\centerline{\includegraphics[width=0.495\textwidth]{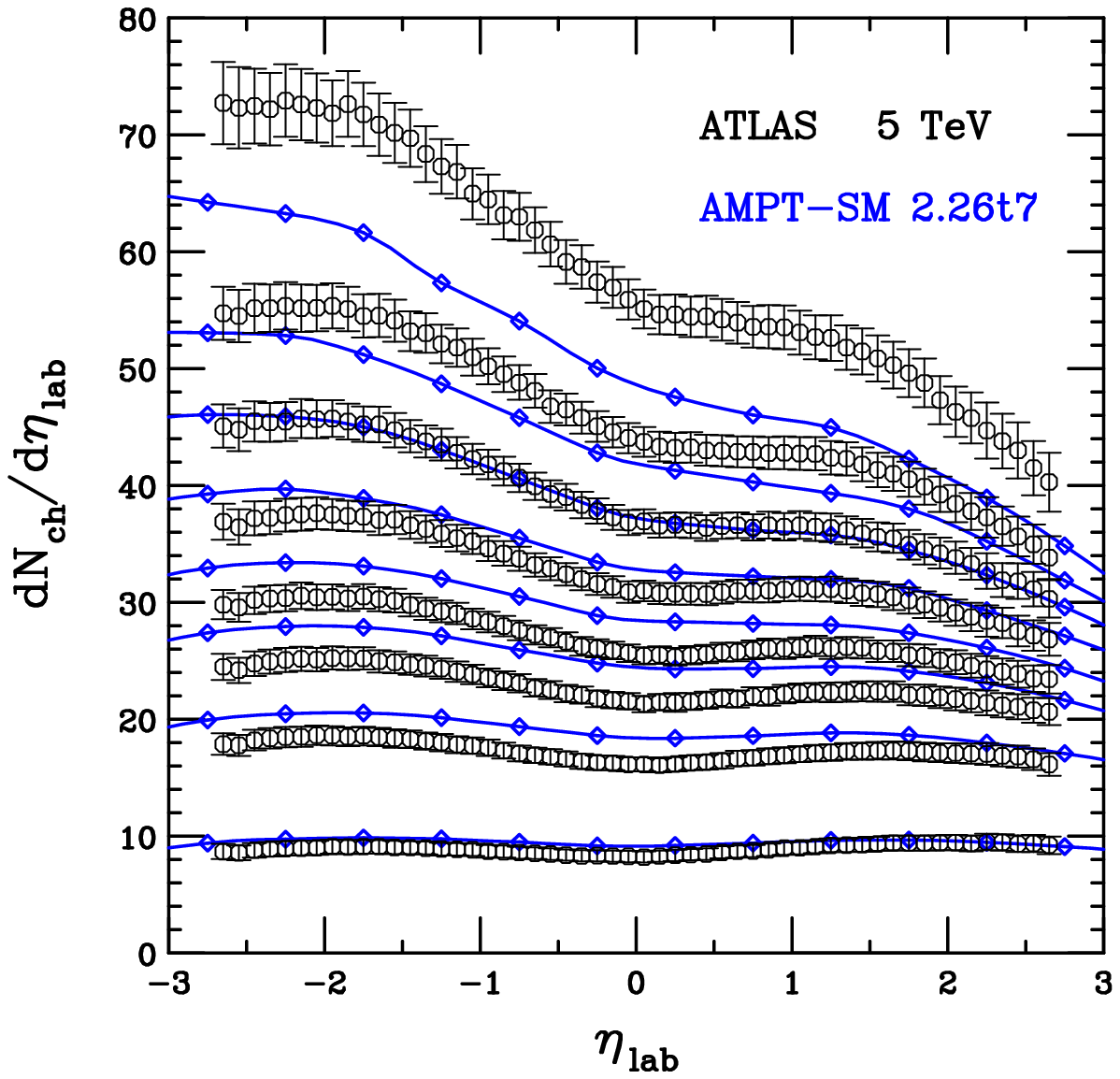}}
\caption[]{(Color online) 
  The new $\mathtt{AMPT-SM}$ charged hadron multiplicity distribution,
  $dN_{\rm ch}/d\eta$, at 5 TeV (in blue) is compared to the ATLAS data (black
  points) \protect\cite{Aad:2015zza} in
  the same centrality bins: (0-1)\%, (1-5)\%, (5-10)\%, (10-20)\%, (20-30)\%,
  (30-40)\%, (40-60)\% and (60-90)\%.
} 
\label{dndeta5+ATLAS}
\end{figure}

In Ref.~\cite{Albacete:2016veq}, calculations from 
the default $\mathtt{AMPT}$ model were compared
to the ATLAS data \cite{Aad:2015zza}.
The $\mathtt{AMPT}$ calculations used the same centrality
bins as the experiment, the same as that given here.  When compared
to the data, the prior version of $\mathtt{AMPT-def}$
showed the same inflection point
near midrapidity but tended to underestimate the multiplicity in the most
central collisions.  The comparison of the current $\mathtt{AMPT-SM}$ version
to the same data in Fig.~\ref{dndeta5+ATLAS} show a similar level of agreement.
Note, however, that the curvature of the calculations in the lead-going
direction is more similar to the data in the new version.

\subsection{Transverse Momentum distributions}
\label{sec:dnchdpt}

Here the transverse momentum distributions are presented.  First, results are
shown for charged hadrons from $\mathtt{AMPT}$ and $\mathtt{kTpQCD\_v2.1}$.
Next, the pion, kaon and
proton $p_T$ distributions from $\mathtt{HIJING++}$ are given.  The
$\mathtt{HIJING++}$ pion results are compared to the  $\mathtt{AMPT}$ and
$\mathtt{kTpQCD\_v2.1}$ results for charged hadrons in $p+p$ and $p+$Pb
collisions.

\subsubsection{Charged and identified hadron $p_T$ distributions
  ($\mathtt{AMPT}$, $\mathtt{kTpQCD\_v21}$)}

\begin{figure}[htb]
\centerline{\includegraphics[width=0.495\textwidth]{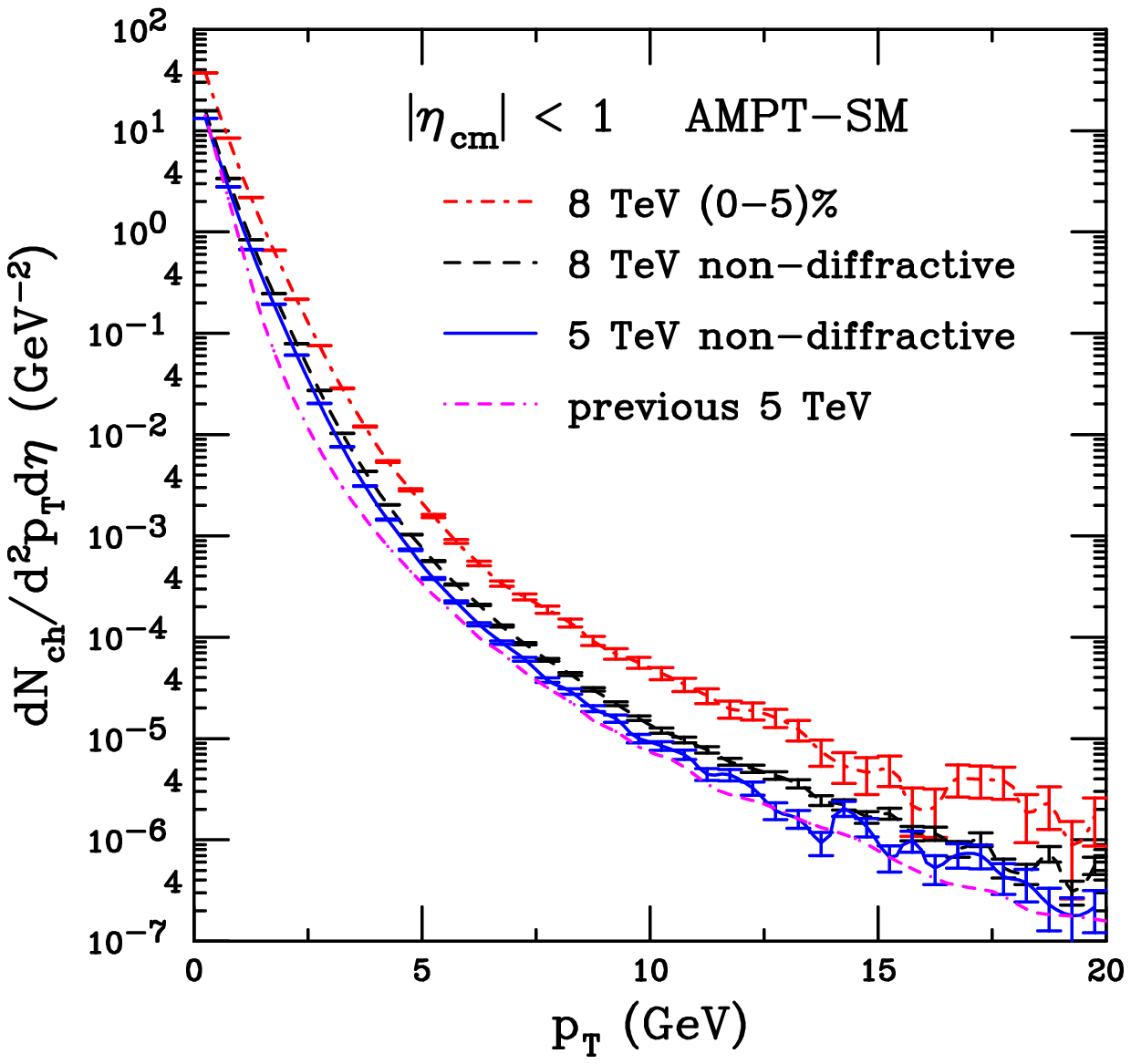}}
\caption[]{(Color online) The $p_T$ spectra of charged hadrons per collision
  in $p+$Pb collisions at both 5 and 8~TeV from $\mathtt{AMPT-SM}$.  The
  previous 5~TeV result is given by the magenta dot-dash-dash-dashed curve while
  the current 5~TeV non-diffractive result is given by the solid blue curve.
  The non-diffractive and (0-5)\% most central results at 8~TeV are given by
  the black dashed and red dot-dashed curves respectively.
}
\label{dndpt_AMPT}
\end{figure}

Figure~\ref{dndpt_AMPT} shows the $p_T$-spectra of charged hadrons per collision
within the center of mass pseudorapidity range $|\eta_{\rm cm}|<1$
for all non-diffractive events and also for the top 5\%
centrality at 8 TeV from $\mathtt{AMPT}$. 
Also shown are the current result for non-diffractive events at 5 TeV 
(obtained with the string melting $\mathtt{AMPT}$ version 2.26t7) and the
previous prediction
\cite{Albacete:2013ei,Albacete:2016veq} for minimum-bias events at 5~TeV 
(obtained with the string melting $\mathtt{AMPT}$ version 2.26t1).
Note that the uncertainties shown are only statistical. 
The $p_T$ spectrum at 8 TeV is obviously harder than that at 5 TeV. 
The current 5 TeV $p_T$ spectrum for non-diffractive events
is enhanced in the intermediate $p_T$ range, $1 \leq p_T \leq 5$~GeV,
relative to the previous prediction for
minimum-bias events. This is mainly due to the small value of the Lund 
parameter, $b$, used in the current parameter set.  The smaller value of $b$
leads to a higher
effective string tension and a harder $p_T$-spectrum for initial
hadrons \cite{Lin:2014tya}.  

\begin{figure}[htb]
\includegraphics[width=0.495\textwidth]{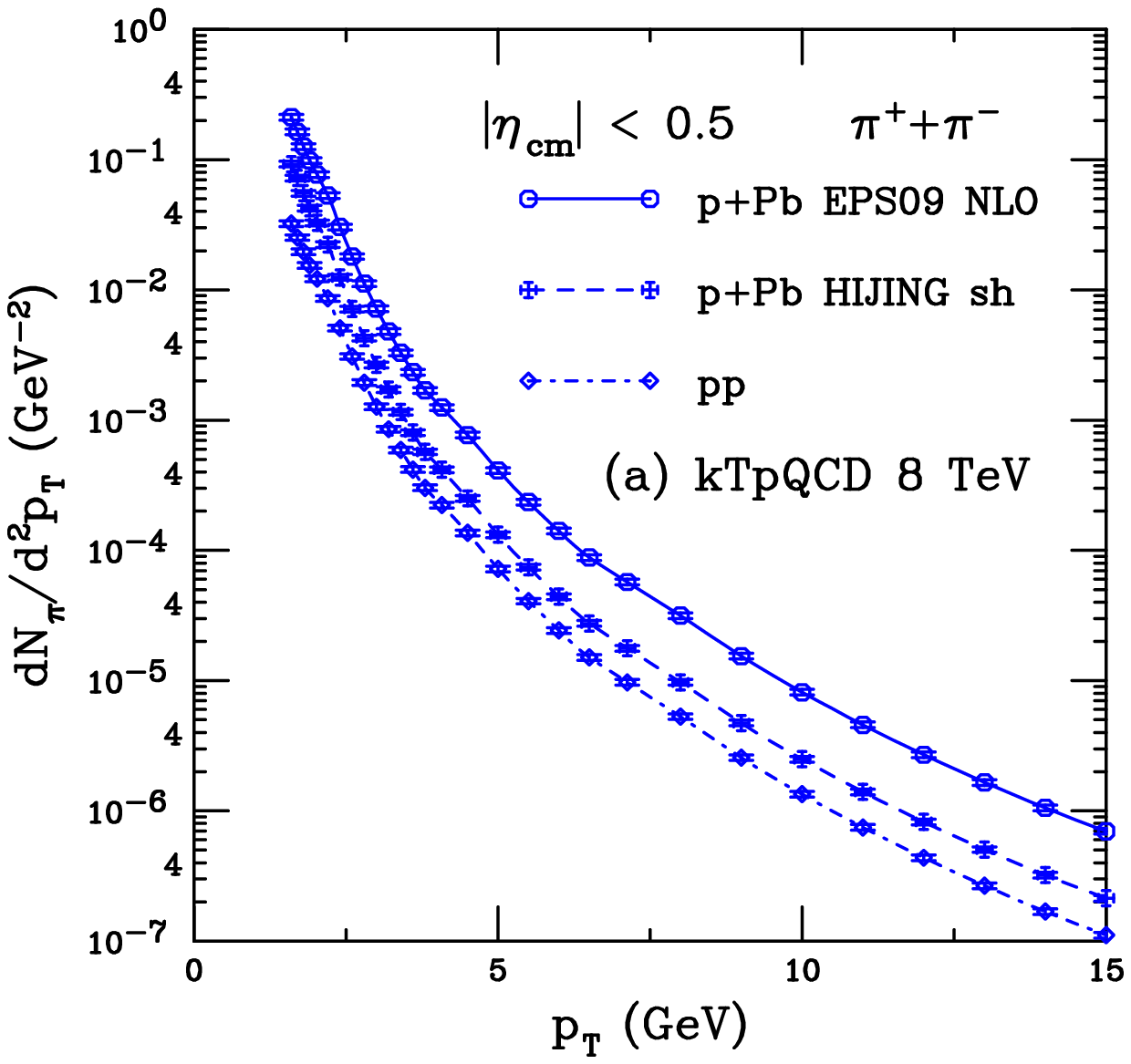} 
\includegraphics[width=0.495\textwidth]{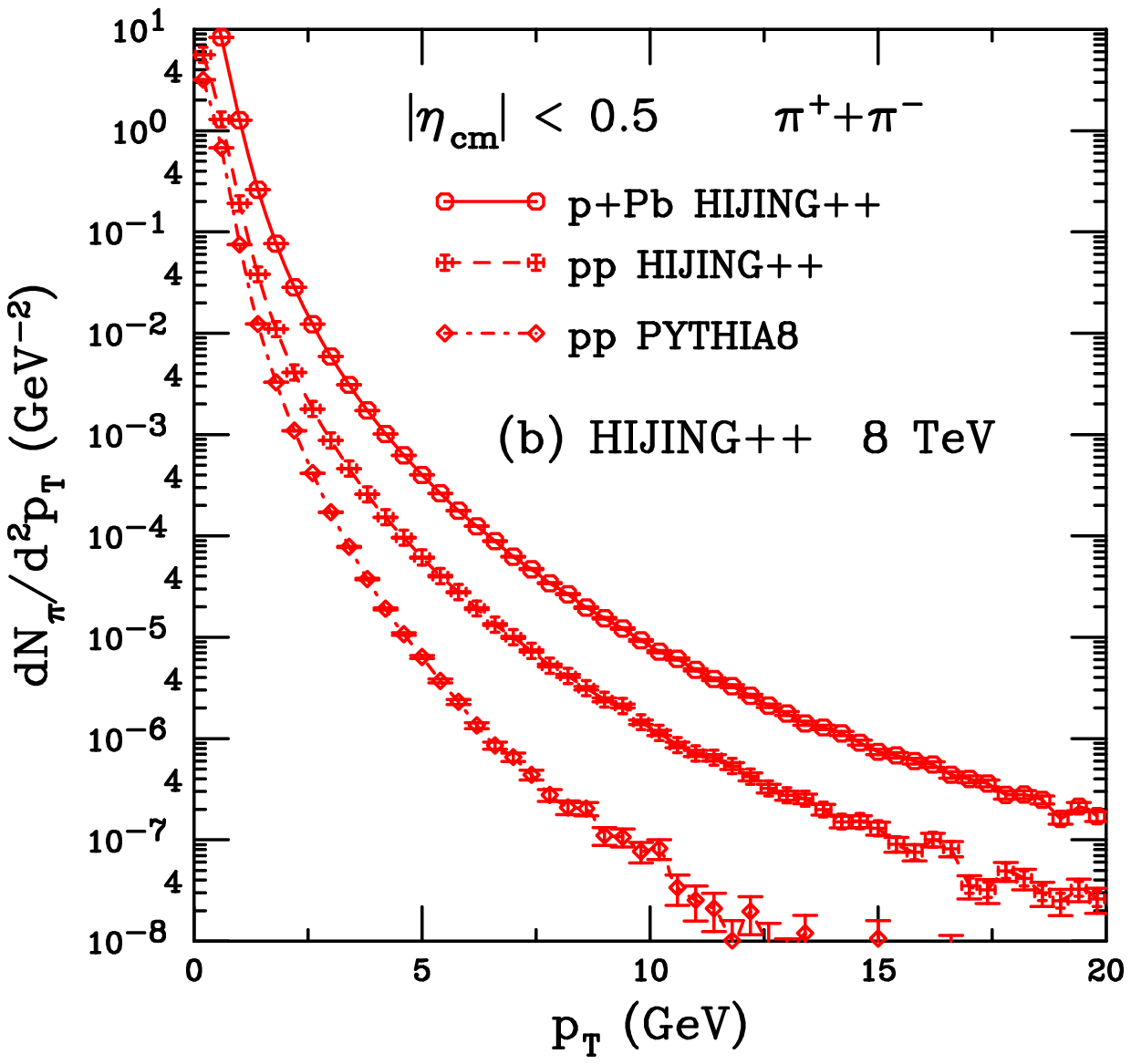} 
\caption[]{(Color online) The $p_T$ spectra of charged pions at 8~TeV from
  $\mathtt{kTpQCD}$ is given in (a) with the $p+$Pb results with EPS09 NLO
  and the HIJING shadowing parameterization given by the solid and dashed
  curves respectively while the $p+p$ result is shown in the dot-dashed curve.
  The $\mathtt{HIJING++}$ 8~TeV results are given in (b) for $p+$Pb (solid
  curve) and $p+p$ (dashed curve) while the $\mathtt{PYTHIA8}$ result is
  shown in the dot-dashed curve.
}
\label{dndpt-pions}
\end{figure}

The predicted spectrum from $\mathtt{kTpQCD\_v21}$ for charge-averaged pions,
$\pi^{\pm}$, is presented in Fig.~\ref{dndpt-pions}(a),
calculated at $\sqrt{s_{_{NN}}}= 8.0$ TeV in $p+p$ and $p+$Pb collisions. The
$\mathtt{HIJING}$~\cite{HIJINGsh} shadowing
parameterization is stronger than
EPS09 NLO~\cite{Eskola:2009uj}.  The difference is significant.  Indeed the
$p+$Pb result with the $\mathtt{HIJING}$ shadowing parameterization is more
compatible with the $p+p$ calculation in $\mathtt{kTpQCD\_v21}$ than the
$p+$Pb result with the EPS09 NLO shadowing parameterization.
Note that these results are shown for $1.6 < p_T <15$~GeV.
The $p+$Pb calculations are for minimum bias collisions.

The $\mathtt{HIJING++}$ result is shown in
Fig.~\ref{dndpt-pions}(b) for $p_T < 20$ GeV.
These calculations were done at 8.16 TeV.  The upper curve is the $p+$Pb
result for minimum bias collisions.
The $p+p$ results with $\mathtt{HIJING++}$ and $\mathtt{PYTHIA8}$
are also shown.  It is clear that the difference between the $p+p$ results
for $\mathtt{HIJING++}$ and $\mathtt{PYTHIA8}$ is large and
increasing with $p_T$.

\begin{figure}[htb]
\includegraphics[width=0.495\textwidth]{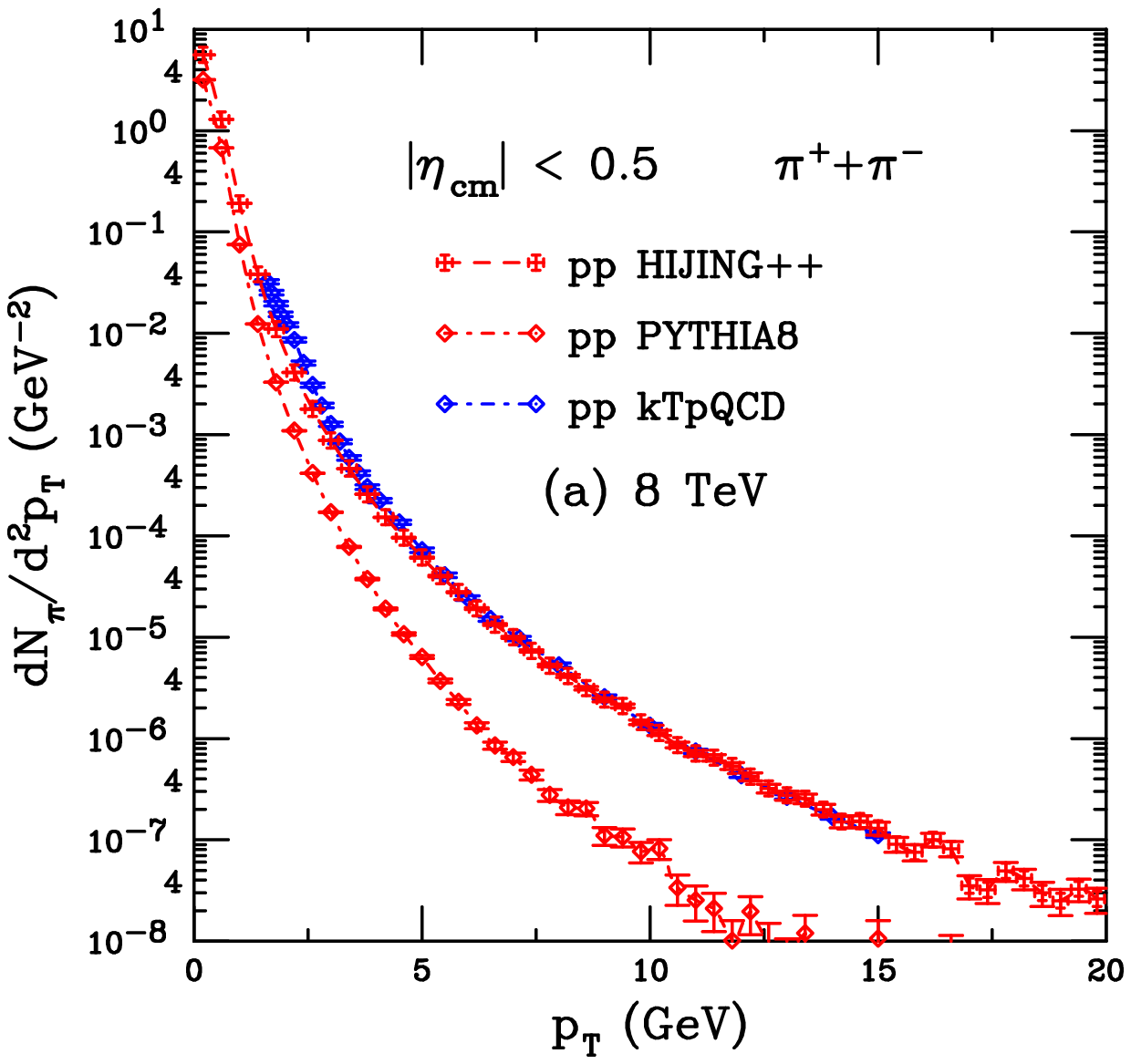} 
\includegraphics[width=0.495\textwidth]{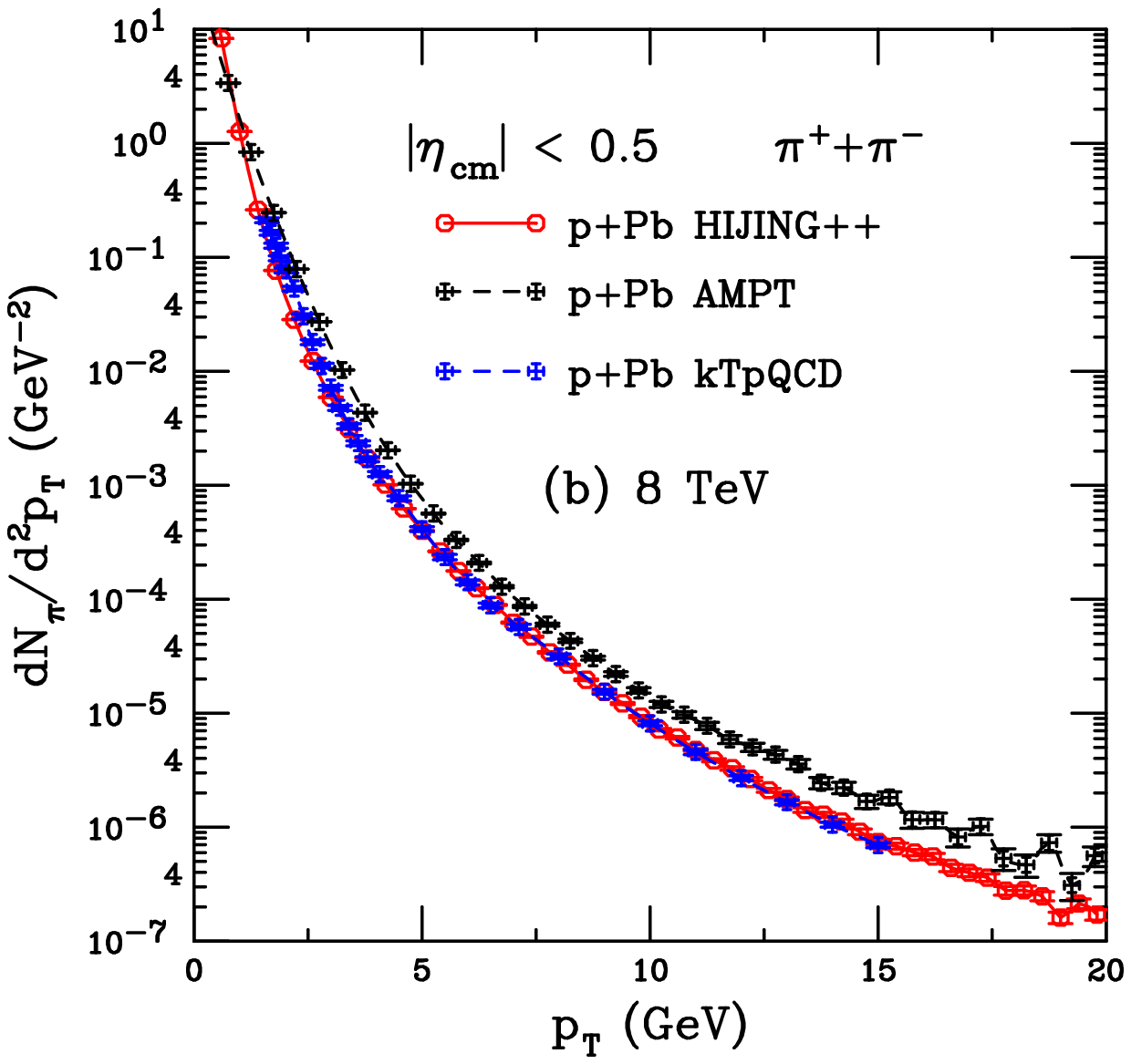} 
\caption[]{(Color online) The $p_T$ spectra of charged pions from
  $\mathtt{kTpQCD}$ and $\mathtt{HIJING++}$ in $p+p$ (a) and $p+$Pb (b)
  collisions, both at 8~TeV.  The $p+p$ results from $\mathtt{HIJING++}$,
  $\mathtt{PYTHIA8}$ and $\mathtt{kTpQCD}$ are given by the red dashed,
  red dot-dashed and blue dot-dashed curves respectively in (a).  The
  $p+$Pb results from $\mathtt{HIJING++}$ and $\mathtt{kTpQCD}$ are given by
  the red solid and blue dashed curves respectively.  The
  $\mathtt{AMPT-SM}$ non-diffractive $p+$Pb result for charged hadrons
  at 8 TeV is given by the black dashed curve in (b).
  }
\label{dndpt_pp_pPb}
\end{figure}

Given the difference in the $p+p$ results, for comparison, the $p+p$
calculation with $\mathtt{kTpQCD\_v21}$
is shown with the $\mathtt{HIJING++}$ and $\mathtt{PYTHIA8}$ curves
in Fig.~\ref{dndpt_pp_pPb}(a).
The perturbative QCD result is in very good agreement
with the $\mathtt{HIJING++}$ calculation even though the two
calculations were done at slightly different energies, 8 TeV for
$\mathtt{kTpQCD\_v21}$ and 8.16 TeV for $\mathtt{HIJING++}$
and $\mathtt{PYTHIA8}$.
The Monash 2013 tune for $\mathtt{PYTHIA8}$ seems to considerably soften
the $p_T$ dependence of light hadron production.

The $p+$Pb result for $\mathtt{HIJING++}$ is also compared with the
$\mathtt{kTpQCD\_v21}$
calculation in Fig.~\ref{dndpt_pp_pPb}(b).
Of the two $\mathtt{kTpQCD}$ results, the one
including the $\mathtt{HIJING}$ shadowing parameterization is shown since this
parameterization is also included in $\mathtt{HIJING++}$.  Again,
the difference in the two results is small.

The $\mathtt{AMPT-SM}$ non-diffractive result at 8 TeV for charged hadrons is
also included in the figure.  There are several differences between the two
generator calculations.  $\mathtt{HIJING++}$ is given in the central rapidity
bin, $|\eta| \leq 0.5$, for charged pions while the $\mathtt{AMPT-SM}$
result is for charged
hadrons in a broader bin, $|\eta| \leq 1$.  Since the charged hadron result is
dominated by pion production and the rapidity bin widths are divided out, these
differences should be negligible.  The largest difference is likely the
overall normalization since $\mathtt{AMPT-SM}$ gives a considerably larger
$p_T$-integrated multiplicity at midrapidity than does $\mathtt{HIJING++}$.
On a logarithmic scale, these differences are rather small.  Thus the two
results are compatible over a broad range of $p_T$, with the $\mathtt{AMPT-SM}$
result becoming somewhat harder for $p_T > 10$~GeV but, overall, the comparison
is good.

\begin{figure}[htb]
\includegraphics[width=0.495\textwidth]{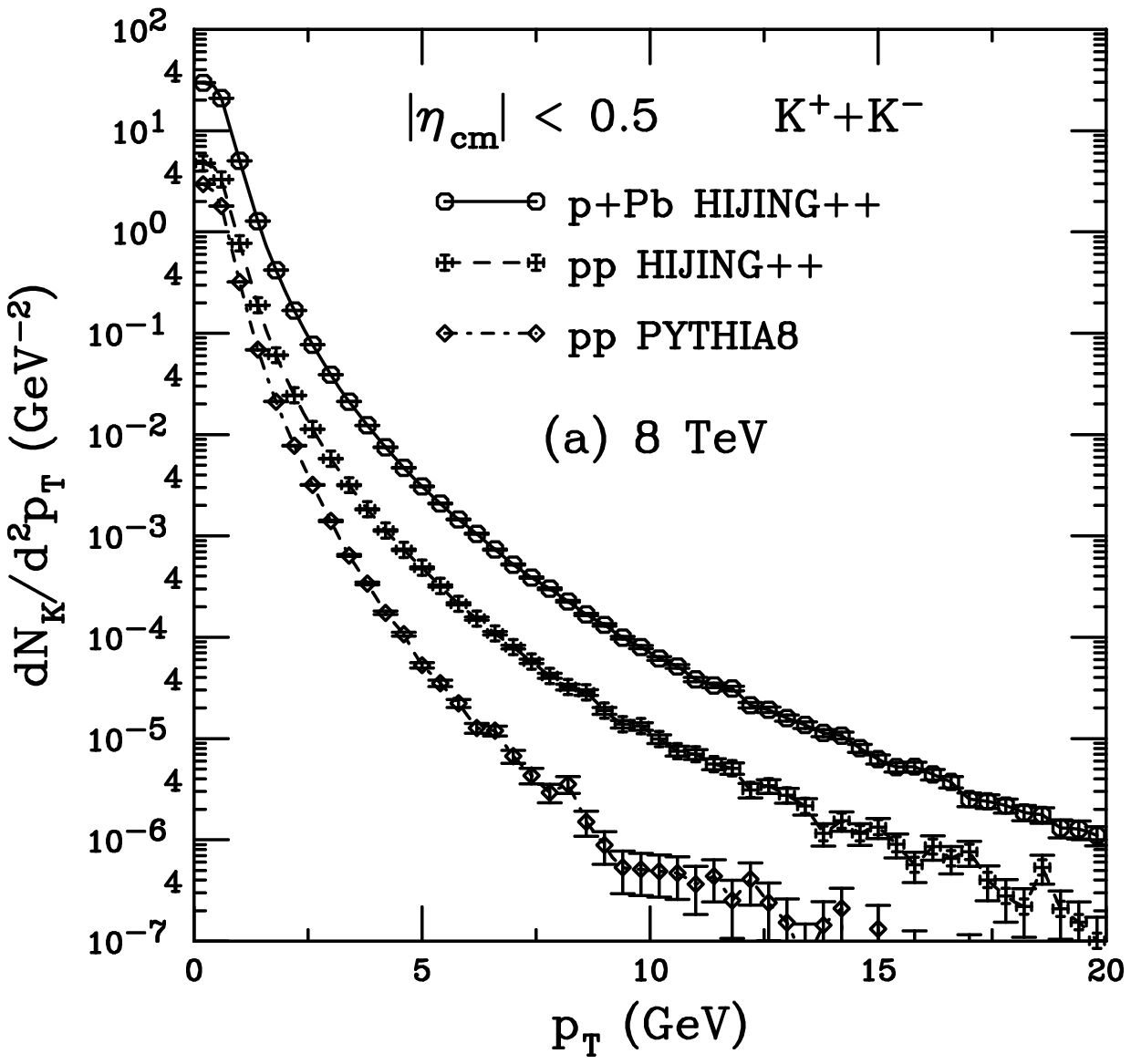} 
\includegraphics[width=0.495\textwidth]{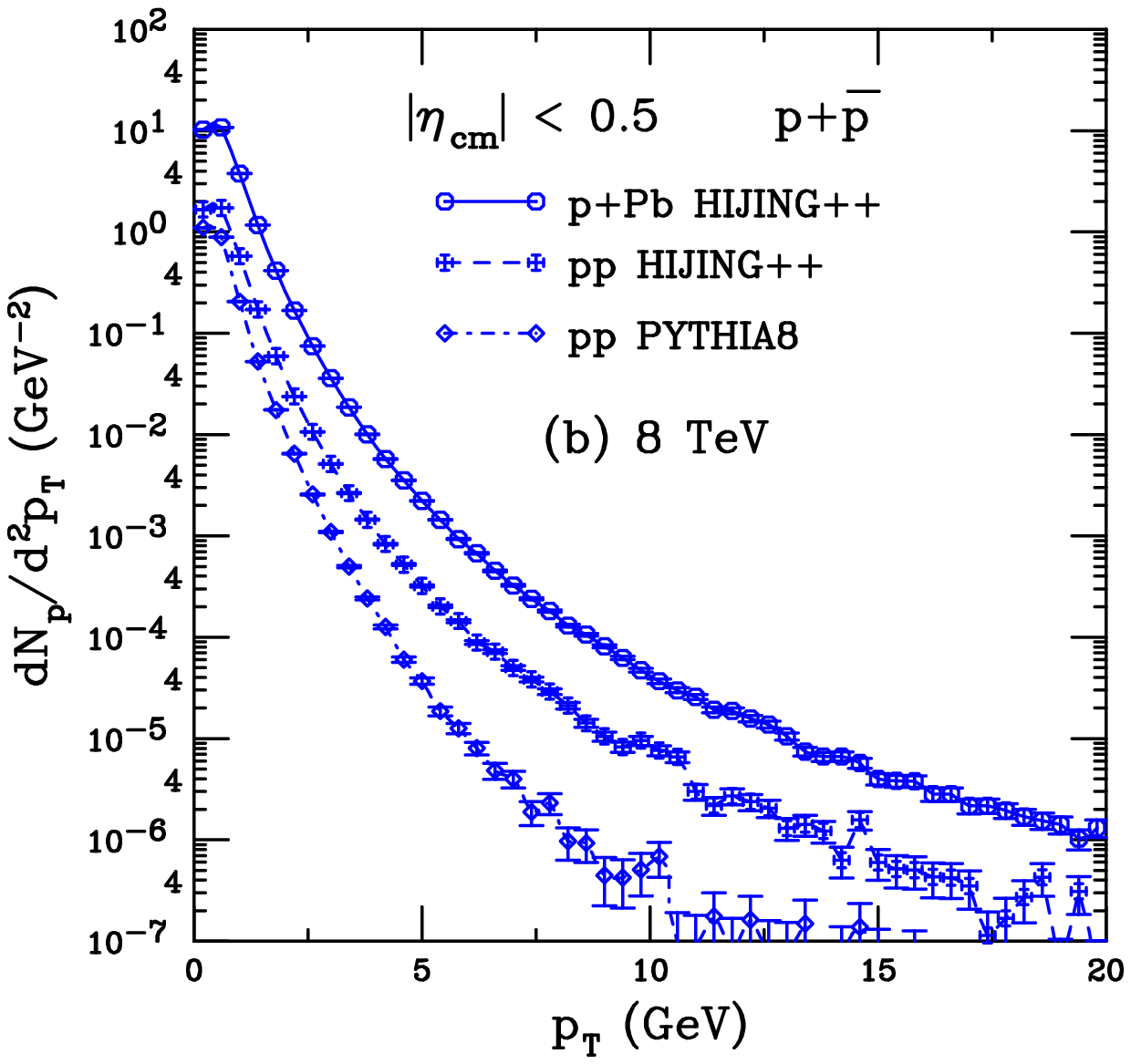} 
  \caption[]{(Color online) The $p_T$ spectra of charged kaons (a) and
    protons (b) in 8~TeV $p+p$ and $p+$Pb collisions.  In both cases the solid
    curves are the $p+$Pb results while the $p+p$ results for
    $\mathtt{HIJING++}$ and $\mathtt{PYTHIA8}$ are given by the dashed and
    dot-dashed curves respectively.
  }
\label{dndpt-HIJING}
\end{figure}

Finally, the $\mathtt{HIJING++}$ results for charged kaons and protons plus
antiprotons are shown in Fig.~\ref{dndpt-HIJING}.  The corresponding $p+p$
results with $\mathtt{HIJING++}$ and $\mathtt{PYTHIA8}$ with the Monash 2013
tune are also given.  The same
difference in the $p+p$ distributions is observed in these cases as well.
Statistical uncertainties, which become larger for the more massive light
hadrons, are shown.

\subsection{Nuclear Suppression Factor $R_{p{\rm Pb}}$}
\label{sec:RpPb}

In this section, calculations of
the nuclear suppression factor are presented.  Results are
shown first at midrapidity for charged hadrons.   The calculations include
initial-state shadowing by Eskola {\it et al.}, cold nuclear matter by Vitev
{\it et al.},
the rcBK results by Lappi and M\"antysaari and Rezaeian,
the $\mathtt{kTpQCD}$
result with two different
shadowing parameterizations, and $\mathtt{HIJING++}$.
The $\mathtt{HIJING++}$ suppression factor for
identified pions, kaons and protons are also shown at midrapidity.   
Finally, the calculations by Vitev {\it et al.}\
and by Lappi and M\"antysaari are shown at
forward rapidity.

\subsubsection{$R_{p{\rm Pb}}(p_T)$ for charged hadrons at $\eta \sim 0$}
\label{sec:RpPb_y0}

\paragraph{EPS09 (K. J. Eskola, I. Helenius, H. Paukkunen)}

The first CMS result for the minimum-bias charged-hadron nuclear modification 
factor ($R_{p \mathrm{Pb}}$) at $\sqrt{s_{_{NN}}} = 5.0$~TeV
showed an enhancement
of $\sim 40$\% at $p_{T} > 20$~GeV \cite{Khachatryan:2015xaa}. 
Such an enhancement would clearly be too large to be accommodated by 
a DGLAP-based nPDF analysis and would thus suggest a violation of 
factorization of the nuclear effects at high $p_T$. A similar behavior
was also seen in the first ATLAS measurement \cite{ATLAS:2014cza} (with some 
cuts on centrality) but in their published result \cite{Aad:2016zif} the $p_{T}$
reach is restricted to $p_{T} \sim 20$~GeV. However, the measurement 
from ALICE \cite{Abelev:2014dsa} was consistent with unity for 
$10 < p_{T} < 50$~GeV. For these early measurements no $p+p$ baseline 
measurement was available at the same collision energy.  During 2015 a short 
$p+p$ run was performed at the LHC with $\sqrt{s} = 5.0$~TeV providing a 
directly measured baseline for $R_{p \mathrm{Pb}}$. Indeed, the new CMS 
measurement of the $R_{p  \mathrm{Pb}}$ \cite{Khachatryan:2016odn} 
show only a moderate enhancement (20\% at most), 
consistent with the nPDF-based calculation when all uncertainties are 
accounted for. Regarding the relevance of the charged-hadron $p+p$ baseline 
calculation (and hence also the ratio $R_{p \mathrm{Pb}}$), the 
independent fragmentation picture is expected to work in the region
$p_T>10$~GeV
where the scale dependence of the computed cross sections is modest and where 
nonperturbative and/or higher-twist effects can be expected to remain small 
\cite{d'Enterria:2013vba}.

\begin{figure}[htb!]
\center
\includegraphics[width=0.7\textwidth]{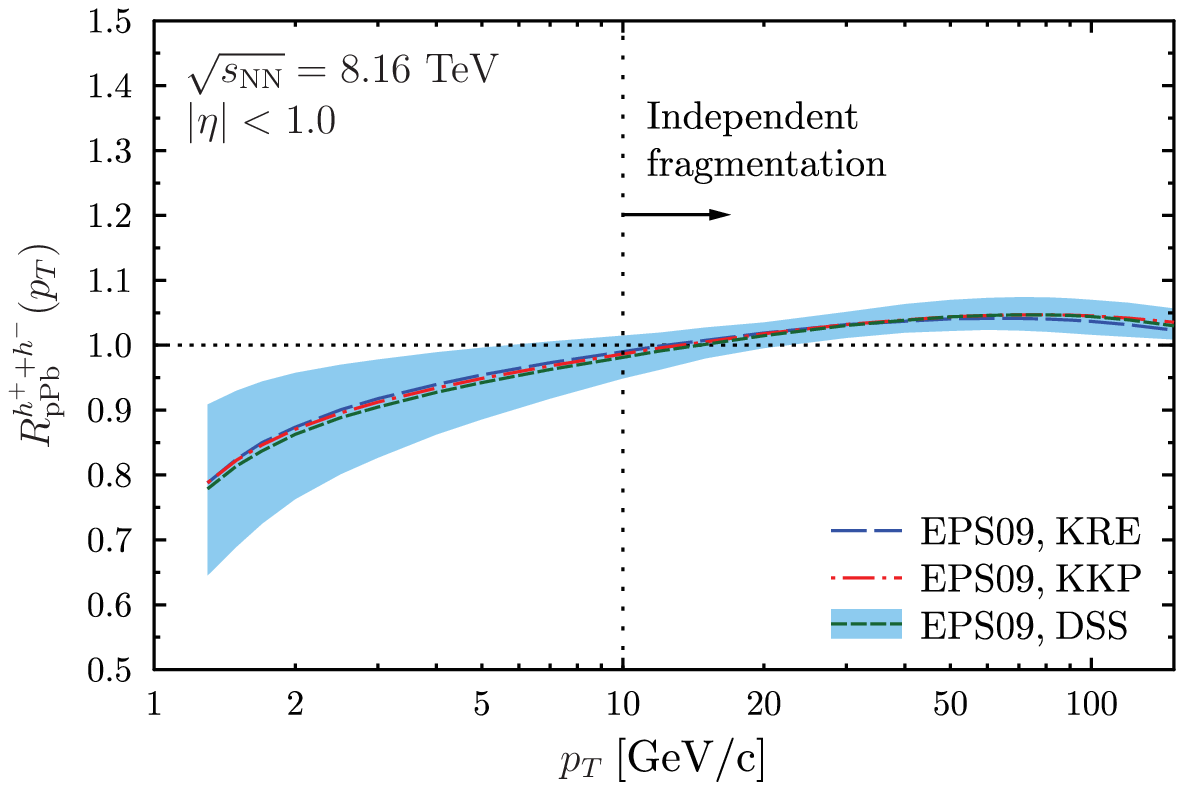}
\caption[]{(Color online) 
Charged-hadron nuclear modification factor for $p$+Pb collisions at 
$\sqrt{s_{_{NN}}} = 8.0$~TeV and $|\eta| < 1.0$. The NLO predictions are computed 
using the CT10 free proton PDFs with EPS09 nuclear modifications and three
fragmentation functions:
Kretzer (blue, long-dashed), KKP (red, dot-dashed) and DSS (green, dashed). The 
uncertainty band is derived from the EPS09 error sets using the DSS
fragmentation function. The dotted 
line with an arrow shows the $p_T$ region where this calculation is expected
to be relevant.  
}
\label{fig:RpPbEPS09}
\end{figure}

Figure~\ref{fig:RpPbEPS09} shows the EPS09-based prediction of the nuclear 
modification factor for charged hadron production in $p$+Pb collisions at 
$\sqrt{s_{_{NN}}} = 8.16$~TeV
at midrapidity ($|\eta|<1$) as a function of $p_{T}$.  The 
calculational framework is the same as in
Refs.~\cite{Albacete:2016veq,Helenius:2012wd}.  The
next-to-leading order (NLO) 
calculations are performed with the \textsc{Incnlo} code 
\cite{Aurenche:1999nz} using the CT10 free proton PDFs \cite{Lai:2010vv} 
and EPS09 NLO nuclear modifications \cite{Eskola:2009uj}.  Three different 
parton-to-hadron fragmentation functions are employed: Kretzer
\cite{Kretzer:2000yf}, KKP \cite{Kniehl:2000fe} and DSS 
\cite{deFlorian:2007hc}.

The theoretical
uncertainties related to scale variations and the proton PDFs cancel out almost 
completely in this ratio so that only uncertainties originating from the
EPS09 NLO sets are 
considered. Also, while the differences between the fragmentation functions
are large \cite{d'Enterria:2013vba}, they also cancel in the ratio.
The behavior is very similar at $\sqrt{s_{_{NN}}} = 5.0$~TeV: some
suppression due to shadowing is seen at small values of $p_{T}$
which turns into a
small enhancement above $p_T \sim 10$~GeV following from the antishadowing
in EPS09 NLO.

Very recently the first nPDF analysis also including data from the
LHC, EPPS16, was completed \cite{Eskola:2016oht}. The central result is very 
similar to the EPS09 NLO fit but, due to increased freedom in the 
parameterization and 
the lack of additional weights on certain data sets, the uncertainties are 
larger. This will result in a somewhat wider uncertainty band than that
shown in Fig.~\ref{fig:RpPbEPS09}.

\paragraph{Other approaches}

\begin{figure}[tb]
\centering
\includegraphics[width=0.495\textwidth]{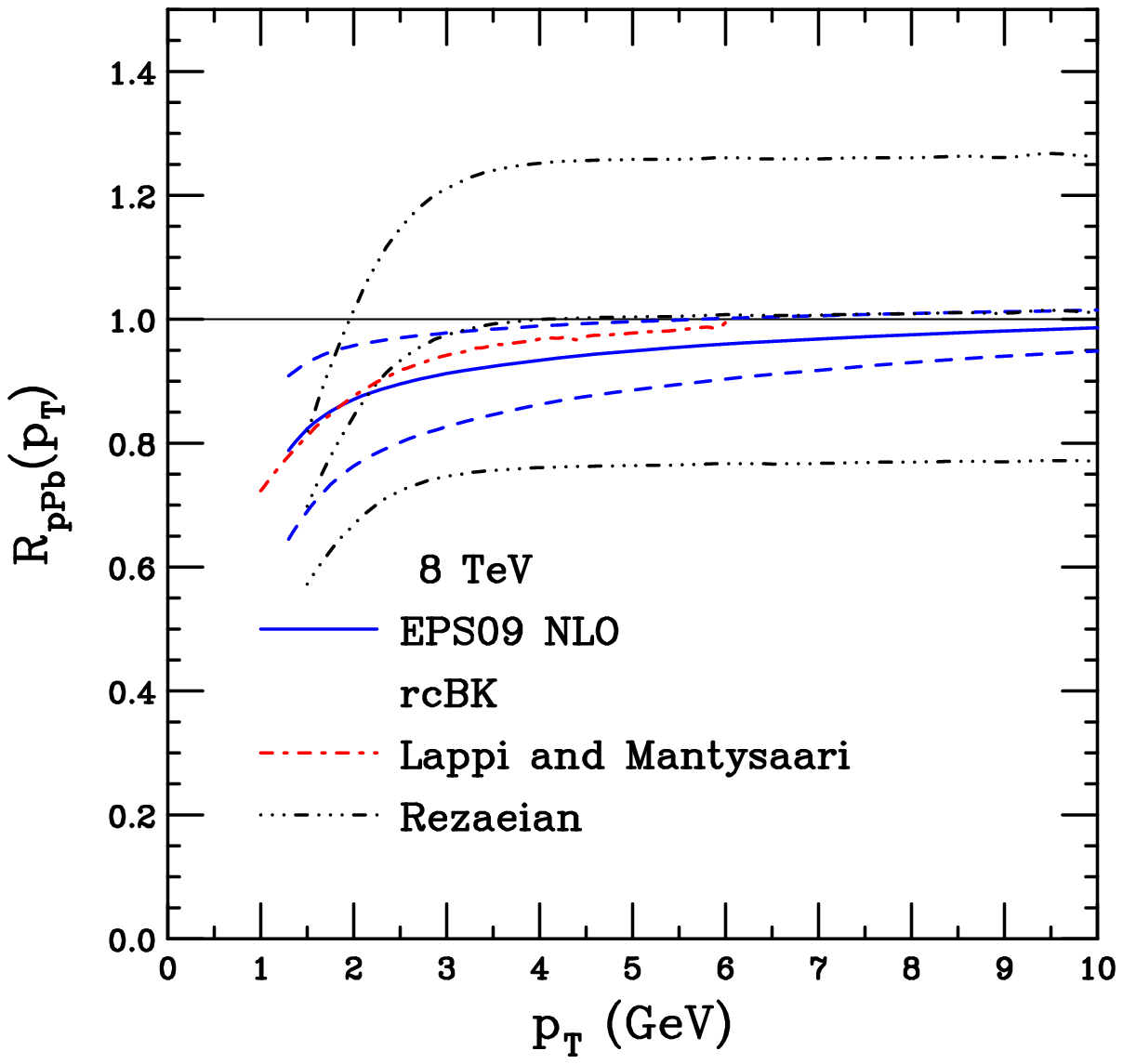} 
\caption[]{(Color online) Charged hadron $R_{p {\rm Pb}}$
  as a function of $p_T$ at 
  midrapidity in 8~TeV collisions.
  The central EPS09 NLO result from Fig.~\protect\ref{fig:RpPbEPS09}
is shown in the solid blue curve.  The dashed blue curves show the uncertainty
in the low $p_T$ region.  The midrapidity CGC calculation by 
Lappi and M\"antysaari is given in the dot-dashed red curve.  The results by
Rezaeian are given in the dot-dot-dot-dashed black curves.}
\label{fig:ch_rpa_mid_CGC}
\end{figure}

Figure~\ref{fig:ch_rpa_mid_CGC} compares the EPS09 NLO central calculation
from Fig.~\ref{fig:RpPbEPS09} with the rcBK results at $y=0$ by Lappi and
M\"antysaari (red curves) and Rezaeian (black curves).  In the
calculations by Lappi and M\"antysaari, the generalization to nuclei is done
using the optical Glauber model, as discussed in Sec.~\ref{sec:sinc} and the
nuclear saturation scale is not a free parameter.
The calculations by Rezaeian are obtained using the hybrid CGC formalism at
leading-order \cite{Dumitru:2005gt} and the solutions of the rcBK
evolution equation \cite{Albacete:2010ad}.
The details of these calculations can be found in 
Ref.~\cite{Rezaeian:2012ye}.  The average initial saturation scale
for the nucleus was
$Q_{0A}^2= 0.168\, N\,\text{GeV}^2$ with the range of $N$ constrained to
$4 \leq N\leq 6$ in
Ref.~\cite{Rezaeian:2012ye}.  The preferred value, $N=5$, corresponds to the
average
value of $Q_{0A}$ extracted from other reactions \cite{Rezaeian:2012ye}.
However, the  exact value of $N$ cannot be determined in the leading-order
approximation \cite{JalilianMarian:2011dt}.
Moreover, the experimental data at small $x$ are
not sufficient to uniquely fix the initial value of the rcBK evolution
equation via a fit \cite{Rezaeian:2012ye}. Therefore, the freedom to choose $N$
in the hybrid factorization formalism introduces rather large uncertainties 
\cite{JalilianMarian:2011dt}.
The LHC data for $R_{pA}$ at 5.02 TeV seem to rule out a
strong Cronin-type peak.  If this feature of the data is verified at higher
energy and thus lower $x$, it can be considered as important
evidence in favor of
small $x$ evolution effects at the LHC \cite{Rezaeian:2012ye}.
Note the average number of binary collisions was assumed to be
$\langle N_{\rm coll}\rangle =6.9$ \cite{dEnterria:2003xac}.
To compare with the LHC data at 8 TeV,
the curves can be rescaled with the experimental value of
$\langle N_{\rm coll}\rangle$.  

In the common $p_T$ range shown,
the rcBK result is quite similar to that of EPS09 NLO at low $p_T$ but rises
toward unity somewhat faster.  Due to the uncertainty in the value of $N$ in
Rezaeian's calculation, that band, although narrower than at 5.02 TeV,
encompasses the EPS09 NLO band and the Lappi and M\"antysaari calculations
for $p_T > 2$~GeV.

\begin{figure}[tb]
\centering
\includegraphics[width=0.495\textwidth]{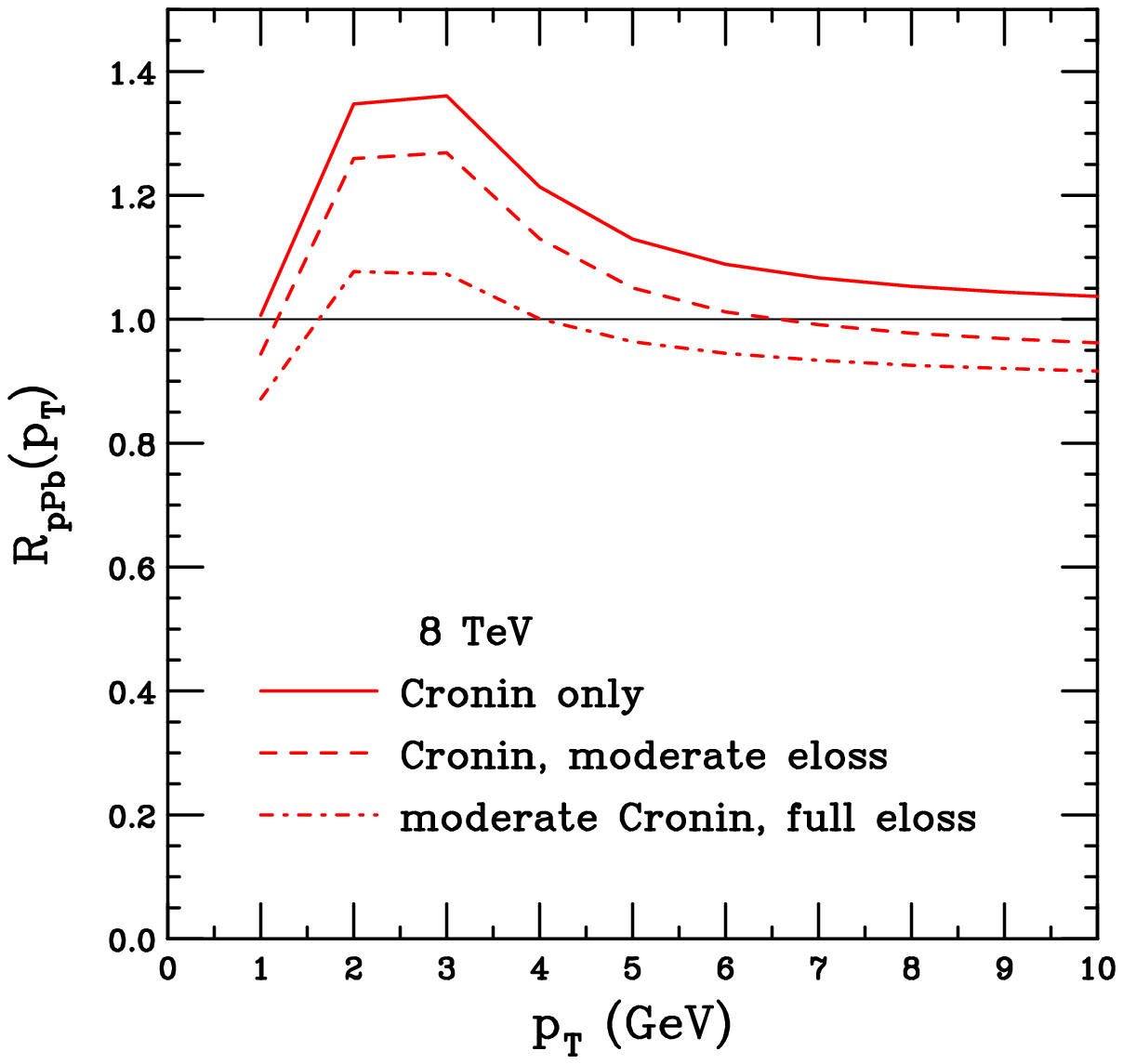} 
\caption[]{(Color online) Charged hadron $R_{p {\rm Pb}}$
  as a function of $p_T$ at 
midrapidity in 8~TeV collisions.  The results by Vitev {\it et al.}\ with
Cronin broadening alone (solid red) and with energy loss (full Cronin and 
moderate energy loss, red dashed, reduced Cronin and stronger energy loss,
red dot dashed) are shown.}
\label{fig:ch_rpa_mid}
\end{figure}

The calculations by Vitev {\it et al.}\
shown in Fig.~\ref{fig:ch_rpa_mid}, on the other hand,
all show an enhancement peaking at $p_T \sim 2-3$ GeV.  The largest enhancement
is with only Cronin broadening.  In this case, $R_{p{\rm Pb}}$ does not drop below
unity for $p_T \leq 10$ GeV.  If the Cronin enhancement is unchanged but
moderate energy loss, with the gluon mean-free path enhanced 50\% over the 1~fm
default value, is included, the enhancement is somewhat reduced.  The smallest
enhancement comes when the default Cronin effect is reduced by a factor of two,
increasing the scattering length from 1~fm to 1.5~fm, while the default
energy loss in cold matter, with a gluon mean-free path of 1~fm, is used.
In this case, the ratio is less than unity
for $p_T > 4$~GeV.

\begin{figure}[tb]
\centering
\includegraphics[width=0.495\textwidth]{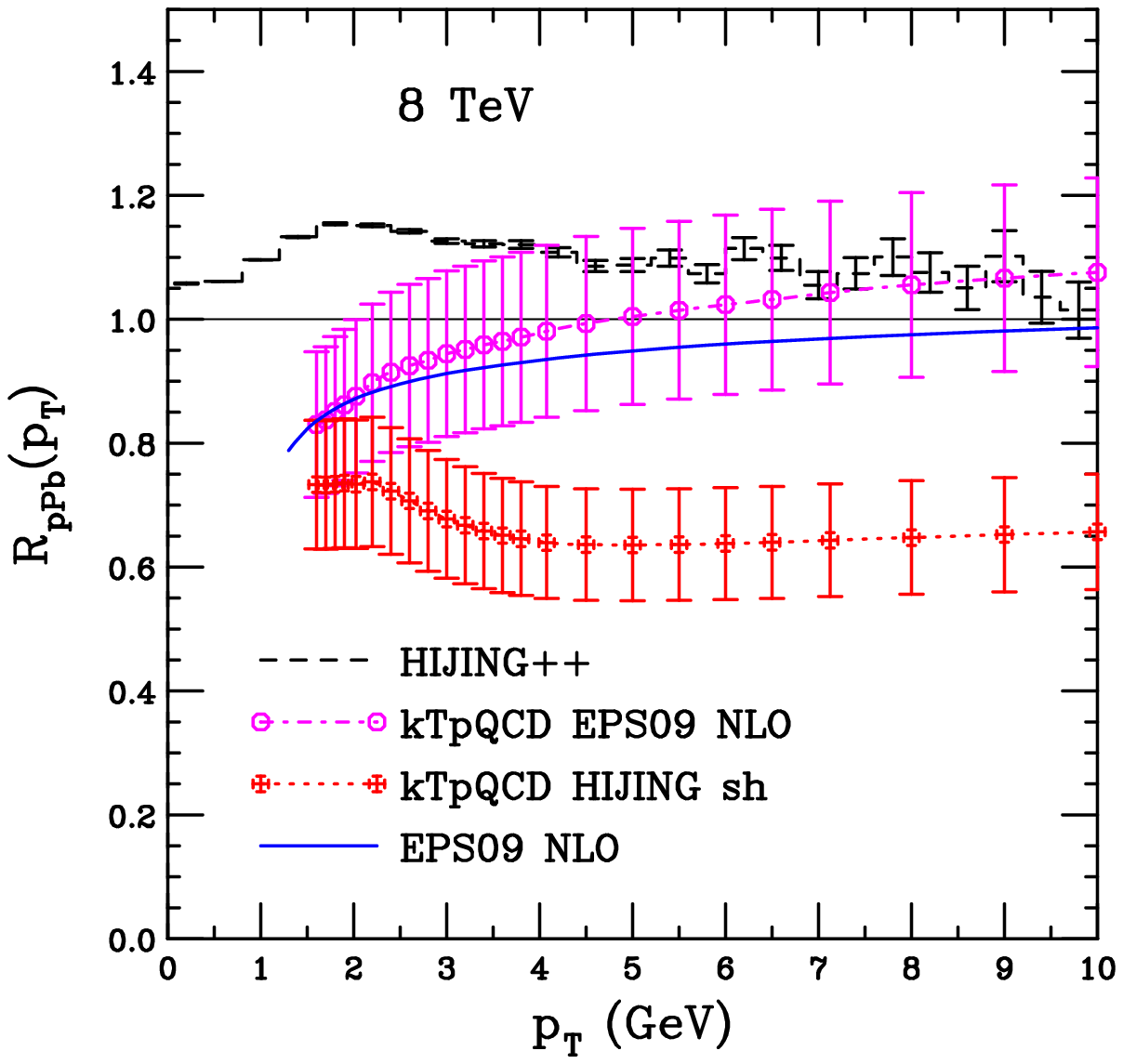} 
\caption[]{(Color online) Charged hadron $R_{p {\rm Pb}}$
  as a function of $p_T$ at 
  midrapidity in 8~TeV collisions
  are shown for $\mathtt{HIJING++}$ (black dashed histogram) and two
  parameterizations in $\mathtt{kTpQCD\_v21}$ (EPS09, magenta curve, and
  the $\mathtt{HIJING}$ shadowing
  parameterization, red curve).  Estimated uncertainties are also shown for
  the last two calculations.  The central EPS09 NLO calculation by
  Eskola (solid blue curve) is also shown.}
\label{fig:ch_rpa_mid2}
\end{figure}

The nuclear modification factors for $\mathtt{HIJING++}$ and $\mathtt{kTpQCD}$
for charged hadrons are shown in Fig.~\ref{fig:ch_rpa_mid2}.  The $p+p$
baseline for the $\mathtt{HIJING++}$ calculation is calculated with
$\mathtt{HIJING++}$ and not $\mathtt{PYTHIA8}$.
As is the case for the cold matter pQCD calculation by Vitev {\it et al.}\
shown in Fig.~\ref{fig:ch_rpa_mid}, the $\mathtt{HIJING++}$ result
is larger than unity
over the $p_T$ range shown.  It shows an enhancement at $p_T \sim 2-3$~GeV
comparable to the dot-dashed curve by Vitev {\it et al.}\
with moderate Cronin and default
energy loss.  On the other hand, the $\mathtt{kTpQCD\_v21}$ calculations show
significant suppression at low $p_T$.  These calculations include an
estimated 10\% uncertainty band to account for uncertainties on the underlying
proton parton density, scale dependence of the perturbative calculation, and
the fragmentation function.  Note that the central EPS09 NLO set is used in
$\mathtt{kTpQCD\_v21}$.
The nuclear PDF uncertainties are not included in the uncertainty
band shown.  The $\mathtt{kTpQCD\_v21}$ result is directly compared to the
central EPS09 NLO calculation.  The two calculations
agree within the $\mathtt{kTpQCD\_v21}$ model uncertainties
although the central $\mathtt{kTpQCD}$ result increases to $R_{p{\rm Pb}}>1$
already at $p_T > 5$~GeV due to the multiple scattering included in this
model.  On the other hand, the calculation with the $\mathtt{HIJING}$
shadowing parameterization decreases
with $p_T$ and seems to saturate for $p_T > 4$~GeV.  The two results only
overlap for $p_T \sim 2-3$~GeV.

\subsubsection{$R_{p{\rm Pb}}(p_T)$ of Identified Particles at  $\eta \sim 0$}
\label{sec:RpPb_pID}

\begin{figure}[tb]
\centering
\includegraphics[width=0.495\textwidth]{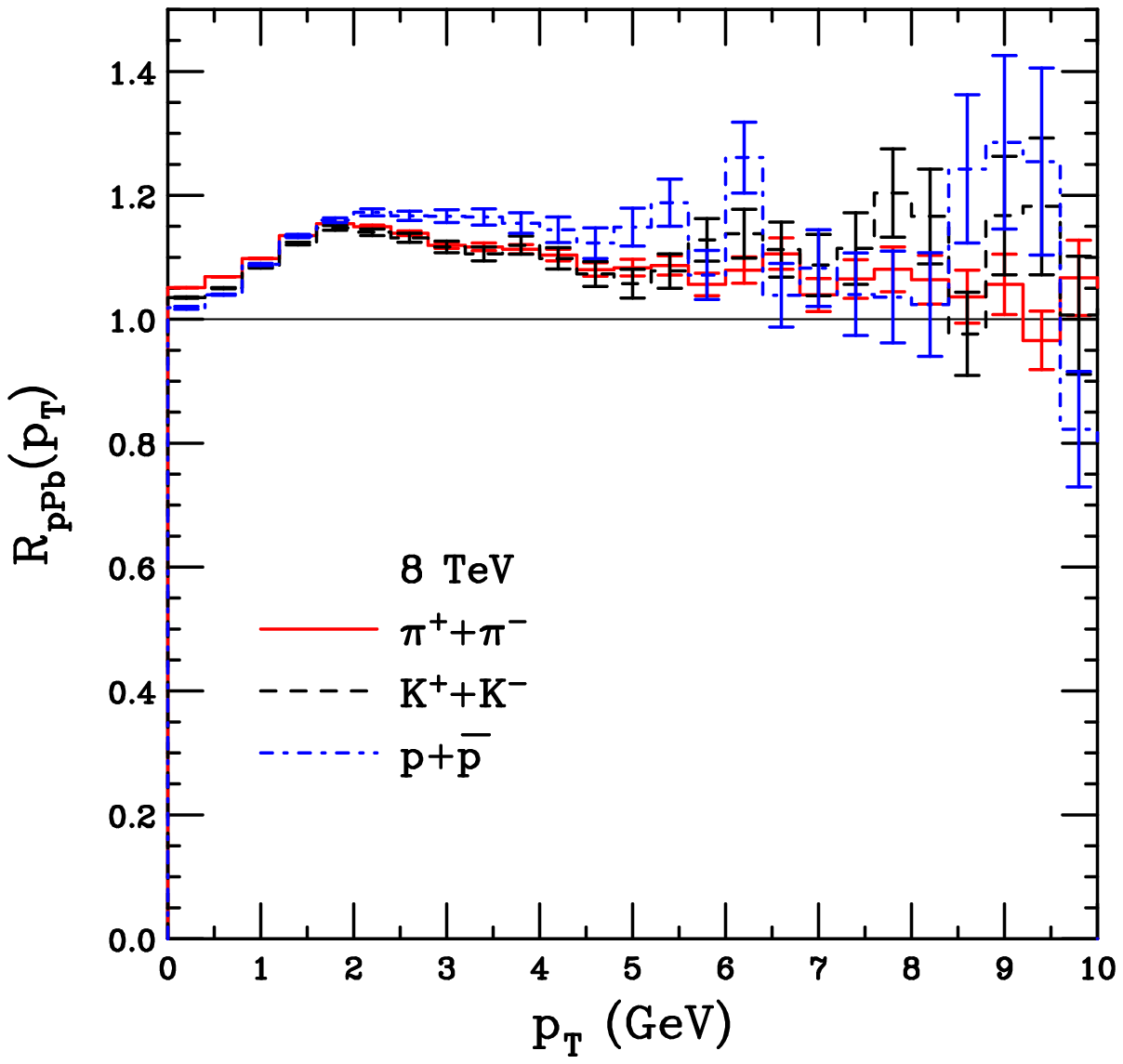} 
\caption[]{(Color online) Charged hadron $R_{pA}$ as a function of $p_T$ at 
  midrapidity in 8~TeV collisions
  are shown for $\mathtt{HIJING++}$ calculations of charged pions
  (solid red histogram), charged kaons (dashed black histogram) and
  protons/antiprotons (dot-dashed blue histogram).}
\label{fig:ch_PID_mid}
\end{figure}

Figure~\ref{fig:ch_PID_mid} shows the $\mathtt{HIJING++}$ calculations of
$R_{p{\rm Pb}}$ for charged pions, charged kaons and protons+antiprotons
formed from the $p+$Pb and $p+p$ calculations with $\mathtt{HIJING++}$ in
Figs.~\ref{dndpt-pions} and \ref{dndpt-HIJING}.  The trend for all three is
similar to that for charged hadrons.  The $p+ \overline p$ ratio has a somewhat
larger enhancement than for $\pi^+ + \pi^-$ and $K^+ + K^-$ in the range
$2 \leq p_T \leq 4$~GeV.  At higher $p_T$, statistical uncertainties become
too large for a meaningful separation.

Recent experimental measurements of identified particle multiplicities in
$p+$Pb collisions~\cite{Abelev:2013haa,Chatrchyan:2013eya} have raised the
possibility that an onset of Cronin-like enhancement might also arise form a
common radial flow-like boost
$\langle \beta_T \rangle$~\cite{Schnedermann:1993ws}. The effect is more
pronounced on particles of larger mass, such as $p$ and $K$ in comparison to
$\pi$. For a quantitative study and discussion of this phenomenon in the
framework of the event generator EPOS, see Ref.~\cite{Werner:2013tya}.
Without radial flow in $\mathtt{HIJING++}$, there is an enhancement, as is also
apparent for charged hadrons in Fig.~\ref{fig:ch_rpa_mid2}, but it has no
significant mass dependence.

\subsubsection{$R_{p {\rm Pb}}(p_T)$ at $|\eta| \neq 0$}
\label{forward_RpPb}

Two results are shown here, the CGC calculation by Lappi and M\"antysaari and
the collinear factorization calculation of cold nuclear matter by Vitev
{\it et al.}.

\paragraph{CGC}

\begin{figure}[tb]
\centering
\includegraphics[width=0.495\textwidth]{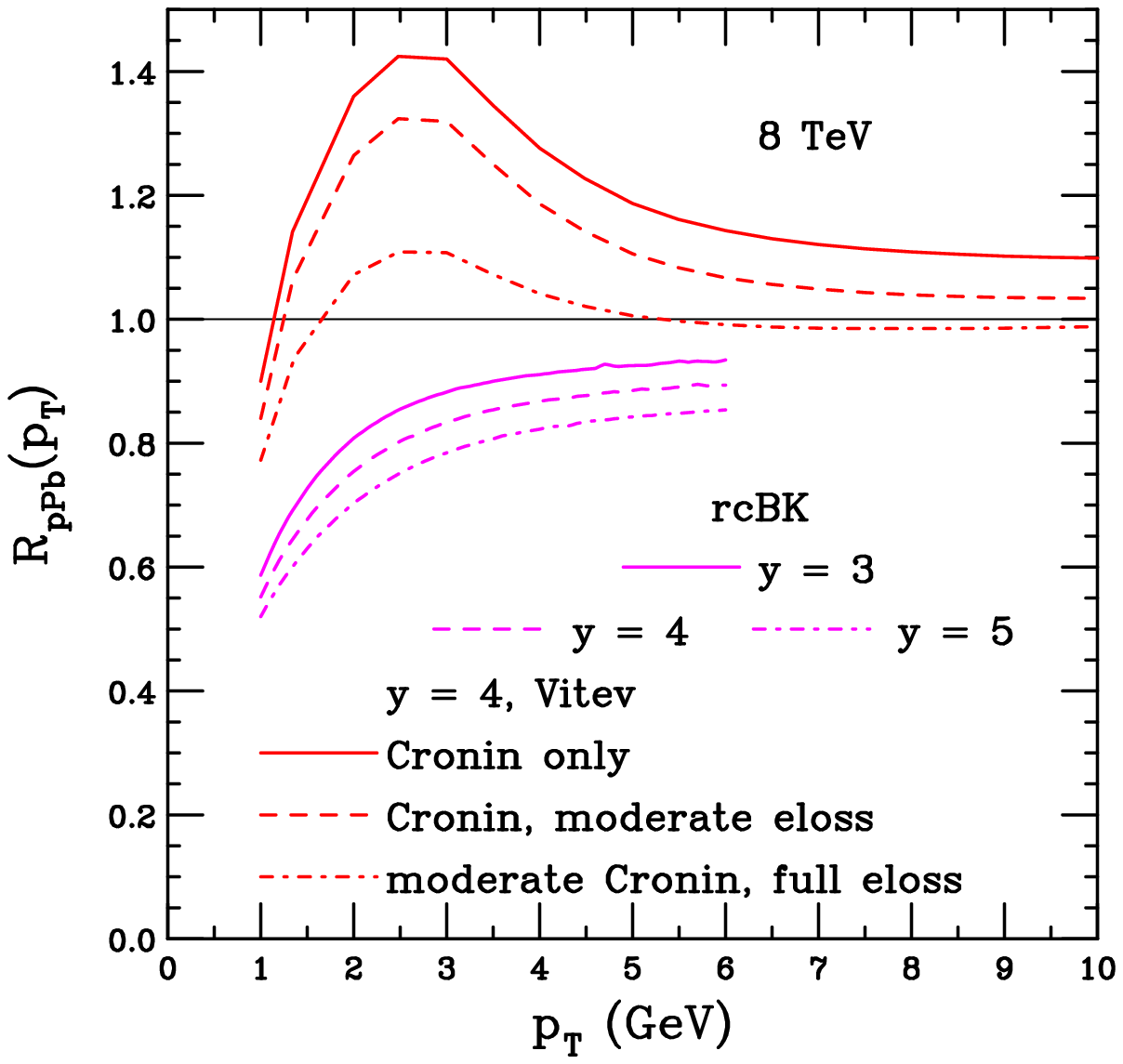} 
\caption[]{(Color online)
  Charged hadron $R_{pA}$ as a function of $p_T$ at forward rapidity in 8~TeV
  collisions.  The
  rcBK results are shown in magenta for $y = 3$ (solid), 4 (dashed) and
  5 (dot-dashed).  The calculations by Vitev {\it et al.}\
  at $y = 4$ are shown in red for
  Cronin only (solid), with moderate energy loss (dashed) and moderate Cronin
  with full energy loss (dot-dashed).
}
\label{fig:ch_rpa}
\end{figure}

The nuclear suppression factor is calculated at midrapidity using $k_T$
factorization while at forward rapidities, $y=3,4$, and 5, the hybrid
formalism is employed.  (See Ref.~\cite{Lappi:2013zma} for a more detailed
comparison of the methods). The results are presented for minimum bias
collisions only as the centrality classes from the Optical Glauber model can
not be expected to match experimental centrality classes defined using
multiplicity distributions. (See also the discussion in
Ref.~\cite{Ducloue:2016pqr}). The predictions for $y=3$, 4 and 5
are shown in
Fig.~\ref{fig:ch_rpa}.

It is emphasized that, in this calculation, there are no free nuclear
parameters except the standard Woods-Saxon distribution. Thus these results
are predictions based only on HERA DIS data.  They show strong suppression for
the rcBK calculation at $y > 0$.  The suppression  factor decreases with
increasing $y$, thus the smaller $x$ region at larger $y$ results in
greater suppression.

\begin{figure}[tb]
\centering
\includegraphics[width=0.495\textwidth]{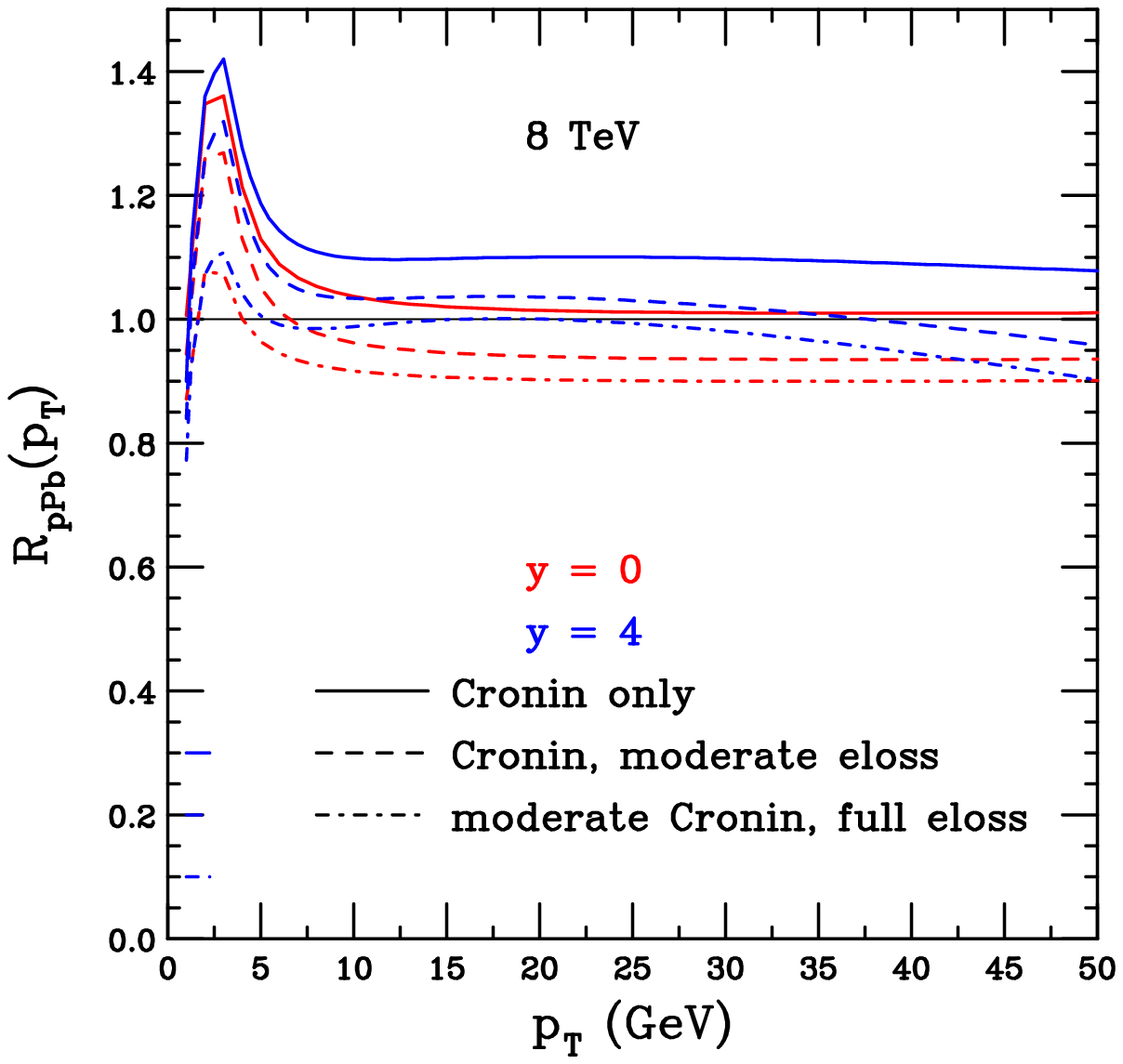} 
\caption[]{(Color online)
  Charged hadron $R_{pA}$ as a function of $p_T$ at mid ($y=0$, blue) and forward
  rapidity ($y=4$, red) in 8~TeV collisions.
  The calculations by Vitev {\it et al.}\ are shown for
  Cronin only (solid), with moderate energy loss (dashed) and moderate Cronin
  with full energy loss (dot-dashed).
}
\label{ch_rpa_Ivan}
\end{figure}

\paragraph{Perturbative QCD}
The calculations by Vitev {\it et al.}\
at $y=4$ are also shown in Fig.~\ref{fig:ch_rpa}.
The results are somewhat higher than those at $y=0$ in the $p_T$ range shown.  
The Cronin effect is slightly enhanced because, at higher rapidity, the
$p_T$ distribution is more steeply falling so that, while
$\langle k_T^2 \rangle_{pp}$ is decreasing, the multiple scattering remains the
same, leading to a somewhat larger enhancement at low $p_T$.  This effect alone
causes the Cronin-only curves at $y = 4$ to be above those at midrapidity.
However, when energy loss is included, the effect at forward rapidity is
slightly stronger than at $y=0$ because the projectile $x$ values are
larger at forward rapidity.  At intermediate $p_T$, the $y=4$ results for
$R_{p{\rm Pb}}$ are above those at $y=0$ but the effect becomes stronger at larger
$p_T$ since the $p_T$ distribution decreases faster at high $p_T$, eventually
causing the high $p_T$ results to drop below those at midrapidity.  The
higher-twist dynamical shadowing plays a negligible role at high $p_T$, even
at $y=4$, because, for massless
particles, the $t$ dependence in Eq.~(\ref{Ivan:HTshad}) results in a decrease
in the shadowing effect as $1/p_T^2$, causing it to become negligible for
$p_T > 4$~GeV.  The difference in the two calculations at $y = 0$ and $y = 4$
are shown in Fig.~\ref{ch_rpa_Ivan} where the $p_T$ dependence is extended to
$p_T \sim 50$~GeV to emphasize the difference between the results for
the two rapidities at higher $p_T$.

\subsubsection{$\mathtt{AMPT}$ Flow Coefficients}
\label{sec:flow}

The $p_T$-dependence of the anisotropy harmonics $v_n$ with $n\!=\!2,3$, and 4
shown here
follows the analysis method of the CMS collaboration \cite{Li:2014zza} where 
\begin{eqnarray}
v_n \{2,|\Delta \eta|>2\}(p_T) =v_{n \Delta}(p_T,p_T^{\rm ref})/
\sqrt {v_{n \Delta} (p_T^{\rm ref},p_T^{\rm ref}}) \, \, .
\end{eqnarray}
The coefficients $v_{n \Delta}(p_T,p_T^{\rm ref})$ are calculated
as $\langle \langle \cos(n \Delta \phi) \rangle \rangle $ 
\cite{CMS:2013bza}, where $\langle \langle ... \rangle \rangle $  
denotes averaging over different charged hadron pairs in each 
event and then averaging over those events. 
Both particles in a pair need to be
within $|\eta_{\rm lab}|<2.4$ and have a minimum separation $|\Delta \eta|$
of 2 units.  The transverse momentum of the reference particle is within 
$0.3 < p_T^{\rm ref} < 3.0$ GeV. 
$\mathtt{AMPT-SM}$ was used earlier to study these observables
in $p+$Pb collisions at 5 TeV and direct comparisons with the 5 TeV $v_2$
and $v_3$ data have shown good overall agreement \cite{Bzdak:2014dia}. 

\begin{figure}[htb]
\centerline{\includegraphics[width=0.495\textwidth]{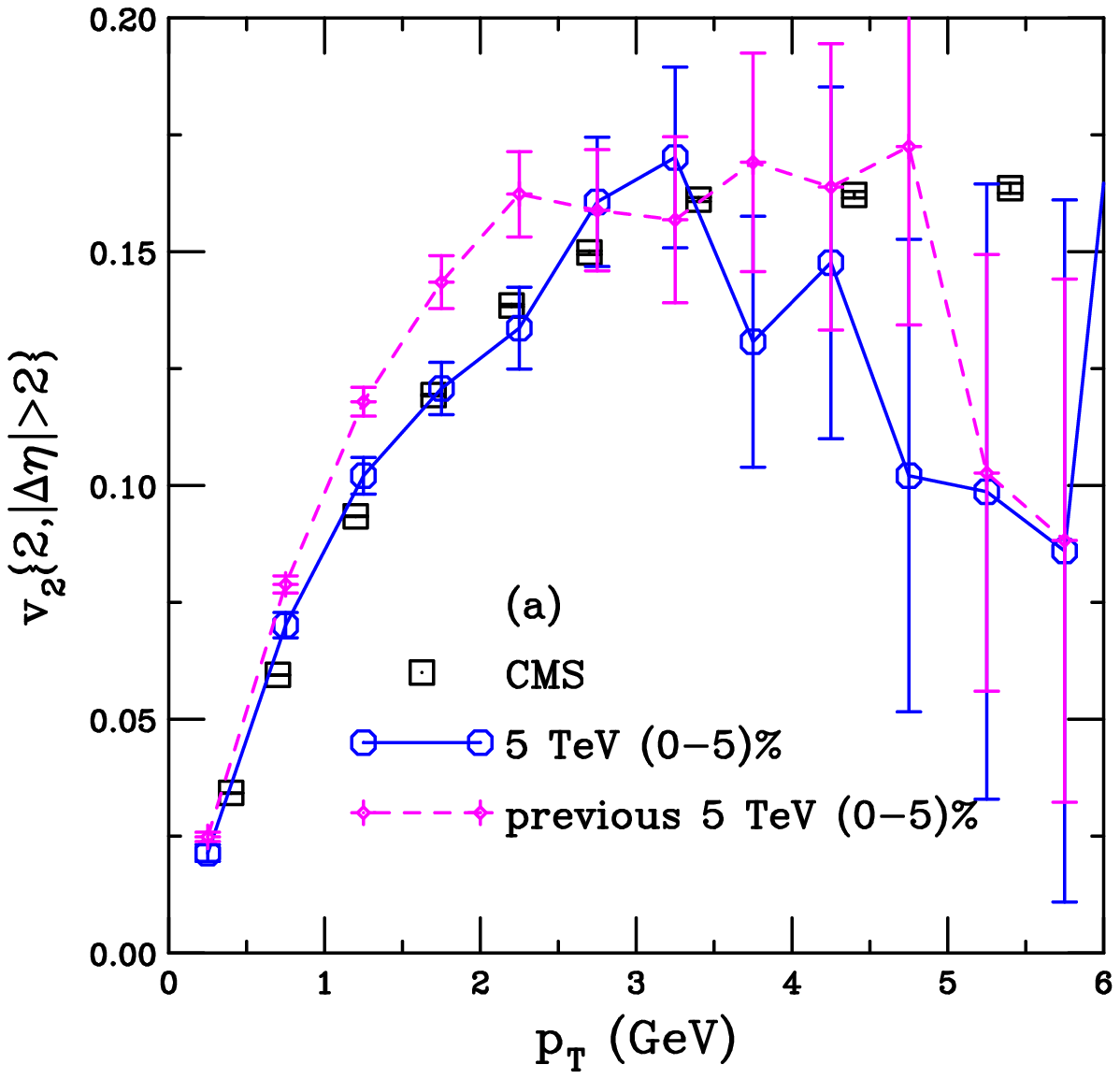}}
\centerline{\includegraphics[width=0.495\textwidth]{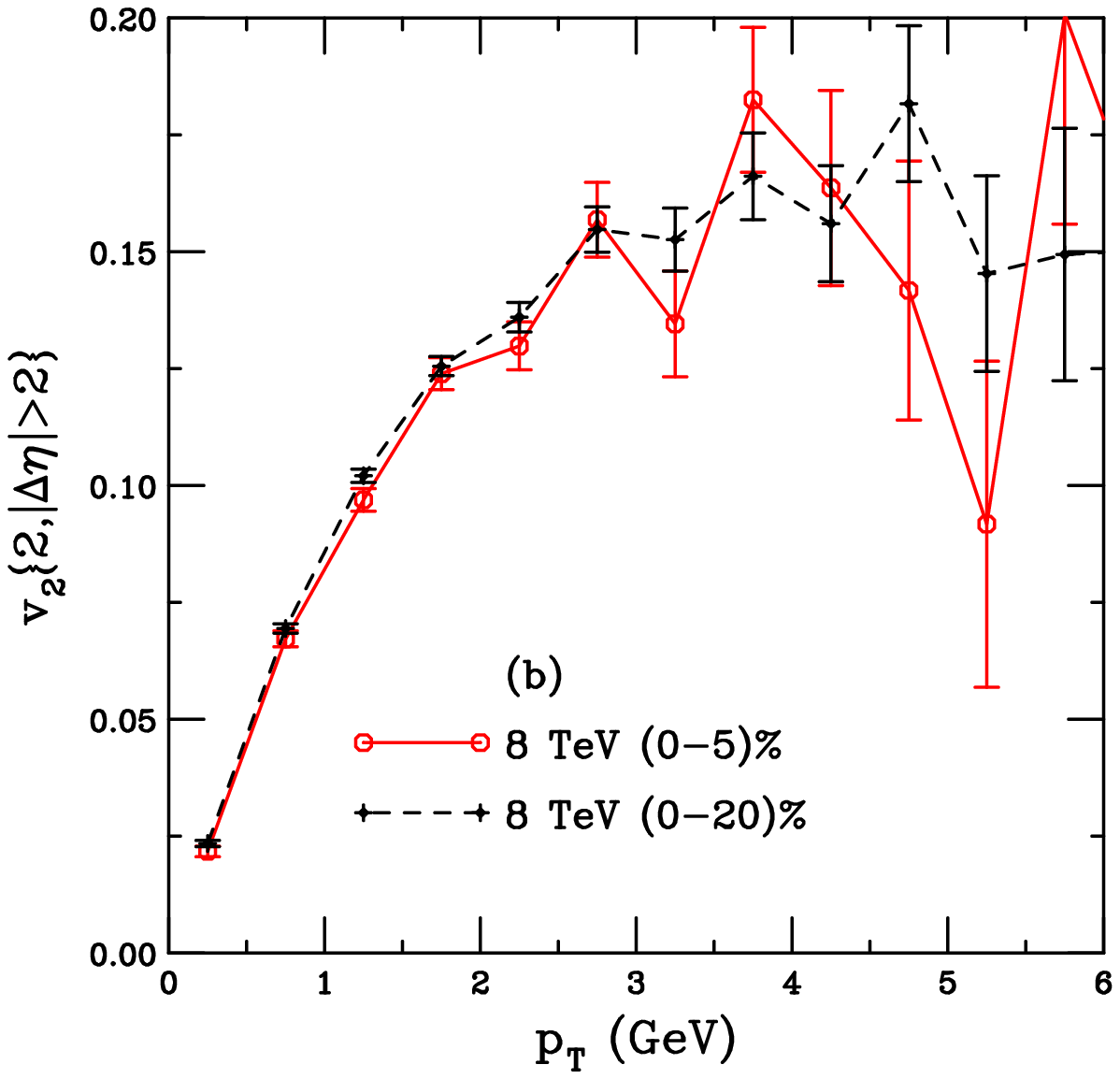}}
\caption[]{(Color online)  The $v_2\{2,|\Delta \eta|>2\}(p_T)$ 
  for $p+$Pb collisions from $\mathtt{AMPT-SM}$ at 5~TeV (a) and 8~TeV (b).
  The 5~TeV results in the (0-5)\% centrality range from $\mathtt{AMPT-SM}$
  versions 2.26t7 (solid blue) and 2.26t1 (dashed magenta) are compared to each
  other and to the CMS 5~TeV $p+$Pb data \protect\cite{Chatrchyan:2013nka}
  in (a).  In (b), the $\mathtt{AMPT-SM}$
  version 2.26t7 calculations at 8~TeV for (0-5)\% (solid red) and (0-20)\%
  (dashed black) are shown.} 
\label{figv2}
\end{figure}

\begin{figure}[htb]
\centerline{\includegraphics[width=0.495\textwidth]{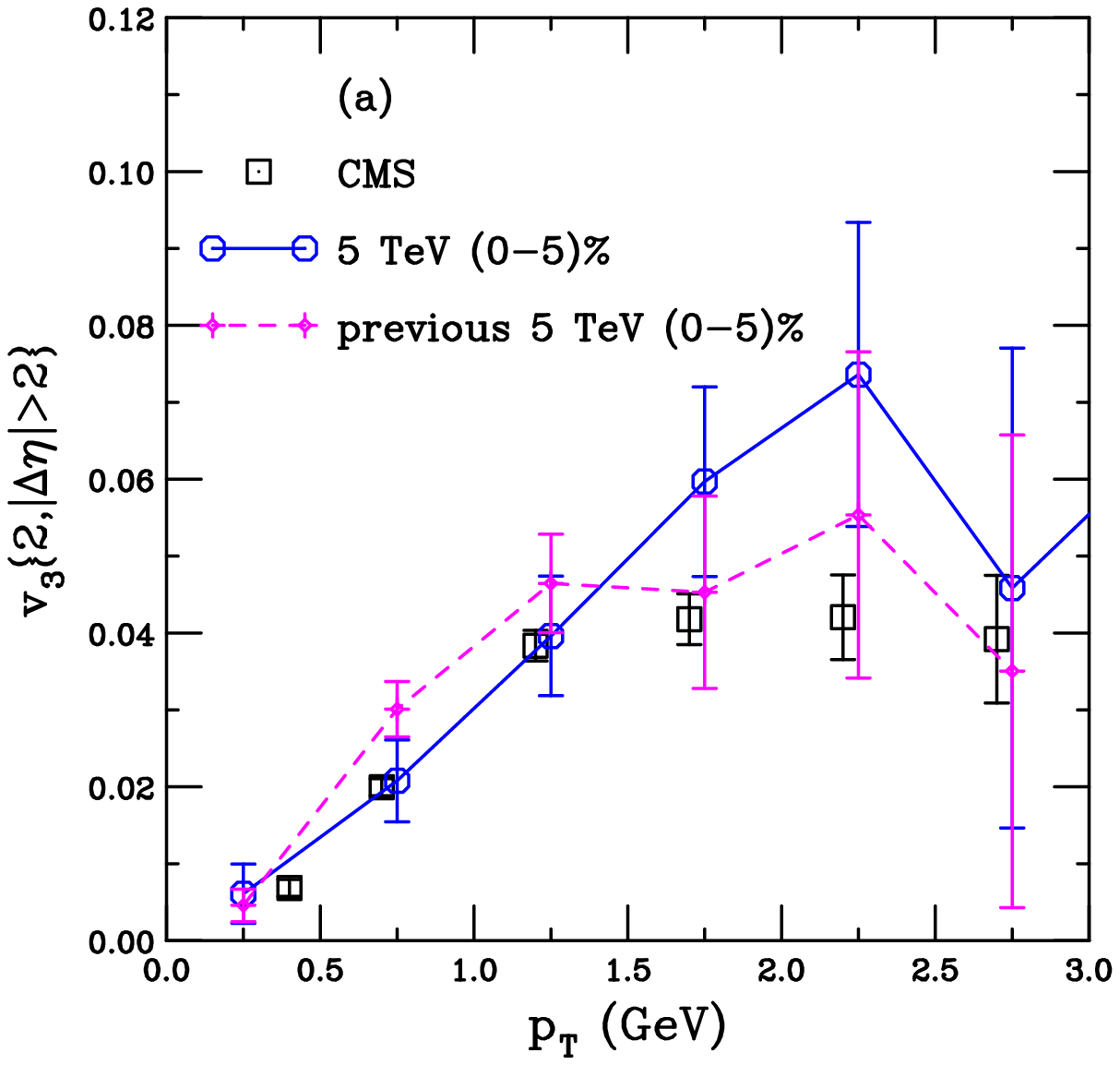}}
\centerline{\includegraphics[width=0.495\textwidth]{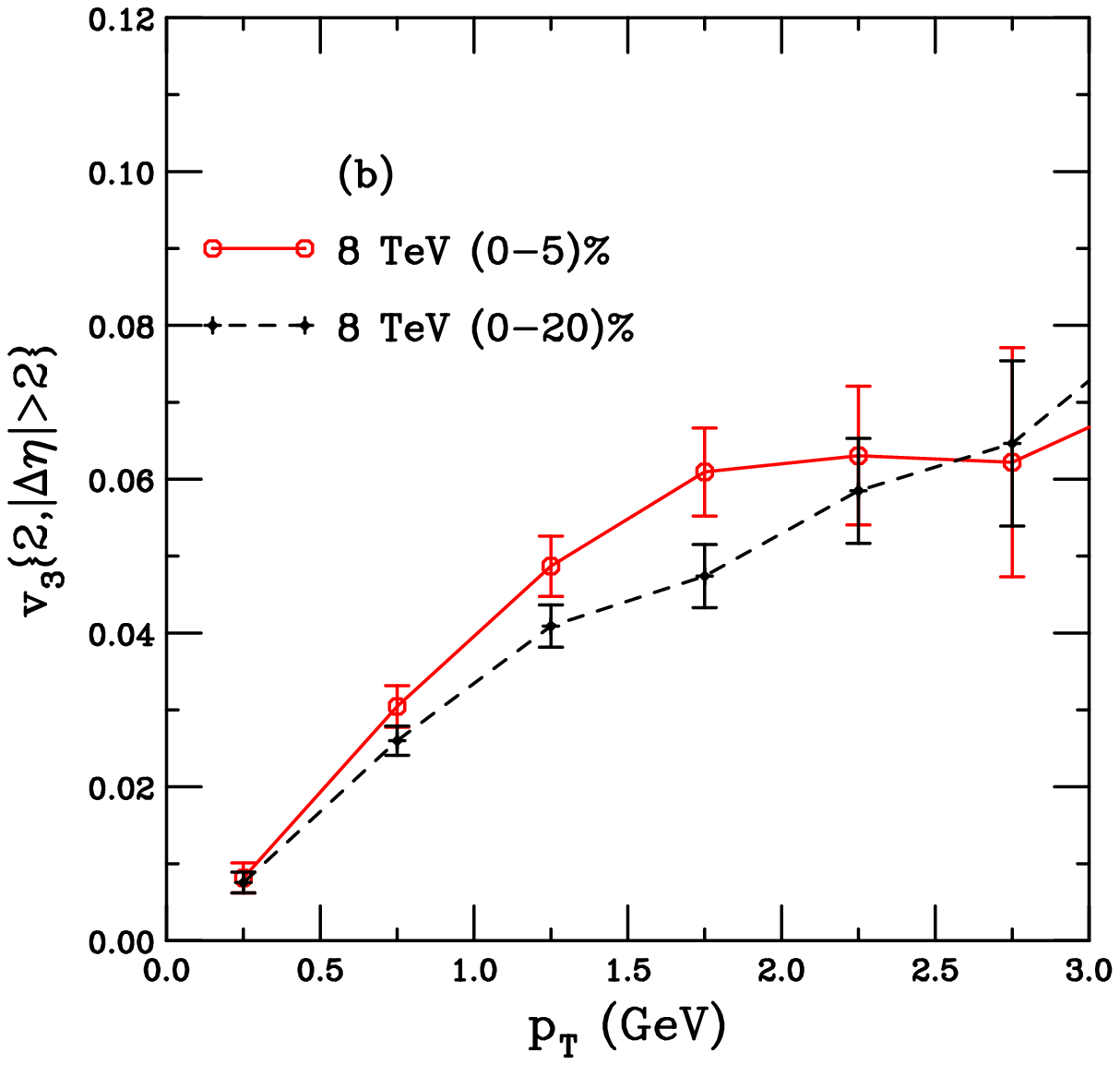}}
\caption[]{(Color online)  The $v_3\{2,|\Delta \eta|>2\}(p_T)$ 
  for $p+$Pb collisions from $\mathtt{AMPT-SM}$ at 5~TeV (a) and 8~TeV (b).
  The 5~TeV results in the (0-5)\% centrality range from $\mathtt{AMPT-SM}$
  versions 2.26t7 (solid blue) and 2.26t1 (dashed magenta) are compared to each
  other and to the CMS 5~TeV $p+$Pb data \protect\cite{Chatrchyan:2013nka}
  in (a).  In (b), the $\mathtt{AMPT-SM}$
  version 2.26t7 calculations at 8~TeV for (0-5)\% (solid red) and (0-20)\%
  (dashed black) are shown.
}  
\label{figv3}
\end{figure}

\begin{figure}[htb]
\centerline{\includegraphics[width=0.495\textwidth]{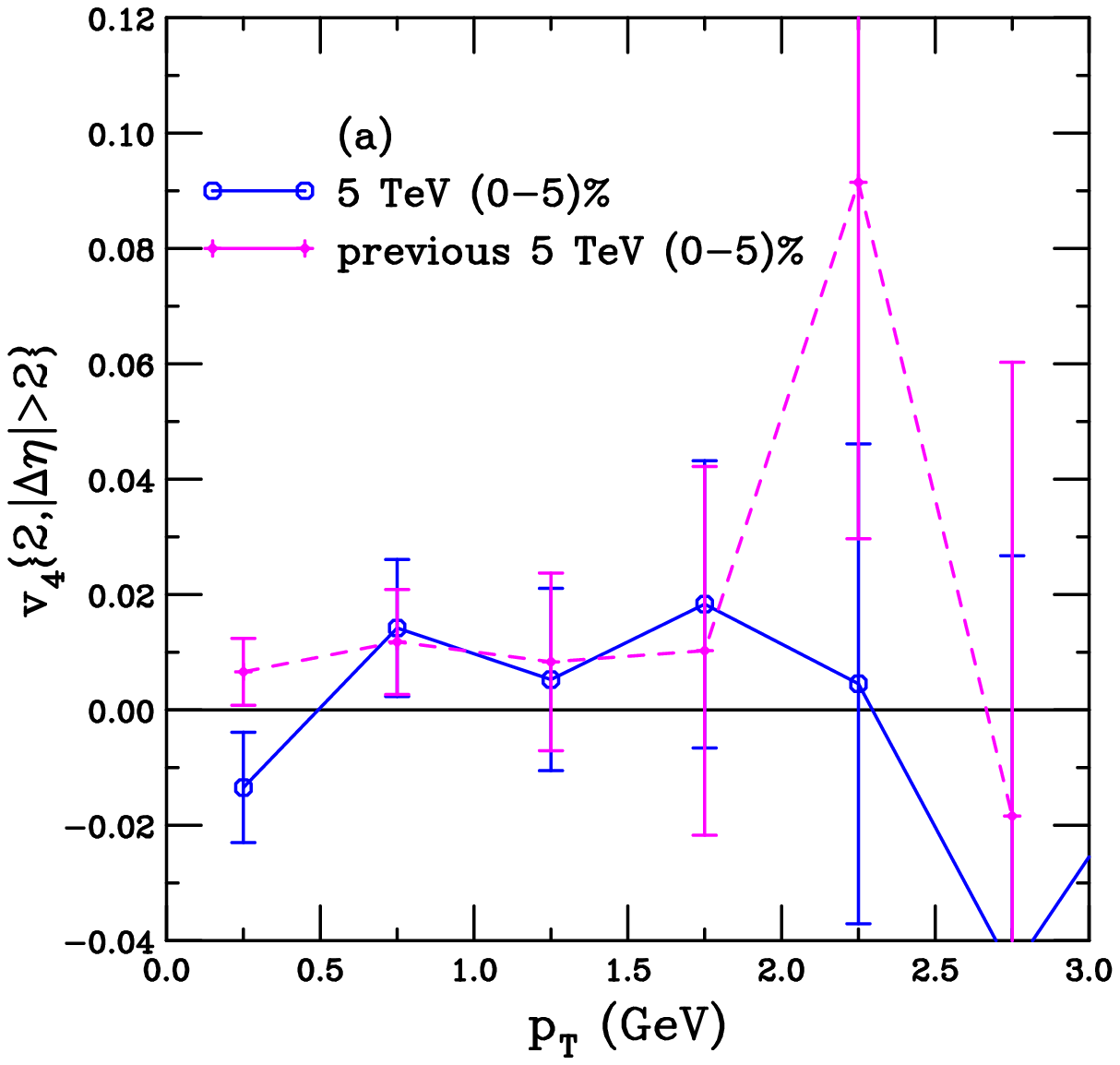}}
\centerline{\includegraphics[width=0.495\textwidth]{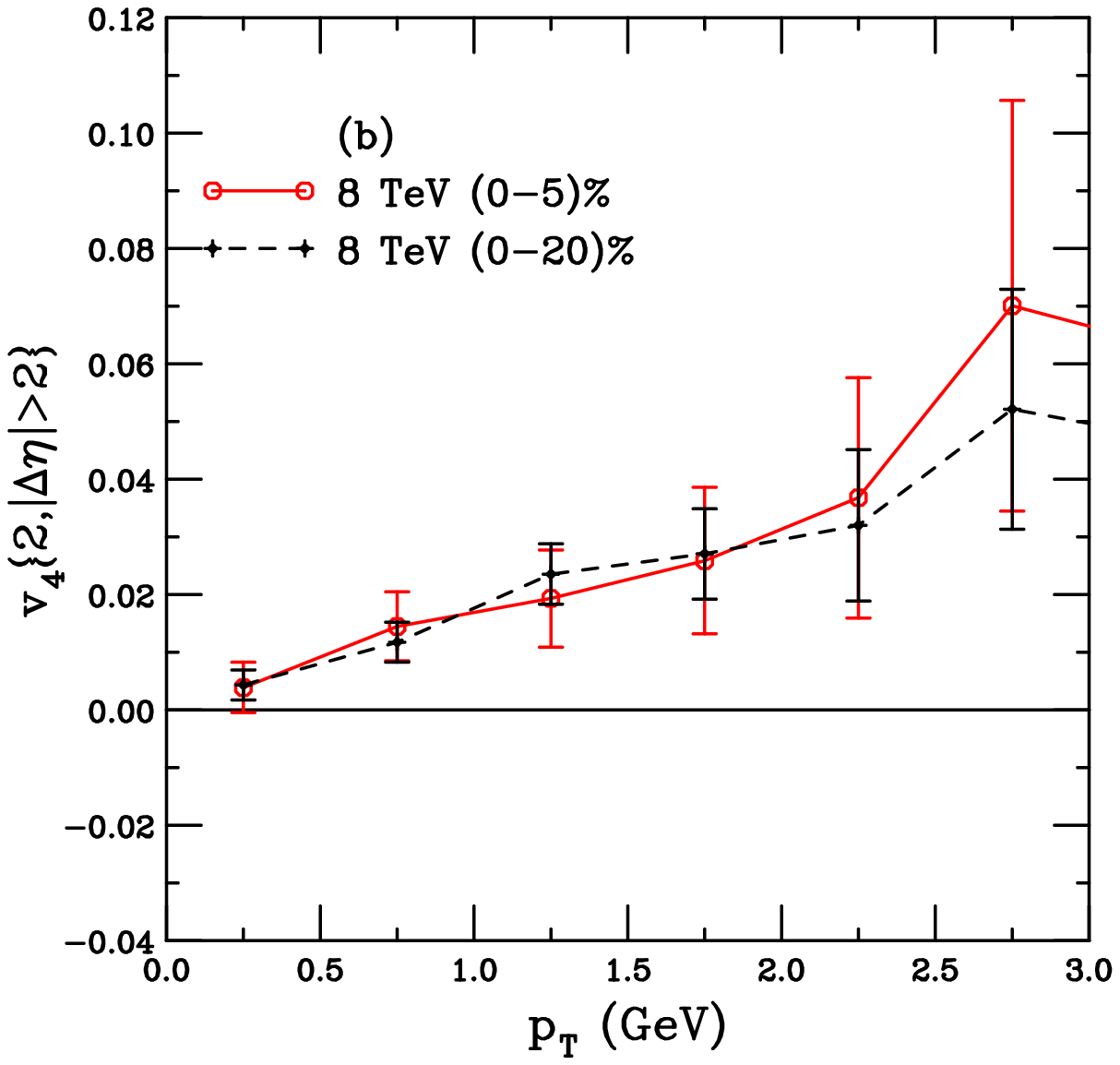}}
\caption[]{(Color online)  The $v_4\{2,|\Delta \eta|>2\}(p_T)$ 
  for $p+$Pb collisions from $\mathtt{AMPT-SM}$ at 5~TeV (a) and 8~TeV (b).
  The 5~TeV results in the (0-5)\% centrality range from $\mathtt{AMPT-SM}$
  versions 2.26t7 (solid blue) and 2.26t1 (dashed magenta) are compared
  in (a).  In (b), the $\mathtt{AMPT-SM}$
  version 2.26t7 calculations at 8~TeV for (0-5)\% (solid red) and (0-20)\%
  (dashed black) are shown.
}  
\label{figv4}
\end{figure}

Figures~\ref{figv2}-\ref{figv4} show the anisotropy harmonics
$v_n\{2,|\Delta \eta|>2\}(p_T)$ for $n=2,3,$ and 4 respectively,
calculated with the two-particle correlation method described above.
Results for the top
5\% centrality 5 TeV $p+$Pb collisions from $\mathtt{AMPT-SM}$ version 2.26t7 in
this study and from the
previous prediction compilation \cite{Albacete:2016veq} using version 2.26t1
are shown in Figs.~\ref{figv2}(a)-\ref{figv4}(a).
Results for the top 5\% centrality and top 20\% centrality 8 TeV $p+$Pb
events are given in Figs.~\ref{figv2}(b)-\ref{figv4}(b).

In Figs.~\ref{figv2}(a) and \ref{figv3}(a), the CMS 5.02~TeV $p+$Pb 
$v_2\{2,|\Delta \eta|>2\}(p_T)$ and $v_3\{2,|\Delta \eta|>2\}(p_T)$ data are
shown for the CMS centrality cut, $120 < N_{\rm trk} < 150$, given in  
Ref.~\cite{Chatrchyan:2013nka}.  The $\mathtt{AMPT}$ results shown in the
magenta curves in these figures, labeled ``previous 5~TeV (0-5)\%'',
employ the same centrality 
definition as in Refs.~\cite{Albacete:2013ei,Albacete:2016veq}, the number of
charged particles in the pseudorapidity window $|\eta_{\rm lab}|< 1$.

As noted in Ref.~\cite{Albacete:2016veq}, this centrality definition is not
identical to that of CMS \cite{Chatrchyan:2013nka}.  Instead, the CMS criteria
$120 < N_{\rm trk} < 150$ roughly corresponds to $0.5-2.5$\% centrality while
the $\mathtt{AMPT-SM}$ results shown in Figs.~\ref{figv2}(a)-\ref{figv4}(a)
are for a $0-5$\% centrality cut at 5~TeV.   Thus, the comparison to 
data here, as in Ref.~\cite{Albacete:2016veq} is inexact.  However, as was also
shown in Ref.~\cite{Albacete:2016veq},
the differences in the flow coefficients between  the (0-5)\% and (0-20)\%
centrality bins was not large.  The difference between the $\mathtt{AMPT-SM}$
centrality selection and that of the CMS data is therefore likely within the
uncertainties of the 5~TeV calculations.

\begin{table}
\caption{Centrality classes of $p+$Pb events from the string melting
  version of $\mathtt{AMPT}$, with centrality determined from the number of
  charged hadrons within $|\eta_{\rm lab}|<2.4$.
  ``All'' refers to all simulated non-diffractive events.}
\begin{tabular}{cccccccc}
\hline
Centrality & $\left <b \right >$(fm) & $b_{\rm min}$(fm) & $b_{\rm
  max}$(fm) & $N_{\rm part}^{\rm Pb}$   & $N_{\rm part-in}^{\rm Pb}$ &
$\langle N_{\rm ch}(|\eta_{\rm lab}|<2.4)\rangle$ \\
\hline \hline 
\multicolumn{7}{c}{$\sqrt{s_{_{NN}}} = 8$~TeV} \\ \hline 
All  & 5.72 & 0.0 & 13.2 & 8.64 & 6.11 & 118.4 \\
0-5\% & 3.41 & 0.0 & 8.1 & 17.55 & 13.35 & 293.1 \\
0-20\% & 3.71 & 0.0 & 9.5 & 15.63 & 11.59 & 237.7 \\ \hline \hline 
\multicolumn{7}{c}{$\sqrt{s_{_{NN}}} = 5$~TeV} \\ \hline 
All  & 5.62 & 0.0 & 13.2 & 8.01 & 5.76 & 97.2 \\
0-5\% & 3.37 & 0.0 & 7.7 & 16.26 & 12.64 & 243.2 \\
0-20\% & 3.67 & 0.0 & 9.3 & 14.42 & 10.91 & 195.4 \\
\hline
\hline
\end{tabular}\\[2pt]
\label{tablev2} 
\end{table}

In the new 5~TeV calculations and the 8~TeV calculations, shown in
Figs.~\ref{figv2}-\ref{figv4}, the centrality selection is closer to the CMS
acceptance at central rapidity, $|\eta|<2.4$.  The characteristics of the 5 and
8~TeV $p+$Pb collisions in the centrality bins shown in this section are given
in Table~\ref{tablev2}.  Note that the broader centrality definition here than
that in Ref.~\cite{Albacete:2016veq} scales the average number of charged
hadrons approximately with the pseudorapidity region while the average number
of participants, both in all collisions and in the inelastic collisions, remains
relatively constant.

Figure~\ref{figv2}(a) compares the 5 TeV results on elliptic flow,
$v_2\{2,|\Delta \eta|>2\}(p_T)$, from version 2.26t1 (labeled
previous 5 TeV) and the updated version 2.26t7 with the CMS data
\cite{Chatrchyan:2013nka}.  The new version of $\mathtt{AMPT-SM}$ gives a lower
$v_2$, in better agreement with the CMS data, albeit for a slightly different
centrality cut, as described above.  However, for $p_T > 3$~GeV, the lower
statistics of the 5~TeV calculations do not allow one to distinguish between
the results.  Figure~\ref{figv2}(b) shows that the elliptic flow
at 8 TeV for the top 5\% centrality is similar to that for the
top 20\% centrality.  The $p_T$ dependence of $v_2\{2,|\Delta\eta|>2\}$ at
8~TeV is also very similar to
the result for the top 5\% centrality at 5 TeV with $\mathtt{AMPT-SM}$
version 2.26t7.

The trends shown in Fig.~\ref{figv3}(a) at 8~TeV are similar to those in
Fig.~\ref{figv2}(a) at 5~TeV.  The updated $\mathtt{AMPT-SM}$ calculation gives
lower $v_3$ at low $p_T$, in better agreement with the CMS data.
In addition, Figs.~\ref{figv3}-\ref{figv4} show that at 8 TeV 
$v_3\{2,|\Delta \eta|>2\}(p_T)$ and $v_4\{2,|\Delta \eta|>2\}(p_T)$ 
for the top 5\% centrality are not very different from those for the top
20\% centrality.  The magnitudes of $v_3$ and $v_4$ at 8 TeV trend somewhat
higher than those at 5 TeV for the top 5\% centrality with $\mathtt{AMPT-SM}$
version 2.26t7.  However, the statistical uncertainties at the lower energy,
especially those on version 2.26t1, shown in Ref.~\cite{Albacete:2016veq},
are rather large. 

\section{Quarkonium and heavy flavor
  (F. Arleo, G. G. Barnaf\"oldi, G. B\'ir\'o, B. Duclou\'e, E. Ferreiro,
  M. Gyulassy, Sz. M. Harangoz\'o, Z.-B. Kang,
  J.-P. Lansberg, T. Lappi, P. L\'evai, G. Ma,
  Y.-Q. Ma, H. M\"antysaari, G. Papp, H.-S. Shao, I. Vitev, R. Venugopalan,
  R. Vogt, H.~Xing, X.-N. Wang, B.-W. Zhang, H.-F. Zhang)}
\label{sec:Onia_HF}

Here, calculations of prompt quarkonium and heavy flavor hadrons are presented
together with the hidden heavy flavor hadron production described first.  The
quarkonium calculations include shadowing in the color evaporation model
(Vogt), a data driven approach studying the effect of shadowing on best-fit
results from proton-proton collisions (Lansberg and Shao),
final-state energy loss (Arleo), comover suppression (Ferreiro) and saturation
effects (Duclou\'e {\it et al.} and Y.-Q.
Ma {\it et al.}).  The heavy flavor calculations include the cold matter
energy loss approach of Vitev {\it et al.}\ and the data-driven shadowing
calculations of Lansberg and Shao.  In addition preliminary
$\mathtt{HIJING++}$ calculations, based on the same model described in
Sec.~\ref{sec:HIJING++} are also shown.  However, in this case, heavy flavor
decays and the associated resonances decays were turned off.

\subsection{Quarkonium}
\label{sec:onia}

The model calculations for prompt quarkonium production
are described here, first for
those calculations based on collinear factorization, including comovers, and
then for the saturation approaches. 

\subsubsection{Collinear Factorization}
\label{sec:onia_pQCD}

\paragraph{EPS09 NLO in the Color Evaporation Model (R. Vogt)}
\label{sec:Vogt}

The predictions for the quarkonium nuclear
suppression factor, considering only shadowing
effects on the parton densities are described here.
The results are
obtained in the color evaporation model (CEM) at next-to-leading order in the
total cross section.
In the CEM, the quarkonium 
production cross section is some fraction, $F_C$, of 
all $Q \overline Q$ pairs below the $H \overline H$ threshold where $H$ is
the lowest mass heavy-flavor hadron,
\begin{eqnarray}
\sigma_C^{\rm CEM}(s)  =  F_C \sum_{i,j} 
\int_{4m^2}^{4m_H^2} ds
\int dx_1 \, dx_2~ f_i^p(x_1,\mu_F^2)~ f_j^p(x_2,\mu_F^2)~ 
\hat\sigma_{ij}(\hat{s},\mu_F^2, \mu_R^2) \, 
\, , \label{sigtil}
\end{eqnarray} 
where $ij = q \overline q$ or $gg$ and $\hat\sigma_{ij}(\hat s)$ is the
$ij\rightarrow Q\overline Q$ subprocess cross section at LO.  At NLO, $ij$
includes $q g$ and $\overline q g$ processes and an additional light parton,
$k$, is emitted in the final state, $ij \rightarrow Q \overline Q k$.
The normalization factor $F_C$ is fit 
to the forward (integrated over $x_F > 0$) 
$J/\psi$ cross section data on only $p$, Be, Li,
C, and Si targets.  In this way, uncertainties due to 
ignoring any cold nuclear matter effects,
which are on the order of a few percent
in light targets, are avoided.  The fits are restricted to the forward cross 
sections only.

The same values of the central charm quark
mass and scale parameters are employed as those 
found in the fits to the open charm total cross section,
$m = 1.27 \pm 0.09$~GeV,
$\mu_F/m = 2.10 ^{+2.55}_{-0.85}$, and $\mu_R/m = 1.60 ^{+0.11}_{-0.12}$ 
\cite{Nelson:2012bc}.  For the CEM calculation, the scales $\mu_F$ and $\mu_R$
are defined as proportional to the transverse mass instead of the quark mass.
The normalization $F_C$ is obtained for the central set,
$(m,\mu_F/m, \mu_R/m) = (1.27 \, {\rm GeV}, 2.1,1.6)$.  
The calculations of the mass and scale uncertainties are
multiplied by the same value of $F_C$ to
obtain the extent of the $J/\psi$ uncertainty band \cite{Nelson:2012bc}.
The results here are based on those of 
Ref.~\cite{Vogt:2015uba} but extended to 8~TeV.
For these calculations, instead of defining
$\mu_F$ and $\mu_R$ relative to the quark mass, as above, they are
defined relative to the transverse mass, 
$\mu_{F,R} \propto m_T = \sqrt{m^2 + p_T^2}$ where 
$p_T$ is that of the $Q \overline Q$ pair, 
$p_T^2 = 0.5(p_{T_Q}^2 + p_{T_{\overline Q}}^2)$.
 
All the 
calculations are NLO in the total cross section and assume that the intrinsic
$k_T$ broadening is the same in $p+p$ as in $p+$Pb.  See
Ref.~\cite{Vogt:2015uba} for details of the calculation.
Note that the effect of the intrinsic $k_T$ on the shape of the quarkonium
$p_T$ distribution can be expected to decrease as 
$\sqrt{s}$ increases because the average $p_T$ also increases 
with energy.  However, the value of $\langle k_T^2 \rangle$ may increase with
$\sqrt{s}$ so that effect remains important at higher energies.  
The energy dependece of $\langle k_T^2 \rangle$ is
$\langle k_T^2 \rangle = 1 + (1/n)\ln(\sqrt{s}/20)$~GeV$^2$
where $n = 12$ for $J/\psi$ and 3 for $\Upsilon$ \cite{Vogt:2015uba}.

The EPS09 band is obtained by calculating the deviations from the central value
for the 15 parameter variations on either side of the central set and adding
them in quadrature.  Only the nPDF uncertainties are shown.  In this approach,
the scale uncertainties defined by the $c \overline c$ total cross section fits,
while reduced relative to changing $\mu_F$ and $\mu_R$ by a factor of two around
a central value of $m_T$, are still larger than those due to the nPDFs.  For
more details, see Ref.~\cite{Vogt:2015uba}.

\paragraph{Data-Driven Models (J.-P. Lansberg and H.-S. Shao)}
\label{sec:Shao_JPL}

In Ref.~\cite{Lansberg:2016deg},
Lansberg and Shao proposed a novel approach to 
evaluate the impact of the nuclear modification of the gluon densities as 
encoded in the nuclear PDFs. It is particularly well-suited for 
quarkonium and open heavy flavor production in proton-nucleus collisions at 
LHC energies, whose leading contributions are to a good approximation from 
$2 \rightarrow 2$ partonic processes. It relies on a $p+p$ data-driven 
parameterization of the partonic scattering amplitude squared which allows 
one to correctly take into account the $2 \rightarrow 2$ kinematics relating 
the momentum of the observed particle and the momentum fraction of the initial 
gluons which enter the evaluation of the nPDFs.  

This method has several advantages.  It can be applied to single quarkonium
and inclusive heavy flavor production with parameters tuned to $p+p$ data on
individual meson production assuming $2 \rightarrow 2$ scattering.  A 
data-driven approach results in a smaller uncertainty on the $p+p$ cross
sections than those from calculations of
theoretical uncertainties since the available $p+p$ data provide 
stringent constraints on the model parameters.  Since the calculation depends
only on a simple, common, parameterization of the amplitude for both open heavy
flavor and quarkonium, it is very efficient.

The functional form of the amplitude in this model is
\begin{eqnarray}
\overline{|\mathcal{A}(k_1 k_2 \rightarrow \mathcal{H} + k_3)|^2} & = &  
\frac{\lambda^2 \, \kappa \, x_1 x_2 s}{M_{\mathcal{H}}^2}
\exp \left[ -\kappa \, {\rm min}(p_T^2, \langle p_T \rangle^2)/M_{\mathcal{H}}^2)
  \right] \\ \label{eq:lansberg_amplitude} & \times &
\left( 1 + \theta(p_T^2 - \langle p_T^2 \rangle^2) \frac{\kappa}{n}
\frac{p_T^2 - \langle p_T \rangle^2}{M_{\mathcal{H}}^2} \right)^{-n} \, \, ,
\nonumber
\end{eqnarray}
where $k_1$ and $k_2$ are the incident partons involved in the hard scattering
to produce final-state particle $\mathcal{H}$ with mass $M_{\mathcal{H}}$
and final-state parton $k_3$
and $x_1$ and $x_2$ are the momentum fractions carried by $k_1$ and $k_2$.
The $\theta$ function ensures that
the second term in the last factor is incorporated
only when $p_T^2 > \langle p_T \rangle^2$.  The amplitude does not include
any dependence on spin or color.  The amplitude
includes four parameters: $\lambda$, $\kappa$, $\langle p_T \rangle$, and $n$.
They are determined from the $p+p$ data after convolution with the proton PDFs,
\begin{eqnarray}
\frac{d\sigma(p + p \rightarrow \mathcal{H} + X)}{d\Phi_2} = \frac{1}{2s}
\int dx_1 dx_2 x_1 f_p(x_1) x_2 f_p(x_2) \overline{|\mathcal{A}(k_1 k_2 
\rightarrow \mathcal{H} + k_3)|^2} \, \, .
\label{eq:lansberg_sigma}
\end{eqnarray} 
The phase space for the $2 \rightarrow 2$ scattering is denoted by $\Phi_2$
and the proton PDFs are denoted by $f_p$.  The factorization scale dependence
of the PDFs is suppressed in Eq.~(\ref{eq:lansberg_sigma}) but the
factorization scale is assumed to
be equal to the transverse mass of the produced particle, 
$m_T = \sqrt{M_{\mathcal{H}}^2 + p_T^2}$.  For the energies considered in 
Ref.~\cite{Lansberg:2016deg},
only the $gg$ contribution to the partonic cross
section is included since this contribution dominates production in the 
kinematic acceptance of the LHC detectors.

By construction, after the parameters are fit to accurately reproduce the
$p+p$ data, the formalism described above can provides reliable $p+A$ cross 
sections after including the nPDF correction factor 
$R_i(x_2) = f_{i,A}(x_2)/Af_{i,p}(x_2)$ in Eq.~(\ref{eq:lansberg_sigma}).
The calculations can then be directly compared to experimental data, either
as individual distributions or as the nuclear modification factor $R_{pA}$ and 
forward-backward asymmetry $R_{\rm FB}$. 

Since the hard 
scattering is parameterized, there is no dependence on either mass or 
renormalization scale: only the factorization scale at which the nPDF is 
evaluated needs to be fixed, introducing an additional uncertainty on top of 
the nPDF uncertainty.  The results can be calculated for any nPDF set included 
in the $\mathtt{LHAPDF5}$ \cite{LHAPDF5} and $\mathtt{LHAPDF6}$ \cite{LHAPDF6}
libraries employing the corresponding version of 
$\mathtt{HELAC-Onia}$~\cite{Shao:2015vga}.    As shown in
Ref.~\cite{Lansberg:2016deg}, 
the nPDF uncertainty is larger than the factorization scale uncertainty found
by varying the central $m_T$ scale by a factor of two, $m_T/2$ and $2m_T$.

Currently predictions are given for the nCTEQ15 \cite{Kovarik:2015cma}
and EPS09 LO and NLO nPDFs sets.  The CT14 NLO proton PDFs 
\cite{CT14} are used with the
nCTEQ15 nPDFs while CT10 NLO proton PDFs \cite{Lai:2010vv} 
are used with the EPS09 LO and NLO nPDFs since the code does not load two
PDF libraries at once \cite{Lansberg:2016deg}.  The gluon distributions in
CT10 NLO and CT14 NLO are quite similar so the resulting difference is not
large. In any case, even though the CTEQ6L1 and CTEQ6M PDFs should be used with
EPS09 LO and EPS09 NLO respectively for consistency, 
Ref.~\cite{Lansberg:2016deg}
notes that the proton PDF is less critical since
EPS09 provides a ratio while nCTEQ15 provides distributions.
Only minimum-bias collisions are considered
since the code has not yet been coupled to a Glauber Monte Carlo. The 
uncertainties are evaluated using the different eigensets provided by the nPDF 
sets.

The value of $n$ in Eq.~(\ref{eq:lansberg_amplitude}) was fixed to 2 for
all the calculations in Ref.~\cite{Lansberg:2016deg}.  The $J/\psi$ average
$p_T$ was fixed to 4.5 GeV for both mid and forward rapidity while the
value of $\langle p_T \rangle$ was fixed to 13.5 GeV for all the $\Upsilon$
data.  Making rapidity-dependent fits is appropriate for the $J/\psi$ because
of the different $p_T$ acceptance at midrapidity ($p_T \geq 6-8$ GeV) and
forward rapidity ($p_T > 0$) for some of the LHC experiments because the large
magnetic fields do not allow detection of leptons from low $p_T$ $J/\psi$
decays.  No acceptance-based fit is required for 
$\Upsilon$ because the larger mass allows all $\Upsilon$ with $p_T > 0$
to be detected, even at midrapidity since the lepton momenta from $\Upsilon$
decays at $p_T = 0$ are above detection threshold.
Thus, for quarkonium, only $\lambda$ and $\kappa$ were fit to data.
Separate fits were made in all cases for CT14 NLO and CT10 NLO.  
While the data used in the fits were typically from Run I data at 7 TeV, 
the parameter values were unchanged for the 8.16 TeV $p+$Pb run.  Note that the
calculations were done for 8 TeV.

\paragraph{Energy Loss (F. Arleo)}
\label{sec:Arleo}

In the coherent energy loss model~\cite{Arleo:2012hn,Arleo:2012rs}, 
the differential $p+A$ production cross section as a function of the 
quarkonium (labeled $\psi$) energy is
\begin{eqnarray}
\label{eq:xspA0-energy}
\frac{1}{A}\frac{d\sigma_{pA}^{\psi}}{d E} \left( E \right)  = 
\int_0^{\varepsilon^{\rm max}} d \varepsilon \,{\cal P}(\varepsilon, E, \ell_{_A}^2) 
\, 
\frac{d\sigma_{pp}^{\psi}}{d E} \left( E+\varepsilon \right) \, ,
\end{eqnarray}
where $E$ ($\varepsilon$) is the energy (energy loss) of the $Q \overline{Q}$ 
pair in the rest frame of nucleus $A$. The upper limit on the energy loss is 
$\varepsilon^{\rm max}=\min(E,E_p-E)$ 
where $E_p$ is the beam energy in that frame.
The energy loss probability distribution, or quenching weight, 
${\cal P}$, is related to the medium-induced, coherent radiation spectrum given 
in Refs.~\cite{Arleo:2012rs,Arleo:2010rb}.  
This result proved to be an excellent 
approximation of the spectrum computed to all orders in the opacity 
expansion~\cite{Peigne:2014uha}.  It depends on the accumulated transverse 
momentum transfer $\ell_{_{\rm A}} = \sqrt{\hat{q} L}$ due to soft rescatterings 
in the nucleus where $L$ is the medium path length obtained from a Glauber 
calculation using realistic nuclear densities and $\hat{q}$ is the transport 
coefficient in cold nuclear matter. The transport coefficient at momentum
fraction $x_2$ is~\cite{Arleo:2012rs}
\begin{eqnarray}
\label{qhat-model}
\hat{q}(x_2) \equiv \hat{q}_0 \left[ \frac{10^{-2}}{x_2} \right]^{0.3}\ ; \ \ \  
x_2 \equiv  \frac{m_T}{\sqrt{s}} \, e^{-y} \, ,
\end{eqnarray}
at small values of $x_2$, $x_2 < 0.01$, 
where $x_2$ is defined in $2 \rightarrow 1$
kinematics.  Here $y$ is the quarkonium rapidity in 
the center-of-mass frame of an elementary proton-nucleon collision, $m_T$ is 
the transverse mass and $\hat{q}_0$ is the only free parameter of the model. 
It is determined by fitting the $J/\psi$ suppression measured by 
the E866 Collaboration~\cite{Leitch:1999ea} in $p+$W relative to $p+$Be 
collisions at $\sqrt{s_{_{NN}}}=38.7$~GeV, see Ref.~\cite{Arleo:2012rs}. 
The fitted value is $\hat{q}_0 =0.075^{+0.015}_{-0.005}$~GeV$^2$/fm.
The $p+p$ production cross section appearing in \eq{eq:xspA0-energy}
is given by the simple parameterization 
\begin{eqnarray}
\frac{d\sigma_{pp}^{\psi}}{ d y}  \propto \left(1- \frac{2 m_T}{\sqrt{s}} 
\cosh{y} \right)^{n(\sqrt{s})}\  \, \, , 
\end{eqnarray}
where the exponent $n$ is obtained from a fit to $p+p$ measurements at
different center-of-mass energies.

\paragraph{$J/\psi$ model comparisons, collinear factorization}
\label{sec:psi}

\begin{figure}[htpb]
\begin{center}
\includegraphics[width=0.495\textwidth]{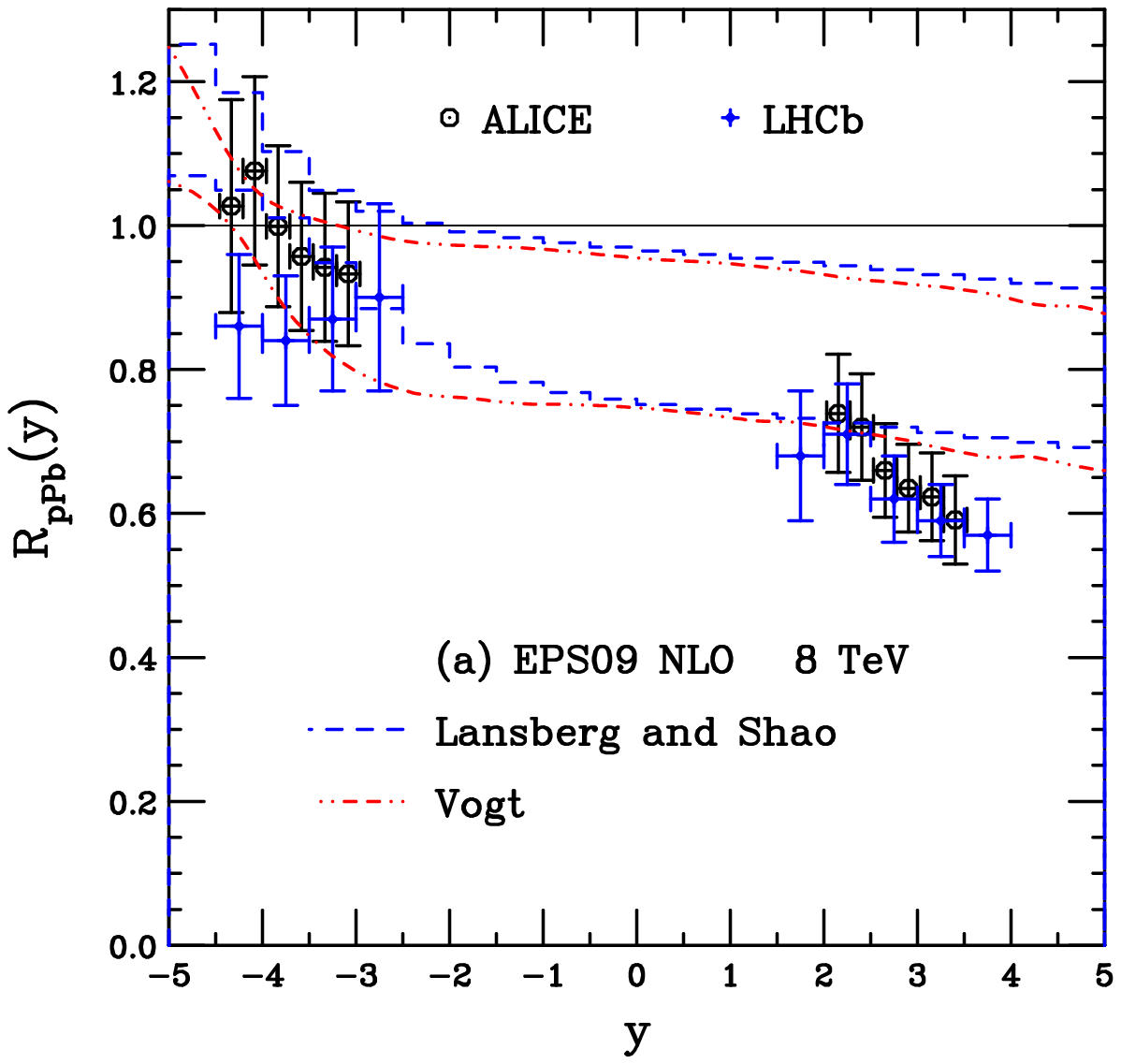} 
\includegraphics[width=0.495\textwidth]{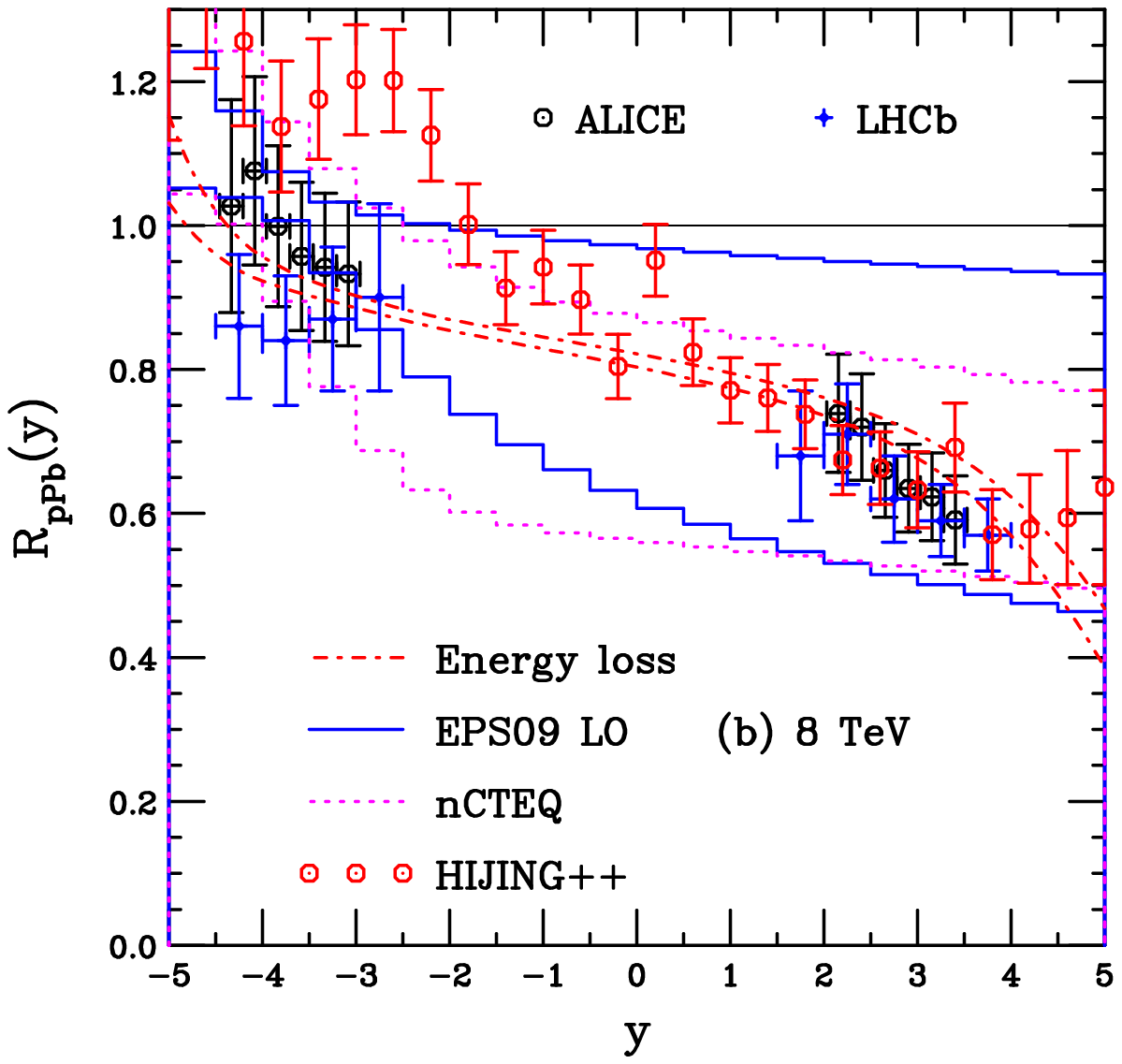} 
\end{center}
\caption[]{(Color online)
The ratio $R_{\rm p{\rm Pb}}$ for $J/\psi$ as a function of rapidity at 8~TeV.
(a) The EPS09 NLO result is compared between the NLO CEM calculation
of Vogt (dot-dot-dash-dashed red curve) and the data-driven result of 
Lansberg and Shao (solid cyan). (b) The data-driven calculation of Lansberg
and Shao for EPS09 LO (solid blue histogram) and nCTEQ (dotted magenta
histogram) is compared to the energy loss only calculation of Arleo (dot-dashed
red curve).  The $\mathtt{HIJING++}$ calculations are shown by the red points.
The ALICE data \protect\cite{Enrico} are shown in black while
the LHCb data \protect\cite{Aaij:2017cqq} are shown in blue.}
\label{fig:Jpsi_y}
\end{figure}

\begin{figure}[htpb]
\begin{center}
\includegraphics[width=0.495\textwidth]{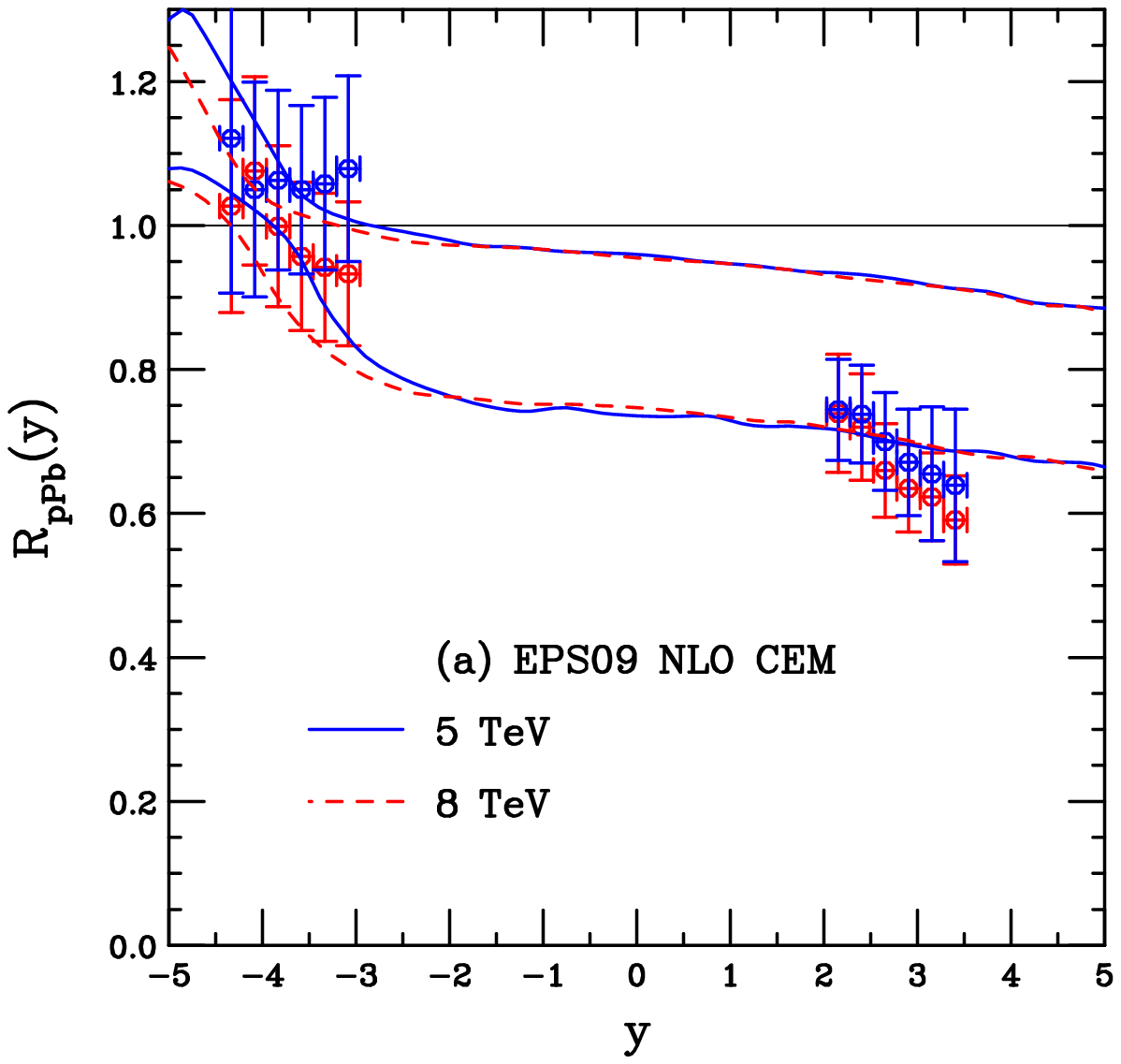} 
\includegraphics[width=0.495\textwidth]{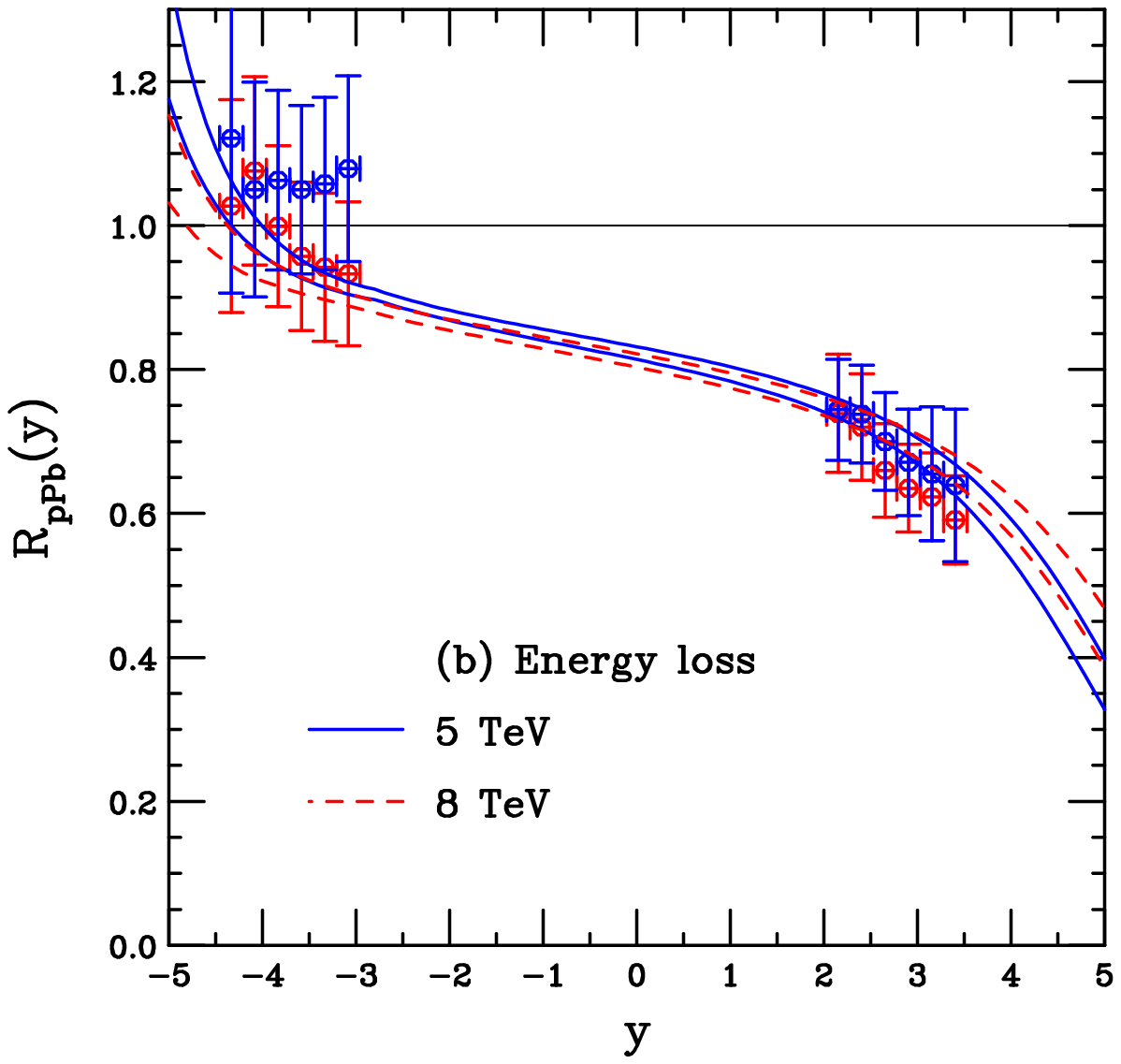} 
\end{center}
\caption[]{(Color online)
  The ratio $R_{\rm p{\rm Pb}}$ for $J/\psi$ as a function of rapidity
  at 5 TeV (blue points and solid blue curves) and 8 TeV
  (red points and dashed red curves) are compared.  The data from ALICE at
  5 TeV \protect\cite{ALICE_psiy5} and 8 TeV \protect\cite{Enrico} are also
  shown.
  (a) The EPS09 NLO CEM result. (b) The energy loss calculation by Arleo.
  }
\label{fig:Jpsi_y_8vs5}
\end{figure}

\begin{figure}[htpb]
\begin{center}
\includegraphics[width=0.495\textwidth]{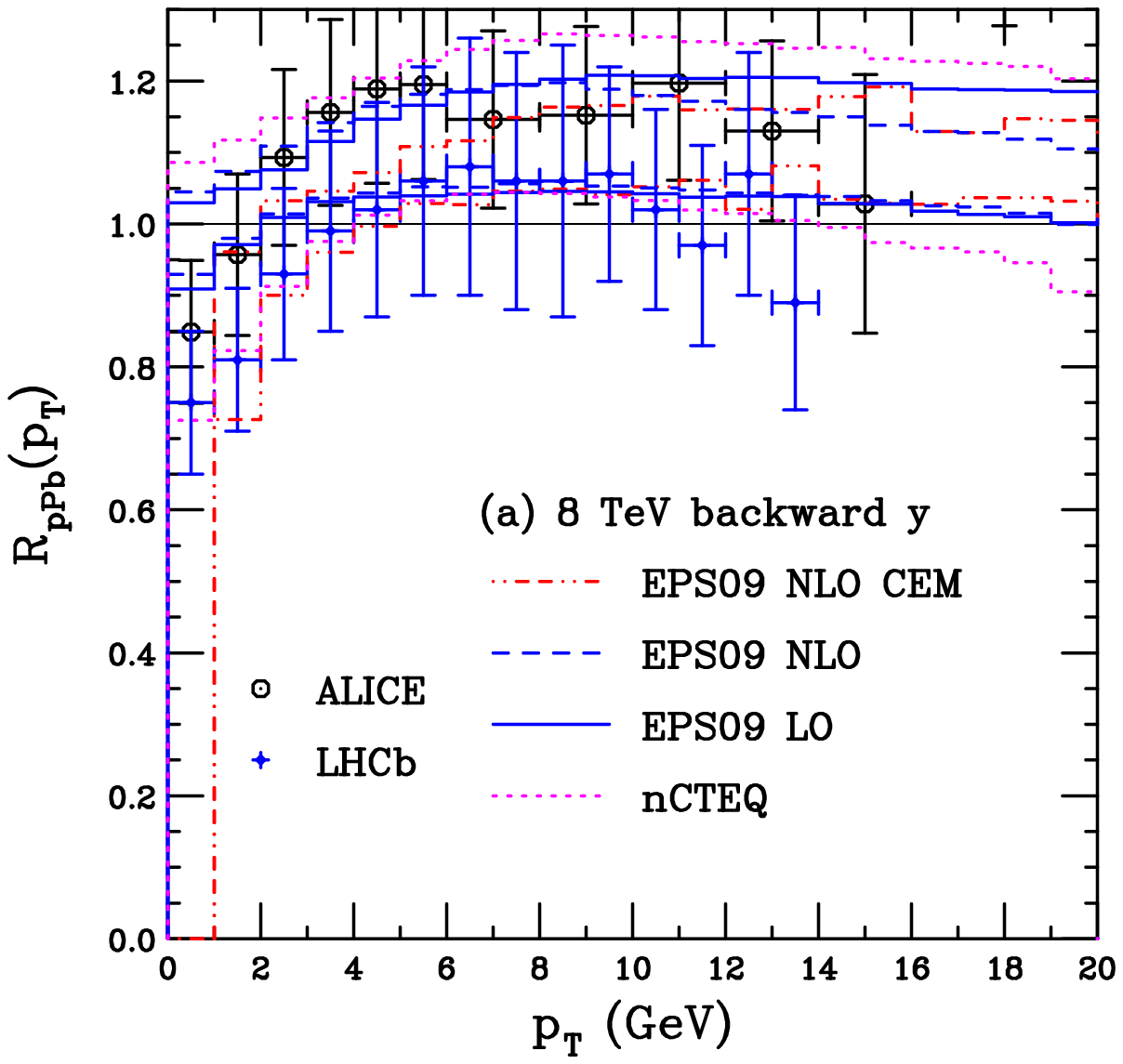} 
\includegraphics[width=0.495\textwidth]{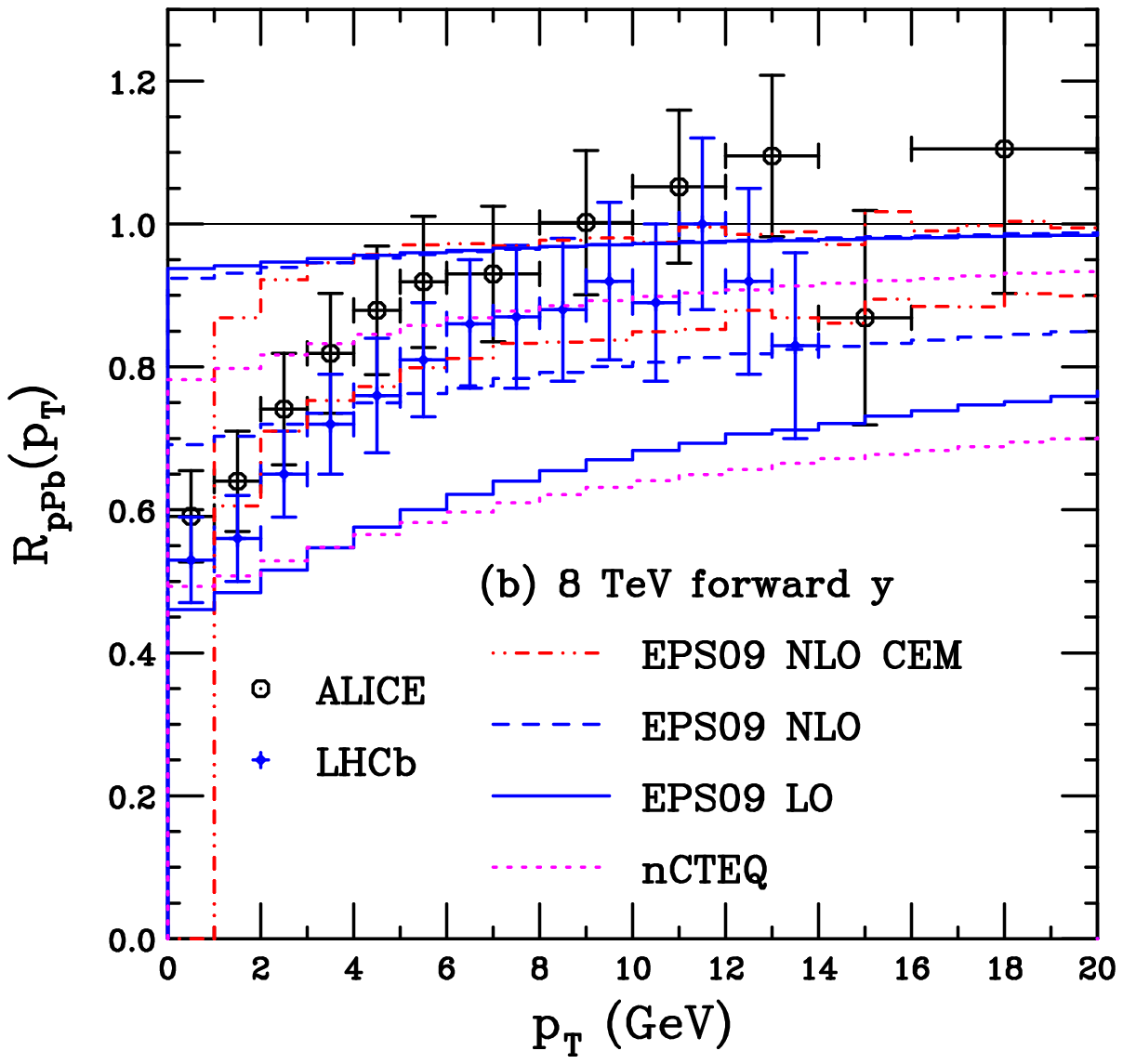} 
\includegraphics[width=0.495\textwidth]{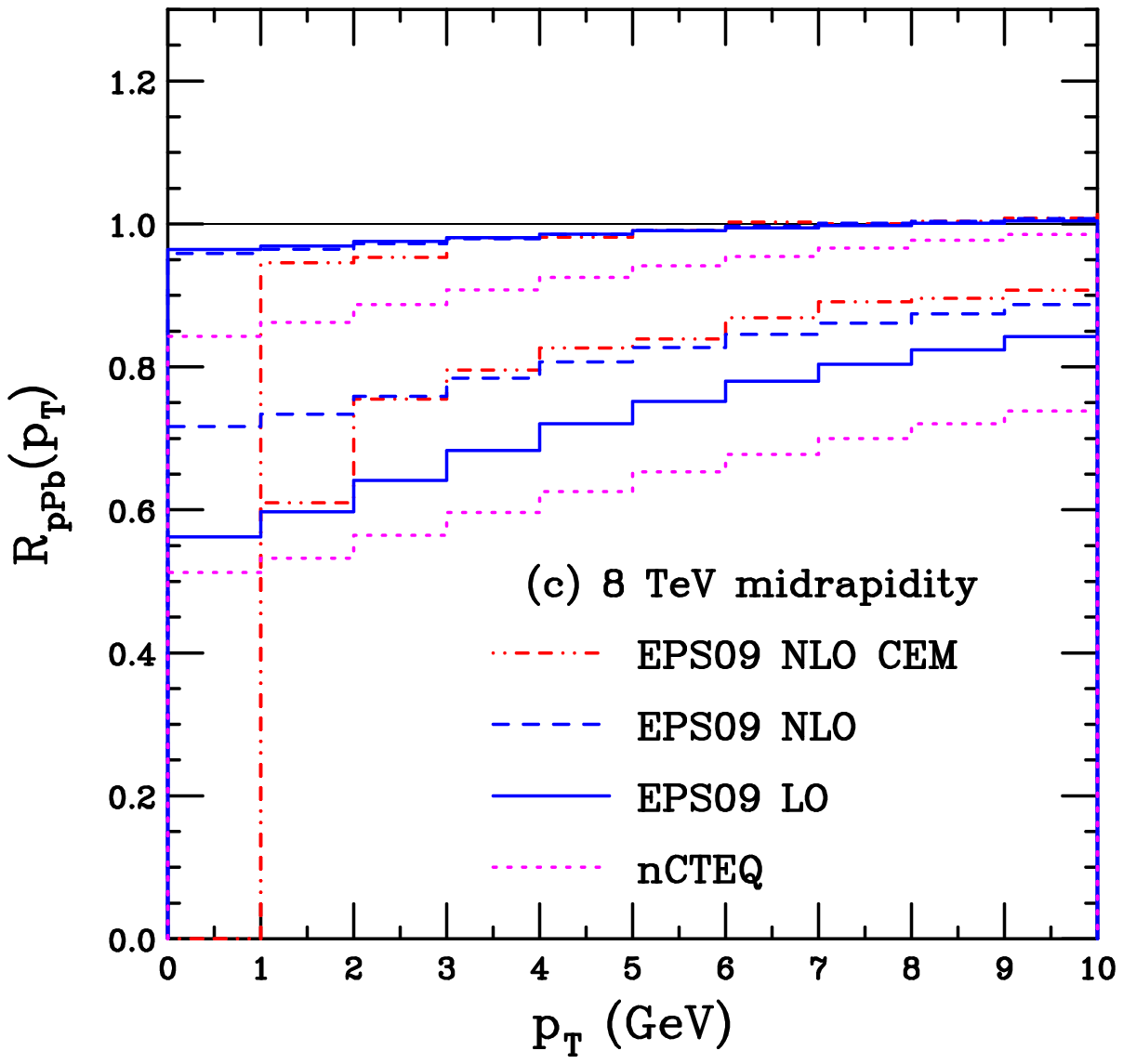} 
\end{center}
\caption[]{(Color online)
  The ratio $R_{\rm p{\rm Pb}}$ for $J/\psi$ as a function of transverse momentum
  at 8~TeV.
The EPS09 NLO results of Vogt (dot-dot-dash-dashed red curve) and 
Lansberg and Shao (solid cyan) are shown with their results for EPS09 LO 
(solid blue histogram) and nCTEQ (dotted magenta histogram) at
backward (a), forward (b), and mid (c) rapidity.
The ALICE data \protect\cite{Enrico} at
backward and forward rapidity are shown in black in (a) and (b) while the LHCb
data \protect\cite{Aaij:2017cqq} are shown in blue.}
\label{fig:Jpsi_pT}
\end{figure}

The predictions for $J/\psi$ suppression in $p+$Pb collisions 
at $\sqrt{s_{_{NN}}}=8$~TeV in the approaches discussed in this section
are shown in Figs.~\ref{fig:Jpsi_y}-\ref{fig:Jpsi_pT}.  The LHCb
\cite{Aaij:2017cqq} and preliminary
ALICE \cite{Enrico} $J/\psi$ data are also
included.  The $\mathtt{HIJING++}$ prediction for $R_{p{\rm Pb}}(y)$ is included
in Fig.~\ref{fig:Jpsi_y}(b).

The values Lansberg and Shao
obtained for $\lambda$ and $\kappa$ for the $J/\psi$ were very similar for the
two proton PDFs.  A clear dependence on rapidity range ($p_T$ acceptance) is
noticeable in the fit parameters with the values of both parameters being
larger at midrapidity for high $p_T$: $\lambda \sim 0.3$ and $\kappa \sim 0.54$
at forward rapidity and $\lambda \sim 0.38$, $\kappa \sim 0.75$ at
midrapidity.

In Fig.~\ref{fig:Jpsi_y}(a), the CEM and data-driven
calculations employing EPS09 NLO are compared.  The data-driven calculations
employ the same parameters, $\langle p_T \rangle$, $\lambda$ and $\kappa$ for
$p+p$ and $p+$Pb.  The 8 TeV energy was run in $p+p$ collisions so no energy
extrapolation is required.  The ratios here and elsewhere do not depend on
$\lambda$ since the normalization is not changed.  The calculations in
Figs.~\ref{fig:Jpsi_y} and \ref{fig:Jpsi_pT} used the parameters for LHCb
(forward rapidity) since these are for low $p_T$ and forward $y$.  These values
were chosen since the $J/\psi$ measurement at ALICE can go to $p_T > 0$ due
to the lower magnetic field.

There is a slight backward shift for the CEM calculation relative to the
data-driven calculation with the same nPDF.  There are several possible reasons
for this.  The EPS09 NLO calculation in the CEM is done for the renormalization
and factorization scales proportional to
$m_T = (0.5(p_{T_Q}^2 + p_{T_{\overline Q}}^2) + m_Q^2)^{1/2}$ with
$m_Q < M_{\mathcal{H}}$.  However, the central value of the factorization scale is
$\mu_F = 2.1 m_T$, larger than the scale used in the data-driven calculation.
The CEM calculation is also dominated by the $2 \rightarrow 3$ contributions
to the $Q \overline Q$ cross section, in particular
$gg \rightarrow Q \overline Q g$, and is thus NLO in the total cross section.
At rapidities larger than $-1$, the two model results with EPS09 NLO are quite
similar.  The EPS09 NLO sets essentially plateau with rapidity for $y > -1$.

While the collinear factorization-based shadowing
calculations at backward rapidity are fully compatible with the ALICE
and LHCb
data, they tend to somewhat underestimate the amount of suppression at forward
rapidity.  This is due to the aforementioned plateau of the calculations at
forward rapidity.  It is noteworthy that the behavior of the gluon suppression
due to shadowing at low $x$ with EPPS16 \cite{Eskola:2016oht} has a shape
similar to that of EPS09 NLO.  However, the number of fit parameters has
increased from 15 in EPS09 NLO to 20 in EPPS16 which mainly influence the width
of the low $x$ shadowing band, especially for the gluons.  Therefore, one might
expect that employing this new set, with its associated uncertainties, would
increase the relative suppression at low $x$ and thus also encompass the
ALICE and LHCb data.
Thus it is important to seek constraints on the gluon density in
the nucleon at low $x$, $x < 0.01$ at least but, in practice, measurements at
even lower $x$ would be preferable to see if the effect saturates at low $x$
or not.

In Fig.~\ref{fig:Jpsi_y_8vs5}, the results that were available for comparison
to the data from the 5 TeV run \cite{Albacete:2016veq},
the EPS09 NLO CEM calculation by Vogt and the
energy loss calculation by Arleo, are compared side-by-side with the nuclear
suppression factor $R_{p{\rm Pb}}(y)$ measured by ALICE at the same energies
\cite{ALICE_psiy5,Enrico}.  The EPS09 LO calculation by Lansberg {\it et al.}
in Ref.~\cite{Albacete:2016veq} used a standard $2 \rightarrow 2$ matrix
element, not the data-driven approach here, and the CGC calculations were
different also.

The ALICE data are remarkably similar so far at the two energies although
the still-preliminary 8.16 TeV data have large uncertainties.
(Note that the same data sets from LHCb are not shown here to make it possible
to distinguish between the data sets at the two energies.  The LHCb data
at 5 \cite{LHCb_psiy5} and 8.16 TeV \cite{Aaij:2017cqq}
are also compatible at the two energies.)
The 5 TeV data in the backward rapidity region do not show a strong rapidity
dependence while there is a decrease with increasing rapidity at 8.16 TeV.
The trends in the data at forward rapidity are very similar although the
slope seems again somewhat larger for 8.16 TeV.  Recall, however, that the
$p+p$ baseline of the 5.02~TeV was obtained from an interpolation between
$p+p$ measurements at higher (7~TeV) and lower (2.76~TeV) energies since there
was no $p+p$ run at 5~TeV until LHC Run II.  It would be interesting to
recalculate the 5~TeV $R_{p{\rm Pb}}$ results for the measured $p+p$ data at the
same energy.

The EPS09 NLO CEM calculation gives essentially identical results for the
two energies for $y > -2$.  At more backward rapidity, the antishadowing
peak has moved to still more negative rapidity in the higher energy
calculation.  The energy loss calculation also shows a shift to more negative
rapidity, the rise at backward rapidity is shifted to the left, toward lower
values of $y$, at the higher energy.  Interestingly, this has the effect of
lifting the 8 TeV result above that of the 5 TeV calculation at positive
rapidity, opposite the trend of the data.  

In both cases the difference in the calculated $R_{p{\rm Pb}}(y)$ will be hard
to distinguish and subtle differences in curvature may not be differentiated
by the data unless the uncertainties are significantly reduced.

In Fig.~\ref{fig:Jpsi_y}(b), the data-driven calculations
with EPS09 LO and nCTEQ are compared with the energy loss calculation.  As has
been observed previously \cite{Vogt:2015uba}, the lower limit of shadowing with
EPS09 LO ($R_{p{\rm Pb}} \sim 1$) is similar to that of EPS09 NLO.  However, the
upper limit of EPS09 LO shows much stronger shadowing (lower $R_{p {\rm Pb}}$) due
to the different low $x$ behavior of CTEQ6M and CTEQ6L1, see
Ref.~\cite{Vogt:2015uba} for
more details.  The nCTEQ calculation shows stronger shadowing at backward
rapidity for the upper limit of shadowing with a steeper rise toward
the antishadowing region.

The energy loss calculation has a different curvature at forward rapidity with
a stronger effect turning on for $y > 3$.  It also has a rise at backward
rapidity since the backward shift in rapidity that causes the drop at forward
rapidity causes a corresponding enhancement at backward rapidity.
Overall, the energy loss model predicts 
rather strong $J/\psi$ suppression at forward rapidity, $y\gtrsim3$, and a 
slight enhancement in the most backward rapidity bins, $y<-4$. 

As can be seen in Fig.~\ref{fig:Jpsi_y}(b), the preliminary
ALICE data are consistent with the shadowing and energy loss
calculations.  In the case of
the data-driven results, the uncertainties in EPS09 LO and nCTEQ are large
enough to encompass the data.  In addition, the maximum achievable shadowing
in these cases is stronger than for EPS09 NLO.  The curvature of the energy loss
calculation is compatible with the decreasing trend of the ALICE data at forward
rapidity.

Finally, the $\mathtt{HIJING++}$ predictions for the $J/\psi$ are also shown
as the red points in Fig.~\ref{fig:Jpsi_y}(b).  The $J/\psi$
are produced in the hard scatterings in the underlying $\mathtt{PYTHIA8}$
generator.  The rather large uncertainties are likely due
to the fact that, in the calculations, all the charmonium channels were
turned on and allowing production in multiple channels can reduce the population
of a specific final quarkonium state.

At rapidities in the range $y > -1$, the $\mathtt{HIJING++}$
calculations are within the uncertainties of the shadowing
models and agree rather well with the forward rapidity data.  
However, for $y < -1$, the calculation gives a significant enhancement, larger
than what one would expect from standard nPDF parameterizations that include
antishadowing.  This may be due to multiple scattering of the final-state
with other particles in the medium.  

Figure~\ref{fig:Jpsi_pT} compares all the shadowing calculations at backward
rapidity ($-4.46<y<-2.96$), forward rapidity ($2.03<y<3.53$), and
midrapidity ($-1.37 <y< 0.44$) as a function of $p_T$.
The backward rapidity region has an antishadowing peak, as the ratio is
larger than unity for all calculations, especially for $p_T > 4$~GeV.  The
level of shadowing at low $p_T$, $p_T \sim 2$~GeV, is similar at forward
rapidity and midrapidity.  This is not surprising because already at $y > -1$
the nPDF calculations are at their maximum $p_T$-integrated shadowing.
However, at midrapidity, the suppression factors increase more rapidly with
$p_T$ than at forward rapidity.  The strongest shadowing at midrapidity is
found with nCTEQ but at forward rapidity nCTEQ and EPS09 LO give comparable
shadowing effects.

The preliminary $p_T$-dependent ALICE data at backward and forward rapidity
are compared to the calculations in Fig.~\ref{fig:Jpsi_pT}.  The LHCb data
are also shown.  Note, however, that these data are in slightly different
rapidity windows, $-5 < y < -2.5$ for backward rapidity and $1.5 < y < 4$ for
forward rapidity \cite{Aaij:2017cqq}.  Despite the difference in rapidity
windows, the data, which include both statistical and systematic uncertainties
added in quadrature, agree rather well in both rapidity regions.  At backward
rapidity, all the calculations overlap with each other as well as with the
measurements.  
On the other hand, at forward rapidity, the data rise faster with
$p_T$ than the calculations which remain less than unity for $p_T \leq 20$~GeV.
Thus the behavior of the data are generally incompatible with the nCTEQ
result for $p_T > 10$~GeV.  However, the data would seem to suggest a faster
evolution with $p_T$ than the current global analyses can account for.

\begin{figure}[htpb]
\begin{center}
\includegraphics[width=0.495\textwidth]{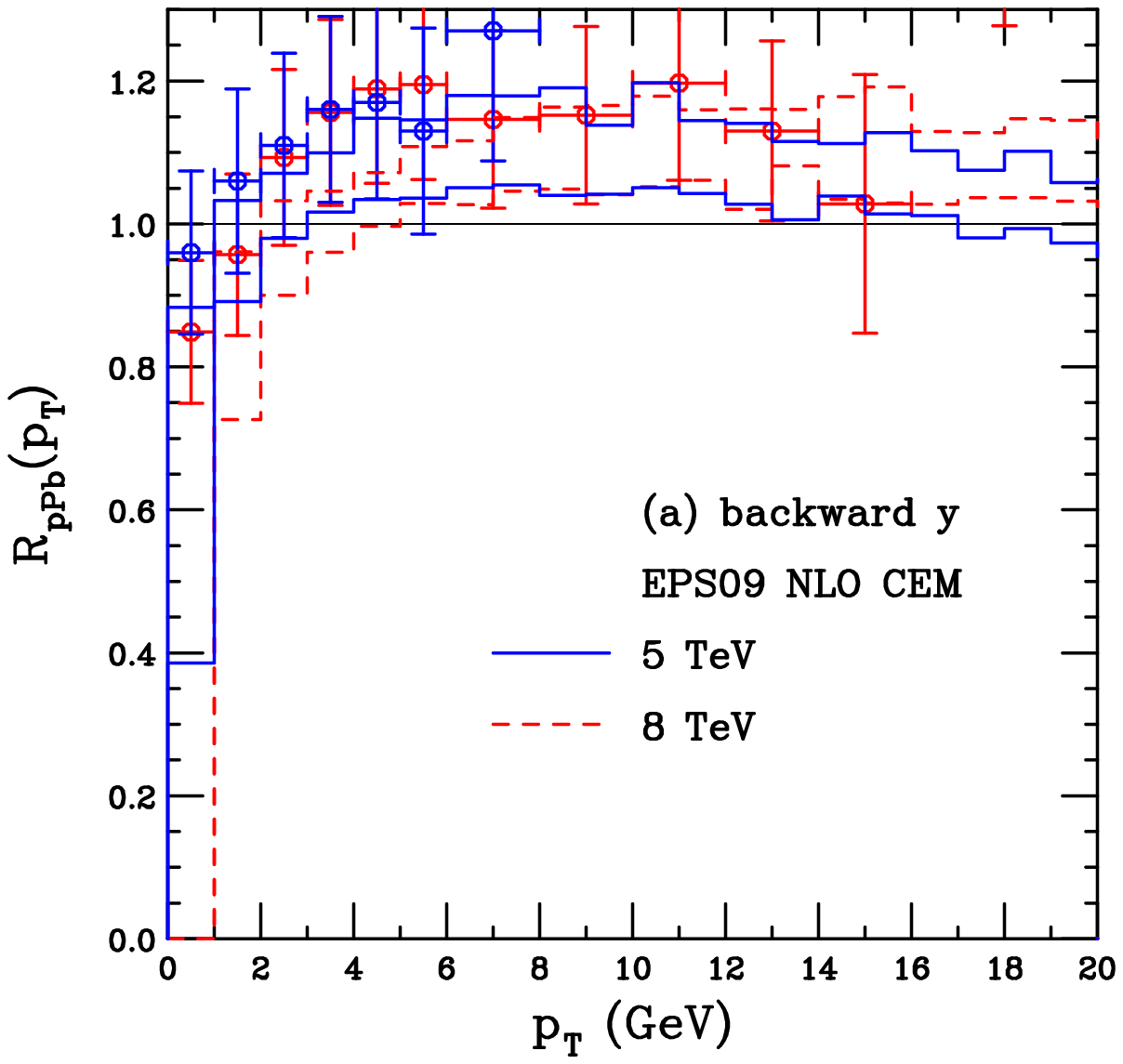} 
\includegraphics[width=0.495\textwidth]{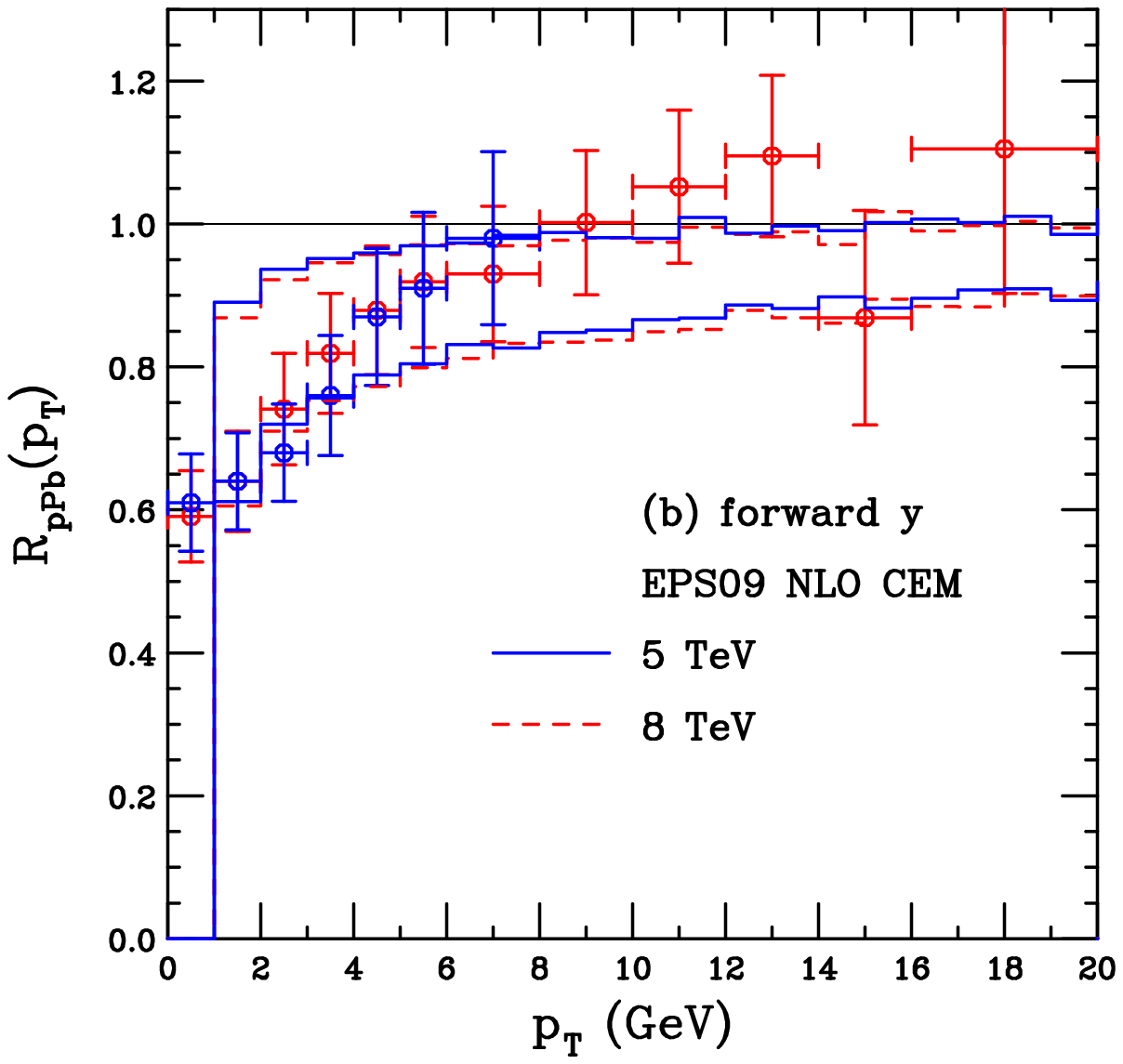} 
\end{center}
\caption[]{(Color online)
  The ratio $R_{\rm p{\rm Pb}}$ for $J/\psi$ as a function of $p_T$ 
  at 5 TeV (blue points and solid blue curves) and 8 TeV
  (red points and dashes red curves) calculated with EPS09 NLO are compared.
  The data from ALICE at
  5 TeV \protect\cite{ALICE_psi_pT} and 8 TeV \protect\cite{Enrico} are also
  shown at backward rapidity (a) and forward rapidity (b).
  }
\label{fig:Jpsi_pT_8vs5}
\end{figure}

Figure~\ref{fig:Jpsi_pT_8vs5} compares the $p_T$ dependence of the suppression
factor at the two energies, both the data and the EPS09 NLO CEM calculation
from Ref.~\cite{Albacete:2016veq}.
The new data at the higher
energy extend the measured $p_T$ range by a factor of two.  At forward
rapidity, the data are very similar where they overlap, not surprising since
they agree rather well in this region of rapidity,
see Fig.~\ref{fig:Jpsi_y_8vs5}.  On the other hand, the 5 TeV data are higher
at low $p_T$ for the backward rapidity region, similar to the rapidity
dependence.  The calculations reflect this: the results are on top of each
other at forward rapidity but there is more low $p_T$ suppression at backward
rapidity.

\paragraph{$\Upsilon$ model comparisons, collinear factorization}
\label{sec:ups}

\begin{figure}[htpb]
\begin{center}
\includegraphics[width=0.495\textwidth]{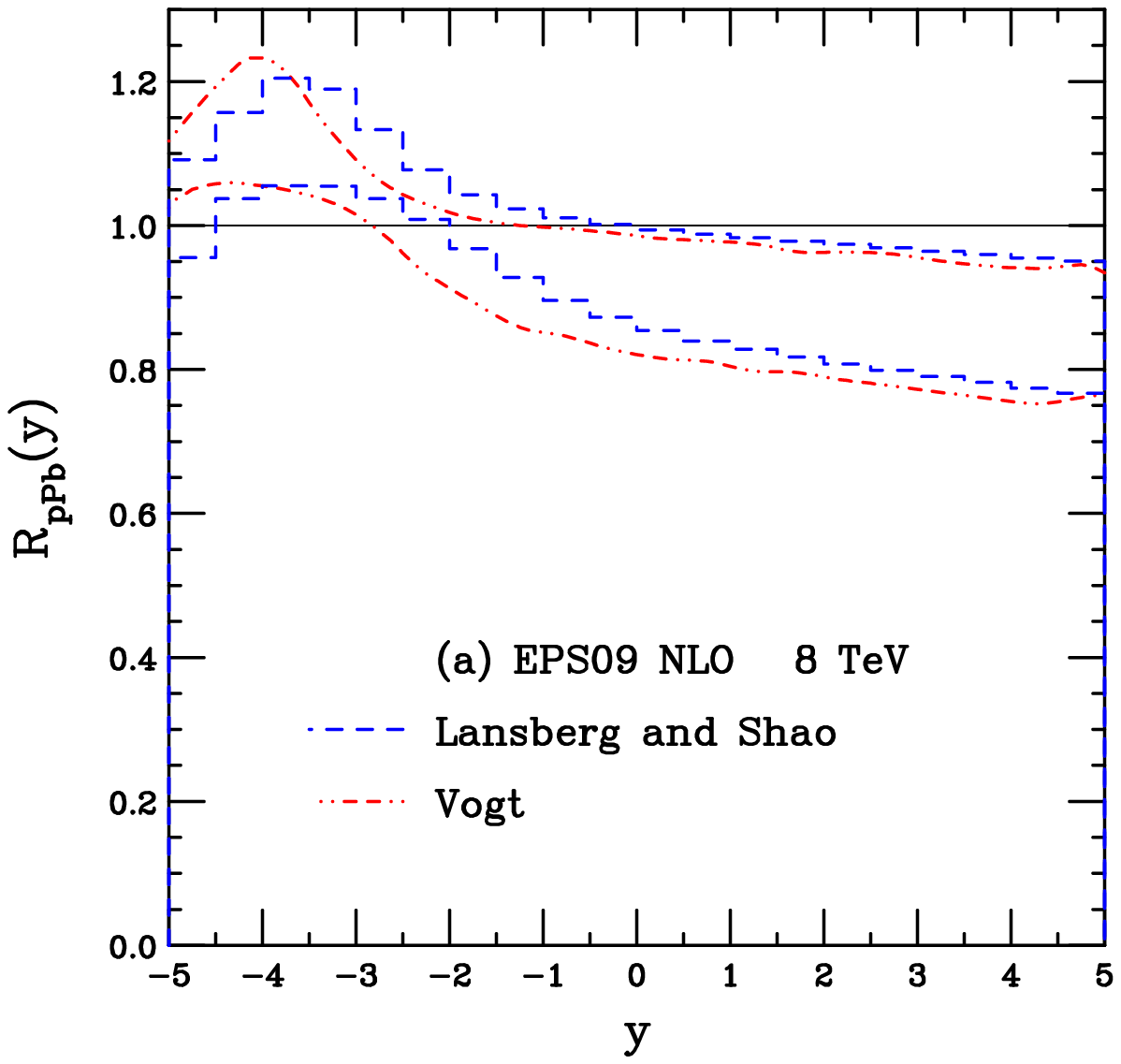} 
\includegraphics[width=0.495\textwidth]{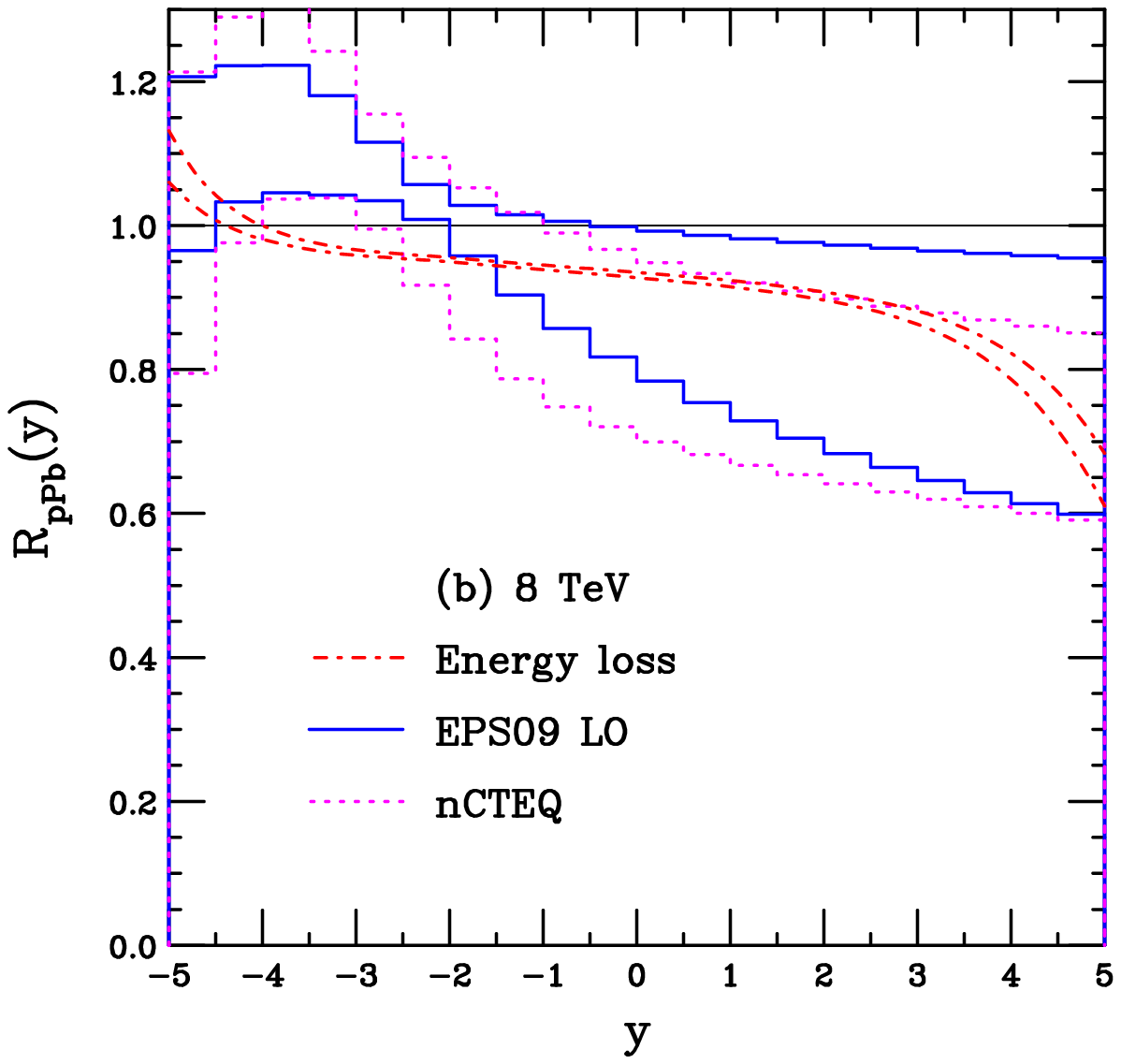} 
\end{center}
\caption[]{(Color online)
The ratio $R_{\rm p{\rm Pb}}$ for $\Upsilon$ as a function of rapidity at 8~TeV.
(a) The EPS09 NLO result is compared between the NLO CEM calculation
of Vogt (dot-dot-dash-dashed red curve) and the data-driven result of 
Lansberg and Shao (solid cyan). (b) The data-driven calculation of Lansberg
and Shao for EPS09 LO (solid blue histogram) and nCTEQ (dotted magenta
histogram) is compared to the energy loss only calculation of Arleo (dot-dashed
red curve).}
\label{fig:Ups_y}
\end{figure}

\begin{figure}[htpb]
\begin{center}
\includegraphics[width=0.495\textwidth]{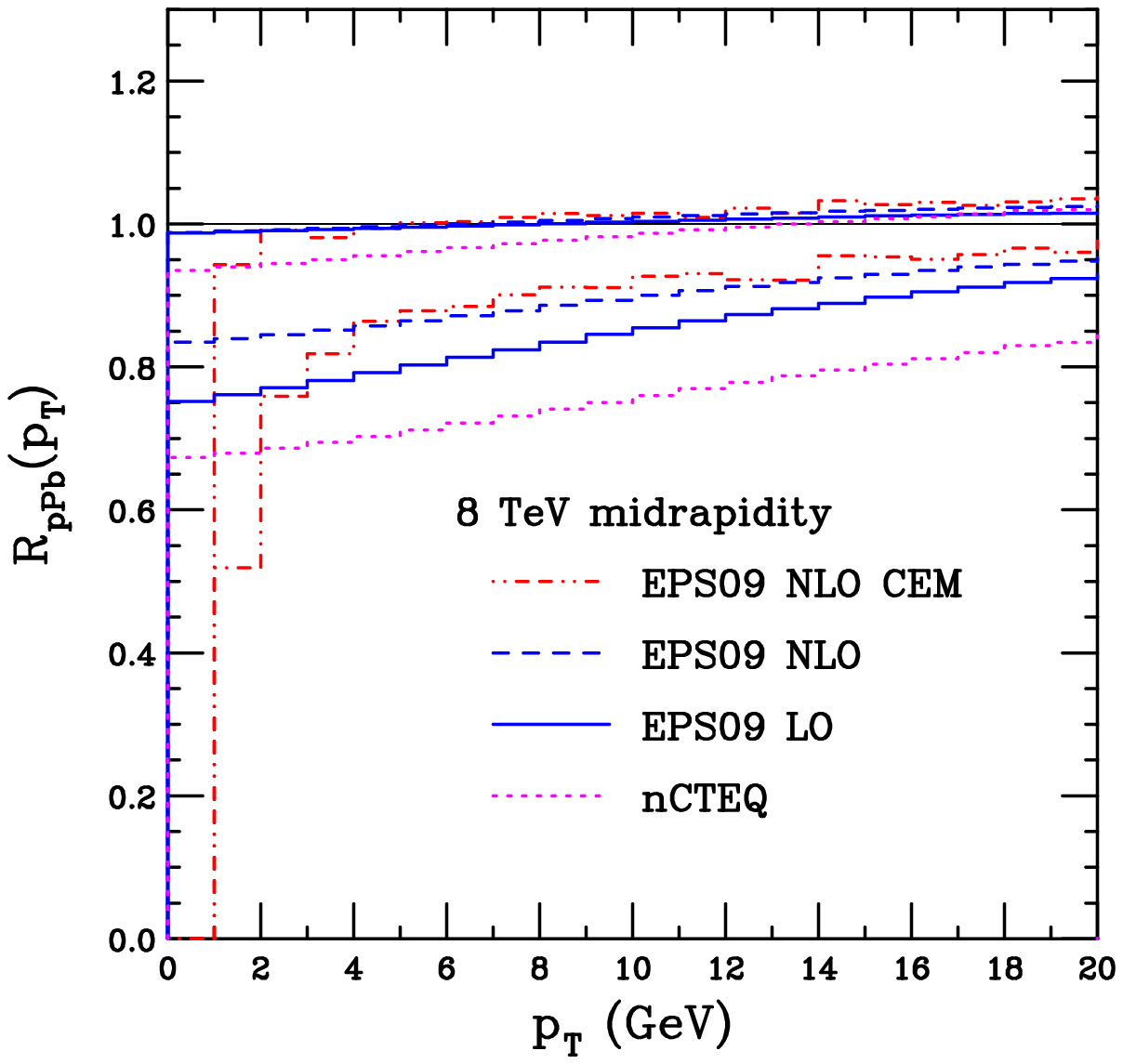} 
\end{center}
\caption[]{(Color online)
  The ratio $R_{\rm p{\rm Pb}}$ for $\Upsilon$ at midrapidity
  as a function of transverse momentum at 8~TeV.
The EPS09 NLO results of Vogt (dot-dot-dash-dashed red curve) and 
Lansberg and Shao (solid cyan) are shown with their results for EPS09 LO 
(solid blue histogram) and nCTEQ (dotted magenta histogram) at
midrapidity.}
\label{fig:Ups_pT}
\end{figure}

The predictions for $\Upsilon$ suppression in $p+$Pb collisions 
at $\sqrt{s_{_{NN}}}=8$~TeV in the approaches discussed in this section
are shown in Figs.~\ref{fig:Ups_y} and \ref{fig:Ups_pT}.
In Fig.~\ref{fig:Ups_y}(a), the CEM and data-driven
calculations employing EPS09 NLO are compared.

For the $\Upsilon$ $p+p$ fits,
Lansberg and Shao found larger values of $\lambda$
while the value of $\kappa$ was smaller.  A stronger dependence on the
proton PDF employed was also noted.  They found $\lambda \sim 0.77$ for CT14 NLO
and $\lambda \sim 0.69$ for CT10 NLO.  The value of $\kappa$ was decreased to
$\kappa \sim 0.085$.  Note that due to the higher average fixed $p_T$ for
$\Upsilon$ relative to $J/\psi$, $\kappa$ can be expected to be different
since $\kappa$ is directly related to $p_T$, see
Eq.~(\ref{eq:lansberg_amplitude}).
On the other hand, $\lambda$ is simply a normalization constant.

The $\Upsilon$ trend is similar to that for $J/\psi$.  However, the larger
$\Upsilon$ mass reduces the shadowing effect for all calculations.  There is a
similar shift between the CEM and data-driven calculations with EPS09 NLO as
seen in Fig.~\ref{fig:Jpsi_y}.  It is less pronounced for $\Upsilon$ than for
$J/\psi$.  The parameters in the CEM calculation for the central
$b \overline b$ fit are $m = 4.65$~GeV, $\mu_F/m_T = 1.6$ and $\mu_R/m_T = 1.1$
\cite{Vogt:2015uba}.
Thus the factorization scale used in this calculation is again
larger than that of the data-driven calculation.  In this case, however, the
difference is not as large and, since the scales are overall larger than
for $J/\psi$, evolution reduces the relative difference.

In Fig.~\ref{fig:Ups_y}(b), the data-driven calculations
with EPS09 LO and nCTEQ are compared with the energy loss calculation.
The shadowing effects are also reduced for the EPS09 LO and nCTEQ calculations
although these still show a stronger effect overall than that with EPS09 NLO.
The suppression due to energy loss predicted for the $\Upsilon$ 
shares the same features as for the $J/\psi$.
However, the suppression is less pronounced than 
that of the $J/\psi$ since 
the (average) coherent energy loss scales as $m_T^{-1}$~\cite{Arleo:2010rb}.

Figure~\ref{fig:Ups_pT} compares all the shadowing calculations at
midrapidity ($-1.37 <y< 0.44$) as a function of $p_T$.  Weaker shadowing is
seen also in the $p_T$ dependence of $\Upsilon$ production at midrapidity.
As was the case for the $J/\psi$, the nCTEQ set has the strongest effect at
midrapidity and low $p_T$.

\paragraph{$J/\psi$ and $\psi'$ Interactions with Comovers (E. G. Ferreiro)}
\label{sec:comovers}

A relative suppression of excited charmonium states as compared to their ground
state has been obtained in d+Au and $p+$Pb collisions by the PHENIX
\cite{Adare:2013ezl}, ALICE  \cite{Abelev:2014zpa,Arnaldi:2014kta} 
and LHCb \cite{Aaij:2016eyl} collaborations.
In particular, stronger
suppression of the  $\psi(2{\rm S})$ relative to the $J/\psi$ 
has been detected. This behavior can be explained by the interactions of the
quarkonium states with a comoving medium \cite{Ferreiro:2014bia}.

In the comover framework, the suppression arises from scattering of the
nascent $\psi$ with the produced particles, the comovers,
that travel along with the $c\bar{c}$ pair \cite{Gavin:1996yd,Capella97}.
The comover suppression affects the $\psi(2{\rm S})$ more strongly than the
$J/\psi$ due to its larger size.
The comover suppression is stronger in regions of phase space where the
comover densities are larger.  Thus the effect is strongest in more central
collisions and, for the asymmetric proton-nucleus collisions, in the
direction of the nucleus.

In the comover interaction model (CIM)
\cite{Capella97,Armesto98,Armesto99,Capella00,Capella05,Capella:2006mb}, the
rate equation that governs the density of charmonium at a given 
transverse coordinate $s$, impact parameter $b$ and rapidity~$y$,
$\rho^{\psi}(b,s,y)$, obeys the expression
\begin{eqnarray}
\label{eq:comovrateeq}
\tau \frac{d \rho^{\psi}}{d \tau} \, \left( b,s,y \right)
\;=\; -\sigma^{{\rm co}-\psi}\; \rho^{\rm co}(b,s,y)\; \rho^{\psi}(b,s,y) \;,
\end{eqnarray}
where $\sigma^{{\rm co}-\psi}$ is the charmonium dissociation cross section
due to interactions with a comoving medium of transverse density
$\rho^{\rm co}(b,s,y)$.

Assuming that the comover density becomes more dilute with time due to the
longitudinal expansion of the medium leads to a $\tau^{-1}$ dependence on
proper time and Eq.~(\ref{eq:comovrateeq}) can be solved analytically. The
result depends only on the ratio $\tau_f/ \tau_0$
of final over initial time. Using the inverse
proportionality between proper time and density, the interaction is
assumed to stop when the density has diluted to the point that the comover
density is equal to the $p+p$ density at the same energy,
$\tau_f/ \tau_0 = \rho^{\rm co}(b, s, y)/\rho_{pp}(y)$.  Thus, the solution of
Eq.~(\ref{eq:comovrateeq}) is given by
\begin{eqnarray}
\label{eq:survivalco}
S^{\rm co}_{\psi}(b,s,y)  \;=\; \exp \left\{-\sigma^{{\rm co}-\psi}
  \, \rho^{\rm co}(b,s,y)\, \ln
\left[\frac{\rho^{\rm co}(b,s,y)}{\rho_{pp} (y)}\right] \right\} 
\end{eqnarray}
where the argument of the logarithm is the interaction time of the $\psi$
with the comovers.

The cross section of charmonium dissociation due to interaction with the
comoving medium, $\sigma^{{\rm co}-\psi}$, was fixed \cite{Armesto98} from fits
to low-energy experimental data to be
$\sigma^{{\rm co}-J/\psi}=0.65$~mb for the $J/\psi$ and
$\sigma^{{\rm co}-\psi(2{\rm S})}=6$~mb for the $\psi(2{\rm S})$. 
These same values were also successfully applied at higher energies to
reproduce the RHIC \cite{Capella:2007jv} and LHC \cite{Ferreiro:2012rq}
$J/\psi$ data in nucleus-nucleus collisions.

The  modification of the gluon parton distribution functions in the nucleus 
is also taken into account in this approach.
Since the effect is identical for the 1S and 2S states \cite{Ferreiro:2012mm},
{\it i.e.} for the $J/\psi$ and the $\psi(2{\rm S})$, it produces an identical
decrease of the $J/\psi$ and the $\psi(2{\rm S})$ yields at mid and forward rapidity
for LHC energies.  However, due to gluon antishadowing, it can induce an
increase of both yields in the backward rapidity region.

The nuclear modification factor is thus
\begin{eqnarray}
\label{eq:ratiopsi}
R^{\psi}_{pA}(b) =
\frac{\int d^2s \, 
  \sigma_{pA}(b) \, n(b,s) \,  S_{\psi}^{\rm sh}(b,s) \, S^{\rm co}_{\psi}(b,s)}
     {\int d^2 s \, \sigma_{pA} (b) \, n(b,s)} \;, 
\end{eqnarray}
where $S^{\rm co}_{\psi}$ is the survival probability due to the comover
interactions and $S_{\psi}^{\rm sh}$ takes into account 
the modification of the nuclear parton distribution functions. 

\begin{figure}[t]
\begin{center}
  \includegraphics[width=0.495\textwidth]{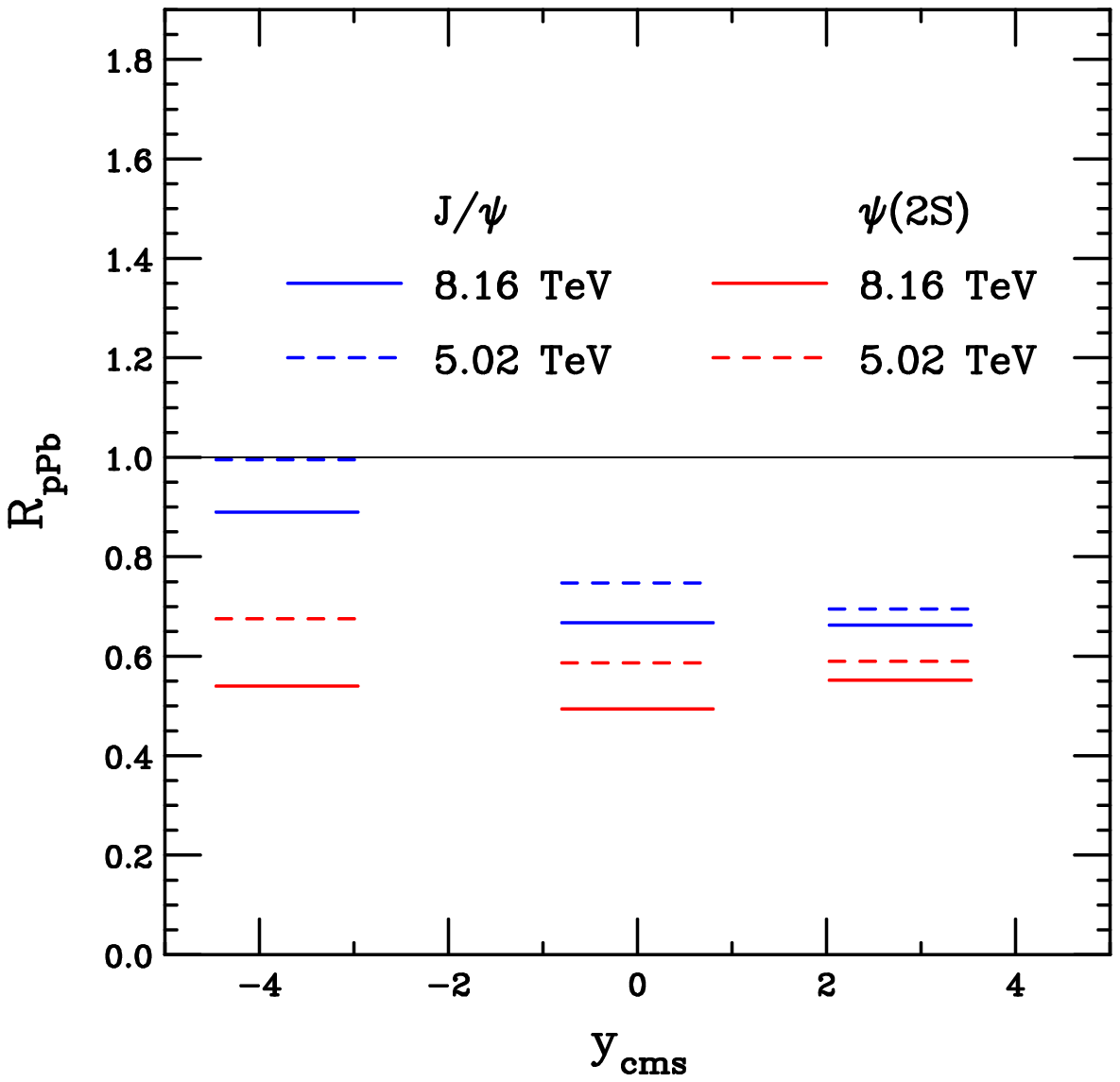}
  \end{center}
\caption[]{\label{fig:figpPby}(Color online) 
The $J/\psi$ (blue lines) and $\psi(2{\rm S})$ (red lines)
nuclear modification factor $R_{p{\rm Pb}}$ as a function of rapidity at 5.02~TeV
(dashed lines) and 8.16~TeV (solid lines). 
}
\end{figure}

Figure~\ref{fig:figpPby} shows the nuclear modification factor $R_{p{\rm Pb}}$
as a function of rapidity in $p+$Pb collisions at $\sqrt{s} = 8.16$~TeV.
Three rapidity intervals are studied: the $p$-going direction, $2.03<y<3.53$;
the Pb-going direction, $-4.46<y<-2.96$; and the midrapidity interval.
EPS09 LO shadowing is assumed \cite{Eskola:2009uj,Ferreiro:2013pua} for both
the $J/\psi$ and the $\psi(2{\rm S})$. 
The interaction with comovers induces a stronger suppression in the backward
rapidity region, the Pb-going direction, due to the higher comover density.
This effect is more important for $\psi(2{\rm S})$ than for $J/\psi$ production since
$\sigma^{{\rm co}-\psi(2{\rm S})} > \sigma^{{\rm co}-J/\psi}$
The effect due to the EPS09 LO shadowing depends on the rapidity interval
considered. It produces additional suppression in the mid and forward rapidity
regions but is compatible with unity in the backward interval accessible to
experiment, see Fig.~\ref{fig:Jpsi_y}(b).

The 8.16 TeV results are compared to those from 5.02 TeV in
Fig.~\ref{fig:figpPby}.  Note the additional suppression at the higher energy,
due to the larger density of produced particles.  The effect is
particularly noticeable at backward rapidity, near the lead nucleus.  At forward
rapidity, in the proton-going direction, the difference is small.

\begin{figure}[t]
\begin{center}
  \includegraphics[width=0.495\textwidth]{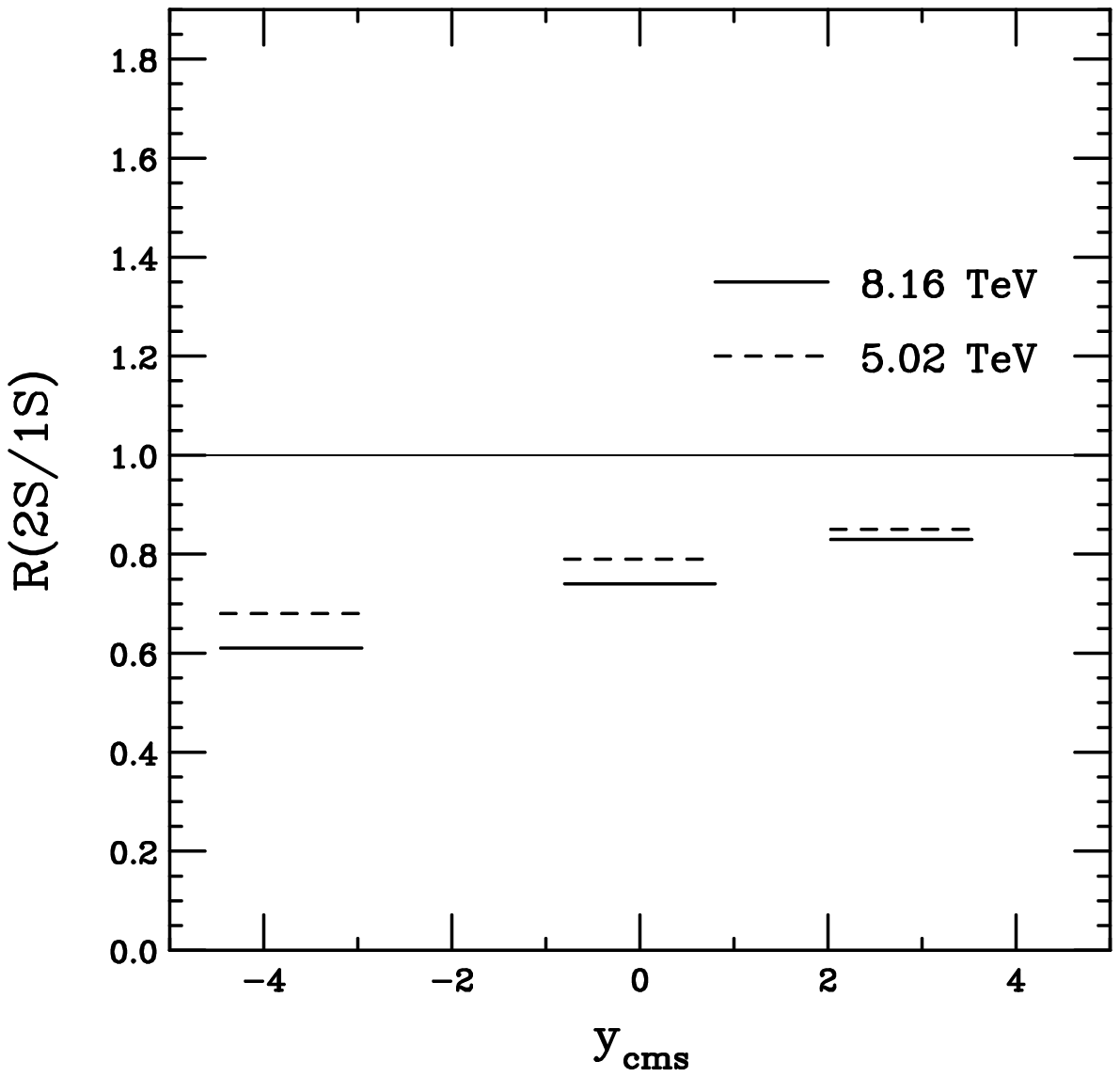}
  \end{center}
\caption[]{\label{fig:figDoubleRpPby}(Color online) 
  The ratio of nuclear modification factors $R_{p{\rm Pb}}(y)$
  for $\psi(2{\rm S})$
  relative to $\psi(1{\rm S})$ are compared at 8.16~TeV (solid) and 5.02~TeV
  (dashed). 
}
\end{figure}

In Fig.~\ref{fig:figDoubleRpPby}, the double ratio
$R_{p{\rm Pb}}(2{\rm S})/R_{p{\rm Pb}}(1{\rm S})$
for $p+$Pb collisions at $\sqrt{s_{_{NN}}}=~8.16$~TeV
is presented and compared with
those at $\sqrt{s_{_{NN}}}=5.02$ \cite{Ferreiro:2014bia}.
The same three rapidity intervals are studied.  The
same trends as in Fig.~\ref{fig:figpPby} are seen.  However, there is a
stronger effect
on the double ratio at backward rapidity than in more the forward rapidity
intervals, away from the nucleus.

\subsubsection{Saturation Approaches}
\label{sec:onia_cgc}

Here results from two saturation approaches are presented.  The first, using
the rcBK approach in Sec.~\ref{sec:sinc}, by
Duclou\'e, Lappi and M\"antysaari employs
the color evaporation model of quarkonium production, also used in
Sec.~\ref{sec:Vogt}.
The second, by Ma, Venugopalan and Zhang, employs nonrelativistic QCD,
NRQCD, as the
baseline quarkonium production model.

\paragraph{CGC+CEM (B. Duclou\'e, T. Lappi and 
H. M\"antysaari)}
\label{sec:lappi_psi}

As discussed in more detail in Refs.~\cite{Ducloue:2016pqr,Ducloue:2015gfa},
the color evaporation model, where a fixed fraction of all $c\overline c$ pairs
produced below the $D$ meson threshold are assumed to become $J/\psi$ mesons,
is employed.   The same CGC framework and same rcBK-evolved parameterization for
the dipole amplitude obtained from DIS fits that is used when calculating
single inclusive particle production in Sec.~\ref{sec:sinc} and in
Ref.~\cite{Lappi:2013zma} is also employed.  Results here are referred to
as CGC+CEM.

The CEM cross section for prompt $J/\psi$ production is written here as
\begin{equation}
  \frac{d \sigma_{J/\psi}}{d^2 p_T d y} = F_{J/\psi} \int_{4m_c^2}^{4m_D^2} d M^2
  \frac{d \sigma_{c\overline c}}{d^2 p_T d y d M^2} 
\end{equation}
where $p_T$ and $y$ are the transverse momentum and rapidity of the produced
$J/\psi$, $m_c$ is the charm mass, and
$m_D=1.864 \gev$ is the $D$ meson mass. The fraction of $c\overline c$ pairs
fragmenting into $J/\psi$ is given by $F_{J/\psi}$ which cancels in $R_{pA}$.
The uncertainty on the calculation is determined by varying $m_c$,
$1.2 \leq m_c \leq 1.5$~GeV.

Only $J/\psi$ production at forward rapidities, where the Bjorken $x$ of the
probe is large and the gluon density in the probe is given by the collinear
parton distribution function $xg(x)$, is considered.  In this region, the
target is probed at small $x$ and the $c\overline c$ production cross section
can be written as
\begin{eqnarray}
  \frac{d \sigma_{c\overline{c}} } {d^2p_T d^2q_T d y_p d y_q} & = & \frac{\alpha_s^2
    N_c}{8\pi^2 d_A}
  \frac{1}{(2\pi)^2}  \\ & & 
  \!\!\!\!\!\!\!\!\!\!\!\!\!\!\!\!\!\!\!\!\!
  \times
\int \frac{d^2 k_T}{(2\pi)^2}
\frac{\Xi_{\rm coll}(p_T + q_T,k_T)}{(p_T + q_T)^2}
\phi_{y_2=\ln{\frac{1}{x_2}}}^{q\overline{q},g}(p_T + q_T,k_T)
x_1 g(x_1,Q^2) \, \, . \nonumber
\end{eqnarray}
Here $d_A=N_c^2-1$ and the $x$ values for the projectile and target,
$x_1$ and $x_2$, are given by
\begin{equation}
x_{1,2} = \frac{\sqrt{p_T^2 + M^2}}{\sqrt{s}} e^{\pm y}.
\end{equation}
The expression for the hard matrix element $\Xi_{\rm coll}$ is given in
Ref.~\cite{Ducloue:2015gfa}.
The propagation of the quark-antiquark
pair through the color field of the target is given by
\begin{equation}
\phi_{_Y}^{q \overline{q},g}(l_T,k_T)=
\int d^2 b_T  \frac{N_c l^2_T}{4 \alpha_s} 
S(k_T) 
S(l_T-k_T) \, \, .
\end{equation}
Here the dipole amplitudes in the Fourier transforms $S(k_T)$ and
$S(l_T-k_T)$ are evaluated at $x_2$. In the
case of proton-proton scattering, the impact parameter dependence is
assumed to factorize and the replacement $\int d^2 b_T \to \sigma_0/2$ is
made. In proton-nucleus collisions the impact parameter integral is performed
using the optical Glauber model as described in Sec.~\ref{sec:sinc}.

\paragraph{CGC+NRQCD (Y.-Q. Ma, R. Venugopalan and H.-F. Zhang)}
\label{sec:Ma}

Here the $J/\psi$ production cross section in $p+p$ and $p+A$ collisions
is calculated within the 
framework of the CGC
\cite{McLerran:1993ni, McLerran:1993ka, Iancu:2003xm, Gelis:2010nm} and the
nonrelativistic QCD approach to quarkonium production~\cite{Bodwin:1994jh}.
The calculations are based on Refs.~\cite{Kang:2013hta,Ma:2015sia}.
The production of $J/\psi$ in $p+p$ collisions within the framework of
CGC+NRQCD was presented in Ref.~\cite{Ma:2014mri}.  

In NRQCD factorization, the production cross section of a quarkonium state
$H$ in the forward region of a $p+A$ collision is expressed
as \cite{Bodwin:1994jh}
\begin{eqnarray} \label{eq:NRQCD}
d\sigma^H_{pA}=\sum_\kappa d\hat{\sigma}_{pA}^\kappa\langle{\cal O}^H_\kappa\rangle
\end{eqnarray}
where $\kappa=\state{2S+1}{L}{J}{c}$ denotes the quantum numbers of the
intermediate $Q\overline{Q}$-pair in the standard spectroscopic notation for angular
momentum.
The superscript $[c]$ denotes the color state of the pair,
which can be either color singlet (CS) with $c=1$ or color octet (CO) with
$c=8$.

For $J/\psi$ production, the most important intermediate
states are $\CScSa$, $\COaSz$, $\COcSa$ and $\COcPj$.
In Eq.~\eqref{eq:NRQCD}, $\langle{\cal O}^H_\kappa\rangle$ are nonperturbative 
universal long-distance matrix elements (LDMEs),
which can be extracted from data and $d\hat{\sigma}^\kappa$ are short-distance 
coefficients (SDCs) for the production of a $Q\overline{Q}$-pair,
computed in perturbative QCD.

To calculate the SDCs in Eq.~\eqref{eq:NRQCD}, CGC effective field 
theory \cite{Gelis:2010nm,Blaizot:2004wv} is applied,
resulting in~\cite{Kang:2013hta,Ma:2014mri}
\begin{align}\label{eq:dsktCS}
\begin{split}
\frac{d \hat{\sigma}_{pA}^\kappa}{d^2\vp d
y}\overset{\text{CS}}=&\frac{\alpha_s (\pi \overline{R}_A^2)}{(2\pi)^{9}
(N_c^2-1)} \underset{\vka,\vk,\vkp}{\int}
\frac{\varphi_{p,y_p}(\vka)}{k_{1T}^2}\\
&\hspace{-1.5cm}\times \mathcal{N}_{Y}(\vk)\mathcal{N}_{Y}(\vkp)\mathcal{N}_{Y}(
\vp-\vka-\vk-\vkp)\,
{\cal G}^\kappa_1
\end{split}
\end{align}
for the color-singlet $\CScSa$ channel and
\begin{align}\label{eq:dsktCO}
\begin{split}
\frac{d \hat{\sigma}_{pA}^\kappa}{d^2\vp d
y}\overset{\text{CO}}=&\frac{\alpha_s (\pi \overline{R}_A^2)}{(2\pi)^{7}
(N_c^2-1)} \underset{\vka,\vk}{\int}
\frac{\varphi_{p,y_p}(\vka)}{k_{1T}^2}\\
&\times \mathcal{N}_Y(\vk)\mathcal{N}_Y(\vp-\vka-\vk)
\,\Gamma^\kappa_8
\end{split}
\end{align}
for the color-octet channels.
Here $\varphi_{p,y_p}$ is the unintegrated gluon distribution of the proton,
\begin{align}\label{eq:unintegrated}
  \varphi_{p,y_p}(\vka)=\pi \overline{R}_p^2 \frac{N_c k_{1T}^2}{4\alpha_s}
  \widetilde{\mathcal{N}}^A_{y_p}(\vka)\,.
\end{align}
The functions ${\cal G}^\kappa_1$ and $\Gamma^\kappa_8$ are calculated
perturbatively \cite{Kang:2013hta,Ma:2014mri}.  
${\cal N}$ ($\widetilde{\mathcal{N}}^A$) are the momentum-space dipole forward
scattering amplitudes with Wilson lines in the fundamental (adjoint)
representation and $\pi \overline{R}_p^2$ ($\pi \overline{R}_A^2$)
is the effective transverse area of the dilute proton (dense nucleus).
These formulas can be used to compute quarkonium production in $p+A$ collisions.
When the nucleus is replaced by a proton, these expressions
can also be used to compute quarkonium
production in  $p+p$ collisions~\cite{Ma:2014mri}.
Note that for d+Au collisions at RHIC,
it is assumed that $\varphi_{d,y_d}(\vka)= 2\, \varphi_{p,y_p}(\vka)$ since gluon
shadowing effects are weak for the deuteron.

In these calculations, the charm quark mass is set to $m_c=1.5\gev$,
approximately half the $J/\psi$ mass.
The CO LDMEs are taken from Ref.~\cite{Chao:2012iv}:
$\mops=1.16/(2N_c) \gev^3$, $\mopa=0.089\pm0.0098 \gev^3$,
$\mopb=0.0030\pm0.0012 \gev^3$ and $\mopc/m_c^2=0.0056\pm0.0021\gev^3$.
Further, as in Ref.~\cite{Ma:2014mri}, ${\cal N}$ and
$\widetilde{\mathcal{N}}^A$ are
obtained by solving the rcBK
equation~\cite{Balitsky:1995ub,Kovchegov:1999yj} in momentum space with
MV initial
conditions~\cite{McLerran:1993ni,McLerran:1993ka}
for the dipole amplitude at the initial rapidity scale $Y_0\equiv\ln(1/x_0)$
(with $x_0=0.01$) for small $x$ evolution.
In Ref.~\cite{Ma:2014mri}, a matching scheme was devised that allowed
interpolation between the collinearly-factorized gluon distribution of the
proton at large $x$ with the unintegrated distribution in
Eq.~(\ref{eq:unintegrated}).
The matching also determined the scale in the collinear gluon
distribution to be $Q=5.1$ GeV and the effective gluon radius of the proton to
be $\overline{R}_p=0.48$~fm.

The initial saturation scale $Q_{s0,A}$ in the nucleus and the effective
transverse nuclear radius $\overline{R}_A$ need to be fixed in $p+A$ collisions.
In this calculation, $Q_{s0,A}^2=2Q_{s0,p}^2$ is adopted.
The radius $\overline{R}_A^2$ is determined from the condition
$R_{pA}\rightarrow 1$ in the high $p_T$ limit.
This results in $\overline{R}_A=\sqrt{A/2}\overline{R}_p$,
giving $\overline{R}_A = 4.9\mathrm{~fm}$ for Pb and
$4.8\mathrm{~fm}$ for Au.

\paragraph{$J/\psi$ model comparisons, saturation}
\label{sec:psi_sat}

\begin{figure}[htpb]
\begin{center}
\includegraphics[width=0.495\textwidth]{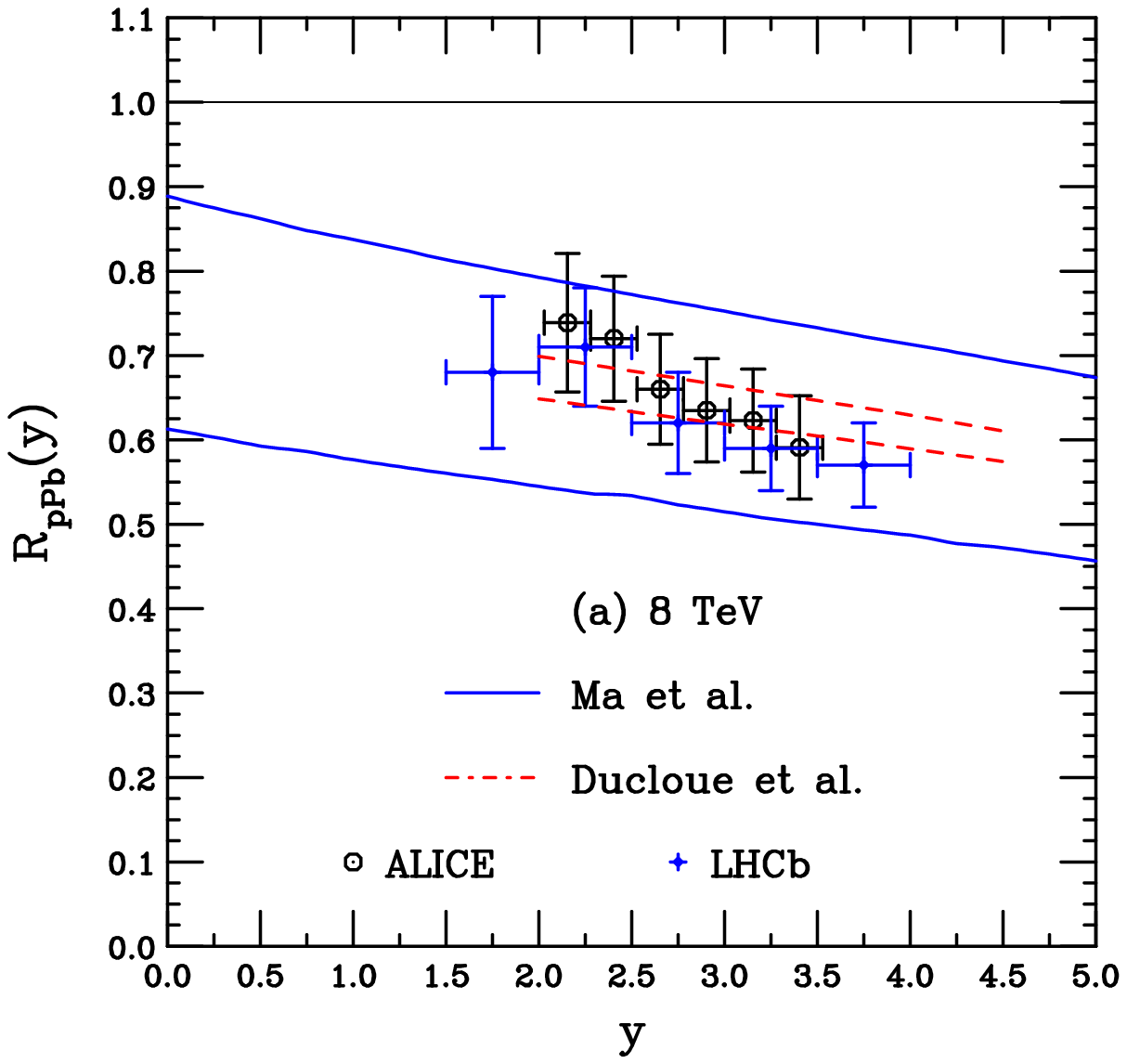} 
\includegraphics[width=0.495\textwidth]{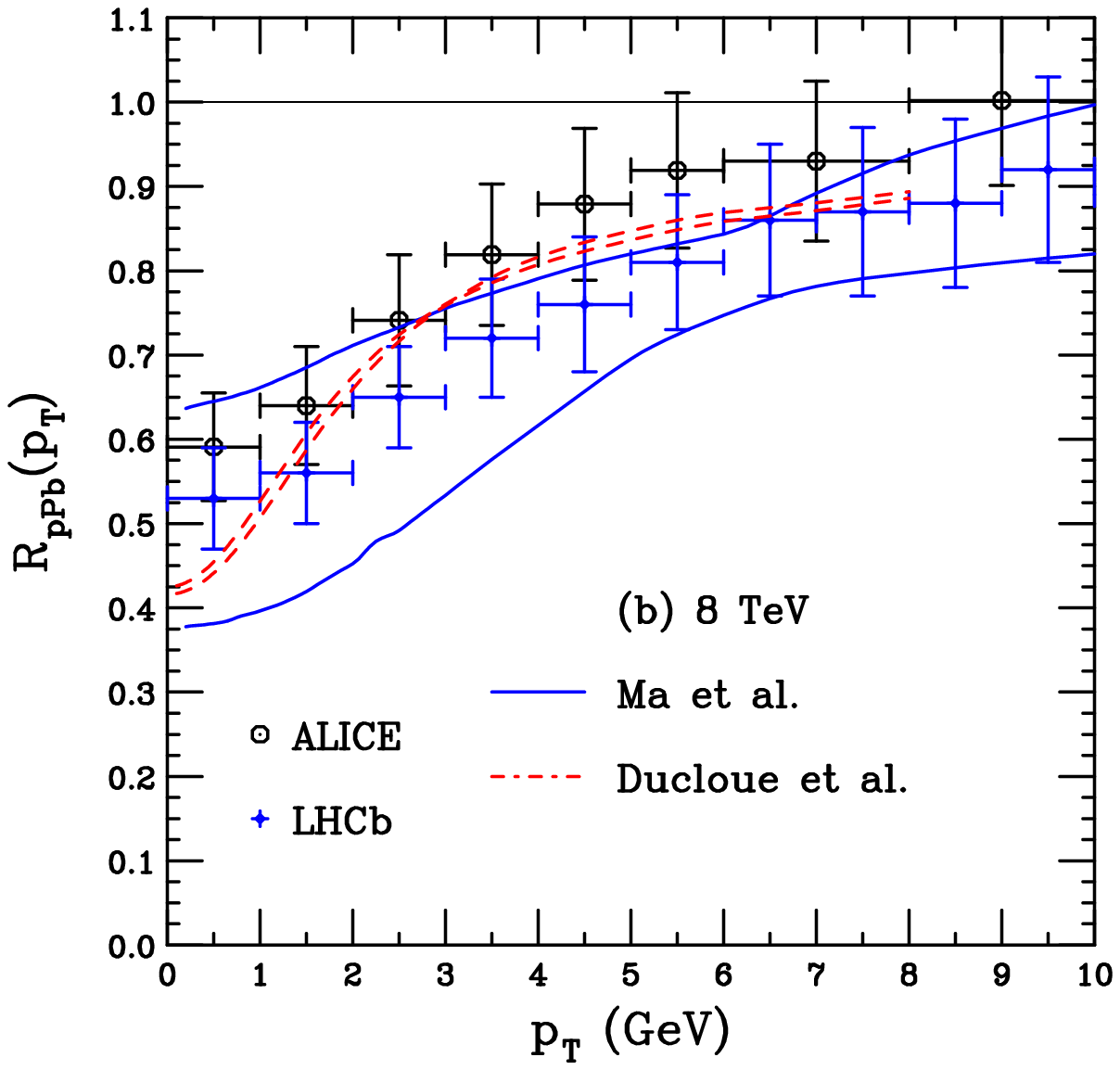} 
\end{center}
\caption[]{(Color online) Nuclear suppression factor for $J/\psi$
  production at 8~TeV
  by Douclou\'e {\it et al.} \protect\cite{Ducloue:2016pqr} (red
  dashed curves) and by Ma {\it et al.} \protect\cite{Ma:2015sia} (solid
  blue curves).  (a) The rapidity dependence.  (b) The
  transverse momentum dependence.
  The ALICE data \protect\cite{Enrico} are shown in
  black while the LHCb data \protect\cite{Aaij:2017cqq} are shown in blue.}
\label{psi_cgc}
\end{figure}

The nuclear suppression factors $R_{pA}$ for the two CGC models are shown in
Fig.~\ref{psi_cgc} as
a function of rapidity (a) and transverse momentum (b).
The calculation of the rapidity dependence by Duclou\'e {\it et al} is shown
only in the rapidity range of the forward ALICE data, $2 < y < 3.5$, while
the Ma {\it et al.} calculation is shown for $y > 0$.  The calculations as
a function of rapidity are integrated over $p_T$ for $p_T > 0$ while the
$p_T$-dependent results are obtained in the forward region, $2 < y < 3.5$ for
both calculations.  The preliminary ALICE \cite{Enrico} and
LHCb \cite{Aaij:2017cqq} are
also shown.

The calculation by Duclou\'e {\it et al.}, which employs the CGC+CEM,
includes uncertainties only due to
the variation of the charm quark mass in the calculation.  Thus the uncertainty
band is rather narrow.  On the other hand, the calculation by Ma and
collaborators, based on CGC+NRQCD, has a broader band.  That is because, in
this calculation, the band corresponds to the range of predictions obtained
by making the ratio $R_{pA}$ for each of the color octet states separately.
In this way, the rather large uncertainties on the individual color octet
matrix elements cancel in the ratios.

In Ref.~\cite{Ma:2015sia}, the authors
noted that if the CGC+NRQCD uncertainty band is based on the individual ratios
of the color octet matrix elements, it should encompass any other calculations
in a similar framework, such as the CGC+CEM calculation of
Ref.~\cite{Ducloue:2016pqr}.
As shown in Fig.~\ref{psi_cgc}, this does indeed
seem to be the case.  The rapidity dependence of the two calculations is nearly
identical.  While the $p_T$-dependent curvature of the two results is somewhat
different, the CGC+CEM calculation is still essentially within the bounds of
the CGC+NRQCD result.

When these results are compared to the collinear factorization calculations with
conventional shadowing and/or final-state energy loss, as in
Figs.~\ref{fig:Jpsi_y} and \ref{fig:Jpsi_pT}, it can be seen that the maximum
CGC+NRQCD suppression as a function of rapidity is similar to that of the
nCTEQ and EPS09 LO suppression factors at forward rapidity.  The minimum
saturation effect
is similar to the minimum effect of nCTEQ.  The energy loss calculation is
within the uncertainty band of the CGC results.

The same general trend is seen as a function of
$p_T$ for both CGC+NRQCD and the calculations with conventional shadowing.
All calculations exhibit $R_{pA}(p_T) \rightarrow 1$ at forward rapidity but the
CGC+NRQCD calculation shows a faster increase with $p_T$ than the
conventional shadowing calculations.  The curvature with CGC+NRQCD is also
somewhat different with a narrowing of the band at $p_T\sim 6$~GeV where there
is a crossover of the $p_T$-dependence of the color octet ratios.  No such
behavior is observed for the collinear factorization calculations.  Nonetheless,
the results from the two approaches
are becoming more similar than earlier CGC calculations shown in
Ref.~\cite{Albacete:2016veq}.

Figure~\ref{psi_cgc} also compares the CGC calculations with the LHCb data
and the preliminary ALICE data
at forward rapidity.  The rapidity dependence of the two approaches
agrees with the data, including the decreasing trend of the data with
rapidity.  In addition, the curvature of the $p_T$ dependence agrees quite
well with the data for the range $p_T \leq 10$~GeV.  While the $p_T$ dependence
is rather different for CGC+CEM and CGC+NRQCD, the data sets cannot distinguish
between the two approaches at this point.

\subsection{Heavy Flavors Hadrons (Z.-B. Kang, J.-P. Lansberg,
  H.-S. Shao, I. Vitev and H. Xing)}
\label{sec:open}

Here cold matter calculations by Vitev and collaborators
are compared to the data-driven
calculations of Lansberg and Shao with shadowing only.  The preliminary
$\mathtt{HIJING++}$ calculations are also shown for $R_{p{\rm Pb}}(y)$.

\subsubsection{$D$ mesons}
\label{sec:Dmesons}

\begin{figure}[hbt]\centering
\includegraphics[width=0.495\textwidth]{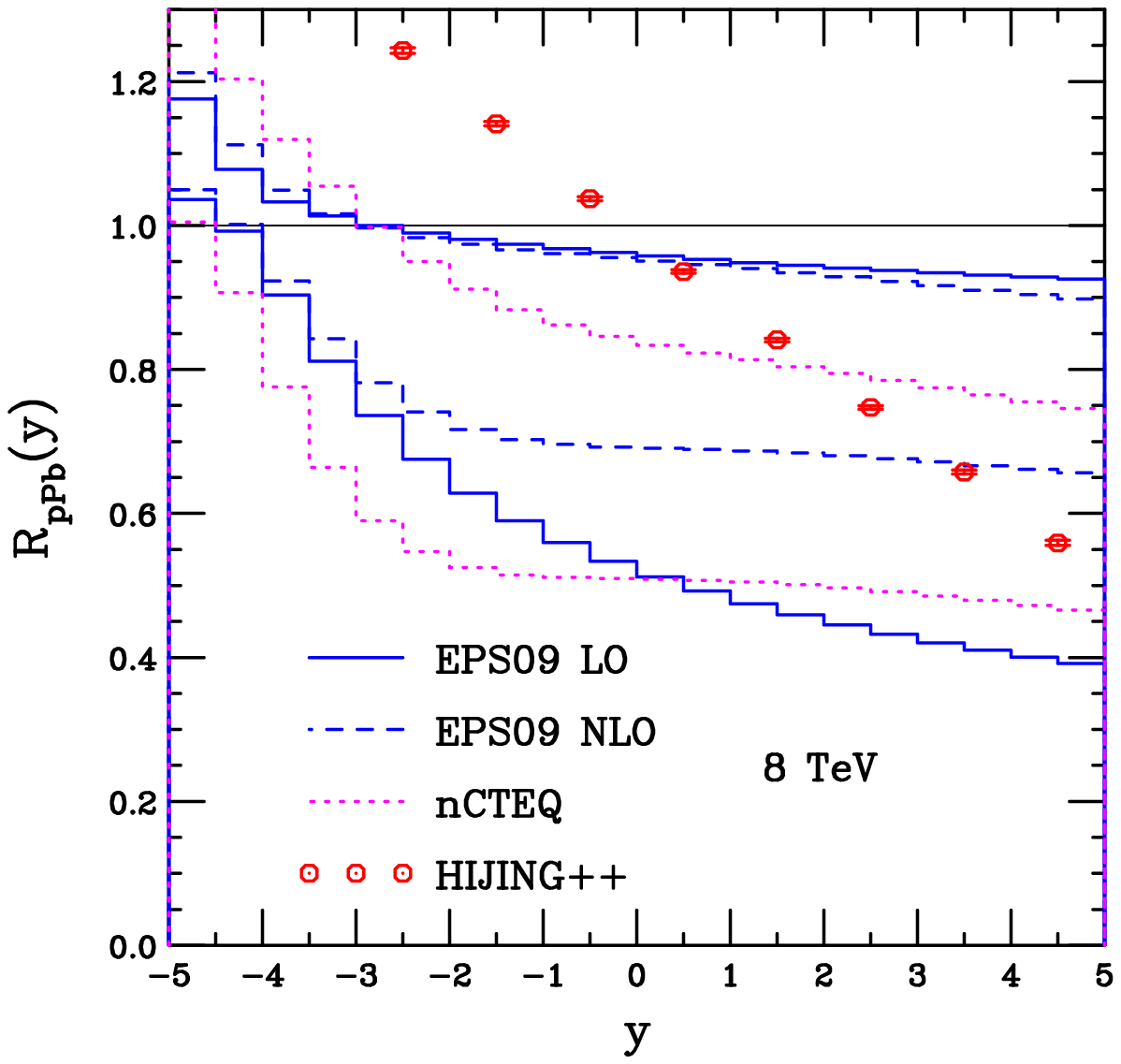}
\caption[]{(Color online)
  Prediction for $D$-meson $R_{p{\rm Pb}}$ in $p+$Pb collisions at 8~TeV
  as a function of rapidity.
The data-driven shadowing results of Lansberg and Shao are shown for EPS09 NLO
(dashed blue histogram), EPS09 LO 
(solid blue histogram) and nCTEQ (dotted magenta histogram).
The $\mathtt{HIJING++}$ calculations are the red points.}
\label{fig:Dmeson_y}
\end{figure}

\begin{figure}[hbt]\centering
\includegraphics[width=0.495\textwidth]{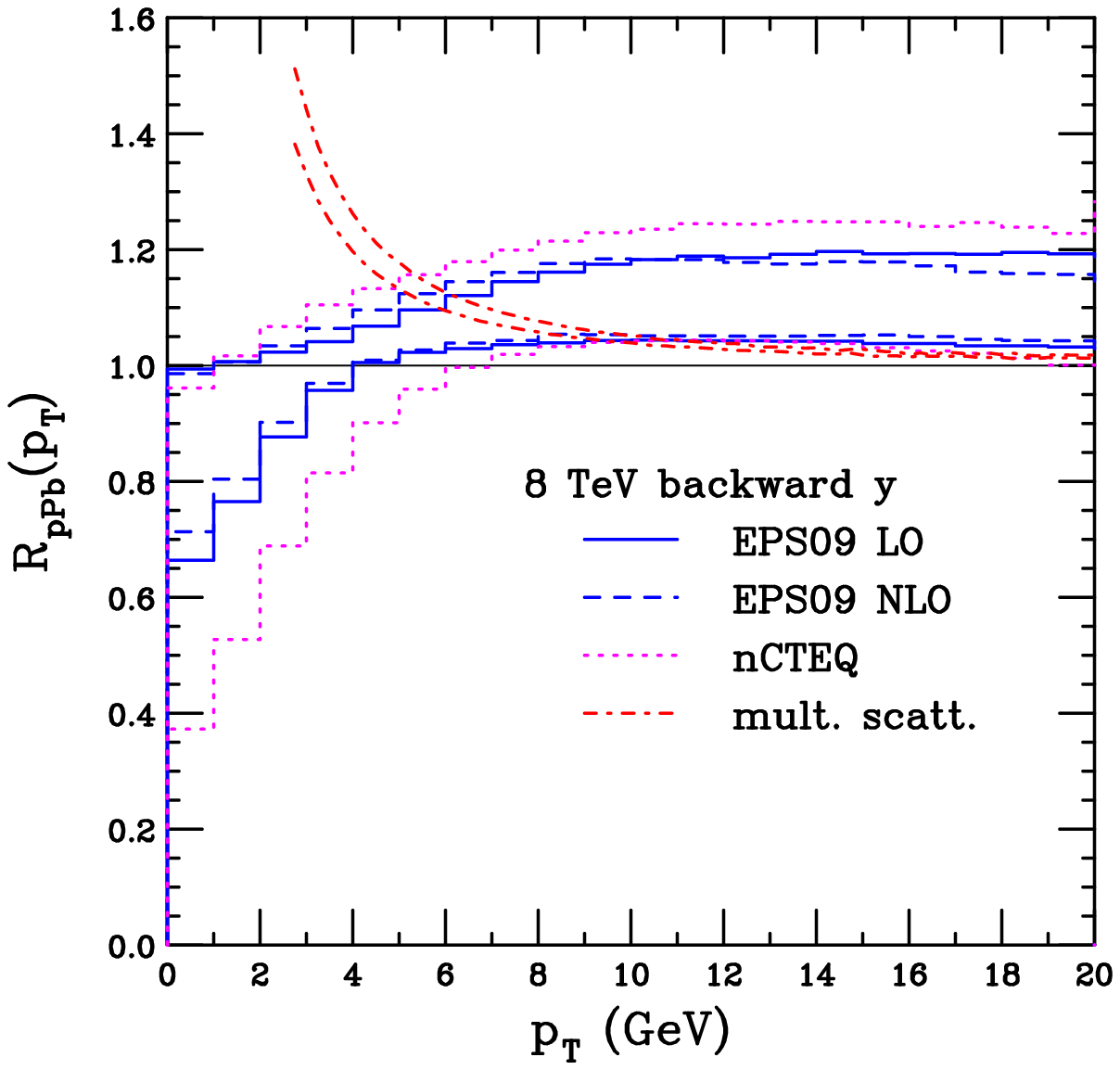}
\caption[]{(Color online)
Prediction for $D$-meson $R_{p{\rm Pb}}$ in $p+$Pb collisions at
$\sqrt{s_{_{NN}}}=8$ TeV with $-4 < y < -2.96$. The red band corresponds to
$0.09 \leq \xi^2 \leq 0.12$~GeV$^2$. The data-driven shadowing results of
Lansberg and Shao at 8 TeV and $-4.46 < y < -2.96$ are shown for EPS09 NLO
(dashed blue histogram), EPS09 LO 
(solid blue histogram) and nCTEQ (dotted magenta histogram).
}
\label{fig:Dmeson_mY}
\end{figure}

\begin{figure}[hbt]\centering
\includegraphics[width=0.495\textwidth]{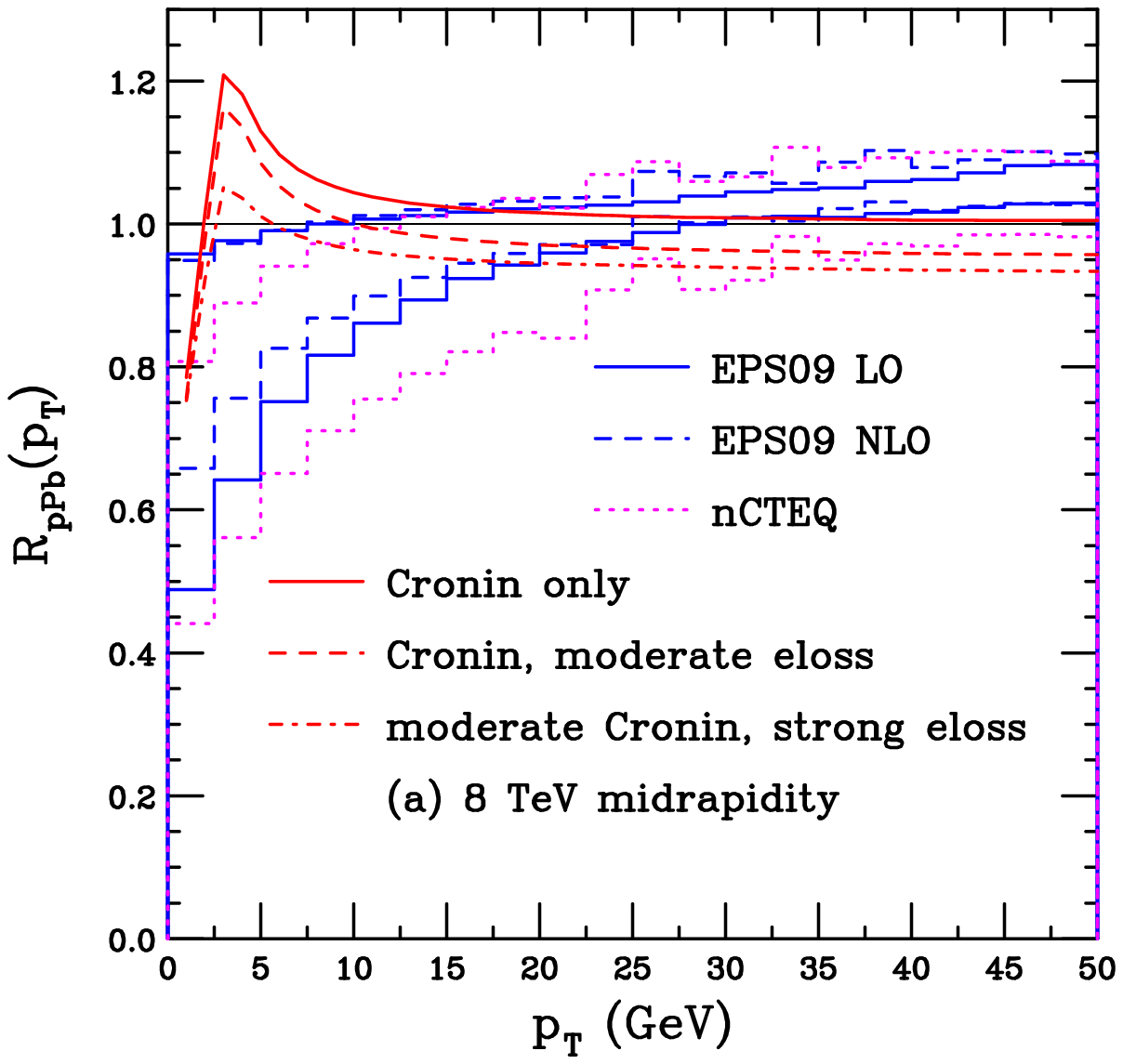}
\includegraphics[width=0.495\textwidth]{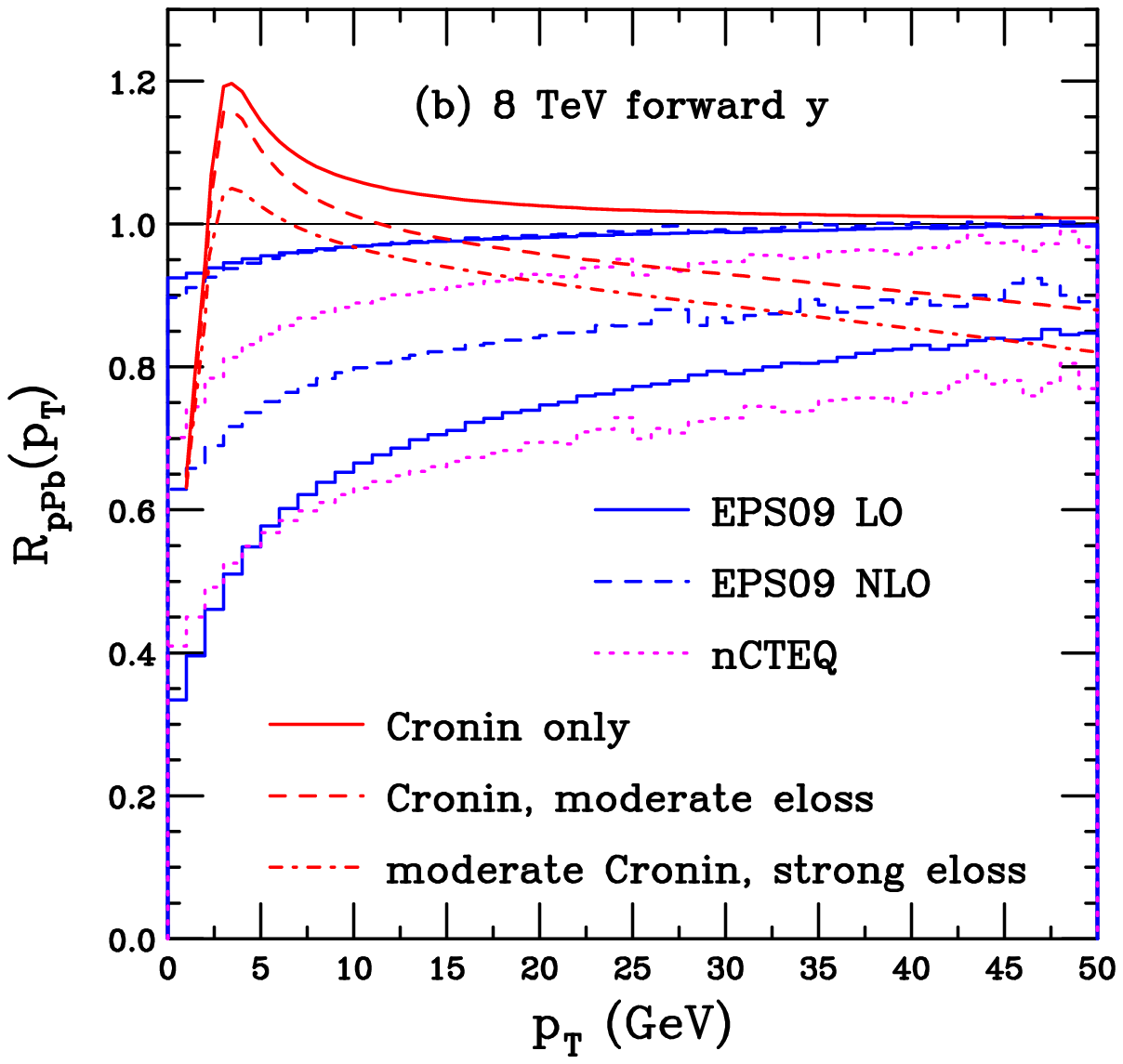}
\caption[]{(Color online)
Prediction for $D$-meson $R_{p{\rm Pb}}$ in $p+$Pb collisions at midrapidity (a)
and forward rapidity (b) at 8~TeV.
The data-driven shadowing results of Lansberg and Shao are shown for EPS09 NLO
(dashed blue histogram), EPS09 LO 
(solid blue histogram) and nCTEQ (dotted magenta histogram).
Results with
Cronin broadening alone (solid red) and with energy loss (full Cronin and 
moderate energy loss, red dashed, reduced Cronin and stronger energy loss,
red dot dashed) are also shown.}
\label{fig:Dmeson_pY}
\end{figure}

The predictions for $D$ meson suppression in $p+$Pb collisions 
at $\sqrt{s_{_{NN}}}=8.16$~TeV in the approaches discussed in this section
are shown in Figs.~\ref{fig:Dmeson_y}-\ref{fig:Dmeson_pY}.

Figure~\ref{fig:Dmeson_y}, compares the data-driven
calculations as a function of rapidity.  In this case, the average $p_T$
parameter in Eq.~(\ref{eq:lansberg_amplitude}) is also allowed to vary
although $n$ is still fixed at $n=2$.  The values of $\langle p_T \rangle$,
$\kappa$ and $\lambda$ are fixed for all rapidity and have a stronger dependence
on proton parton density for $D$ mesons than for $J/\psi$, perhaps because of
the fixed, higher average $J/\psi$ $p_T$, $\langle p_T \rangle = 4.5$~GeV.
Here, for CT14 NLO and nCTEQ, $\kappa = 1.01$, $\lambda = 2.29$, and
$\langle p_T \rangle = 0.88$~GeV, while for CT10 with EPS09 LO and EPS09 NLO,
$\kappa = 1.62$, $\lambda = 2.38$, and $\langle p_T \rangle = 0.52$~GeV.  The
three results for $D^0$ mesons as a function of rapidity look similar to those
of Fig.~\ref{fig:Jpsi_y}.  However, a comparison of Fig.~\ref{fig:Dmeson_y}
with Fig.~\ref{fig:Jpsi_y} shows that the shadowing effect
on $D$ mesons is larger
than for $J/\psi$ at forward rapidity.  Recall that no other nuclear effects
are included.

The $\mathtt{HIJING++}$ calculation is also shown in Fig.~\ref{fig:Dmeson_y}.
The statistics are much higher than in Fig.~\ref{fig:Jpsi_y}(b) since the
calculation accounts for all $D^0$ and $\overline{D}^0$ mesons produced in
hard scatterings with none lost to decays.  Note that while the results at
forward rapidity lie within the large uncertainties of the shadowing
calculations, the curvature is very different.  Indeed, the $\mathtt{HIJING++}$
result is essentially linearly rising as one goes from forward to backward
rapidity, resulting in a relatively large enhancement at backward rapidity.
This enhancement may be due to multiple scattering closer to the lead nucleus.

Figure~\ref{fig:Dmeson_mY} compares the shadowing calculations at backward
rapidity, $-4.46 < y < -2.96$, as a function of $p_T$.  As was the case for
$J/\psi$, in this region there is antishadowing for $p_T > 6$~GeV.   The
maximum shadowing effect is fairly strong in all three cases.  Note also that
here the calculations with EPS09 LO and EPS09 NLO are very similar and only
become distinct at more forward rapidity, see
Fig.~\ref{fig:Dmeson_pY}.

A higher-twist multiple scattering
calculation based on Refs.~\cite{Kang:2014hha,Kang:2013ufa},
within the generalized higher-twist
factorization formalism \cite{Luo:1994np}, is also shown.
The prediction for incoherent multiple scattering on heavy meson 
production is given in the backward rapidity region.
The double scattering contributions to the $D$-meson differential cross
sections are calculated explicitly by taking into account 
both initial-state and final-state interactions. The final result depends on 
both the parameterized twist-4 quark-gluon and gluon-gluon correlation
functions.  Only one parameter, $\xi^2$, characterizing the strength
of parton multiple scattering needs to be determined.

As shown in 
Fig.~\ref{fig:Dmeson_mY}, 
the band corresponds to $0.09 \leq \xi^2 \leq 0.12$~GeV$^2$, 
extracted from DIS data \cite{Qiu:2003vd}.  The double scattering contribution
at $\sqrt{s_{_{NN}}} = 8$~TeV in the backward rapidity region is predicted to
lead to a Cronin-like enhancement in the intermediate $p_T$
range. This feature is understood as the incoherent 
multiple scattering of hard partons in the large nucleus 
\cite{Kang:2014hha,Kang:2013ufa}. Such a feature has already been observed by
the recent measurements at RHIC and the LHC. The backward rapidity
measurements of heavy 
meson production in future LHC $p+A$ programs will provide an excellent 
opportunity to investigate perturbative QCD dynamics and test the 
predictive power of the higher-twist formalism, as well as further constrain 
the properties of cold nuclear matter.

Figure~\ref{fig:Dmeson_pY} presents the results for cold nuclear matter
calculated by Vitev {\it et al.}\
including Cronin broadening and cold matter energy loss along with
the shadowing calculations by Lansberg and Shao.  As was the case for backward
rapidity, the shadowing calculations show a stronger dependence on $p_T$ than
the $J/\psi$.  The effect is larger at lower $p_T$ due to the lower overall
scale, $m_T \sim M_D$ at $p_T \sim 0$ rather than $m_T \sim M_{J/\psi}$.  Note
also the larger $p_T$ scale in Figs.~\ref{fig:Dmeson_mY} and
\ref{fig:Dmeson_pY}.  At midrapidity, there is some antishadowing seen for
$p_T > 20$~GeV.  The Cronin effect included in the cold nuclear matter 
calculation by Vitev {\it et al.}\
results in a low $p_T$ enhancement.  The chosen assumptions
for the two different calculations (Cronin and energy loss vs. shadowing
only) result in the largest difference for
$p_T < 10$~GeV.  At higher $p_T$, even very precise data can likely not
distinguish between the approaches unless there is a clear trend with $p_T$
and, even in this case, it could be difficult.

The LHCb Collaboration has recently released data on $R_{p{\rm Pb}}$ in $p+$Pb
collisions at 5~TeV \cite{LHCbD0}.  They have determined the nuclear
modification factor as a function of rapidity and as a function of transverse
momentum at forward and backward rapidity.  These data agree with the
shadowing only calculations presented here and do not exhibit any low $p_T$
enhancement as might be expected from Cronin enhancement, see
Ref.~\cite{LHCbD0}.

\subsubsection{$B$ mesons}
\label{sec:Bmesons}

\begin{figure}[hbt]\centering
\includegraphics[width=0.495\textwidth]{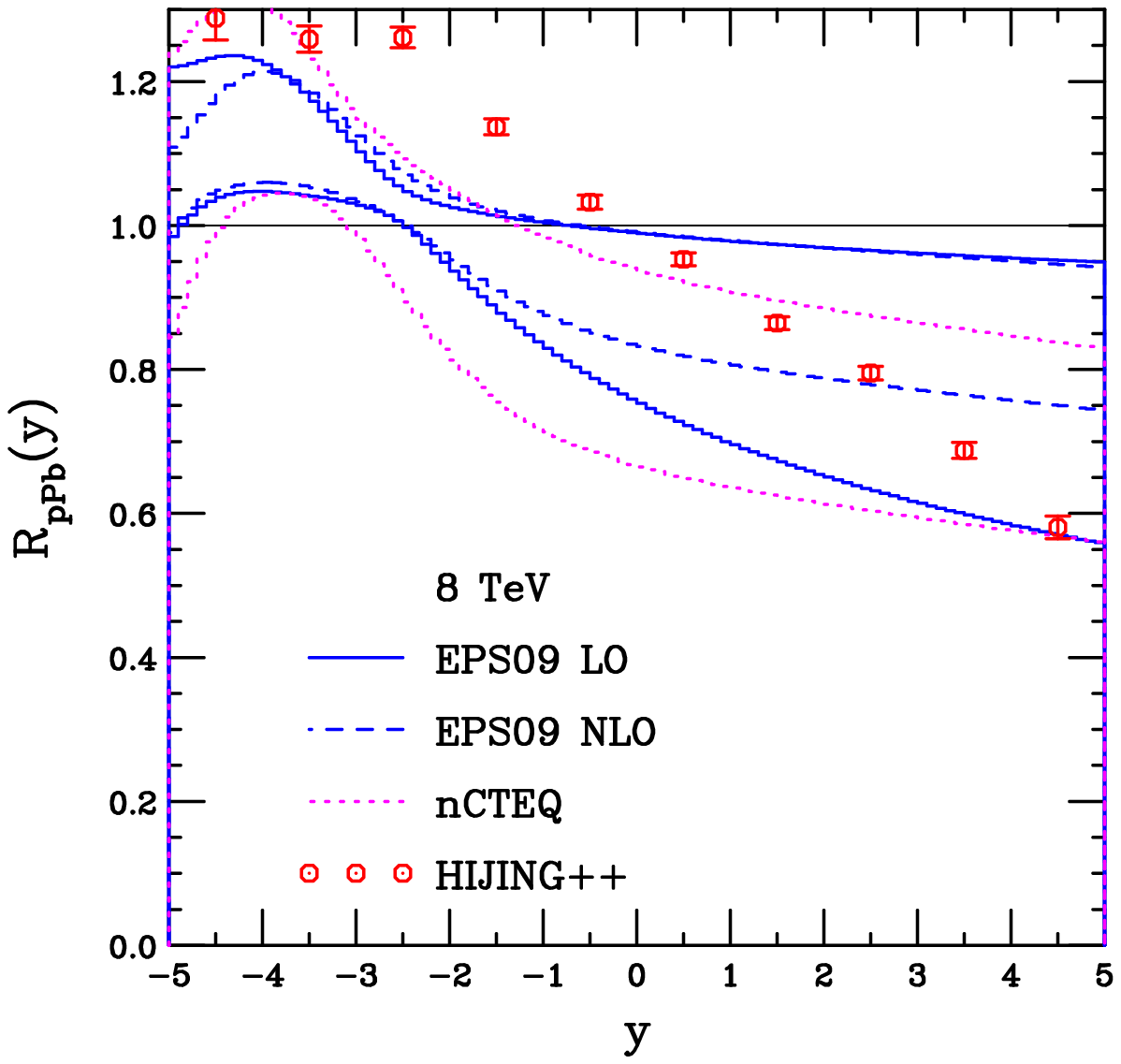}
\caption[]{(Color online)
Prediction for $B$-meson $R_{p{\rm Pb}}$ in $p+$Pb collisions as a function of
rapidity at 8~TeV.
The data-driven shadowing results of Lansberg and Shao are shown for EPS09 NLO
(dashed blue histogram), EPS09 LO 
(solid blue histogram) and nCTEQ (dotted magenta histogram).
The $\mathtt{HIJING++}$ calculations are the red points.}
\label{fig:Bmeson_y}
\end{figure}

\begin{figure}[hbt]\centering
\includegraphics[width=0.495\textwidth]{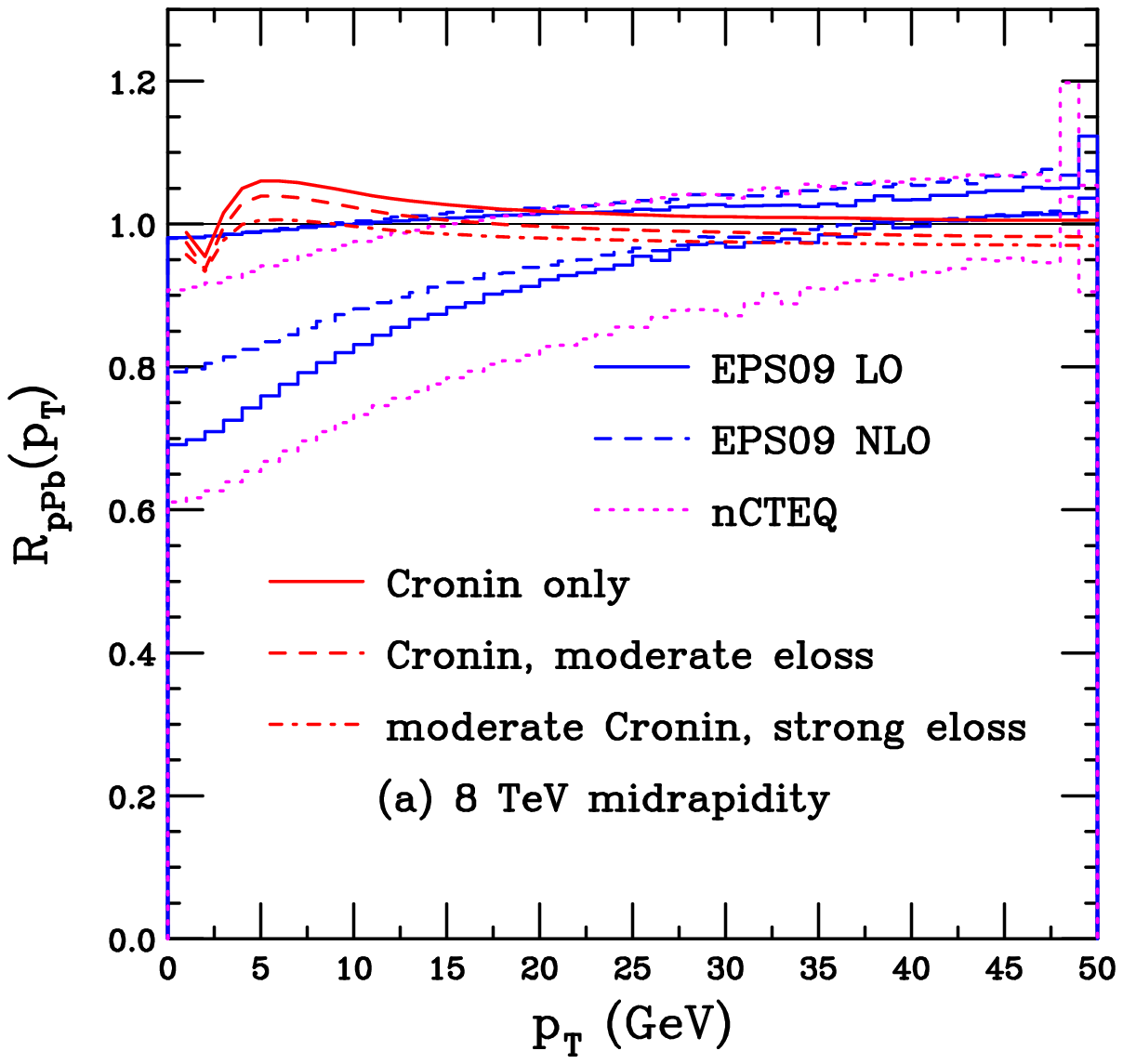}
\includegraphics[width=0.495\textwidth]{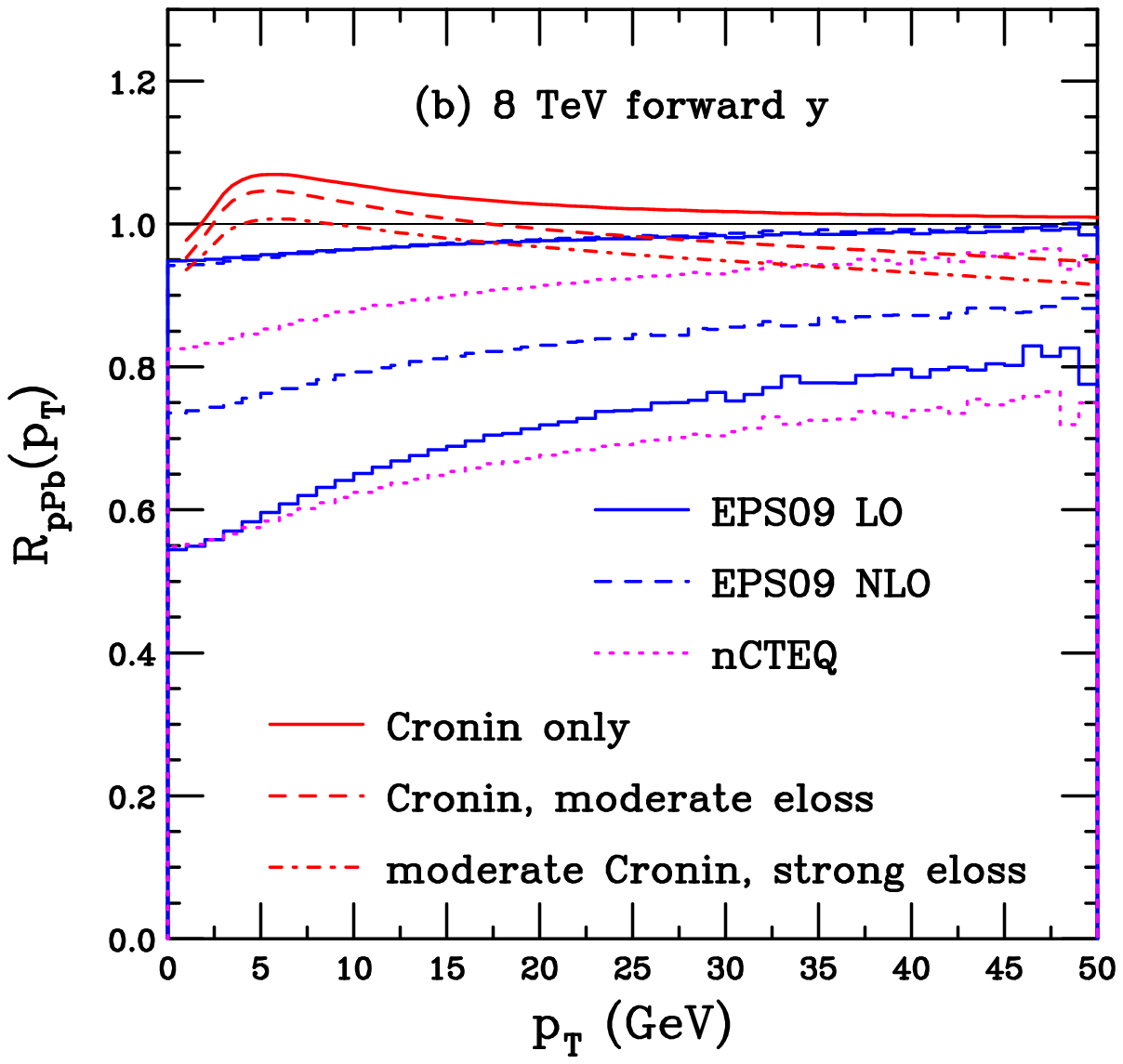}
\caption[]{(Color online)
Prediction for $B$-meson $R_{p{\rm Pb}}$ in $p+$Pb collisions at midrapidity (a)
and forward rapidity (b) at 8~TeV.
The data-driven shadowing results of Lansberg and Shao are shown for EPS09 NLO
(dashed blue histogram), EPS09 LO 
(solid blue histogram) and nCTEQ (dotted magenta histogram).
Results with
Cronin broadening alone (solid red) and with energy loss (full Cronin and 
moderate energy loss, red dashed, reduced Cronin and stronger energy loss,
red dot dashed) are also shown.}
\label{fig:Bmeson_pT}
\end{figure}

The predictions for $B$-meson suppression in $p+$Pb collisions 
at $\sqrt{s_{_{NN}}}=8.16$~TeV 
are shown in Figs.~\ref{fig:Bmeson_y} and \ref{fig:Bmeson_pT}.

In Fig.~\ref{fig:Bmeson_y}, the data-driven calculations
with EPS09 LO, EPS09 NLO and nCTEQ are compared.  The parameters for these
calculations were not included in Ref.~\cite{Lansberg:2016deg} but were
determined for this report.  They were obtained by fitting the LHCb data
at 7 TeV \cite{Aaij:2013noa}.
The power $n$ in Eq.~(\ref{eq:lansberg_amplitude})
was again kept fixed at $n= 2$ but $\langle p_T \rangle$, $\kappa$ and $\lambda$
were fit.  The values were similar for the two proton parton densities:
$\langle p_T \rangle = 5.51$~GeV, $\kappa = 0.56$ and $\lambda = 1.05$ for
CT10 NLO and EPS09 LO, EPS09 NLO while 
$\langle p_T \rangle = 4.96$~GeV, $\kappa = 0.58$ and $\lambda = 1.02$ for
for CT14 NLO and nCTEQ.
There is significantly less suppression due to shadowing
than for $D^0$ mesons but somewhat stronger shadowing than for $\Upsilon$
production in Fig.~\ref{fig:Ups_y}.

The $\mathtt{HIJING++}$ calculation is shown by the points in
Fig.~\ref{fig:Bmeson_y}.  As with the $J/\psi$ and $D^0$ results, the
rapidity dependence of the $B^+$ ratio for $\mathtt{HIJING++}$ is within
the uncertainties of the nPDF results at forward rapidity.  In this case,
however, the enhancement at backward is not as large and, for $y < -3$, the
result is compatible with the nCTEQ uncertainty.

A comparison of the $\mathtt{HIJING++}$ calculations in
Figs.~\ref{fig:Jpsi_y}(b), \ref{fig:Dmeson_y} and \ref{fig:Bmeson_y} shows that,
at forward rapidity, the results are all compatible and exhibit a linear
dependence on rapidity that is typically stronger than all the calculations
with nPDF modifications alone.  This is likely because of the strong low-$x$
shadowing of the parameterization used in $\mathtt{HIJING++}$.  This
parameterization does not yet include any scale evolution.  Thus the only
scale dependence in the $\mathtt{HIJING++}$ calculation arises from that of
the proton PDF.  There are differences at more backward rapidity which likely
arise from the hadronization model of $\mathtt{PYTHIA8}$ and multiple
scatterings near the target rapidity region.

Figure~\ref{fig:Bmeson_pT} compares all the shadowing calculations at
midrapidity and forward rapidity as a function of $p_T$ with the Cronin
and energy loss cold matter calculations by Vitev {\it et al.}.
The suppression factor
at $p_T \sim 0$ is smaller than for $\Upsilon$ and also has more curvature
at midrapidity.  The effect due to the nuclear parton density decreases
faster with $p_T$
than for $\Upsilon$.  The Cronin peak here is much
smaller, less than 10\% even at forward rapidity.   Note that although these
results show a small enhancement as a function of $p_T$, the effect is reduced
relative to $D$ mesons because the $B$ calculation uses a larger factorization
scale.

\section{Drell-Yan Production (F. Arleo)}
\label{sec:DY}

Measurements of $J/\psi$ production in $p+$Pb collisions at the LHC at
$\sqrt{s_{_{NN}}} =5.02$~TeV by ALICE~\cite{Abelev:2013yxa} and
LHCb~\cite{Aaij:2013zxa} has triggered an intense debate on the origin of the
reported nuclear suppression, which could be attributed to either modifications
of the nuclear parton
distribution~\cite{Albacete:2013ei,Vogt:2015uba,Ferreiro:2013pua} or coherent
energy loss effects~\cite{Arleo:2012hn,Arleo:2012rs,Peigne:2014uha}, see
Sec.~\ref{sec:onia}. It was suggested in Ref.~\cite{Arleo:2015qiv} that the
Drell-Yan process could play a key role in clarifying the origin of the
quarkonium suppression reported in $p+$Pb collisions at the LHC since no
coherent energy loss is expected on Drell-Yan production in this framework.

Predictions of
the nuclear modification factor, $R^{\rm DY}_{p {\rm Pb}}$, of low-mass Drell-Yan
lepton pairs in $p+$Pb collisions at $\sqrt{s_{_{NN}}}=8.16$~TeV are given.
Using the $\mathtt{DYNNLO}$~\cite{Catani:2007vq,Catani:2009sm}
Monte Carlo program, the
NLO single differential cross section ${\rm d}\sigma/{\rm d}y$ is computed in
$p+p$ and $p+$Pb collisions at $\sqrt{s_{_{NN}}}=8.16$~TeV and
$R_{p {\rm Pb}}^{\rm DY}(y)$ is calculated.
The MSTW NLO~\cite{Martin:2009iq} proton PDFs are used with factorization and
renormalization scales equal to the lepton pair mass, $M_{\rm DY}$. The $p+$Pb
calculations were carried out using the NLO nPDF sets
EPS09~\cite{Eskola:2009uj}, DSSZ~\cite{deFlorian:2011fp} and
nCTEQ15~\cite{Kovarik:2015cma}. For completeness, the DY cross section was
also computed in $p+$Pb collisions assuming no nPDF corrections. The lepton
pair mass range considered in this calculation is $10.5 < M_{\rm DY} < 20$~GeV.

\begin{figure}[t]
\begin{center}
    \includegraphics[width=8.8cm]{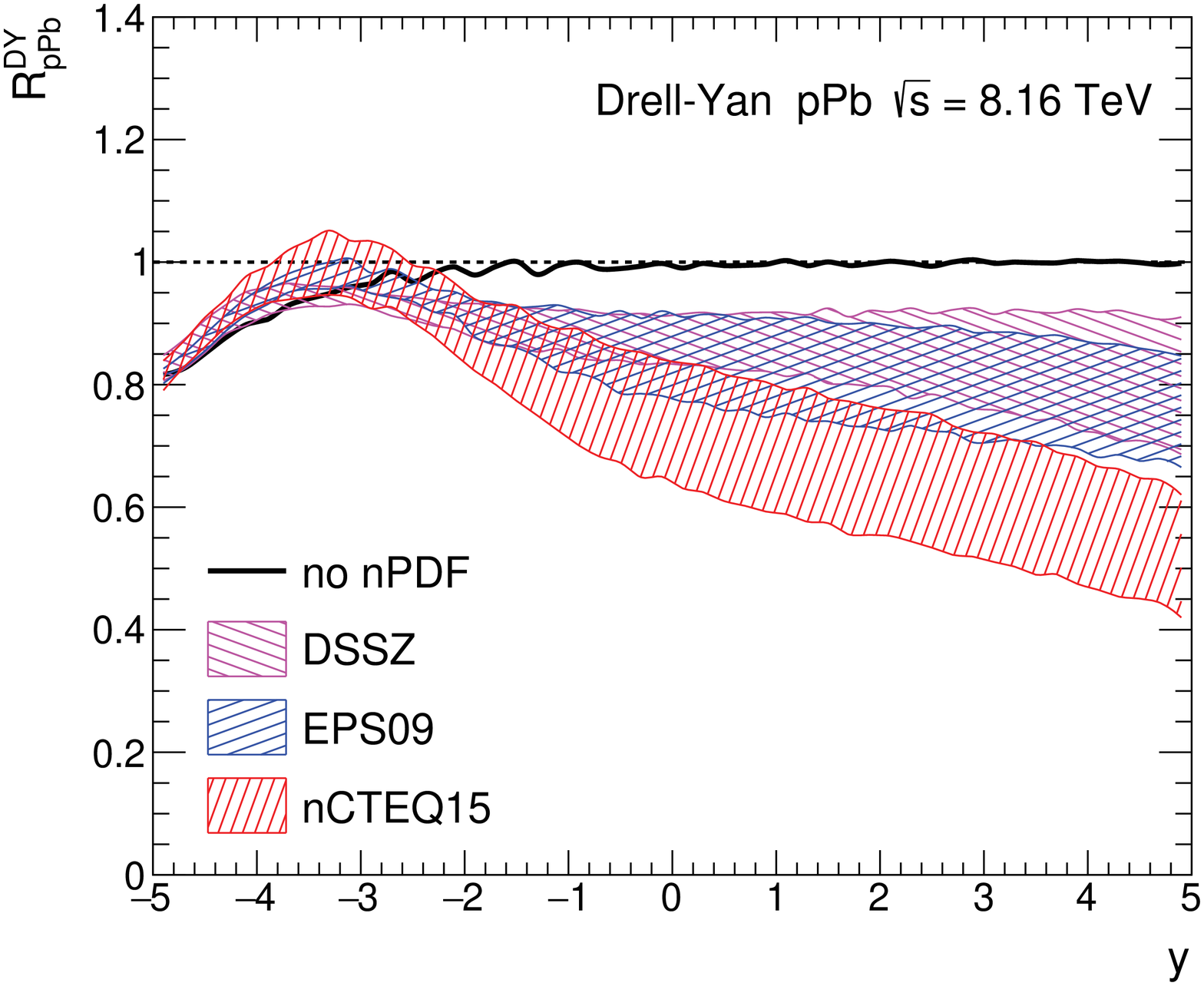}
  \end{center}
\caption[]{(Color online) The Drell-Yan nuclear suppression factor in $p+$Pb
  collisions 
  at $\sqrt{s_{_{NN}}}=8.16$~TeV for the DSSZ (magenta), EPS09 (blue), nCTEQ15
  (red) and isospin only (black line)
  \protect\cite{Arleo:2015qiv}.}
  \label{fig:psi_dy}
\end{figure}

The Drell-Yan
suppression in $p+$Pb collisions is shown in Fig.~\ref{fig:psi_dy} as a
function of the lepton pair rapidity. In the most forward bins,
$3 \lesssim y \lesssim 5$ (corresponding to
$10^{-5} \lesssim x_2 \lesssim 10^{-4}$ using
$x_2 = {M_{\rm DY}}\,e^{-y}/{\sqrt{s_{NN}}}$), the suppression is quite strong,
$R_{p {\rm Pb}}^{\rm DY} \simeq 0.4$--$0.7$,
using nCTEQ15.  It is less pronounced using
DSSZ or EPS09, giving $R_{\rm pPb}^{\rm DY}\simeq 0.7$--$0.9$. These calculations
demonstrate the discriminating power of low-mass Drell-Yan production in $p+$Pb
collisions at the LHC and could set tight constraints on antiquark shadowing
at very small $x$.
In the backward region ($y<0$) the depletion of Drell-Yan production in $p+$Pb
with respect to $p+p$ collisions is due to isospin effects~\cite{Arleo:2015qiv}.

\section{Jets}
\label{sec:jets}

Two results are presented here.  The first, by Vitev, focuses on cold matter
energy loss.  The second, by Kotko {\it et al.}, discusses saturation in
forward-forward dijet production.

\subsection{Jet $R_{p{\rm Rb}}$ (I. Vitev)}
\label{sec:Vitev_jets}

\begin{figure}[hbt]\centering
\includegraphics[width=0.495\textwidth]{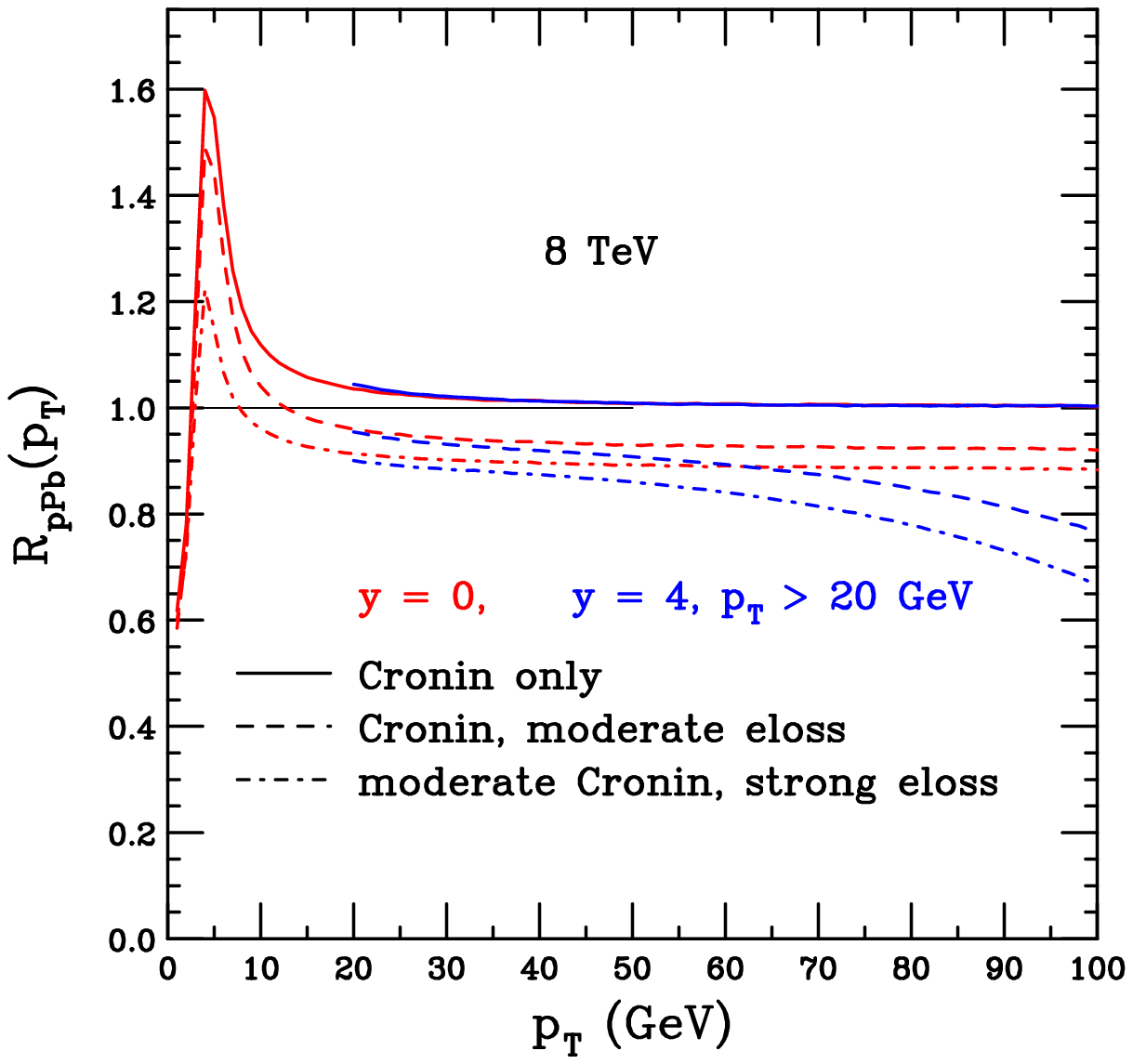}
\caption[]{(Color online)
Prediction for single jet $R_{p{\rm Pb}}$ in $p+$Pb collisions at mid ($y=0$)
and forward ($y=4$) rapidity at 8~TeV.
Results with
Cronin broadening alone (solid) and with energy loss (full Cronin and 
moderate energy loss, dashed, and reduced Cronin and stronger energy loss,
dot dashed) are shown.  The midrapidity ratios are given in red and the
$y = 4$ results for $p_T > 20$ GeV are shown in blue.}
\label{fig:jet_pT}
\end{figure}

Calculations with cold matter energy loss and Cronin broadening are
shown in Fig.~\ref{fig:jet_pT}.  The results are calculated at $y=0$ and
$y=4$.  At forward rapidity, only the results for $p_T > 20$~GeV are shown
since they match those at lower $p_T$.  The results without energy loss are
on top of each other over all $p_T$.  When energy loss is included, since the
effect is stronger at high $p_T$, the curves at $y = 0$ and $y=4$
start to deviate at this value.  By $p_T \sim 100$~GeV, they differ by 20\%
for moderate energy loss and 33\% for strong energy loss.
Note also that the Cronin peak is large for `jets' with $p_T < 10$~GeV, even
larger than for light hadrons due to the massless parton initiating the jet. 
Jets at low transverse momentum cannot be reliably reconstructed.  The Cronin
enhancement shown here will manifest itself in the Cronin enhancement of light
hadrons which will be lower and shifted in $p_T$ due to fragmentation.

\subsection{Forward jets (P. Kotko, K. Kutak and S. Sapeta)}
\label{sec:KKS}

A summary of results predicting
\cite{Albacete:2016veq} saturation \cite{Gribov:1984tu}
effects in production of forward-forward dijets in $p+$Pb scattering at the LHC 
at 8 TeV is presented here. In particular, the prescription proposed in
Ref.~\cite{Kutak:2014wga} to include the hard 
scale dependence in the small~$x$ gluon evolution equations using the Sudakov 
form factor is applied.  (For other approaches, see
Refs.~\cite{vanHameren:2014ala,Mueller:2013wwa}).  
In these calculations, the high energy factorization
formalism~\cite{Catani:1990eg}, which accounts for both the high energy scale of
the scattering and the hard momentum scale $p_T$ provided by the produced hard
system, is applied. Furthermore, the formalism is a very good approximation
of the predictions of the
nuclear modification factors obtained recently within the improved transverse
momentum dependent
formalism \cite{Kotko:2015ura,vanHameren:2016ftb}.
In the asymmetric configuration, the high energy factorization formula 
is \cite{Deak:2010gk}
\begin{eqnarray}
  \frac{d\sigma}{dy_1dy_2dp_{T,1}dp_{T,2}d\Delta\phi} 
  & = &
  \sum_{a,c,d} 
  \frac{p_{T,1}p_{T,2}}{8\pi^2 (x_1x_2 S)^2}{\cal M}_{ag^*\rightarrow cd} \nonumber \\
  & \times & x_1 f_{a/A}(x_1,\mu^2)\,{\cal F}_{g/B}(x_2,k^2_T,\mu)
  \frac{1}{1+\delta_{cd}}\,,
\label{eq:cs-fac}
\end{eqnarray}
where
\begin{eqnarray}
  k^2 = p_{T,1}^2 + p_{T,2}^2 + 2p_{T,1}p_{T,2} \cos\Delta\phi\,.
\end{eqnarray}
It is assumed that $x_1\simeq 1$ and
$x_2 \ll 1$ where
$\Delta\phi$ is the azimuthal distance between the outgoing
partons. The squared matrix element, ${\cal M}_{ag^*\rightarrow cd}$,
includes $2\to 2$ scattering with one off-shell initial state gluon, $g^*$,
and three on-shell
partons, $a,c$, and $d$. On the side of the off-shell gluon in
Eq.~(\ref{eq:cs-fac}), the unintegrated gluon density, 
${\cal F}_{g/B}(x_2,k^2,\mu^2)$~\cite{Kutak:2003bd,Kutak:2004ym,Kutak:2012rf}, 
is employed.  It depends on the
longitudinal momentum fraction, $x_2$,
the transverse momentum of the off-shell gluon, $k_T$, and hard scale $\mu$,
taken to be, for example, the average transverse momentum of the two leading
jets. 
On the side of the on-shell parton, probed at high values of
$x_1$, it is legitimate to use the collinear
parton density~$f_{a/A}(x_1,\mu^2)$.

\begin{figure}[t!]
  \begin{center}
\stackinset{l}{0.215\textwidth}{b}{0.095\textheight}{(a)}{%
  \includegraphics[width=0.495\textwidth, angle=-90]{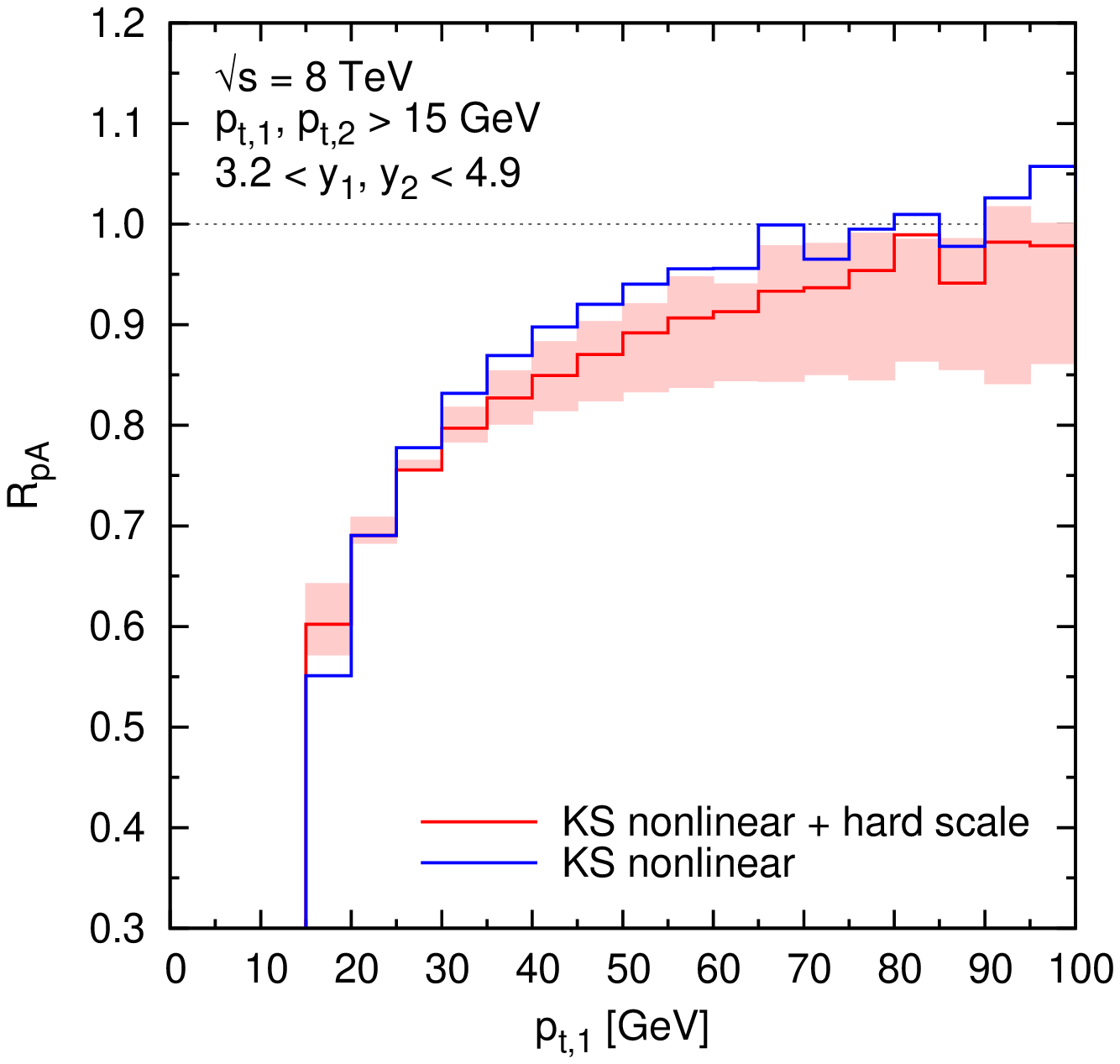}%
}
\stackinset{l}{0.1\textwidth}{b}{0.18\textheight}{(b)}{%
  \includegraphics[width=0.495\textwidth, angle=-90]{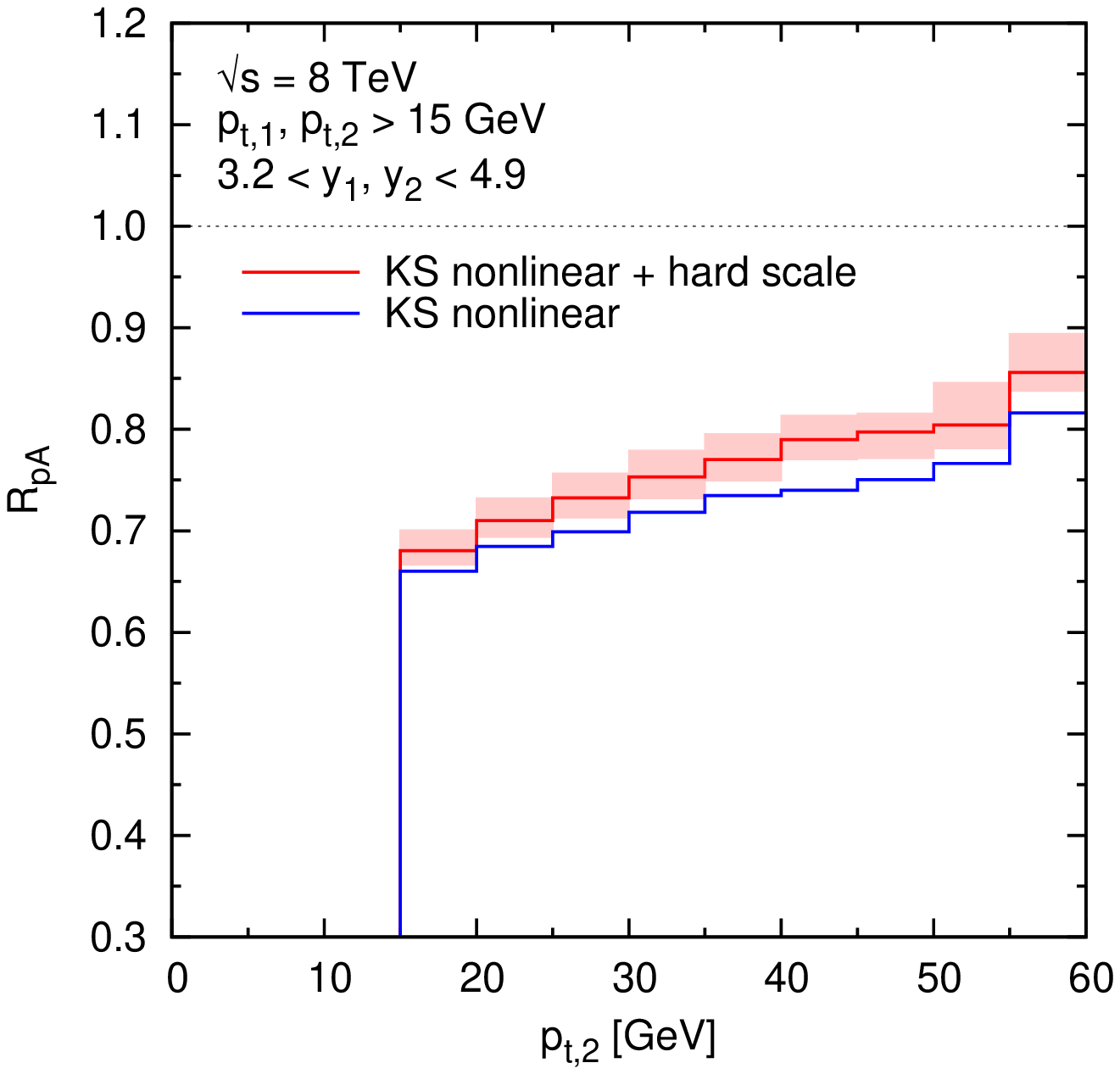}%
}
\stackinset{l}{0.12\textwidth}{b}{0.085\textheight}{(c)}{%
  \includegraphics[width=0.495\textwidth, angle=-90]{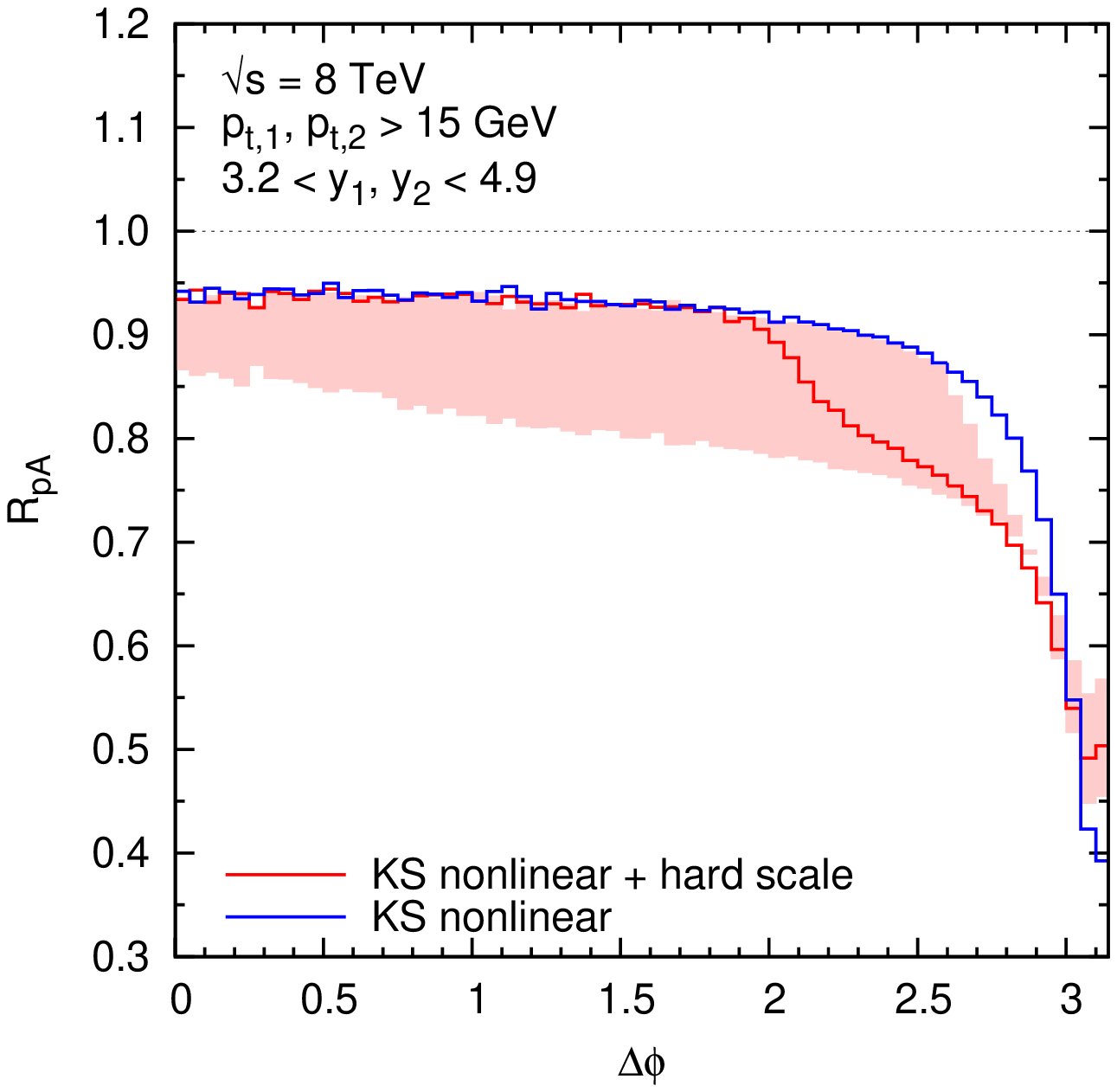}%
}
  \end{center}
  \vspace{-10pt}
  \caption[]{(Color online)
The blue lines correspond to predictions obtained with 
the KS gluon density~\protect\cite{Kutak:2012rf} while the red lines are
predictions using a hard-scale-dependent gluon density
\protect\cite{Kutak:2014wga} at 8~TeV. In both
cases, the renormalization and factorization scales are set to 
$\mu= (p_{T,1}+p_{T,2})/2$.
The light red bands illustrate the scale variation by factors of 0.5 and 2
for the ``KS nonlinear + hard scale'' result.
The analogous variation for the pure KS gluon gives a negligible effect.
The suppression factor for the hardest jet is shown in (a) while that for
the subleading jet is shown in (b).  The nuclear
modification factor as a function of azimuthal angle between jets is given
in (c).
  }
\label{fig:decor-p-Pb}
\end{figure}

At 8~TeV, $x_2$ can be as low as $10^{-5}$ in the forward-forward jet
configuration so that fairly strong suppression at low $p_T$ for the hardest
jet and large azimuthal separation may be observed.

In Fig.~\ref{fig:decor-p-Pb}(a), the ratios of the
differential
cross sections for forward-forward dijet production in $p+$Pb relative to
$p+p$ collisions is shown as a
function of the $p_T$ of the hardest jet.  
Figure~\ref{fig:decor-p-Pb}(b) shows the spectral ratio of the subleading
(second hardest) jet.

All results were obtained with the CT10
NLO PDFs~\cite{Lai:2010vv} on the side of the projectile.
The blue histograms correspond to the KS gluon
density~\cite{Kutak:2012rf} while the red histograms include
Sudakov resummation effects as well as the
KS gluon density~\cite{Kutak:2014wga}.  The light red band around the
hard scale result shows the effect of varying the scale,
$\mu = (p_{T,1}+p_{T,2})/2$, by a factor of two on either side: $\mu/2$ and
$2\mu$.  The effect of the scale variation is negligible for the KS gluon
density alone (blue curves).

\section{Photons (I. Vitev)}
\label{sec:photons}

\begin{figure}[hbt]\centering
\includegraphics[width=0.495\textwidth]{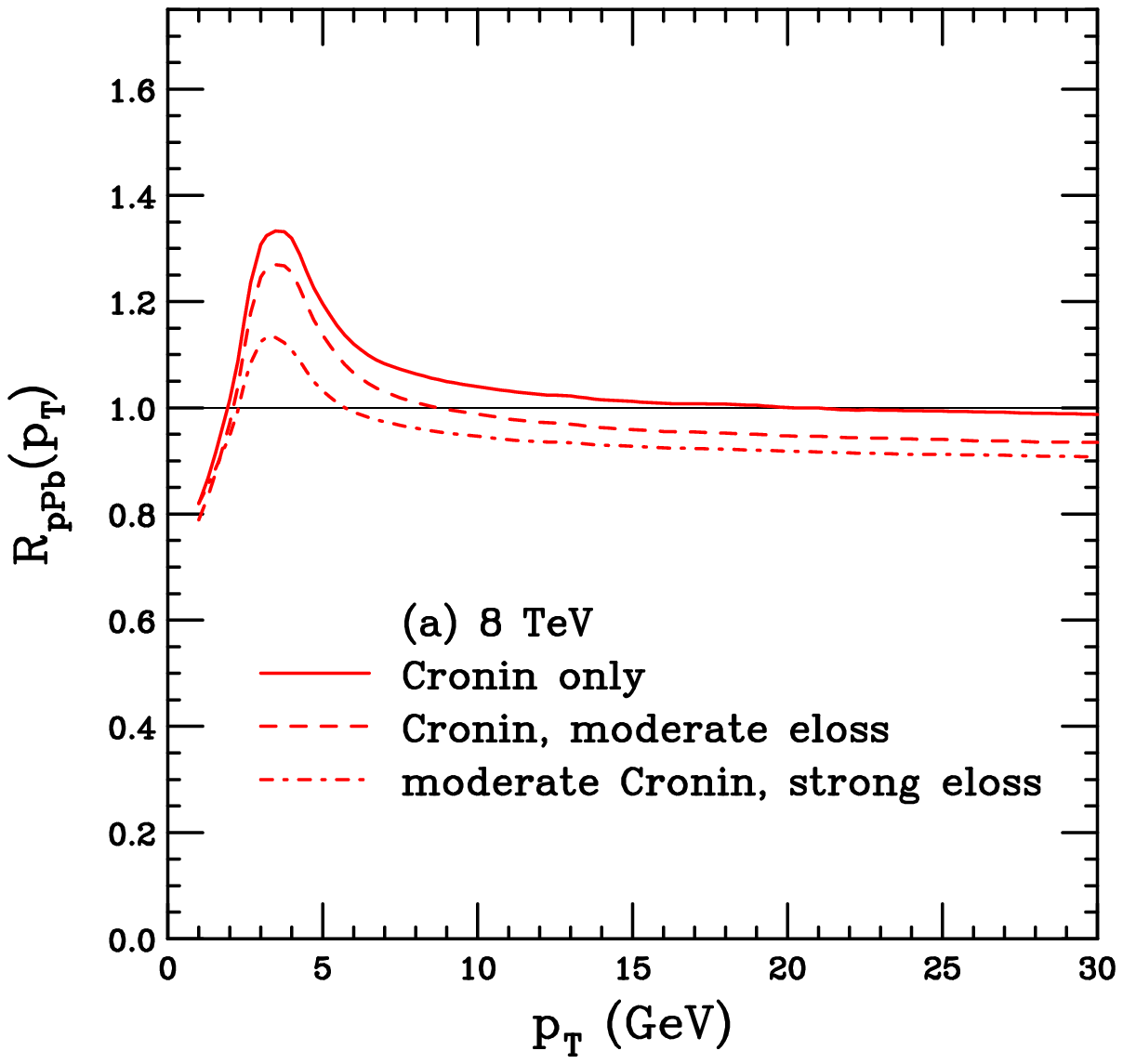}
\includegraphics[width=0.495\textwidth]{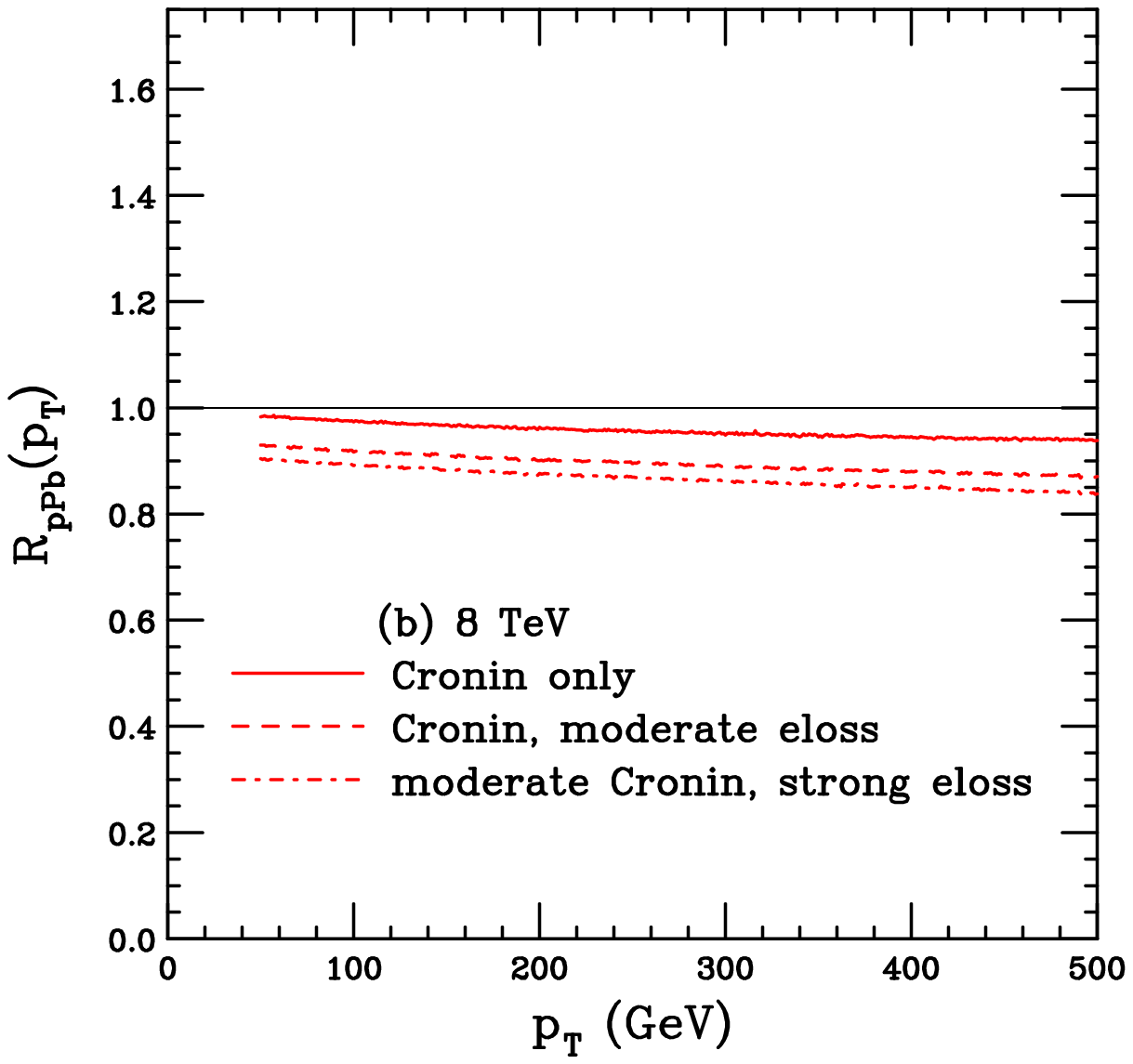}
\caption[]{(Color online)
Prediction for direct photon $R_{p{\rm Pb}}$ in $p+$Pb collisions at 
midrapidity for $p_T < 30$~GeV (a)
and $p_T > 50$~GeV (b) in 8~TeV collisions.
Results with
Cronin broadening alone (solid) and with energy loss: full Cronin and 
moderate energy loss (dashed) and reduced Cronin but stronger energy loss
(dot-dashed), are shown.}
\label{fig:photons}
\end{figure}

The direct photon $R_{p{\rm Pb}}(p_T)$ calculated by Vitev is shown in
Fig.~\ref{fig:photons}.  The results are divided into two parts to emphasize
the different regions: a low $p_T$
part, $p_T < 50$~GeV, shown in (a) and a high $p_T$ part,
$p_T > 50$~GeV, shown in (b).  Both results are shown at
midrapidity.

At low $p_T$, there is
significant enhancement for $p_T < 7$~GeV due to the Cronin effect.  At high
$p_T$, isospin dominates the solid curve labeled Cronin only and is the main
contribution to the decrease at high $p_T$.  The energy loss increases the
high $p_T$ suppression, similar to the effect at low $p_T$.

\section{Gauge Bosons (P. Ru, S. A. Kulagin, R. Petti, E. Wang, B.-W. Zhang
  and W.-N. Zhang)}
\label{sec:WZ_prod}

The predictions for weak gauge boson production in proton-lead collisions at 
$\sqrt{s_{_{NN}}}=8$~TeV are made within the framework of pQCD.
The numerical results, at NLO accuracy, are 
calculated using $\mathtt{DYNNLO}$~\cite{Catani:2007vq,Catani:2009sm}, 
incorporating nuclear parton distribution functions. In this study, three sets 
of nPDFs are used in the numerical 
simulations: the central
EPS09 NLO~\cite{Eskola:2009uj} and
nCTEQ~\cite{Schienbein:2009kk,Kovarik:2010uv} sets are
both matched with the CT10 NLO~\cite{Lai:2010vv,Gao:2013xoa} proton PDFs while
the KP~\cite{Kulagin:2004ie,Kulagin:2014vsa} set is matched with the 
ABMP15~\cite{Alekhin:2015cza} proton PDFs.
The baseline results mentioned in the following refers to
the results obtained with isospin alone,
without any other cold nuclear matter effects.
Thus the baseline results for the EPS09 and nCTEQ are
calculated with CT10 proton PDFs and isospin while those for KP are
calculated with ABMP15 and isospin.
Both the factorization and renormalization scales are set to the
gauge boson mass.
More details of the calculations can be found in 
Refs.~\cite{Albacete:2016veq,Ru:2014yma,Ru:2016wfx}.

\subsection{$W^\pm$ production}

\subsubsection{Charged lepton pseudorapidity distribution}
\begin{figure}[h]
  \stackinset{l}{0.21\textwidth}{b}{0.4\textheight}{(a)}{%
    \stackinset{l}{0.67\textwidth}{b}{0.4\textheight}{(b)}{%
      \stackinset{l}{0.21\textwidth}{b}{0.065\textheight}{(c)}{%
        \stackinset{l}{0.67\textwidth}{b}{0.065\textheight}{(d)}{%
  \includegraphics[width=\textwidth]{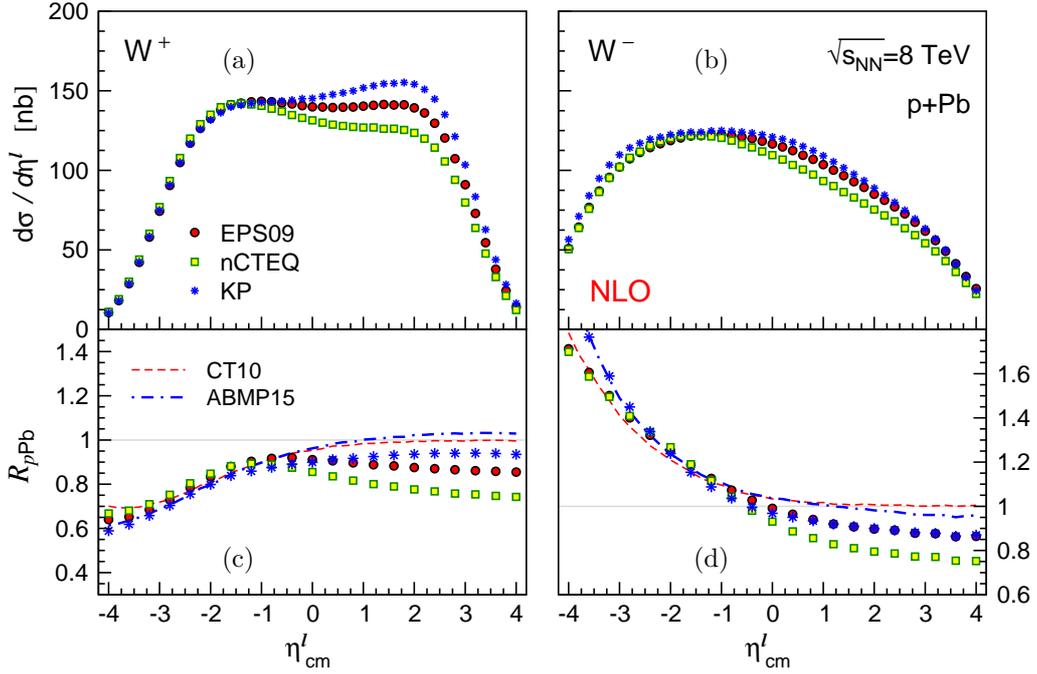}%
  }}}}
\caption[]{(Color online)
The differential cross section as a function of the charged lepton
pseudorapidity for $W^+$~(a) and $W^-$~(b) production in $p+$Pb collisions at
8~TeV for EPS09 (red circles), nCTEQ (green squares) and KP (blue stars).
The corresponding $R_{p{\rm Pb}}$ calculations are 
shown in (c) and (d) respectively, along with the nuclear modification factor
due to isospin alone, with CT10 (red dashes) and ABMP15 (blue dot-dashed).
The range of the charged-lepton transverse momentum is 
$p_T^l>25$~GeV~\protect\cite{Khachatryan:2015hha}.}
\label{fig:W-eta@pPb}
\end{figure}

The differential cross sections for $W^+$ and $W^-$ production are shown
as a function of the charged lepton pseudorapidity
in the center of mass frame. The final-state cut on the charged 
lepton transverse momentum, $p_T^l>25$~GeV,
is used according to the CMS measurement at 
$\sqrt{s_{_{NN}}}=5.02$~TeV ~\cite{Khachatryan:2015hha}.
The results are shown in Fig.~\ref{fig:W-eta@pPb}(a) and (b) for $W^+$ and
$W^-$ respectively.
Obvious differences among the three nPDFs can be seen in the forward direction, 
especially for $W^+$ production.

To make a better comparison of the different nuclear
modifications, Fig.~\ref{fig:W-eta@pPb}(c) and (d)
show the nuclear modification factors, $R_{p{\rm Pb}}$ as a function of charged
lepton pseudorapidity.  The points show the results including the nPDFs while
the curves show the CT10 and ABMP15 sets with isospin effects alone.  The
shape of $R_{p{\rm Pb}}$ in the backward pseudorapidity region is dominated by the
isospin effect since this is the region where the lead nucleon momentum fraction
is large (the antishadowing region) while shadowing effects should dominate
for small $x$ in the lead nucleus in the forward region.

The KP nPDF has very small antishadowing
at these high scales since the ratio agrees well with the ABMP15 curve for
$\eta_{\rm c.m.}^l < 0$.   The EPS09 and nCTEQ sets exhibit antishadowing in this
region so there is some deviation from the CT10 isospin curve.  In the forward
pseudorapidity region, however, there are clear deviations from the curves with
isospin alone.  Note that at forward rapidity, the CT10 ratios go to unity
for both $W^+$ and $W^-$ production.  However, the ABMP15 ratio is $\sim 5$\%
enhanced for $W^+$ and $\sim 5$\% suppressed for $W^-$.  In the case of $W^+$
production, there is a clear separation between the three nPDFs with the KP
set giving the smallest effect and nCTEQ the largest.  With $W^-$ production,
the difference in the isospin ratio at forward pseudorapidity for the two
proton PDFs causes the KP and
EPS09 ratios to overlap at $\eta_{\rm c.m.}^l > 0$ while the nCTEQ result shows
more suppression.



Note that the isospin effects are actually quite large at negative
pseudorapidity and dominate the behavior of $R_{p{\rm Pb}}$ in that region.
Thus a depletion of $\sim 20$\% is predicted for $W^+$ at $\eta_{\rm cm}^l = -2$
and a $\sim 20$\% enhancement for $W^-$ at $\eta_{\rm cm} = 2$.  This behavior
can be expected because the neutron excess in the lead nucleus causes a
relative reduction in $u$ quarks for $W^+$ production on the order of 0.7 in
$p+$Pb collisions, relative to $p+p$, near the edge of phase space at
backward pseudorapidity.  On the
other hand, the enhancement of $d$ quarks in the lead nucleus gives an
ehancement in $W^-$ production in the same direction by a factor of $\sim 1.6$,
as reflected by the isospin results shown in the curves in
Fig.~\ref{fig:W-eta@pPb}.

\subsubsection{Charged lepton asymmetry}
\begin{figure}[h]
\begin{center}
\stackinset{l}{0.3\textwidth}{b}{0.425\textheight}{(a)}{%
\stackinset{l}{0.3\textwidth}{b}{0.225\textheight}{(b)}{%
\stackinset{l}{0.3\textwidth}{b}{0.12\textheight}{(c)}{%
  \includegraphics[width=0.5\textwidth]{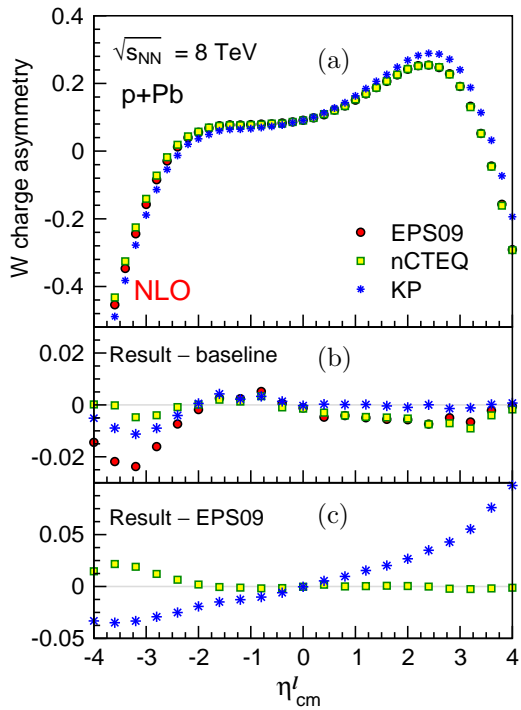}%
  }}}
  \end{center}
\caption[]{(Color online)
(a) The $W^+/W^-$ charge asymmetry as a function of the charged lepton 
pseudorapidity in $p+$Pb collisions at 8~TeV.
(b) The differences between each result and its corresponding baseline as a
function of charged lepton pseudorapidity.  In both (a) and (b), the
calculations with CT10+EPS09 are shown by red circles, CT10+nCTEQ by green
squares and ABMP15+KP by blue stars. (c)
The differences between nCTEQ and KP shadowing and EPS09 NLO are shown by the
green squares and blue stars respectively.
}
\label{fig:W-cas@pPb}
\end{figure}

The asymmetry between $W^+$ and $W^-$ production is mainly due to the
asymmetric $u$ and $d$ quark distributions
in the proton and the lead nucleus, as can be seen in 
Fig.~\ref{fig:W-eta@pPb}.
The related observable, the charge asymmetry, shown in
Fig.~\ref{fig:W-cas@pPb}(a), is
\begin{equation}
\label{cas}
\mathcal{A}(\eta^l)=\frac{d\sigma(W^+)/d\eta^l-d\sigma(W^-)/d\eta^l}{d\sigma(W^+
)/d\eta^l+d\sigma(W^-)/d\eta^l} \, \, .
\end{equation} 
The differences between the $p+$Pb result with shadowing and its corresponding
baseline is shown in Fig.~\ref{fig:W-cas@pPb}(b).
The effect of shadowing for all three
nPDFs is significantly reduced in the charge asymmetry.
A slight suppression in the backward region is predicted by EPS09 NLO, likely
the result of the asymmetric nuclear
modifications of the valence and sea quarks~\cite{Ru:2016wfx}.
The differences between the KP and nCTEQ results with respect to EPS09 NLO are
shown in (c).  This is done to illustrate where the nPDFs differ
most since the difference between each set and its corresponding baseline
is small.  Relative to both EPS09 NLO and  nCTEQ, the KP nPDFs predict a
positive difference
at forward rapidity and a negative difference at backward rapidity,
largely due to the different large-$x$ $u/d$ ratio and small-$x$ 
$\overline{u}/\overline{d}$
ratio in the underlying proton PDFs~\cite{Ru:2016wfx}.

\subsection{$Z^0$ production}

\subsubsection{$Z^0$ rapidity distribution}
\begin{figure}[h]
\begin{center}
\stackinset{l}{0.2\textwidth}{b}{0.40\textheight}{(a)}{%
\stackinset{l}{0.2\textwidth}{b}{0.1\textheight}{(b)}{%
  \includegraphics[width=0.5\textwidth]{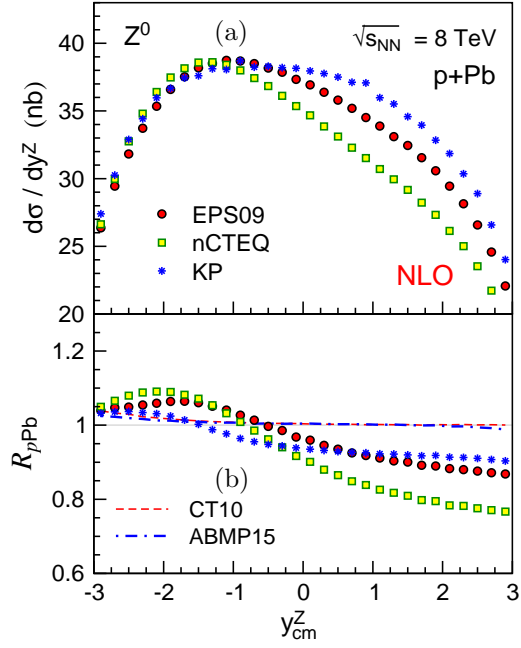}%
  }}
  \end{center}
\caption[]{(Color online)
  (a) The differential cross section as a function of $Z^0$ rapidity in $p+$Pb
  collisions at 8~TeV for EPS09 (red circles),
  nCTEQ (green squares) and KP (blue stars).
  The corresponding $R_{p{\rm Pb}}$ calculations are 
  shown in (b), along with the nuclear modification factor
  due to isospin alone, with CT10 (red dashed) and ABMP15 (blue dot-dashed).
  The $Z^0$ mass window used in the calculation is 
  $60<m_Z<120$~GeV~\cite{CMS:2014sca}.}
\label{fig:Z-y@pPb}
\end{figure}

The differential cross section as a function of the $Z^0$ rapidity is 
shown in Fig.~\ref{fig:Z-y@pPb}(a).
The $Z^0$ mass interval, $60<m_Z<120$~GeV~\cite{CMS:2014sca}, is used.
Asymmetric forward-backward distributions are predicted by the three 
nPDFs.


The nuclear suppression factors are shown in
Fig.~\ref{fig:Z-y@pPb}(b).  As in Fig.~\ref{fig:W-eta@pPb}(c) and (d), the
results are shown by the points for calculations with isospin and shadowing
and by the lines for isospin alone.  The isospin effect is very small for $Z^0$
production with deviations from unity on the percent level or less over most of
phase space.  The isospin effect is small because $Z^0$ production proceeds
through both $u$ and $d$ quarks.
Thus the suppression factors for the calculations with isospin
and nPDF modifications more clearly reflect the shape of the shadowing
parameterizations themselves.  The higher scale, $\mu_F = m_Z$, and lower $x$
conspire to both lower the antishadowing peak and move it closer to midrapidity,
compare Fig.~\ref{fig:Z-y@pPb}(b) and Fig.~\ref{fig:Bmeson_y} for $B$ mesons..

\subsubsection{Forward-backward asymmetry}
\begin{figure}[h]
\begin{center}
  \includegraphics[width=0.5\textwidth]{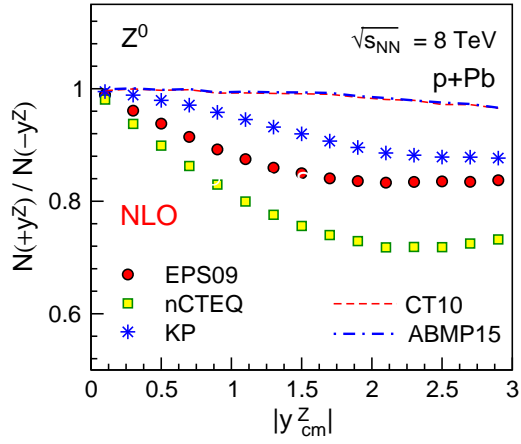}
  \end{center}
\caption[]{(Color online)
  The forward-backward asymmetry for $Z^0$ production at 8~TeV.
  The nPDF esults are shown for EPS09 (red circles),
  nCTEQ (yellow squares) and KP (blue stars) while calculations with isospin
  alone are shown for CT10 (red dashes) and ABMP15 (blue dot-dashed).}
\label{fig:Z-fb@pPb}
\end{figure}

The $Z^0$ forward-backward asymmetry, defined as
\begin{equation}
\label{fb}
R_{FB}(y^Z)=N(+y^Z)/N(-y^Z) \, ,
\end{equation}
is given in Fig.~\ref{fig:Z-fb@pPb}.
The baseline results with isospin alone, for both CT10 and ABMP15, show an
almost symmetric forward-backward ratio.  The deviation at forward rapidity in
both cases is only $\sim 2.5$\%, due to the small enhancement in $R_{p{\rm Pb}}$
at negative rapidity seen in Fig.~\ref{fig:Z-y@pPb}(b).

However, calculations with the three nPDFs predict different forward-backward
asymmetries.  The strongest predicted effect is with nCTEQ.
The results demonstrate that the forward-backward
asymmetry shows a strong sensitivity to the
shadowing parameterization since the asymmetry due to the underlying proton
PDFs is negligible.
This measurement at the LHC may provide more valuable constraints on the nPDFs,
especially the quark distributions~\cite{Ru:2016wfx}.

\subsubsection{$Z^0$ transverse momentum distribution}
\begin{figure}[h]
  \begin{center}
    \stackinset{l}{0.17\textwidth}{b}{0.415\textheight}{(a)}{%
    \stackinset{l}{0.17\textwidth}{b}{0.22\textheight}{(b)}{%
    \stackinset{l}{0.17\textwidth}{b}{0.12\textheight}{(c)}{%
      \includegraphics[width=0.5\textwidth]{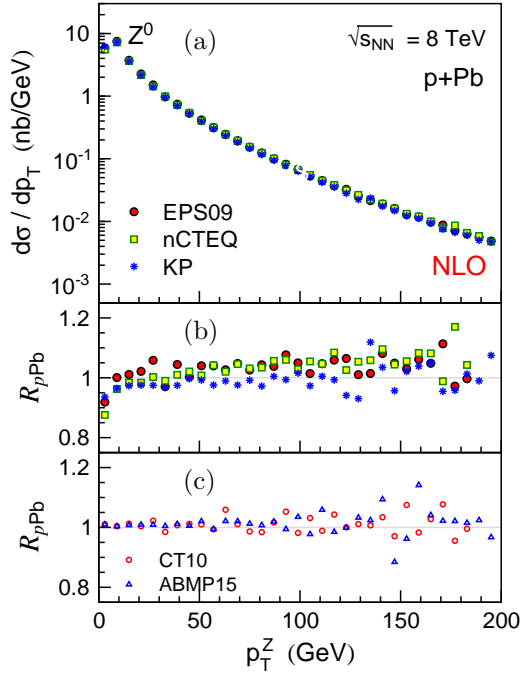}%
      }}}
    \end{center}
\caption[]{(Color online)
  (a) The differential cross section as a function of $Z^0$ transverse momentum
  in $p+$Pb collisions at 8~TeV for EPS09 (red circles),
  nCTEQ (yellow squares) and KP (blue stars).
  The corresponding $R_{p{\rm Pb}}$ calculations are 
  shown for the nPDF effects in (b) while the nuclear modification factor
  due to isospin alone is shown in (c) for CT10 (open red circles) and
  ABMP15 (blue triangles).}
\label{fig:Z-pt@pPb}
\end{figure}

The differential cross section as a function of $Z^0$ transverse momentum 
is shown in Fig.~\ref{fig:Z-pt@pPb}(a) for the $Z^0$ mass
window $60<m_Z<120$~GeV and rapidity interval
$-2.5<y^Z_{\rm cm}<1.5$~\cite{CMS:2014sca}.
The nuclear suppression factors with nuclear PDFs, shown in
Fig.~\ref{fig:Z-pt@pPb}(b), show some scatter but, overall, EPS09 and nCTEQ
predict a small enhancement, $\sim5$\%, at large $p_T$ due to antishadowing
of the nuclear gluon distributions~\cite{Kang:2012am,Ru:2015pfa}.
The nuclear modification of the KP nPDFs shows a slight suppression at large
$p_T$ relative to EPS09 NLO and nCTEQ.  The results with isospin alone, shown
in Fig.~\ref{fig:Z-pt@pPb}(c), display a relatively reduced amount of scatter
consistent with unity.

\section{Top quark production cross sections (D. d'Enterria)}
\label{sec:topQ}

\begin{figure*}[htbp]
\centering
\includegraphics[width=0.6\textwidth]{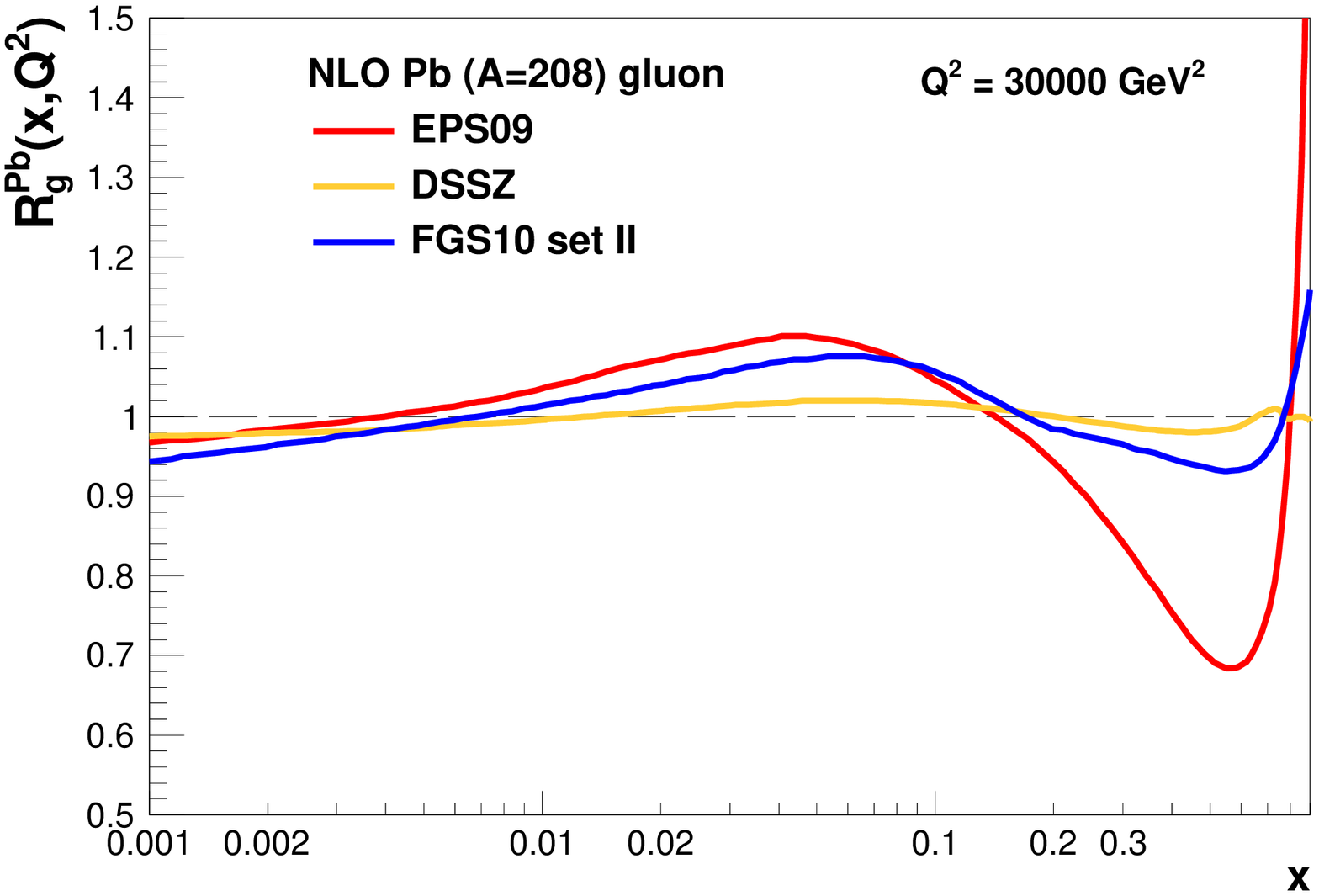}
\caption{(Color online) Ratio of the lead-to-proton gluon densities in the
  antishadowing
  ($x \approx 0.05-0.1$) and EMC ($x\approx 0.1-0.6$) regions probed by
  $t \overline t$
  production at virtualities $Q^2= m_t^2\approx 3\times 10^{4}$~GeV$^2$
  in $p+$Pb collisions at the LHC, for three different NLO nuclear PDF sets:
  EPS09~\cite{Eskola:2009uj} (red), DSSZ~\cite{deFlorian:2011fp} (yellow), and
  FGS10~\cite{Frankfurt:2011cs} (blue).}
\label{fig:nPDFs} 
\end{figure*}

The top quark is the heaviest elementary particle in the Standard Model 
and remains unobserved so far in nuclear collisions. 
Its cross section in hadronic collisions is dominated
by pair production in gluon-gluon fusion ($g\,g \rightarrow t \overline t+X$), 
which is theoretically calculable today with great accuracy via perturbative
quantum chromodynamics methods.
(At NLO, more than 85\% of the $t \overline t$ cross section at 8.16~TeV
involves initial-state gluons from the colliding nucleons.)
Calculations at next-to-next-to-leading-order
(NNLO) including next-to-next-to-leading-log (NNLL) soft-gluon resummation
are available using {\it e.g.} $\mathtt{Top++}$~\cite{Czakon:2013goa}.
Differential $t \overline t$
cross sections are also available at NLO accuracy using the $\mathtt{mcfm}$
code~\cite{mcfm1,mcfm2}.
The study of the $t \overline t$ cross section modifications in
proton-nucleus compared to $p+p$ 
collisions at the same nucleon-nucleon center-of-mass energy provides a novel,
well-calibrated probe of the nuclear gluon density at the
LHC~\cite{d'Enterria:2015mgr}, in particular in the unexplored high-$x$ region 
($x \gtrsim 2\, m_t/\sqrt{s_{_{NN}}} \approx 0.05$) where ``antishadowing''
and ``EMC'' effects are expected to modify its shape compared to the free
proton case, see Fig.~\ref{fig:nPDFs}.

The study of top quark production in $p+$Pb collisions provides information on 
the nuclear PDF that is complementary to that from similar studies with
electroweak
bosons~\cite{Khachatryan:2015hha,Khachatryan:2015pzs,Aad:2015gta,Alice:2016wka}.
The cross sections of the latter are more sensitive to quark, rather than gluon,
 densities, at $x$ values about a factor of two smaller. 
 In addition, a good understanding of top quark production in proton-nucleus
 collisions is crucial as a baseline for upcoming studies of heavy-quark
 energy loss in the quark-gluon-plasma formed in nucleus-nucleus
 collisions~\cite{d'Enterria:2015mgr,Dainese:2016gch,Apolinario:2017}.

The top quark decays very rapidly, 
$\tau_0 = \hbar/\Gamma_{t} \approx 0.15$~fm$/c$, before hadronizing into 
$t \rightarrow W\,b$ with a $\sim 100$\% branching ratio, with the $W$ bosons
themselves decaying either leptonically 
($t\rightarrow W(\ell\,\nu)\,b$, 1/3 of the time) or hadronically
($t\rightarrow W(q \overline q)\,b$, 2/3 of the time)~\cite{Olive:2016xmw}. 
 In Pb+Pb collisions, the charged leptons $\rm \ell = e,\mu$ from the
 fully-leptonic final-state ($t \overline t \rightarrow b \overline b \,
 2\ell\,2\nu$)
 are totally unaffected by final-state interactions, thereby providing the
 cleanest channel for its observation in the complicated heavy-ion environment
 \cite{d'Enterria:2015mgr}, though at the price of a relatively low branching
 ratio (BR~$\approx$~4\% for the $ee,\,e\mu$ and $\mu\mu$ modes combined). 
 In the $p+$Pb case, thanks to the lower backgrounds and the absence of
 final-state effects for jets compared to Pb+Pb collisions, 
 the leptons+jets final state
 ($t \overline t\rightarrow b \overline b\,\ell\,\nu\,2j$) is easily
 measurable and has a much larger branching ratio (BR~$\approx 30$\%) than the
 purely leptonic decay.  Predictions are presented for the total, 
 fiducial, and differential (for the $\ell$+jets channel) cross sections for
 $t \overline t$ production in $p+$Pb at $\sqrt{s_{_{NN}}} = 8.16$~TeV. 

\subsection{Total and fiducial $t \overline t$ cross sections}
\label{sec:sigma_tot}

The total and differential $p + {\rm Pb} \rightarrow t \overline t+X$
cross sections are computed first
at NLO accuracy with $\mathtt{mcfm}$~v8.0~\cite{mcfm1,mcfm2},
using the CT10 NLO~\cite{Gao:2013xoa} and CT14 NLO~\cite{CT14} proton parton
distribution functions and the nuclear modifications
for Pb given by the EPS09~\cite{Eskola:2009uj} and
EPPS16~\cite{Eskola:2016oht} nPDF sets.  A 
$K$-factor, $K=\sigma({\rm NNLO+NNLL})/\sigma({\rm NLO})\approx 1.20$,
is then computed with $\mathtt{Top++}$v2.0~\cite{Czakon:2013goa} using the
NNLO CT10 and CT14 PDFs alone, in order to scale up the NLO $\mathtt{mcfm}$
cross section to NNLO~+~NNLL accuracy. 
The $\mathtt{Top++}$ and $\mathtt{mcfm}$ codes are run with $N_f = 5$ flavors,
the top pole mass set to $m_t=172.5$~GeV, 
default renormalization and factorization scales set to $\mu_R =\mu_F = m_t$,
and the strong coupling set to $\alpha_s = 0.1180$. 
All numerical results have been obtained with the latest standard model
parameters for particle masses, widths and couplings~\cite{Olive:2016xmw}.
The PDF uncertainties include those from the proton and nuclear PDFs
combined in quadrature, as obtained from the corresponding 
56~+~96 (52~+~32) eigenvalues of the CT14~+~EPPS16 (CT10~+~EPS09) sets.
The theoretical 
uncertainty arising from the scale choice is estimated by modifying
$\mu_R$ and $\mu_F$ within a factor of two with respect to their default
values. In the $p+p$ case, such a NNLO+NNLL calculation yields predicted cross
sections in very good agreement with the experimental data at
$\sqrt{s} = 7$, 8, and 13~TeV at the LHC
\cite{Chatrchyan:2011ew,Chatrchyan:2011nb,ATLAS:2012aa,Chatrchyan:2013kff,Aad:2015pga,Khachatryan:2015uqb}. 
The computed nucleon-nucleon cross sections are then scaled by the Pb mass
number ($A= 208$) to obtain the corresponding $p+$Pb cross sections. 

\begin{table*}[htbp!]
\renewcommand\arraystretch{1.4}%
\caption{Total and fiducial (in the $\ell$+jets channel, after typical
  acceptance cuts) cross sections for $t \overline t$ production in $p+p$ and
  $p+$Pb
  collisions at $\sqrt{s_{_{NN}}} = 8.16$~TeV at NNLO~+~NNLL
  accuracy with different
  proton (CT10 and CT14) and ion (EPS09 and EPPS16) PDFs. The first and second
  quoted uncertainties correspond to the PDF and scale uncertainties. 
\label{tab:sigma_ttbar}}\vspace{-0.5cm}
\begin{center}
\begin{tabular}{lccc}\hline
  & \multicolumn{2}{c}{$\sigma(t \overline t)$ total} &
  $\sigma(t \overline t\rightarrow b \rightarrow b\,\ell\nu\,2j)$
  fiducial\\ \hline
PDF sets  & CT10 & CT14 & CT14 \\ \hline
$p+p$  & 265.8 $\,^{+17.4}_{-14.3}${\footnotesize(PDF)}$\,^{+6.9}_{-9.3}$ pb
& 272.6 $\,^{+17.2}_{-15.3}${\footnotesize(PDF)}$\,^{+7.0}_{-9.5}$ pb
& 31.53 $\,^{+2.00}_{-1.77}${\footnotesize(PDF)}$\,^{+0.81}_{-1.10}$ pb \\ \hline
\hline
PDF sets  & CT10~+~EPS09 & CT14~+~EPPS16 & CT14~+~EPPS16 \\ \hline
$p+$Pb & 57.5 $\,^{+4.3}_{-3.3}${\footnotesize(PDF)}$\,^{+1.5}_{-2.0}$ nb & 
59.0 $\pm$ 5.3{\footnotesize(PDF)} $\,^{+1.6}_{-2.1}$ nb
& 6.82 $\pm$ 0.61{\footnotesize(PDF)} $\,^{+0.18}_{-0.24}$ nb \\\hline
$R_{p {\rm Pb}}$ & 1.04 $\,^{+0.04}_{-0.02}${\footnotesize(PDF)}
& 1.04 $\pm$ 0.07{\footnotesize (PDF)}
& 1.04 $\pm$ 0.07{\footnotesize (PDF)}\\\hline
\end{tabular}
\end{center}
\end{table*}

The total $t \overline t$ cross sections for $p+p$ and $p+$Pb collisions for
various proton and lead PDFs are listed in the first two columns of
Table~\ref{tab:sigma_ttbar}, as well as the nuclear modification factor
$R_{p{\rm Pb}} = \sigma_{p{\rm Pb}}/(A\sigma_{pp})$.
For $p+$Pb, the CT14~+~EPPS16 calculations give a central $t \overline t$
cross section 2.6\% larger than that computed with CT10~+~EPS09.
The cross section uncertainties linked to the PDF choice 
are $\pm 9$\% for CT14~+~EPPS16, and $+7.5\%/-5.8\%$ for CT10~+~EPS09. The
theoretical $\mu_F$, and $\mu_R$ scale uncertainties
amount to $+2.5\%/-3.5\%$. Compared to the corresponding $p+p$ results, a
small net overall antishadowing effect increases the total top-quark cross
section by 4\% for both the EPPS16 and EPS09 sets,
$R_{p{\rm Pb}} = 1.04 \,^{\pm 0.07({\rm EPPS16})}_{\pm0.03({\rm EPS09})}$,
where the proton PDF and theoretical scale uncertainties cancel out in the
ratio.

Fiducial top-pair production cross sections can be measured in the $\ell$+jets
channel at the LHC taking into account their decay branching ratio
(BR~$\approx~30$\%), the basic ATLAS/CMS detector acceptance 
constraints, and standard final-state selection criteria applied to remove
$W$+jets and QCD multijet 
backgrounds~\cite{Chatrchyan:2011ew,Chatrchyan:2013kff,Aad:2015pga}, such as:
\begin{itemize}
\item One isolated charged lepton ($\ell = e, \mu$) with $p_T > 30$~GeV,
  $|\eta|<2.5$, and $R_{\rm isol} = 0.3$;
\item Four jets (reconstructed employing the anti-$k_T$ algorithm with $R=0.5$)
 with $p_T > 25$~GeV and $|\eta|<3.0$;
\item Lepton-jets separation of $\Delta R (\ell,j) > 0.4$.
\end{itemize}
Often such cuts are sufficient to carry out the $t \overline t$ measurement
although, if needed, a threshold on the missing transverse momentum from the
unobserved neutrino can be added.

The impact of such cuts, evaluated with $\mathtt{mcfm}$,
indicates a 39.5\% acceptance for 
the total cross section with a very small dependence on the underlying PDF
(the maximum difference in acceptances using the proton and ion
PDFs amounts to $\pm 0.7$\% on the final cross section). The events that pass
such selection criteria 
are then often required in addition to having two $b$-tagged jets. For an
intermediate $b$-tagging efficiency of 70\%, 
this results in a final combined 
acceptance$\times$efficiency of $\sim 20$\% for a $t \overline t$-enriched
sample consisting of one isolated charged lepton, two light-quark jets, and two
$b$-jets. Taking into account the $\ell$+jets branching ratio 
(BR~$\approx 0.3$), the aforementioned acceptance and efficiency, and the
180~nb$^{-1}$ integrated luminosities collected by ATLAS and CMS in $p+$Pb
collisions at 8.16~TeV, $\sim 600$ top-quark pair
events may be expected to be reconstructed in the decay channel.

\subsection{Differential $t \overline t\rightarrow \ell + \, {\rm jets}$
  distributions}
\label{sec:sigma_diff}

\begin{figure*}[htbp]
\centering
    \stackinset{l}{0.65in}{b}{1.425in}{(a)}{%
\includegraphics[width=0.49\textwidth]{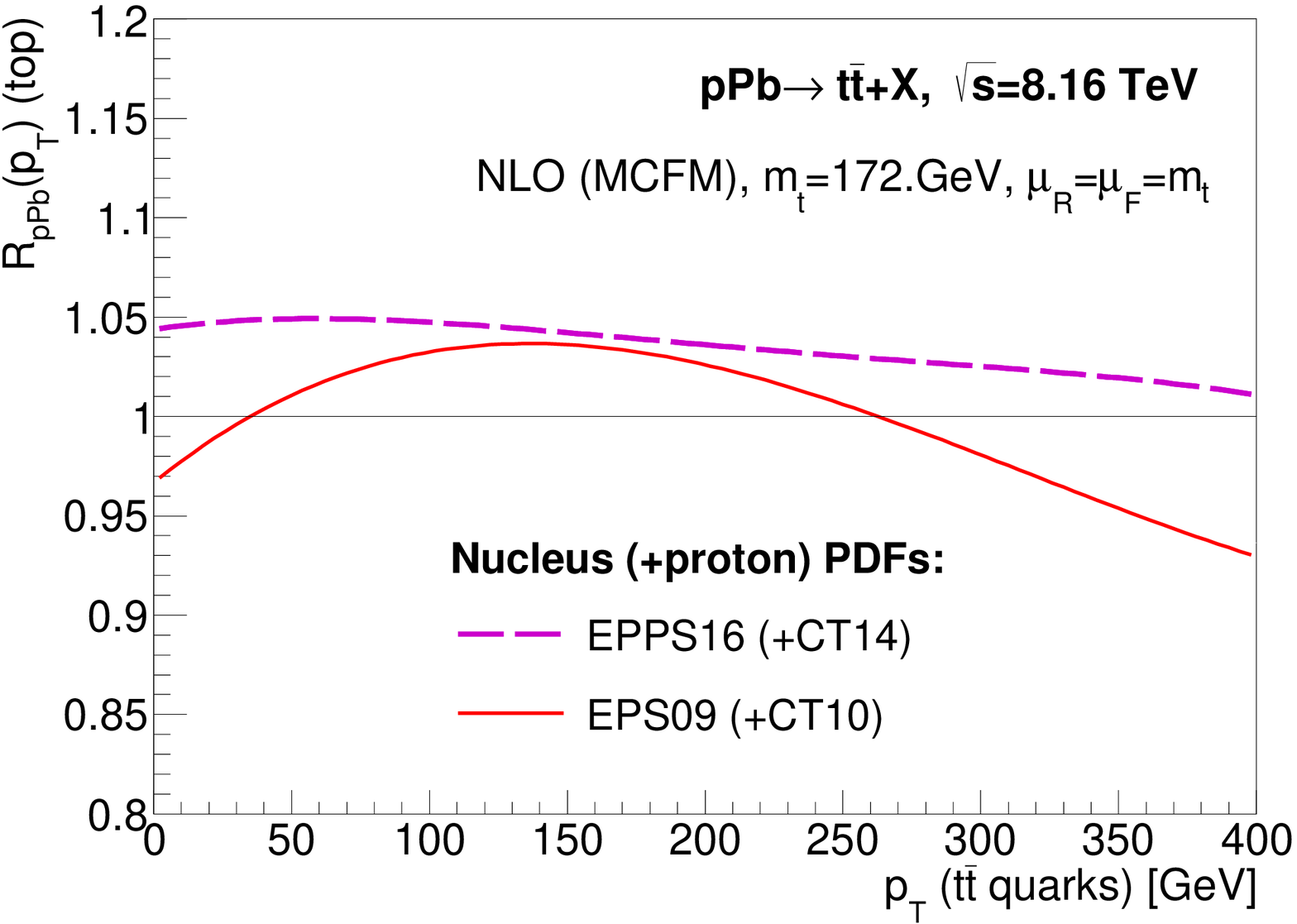}}
    \stackinset{l}{0.65in}{b}{1.45in}{(b)}{%
\includegraphics[width=0.49\textwidth]{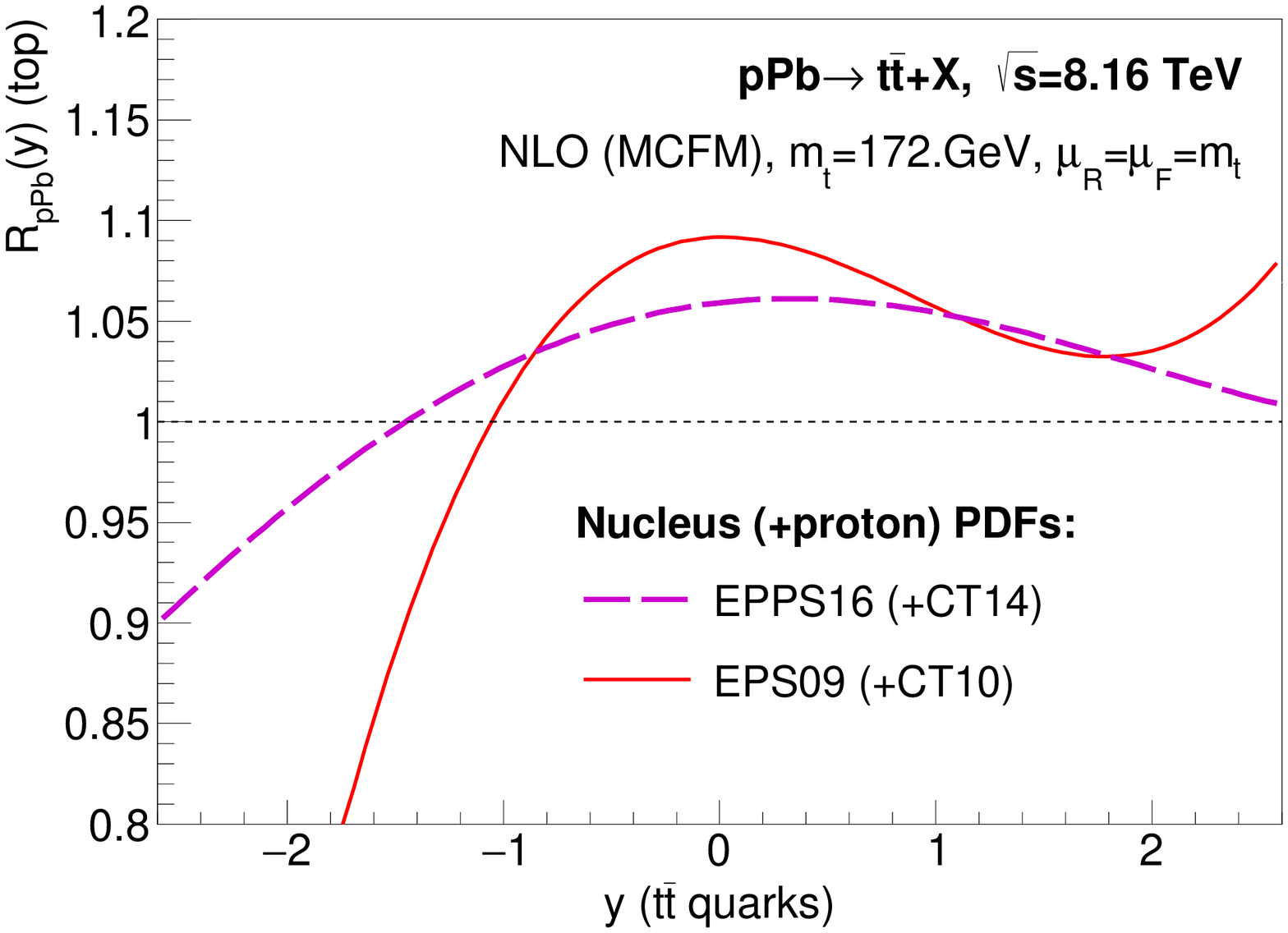}}
    \stackinset{l}{0.65in}{b}{1.45in}{(c)}{%
\includegraphics[width=0.49\textwidth]{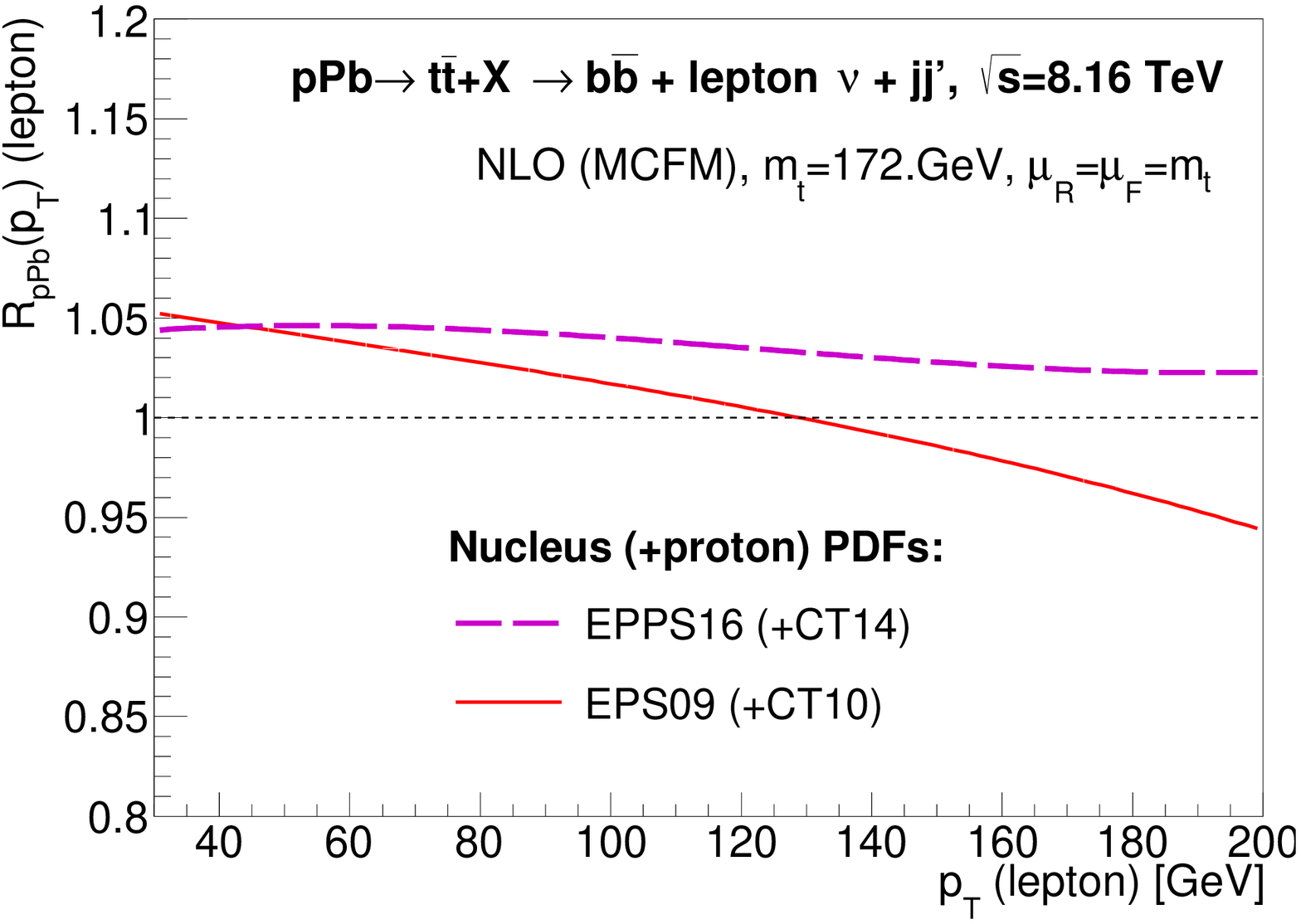}}
    \stackinset{l}{0.65in}{b}{1.45in}{(d)}{%
\includegraphics[width=0.49\textwidth]{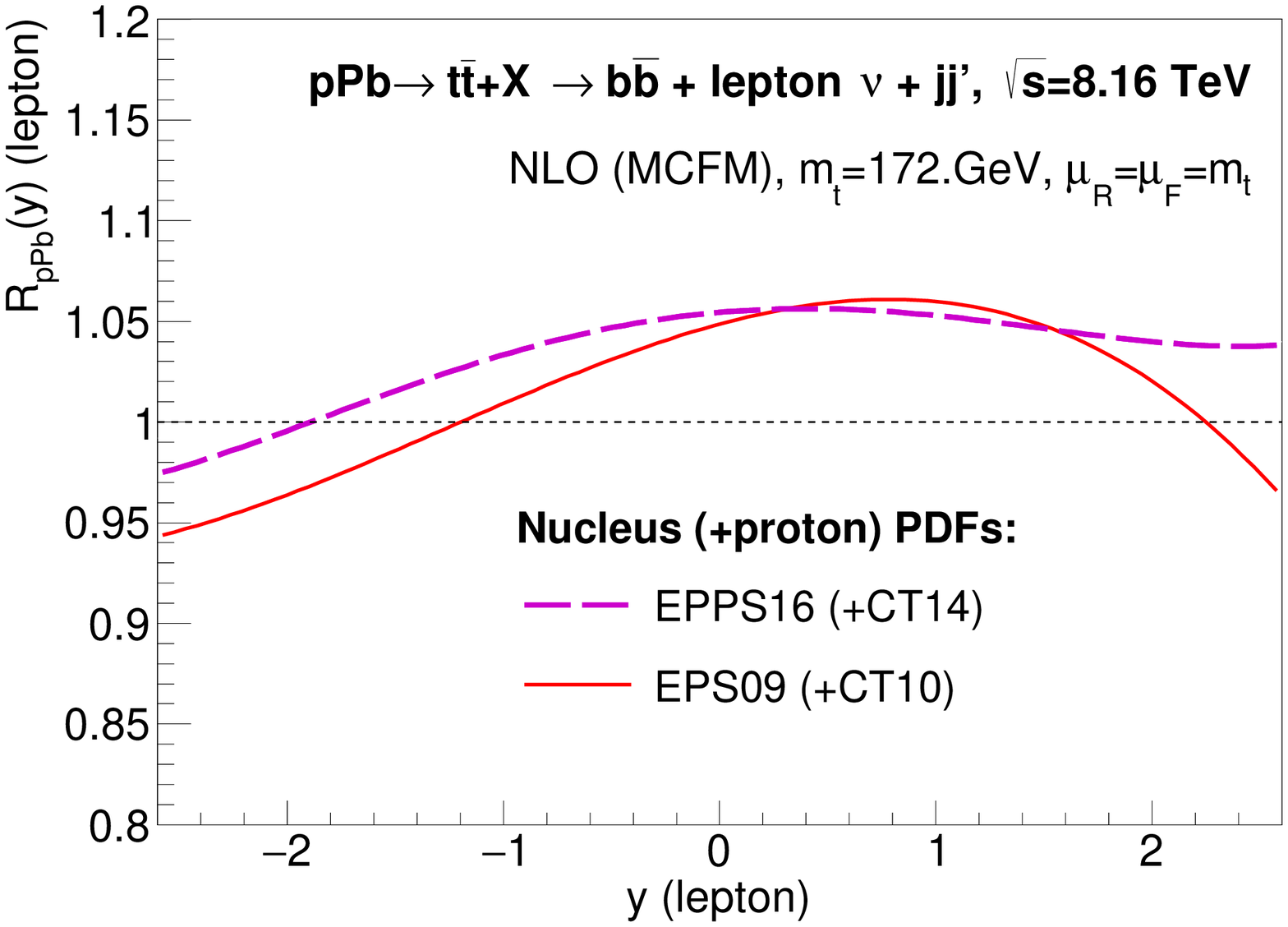}}
    \stackinset{l}{0.65in}{b}{1.45in}{(e)}{%
\includegraphics[width=0.49\textwidth]{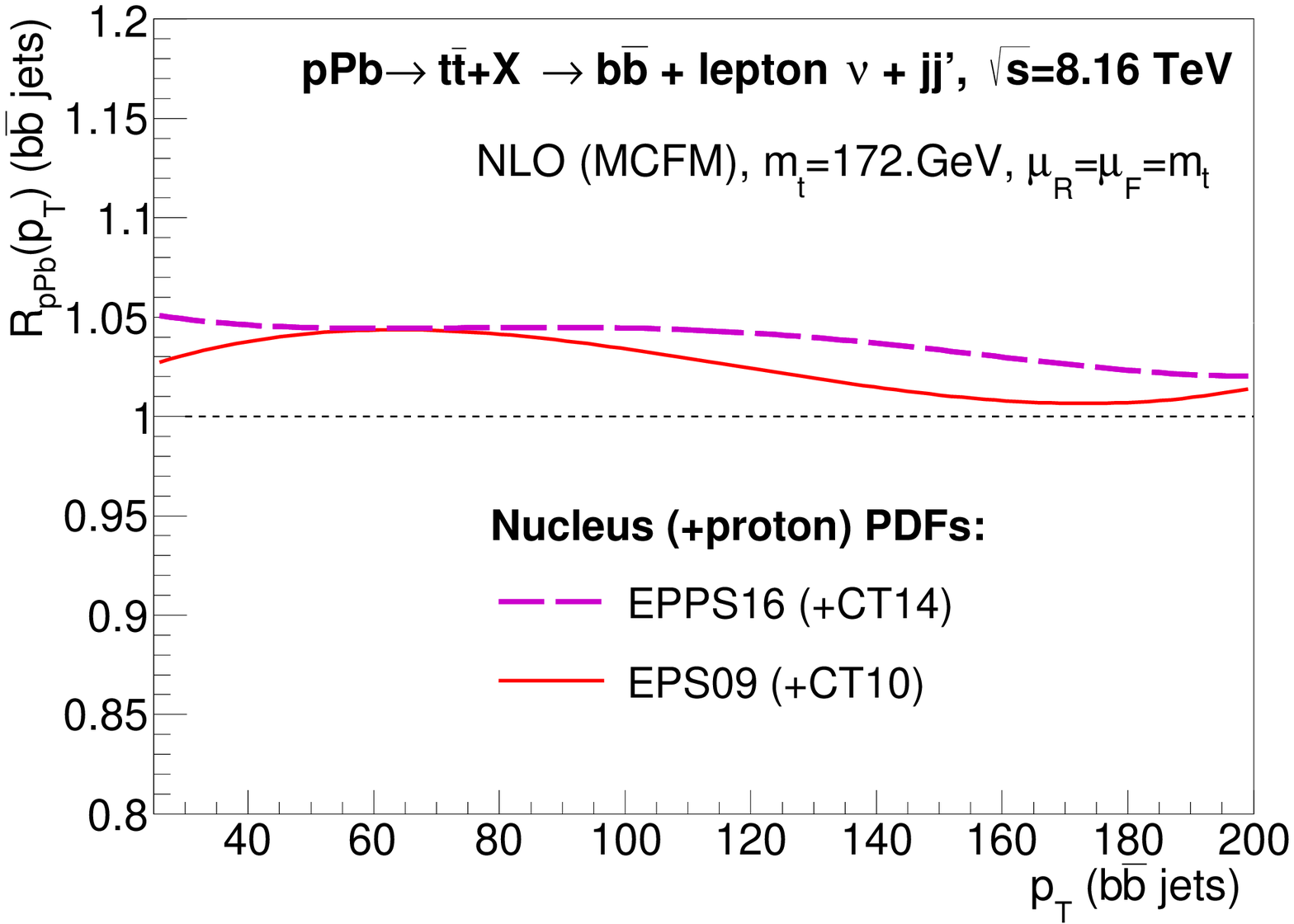}}
    \stackinset{l}{0.65in}{b}{1.46in}{(f)}{%
\includegraphics[width=0.49\textwidth]{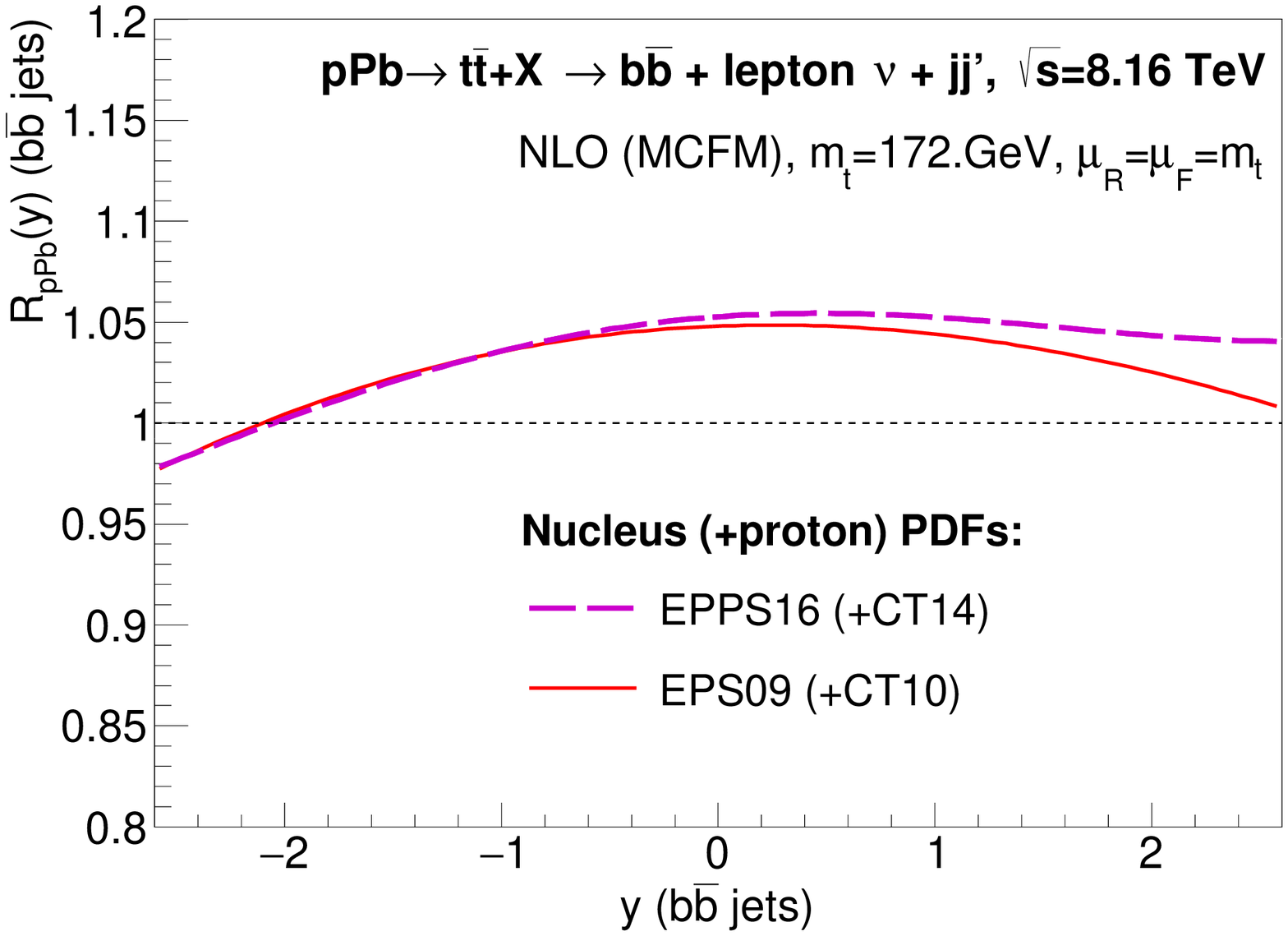}}
    \caption{(Color online)
      Nuclear modification factors as a function of transverse momentum
  (left) and rapidity (right) for $t \overline t$ production in the $\ell$+jets
  channel at $\sqrt{s_{_{NN}}} = 8.16$~TeV for: the produced top quarks, (a) and
  (b), their isolated decay leptons, (c) and (d), and their $b$-jet decays,
  (e) and (f),
  obtained at NLO accuracy with the central sets of CT14+EPPS16 (dashed curves) 
  and CT10+EPS09 (solid curves).}
\label{fig:RpPb_ttbar} 
\end{figure*}

As seen in the previous section, the total integrated $t \overline t$ cross
sections are modified by only a few percent by nPDF effects in $p+$Pb
compared to $p+p$ collisions at 8.16~TeV, giving $R_{p {\rm Pb}} = 1.04$.
However, Fig.~\ref{fig:nPDFs} indicates that $gg \rightarrow t \overline t$
processes at different $x$ values, {\it i.e.} probed at different rapidities
and/or transverse momenta of the produced top quarks, should be much more
sensitive to the underlying positive (antishadowing) and negative (EMC and
shadowing) modifications. This
was quantitatively confirmed in Ref.~\cite{d'Enterria:2015mgr} which showed 
that rapidity distributions of the isolated leptons in the fully-leptonic
$t \overline t$ decay mode are indeed sensitive to the underlying nPDF and can
be used to reduce the uncertainties of the EPS09 nuclear gluon density. 
A similar study is presented here, but for the $\ell$+jets channel,
$t \overline t \rightarrow b \overline b \,\ell\nu\,2j$, and using the recent
EPPS16 nPDF set.  Figure~\ref{fig:RpPb_ttbar} shows the nuclear modification
factors,
$R_{p{\rm Pb}}(X) = (d\sigma_{p{\rm Pb}}/dX)/(A\,d\sigma_{pp}/dX)$,
as a function of transverse momentum ($X=p_T$, left panels) and rapidity
($X=y$, right panels)
for the produced top quarks, (a) and (b), their isolated decay leptons,
(c) and (d), and 
their $b$-jet decays, (e) and (f)
as obtained with EPPS16 (dotted curves) and EPS09 (solid curves).
Any effect related to the choice of the proton PDF (CT10 or CT14) mostly
cancels in the $p+$Pb/$p+p$ ratio.  This ratio is then most sensitive to
modifications of the nuclear gluon densities alone. 
The effect of antishadowing (shadowing or EMC) in the nPDF results in small
enhancements (deficits) in the distributions at lower (higher) $p_T$ as well
as at central (forward and backward) rapidities 
$y\approx 0$ ($|y|\gtrsim 2$). In general, the effects are larger for the
initially-produced 
top quarks than for their decay products (isolated leptons and $b$-jets),
but are nonetheless also visible for the latter.

\subsection{Summary of top quark production}
\label{sec:summary}

The total, fiducial, and differential cross sections for top-quark pair
production in $p+$Pb collisions at $\sqrt{s_{_{NN}}} = 8.16$~TeV have been
computed at up to NNLO+NNLL accuracy using
the CT14 and CT10 proton PDFs and the EPPS16 and EPS09 nuclear PDFs.
The total cross sections are
$\sigma(p {\rm Pb} \rightarrow t \overline t+X) = 59.0 \pm 5.3$
{\footnotesize \sc (CT14+EPPS16)}$\,^{+1.6}_{-2.1}$ (scale)~nb and
$57.5 \pm \,^{+4.3}_{-3.3}$ {\footnotesize\sc (CT10+EPS09)}$\,^{+1.5}_{-2.0}$
(scale)~nb, a
few percent modification with respect to the
result obtained using only the free proton PDFs,
$R_{p{\rm Pb}} = 1.04 \,^{\pm 0.07({\rm EPPS16})}_{\pm0.03({\rm EPS09})}$. In the
lepton+jets decay mode, $t \overline t \rightarrow b \overline b \,
W(\ell\nu)\,W(q \overline q')$, one expects 600~$t \overline t$ events in 
the 180~nb$^{-1}$ integrated luminosity collected at the LHC, after typical
ATLAS/CMS acceptance cuts and efficiency losses. 
The ratios of the $t \overline t$ differential cross sections in $p+$Pb
relative to those in $p+p$ collisions as functions of the $p_T$ and rapidity
of the charged decay leptons and of the $b$-jets are sensitive to
the antishadowing and EMC gluon density modifications at high
virtualities in the nucleus. Precise differential measurements of top-quark
pair production thus provide a novel tool for studying the nuclear parton
distribution functions in a so-far unexplored kinematic regime.

\section{Summary}

Theoretical predictions for the production cross sections and yields of charged
hadrons, identified light hadrons, quarkonium and heavy-flavor hadrons,
Drell-Yan
dileptons, jets, photons, gauge bosons, and top quarks in $p+$Pb collisions at
$\sqrt{s_{_{NN}}} = 8.16$~TeV, of relevance for the November 2016 $p+$Pb
run at the LHC, have been compiled. The corresponding predictions of the
normalized $p+$Pb to $p+p$ cross section ratios have been presented for most
of the observables, providing new insights on
the expected role of cold nuclear matter effects over a wider region of phase
space than ever before available.

When multiple predictions for the same observable are presented, the results
are compared.  These results are generally compatible with each other.
Exceptions include the charged
hadron multiplicity distributions in Sec.~\ref{sec:dnchdeta}
and the $D$ and $B$ meson calculations in Sec.~\ref{sec:open}.
The main difference between the heavy flavor
results is the inclusion of the Cronin effect in the calculations by Vitev
{\it et al.}\ and
its absence in the data-driven shadowing only calculations by Lansberg and
Shao.

The $J/\psi$ data so far available from ALICE and LHCb in Sec.~\ref{sec:psi}
highlight the need for measurements that
can more directly and more stringently constrain the low $x$ gluon distribution
in the nucleus.  A future electron-ion collider
can only provide a partial answer since the
low $x$ reach is not as great at the proposed US facilities as that covered
by the LHC experiments.  Future global analyses that employ the 5.02 and 8.16
TeV $p+$Pb lead results will be helpful but perhaps not sufficient to provide
a definitive answer.

The predictions shown here, in large
part, focus on minimum bias collisions and hard processes.  It is worth noting
that the high multiplicity $p+$Pb and $p+p$ data show very interesting results
suggesting collective phenomena akin to that observed in nucleus-nucleus
collisions in these much smaller systems.  While the phenomenology of these
systems is still in development and not as mature as the cold matter effects
discussed here, many exciting results in this area can be expected in the
future.  For example, double (and even triple) parton interactions,
although not discussed here,
have been shown to play a role at the LHC, for example, in like-sign charm
production, as observed by LHCb \cite{Aaij:2012dz}.  Since these are included
in $\mathtt{PYTHIA8}$, they will also be part of $\mathtt{HIJING++}$.

\section{Acknowledgments}

T. L. and B. D. are supported by the European Research Council grant 
ERC-2015-CoG-681707 and by the Academy of
Finland, Grants No. 267321,  No. 273464 and No. 303756. 
The research of K. J. E. and H. P. 
was supported by the Academy of Finland, Project 297058.
The work of K. K. has been supported by Narodowe Centrum Nauki with Sonata Bis 
grant DEC-2013/10/E/ST2/00656. 
P. K. acknowledges the support of grants DE-SC-0002145 and DE-FG02-93ER40771.
H. M., B. P. S., P. T. and R. Venugopalan are supported under DOE Contract No.
DE-SC0012704.
S. S. is supported by DOE Award No. DE-FG02-97ER410014.
The research of B. P. S., S. S., P. T. and R. Venugopalan used resources of the
National Energy Research Scientific Computing Center, supported by the Office
of Science of the U. S. Department of Energy under Contract No.
DE-AC02-05CH11231.
The research of B. D., T. L. and H. M. used the computing resources of 
CSC—IT Center for Science in Espoo, Finland.
The work of E. G. F. is supported by the
Ministerio de Ciencia e Innovacion of Spain under project
FPA2014-58293-C2-1-P.
The work of J. P. L. is supported in part by the French CNRS via the LIA FCPPL 
(Quarkonium4AFTER) and the D\'efi Inphyniti-Th\'eorie LHC France.
The work of A. R. was supported in part
 by Fondecyt grant 1150135,  Anillo ACT1406, Conicyt PIA/Basal FB0821 and  ECOS-
Conicyt C14E01.
Z. W. L. acknowledges the support of the NSFC of China under grant no. 11628508.
H. S. S. is supported by the ILP Labex (ANR-11-IDEX-0004-02, ANR-10-LABX-63).
The research of R. Vogt was supported in part by the U. S. Department of Energy
under Contract No. DE-AC52-07NA27344. 
The work of P. R., B.-W. Z, and E. W. is supported in part by the Ministry
of Science and Technology 
in China under Project Nos. 2014CB845404, 2014DFG02050, and 
by NSFC of China with Project Nos. 11322546, 11435004, and 11521064.
The work on $\mathtt{HIJING++}$ and $\mathtt{kTpQCD}$
was supported by the Hungarian-Chinese cooperation grant No T\'eT 12 
CN-1-2012-0016 and No. MOST 2014DFG02050, Hungarian National Research Fund 
(OTKA) grants K123815 and K120660. 
G. G. B. also thanks the J\'anos Bolyai Research Scholarship of the
Hungarian Academy of Sciences and acknowledges the support of the Wigner
GPU laboratory.
I. V. is supported by the U.S. Department of Energy, Office of Science under
Contract No. DE-AC52-06NA25396 and by the DOE Early Career Program.
The research of Z. K. is funded by the DOE Office of Science Contract No.
DE-AC52-06NA25396.
H. X. is supported under Contract DE-FG02-91ER40684 of the
DOE Office of Science.
D. d'E. acknowledges discussions with H. Paukkunen on the interface of EPPS16
to $\mathtt{mcfm}$.


\begin{thebibliography}{999}

\bibitem{Enrico} ALICE Collaboration, CERN-ALICE-PUBLIC-2017-001.

\bibitem{Aaij:2017cqq}
  R.~Aaij {\it et al.} [LHCb Collaboration],
  arXiv:1706.07122 [hep-ex].
  
\bibitem{Albacete:2013ei}
  J.~L.~Albacete {\it et al.},
  Int.\ J.\ Mod.\ Phys. E {\bf 22} (2013) 1330007.

\bibitem{Albacete:2016veq} 
  J.~L.~Albacete {\it et al.},
  Int.\ J.\ Mod.\ Phys. E {\bf 25} (2016) 1630005.

\bibitem{Eskola:2016oht}
  K.~J.~Eskola, P.~Paakkinen, H.~Paukkunen and C.~A.~Salgado,
  Eur. Phys. J. C {\bf 77} (2017) 163.

\bibitem{Loizides:2016tew}
  C.~Loizides,
  Nucl.\ Phys.\ A {\bf 956} (2016) 200.

\bibitem{Dusling:2015gta}
K.~Dusling, W.~Li and B.~Schenke,
  Int.\ J.\ Mod.\ Phys.\ E {\bf 25} (2016) 1630002.
  
\bibitem{Romatschke:2016hle}
  P.~Romatschke,
  Eur.\ Phys.\ J.\ C {\bf 77} (2017) 21.

\bibitem{Berges:2017eom}
  J.~Berges, K.~Reygers, N.~Tanji and R.~Venugopalan,
  Phys.\ Rev.\ C {\bf 95} (2017) 054904.

\bibitem{Mantysaari:2016ybx}
  H.~Mäntysaari and B.~Schenke,
  Phys.\ Rev.\ Lett.\  {\bf 117} (2016) 052301.

\bibitem{Mantysaari:2017cni}
  H.~Mäntysaari, B.~Schenke, C.~Shen and P.~Tribedy,
  Phys.\ Lett.\ B {\bf 772} (2017) 681.
  
\bibitem{Schenke:2017bog}
  B.~Schenke,
  Nucl.\ Phys.\ A {\bf 967} (2017) 105.

\bibitem{Adare:2014keg}
  A.~Adare {\it et al.} [PHENIX Collaboration],
  Phys.\ Rev.\ Lett.\  {\bf 114} (2015) 192301.

\bibitem{Adare:2015ctn}
  A.~Adare {\it et al.} [PHENIX Collaboration],
  Phys.\ Rev.\ Lett.\  {\bf 115} (2015) 142301.

\bibitem{Schenke:2014gaa}
  B.~Schenke and R.~Venugopalan,
  Nucl.\ Phys.\ A {\bf 931} (2014) 1039.

\bibitem{Romatschke:2015gxa}
  P.~Romatschke,
  Eur.\ Phys.\ J.\ C {\bf 75} (2015) 305.
  
\bibitem{Chatrchyan:2014hqa}
  S.~Chatrchyan {\it et al.} [CMS Collaboration],
  Eur.\ Phys.\ J.\ C {\bf 74} (2014) 2951.

\bibitem{Khachatryan:2015hha}
  V.~Khachatryan {\it et al.} [CMS Collaboration],
  Phys.\ Lett.\ B {\bf 750} (2015) 565.

\bibitem{Khachatryan:2015pzs}
  V.~Khachatryan {\it et al.} [CMS Collaboration],
  Phys.\ Lett.\ B {\bf 759} (2016) 36.

\bibitem{Aad:2015gta}
  G.~Aad {\it et al.} [ATLAS Collaboration],
  Phys.\ Rev.\ C {\bf 92} (2015) 044915.  

\bibitem{Onengut:2005kv}
  G.~Onengut {\it et al.} [CHORUS Collaboration],
  Phys.\ Lett.\ B {\bf 632} (2006) 65.
    

\bibitem{Lappi:2013zma}
T.~Lappi and H.~M{\"a}ntysaari, 
 Phys. Rev. D {\bf 88} (2013) 114020. 

\bibitem{Albacete:2010sy}
J.~L. Albacete, N.~Armesto, J.~G. Milhano, P.~Quiroga-Arias and C.~A. Salgado,
 Eur. Phys. J. C {\bf 71} (2011) 1705.

\bibitem{Aaron:2009aa}
F.~Aaron {\it et.~al.} [H1 and ZEUS Collaborations], 
 JHEP {\bf 1001} (2010) 109. 

\bibitem{Gribov:1984tu}
L.~Gribov, E.~Levin and M.~Ryskin, 
 Phys. Rept. {\bf 100} (1983) 1.

\bibitem{Kovchegov:2001sc}
Y.~V. Kovchegov and K.~Tuchin, 
Phys. Rev. D {\bf 65} (2002) 074026.

\bibitem{deFlorian:2007aj}
D.~de~Florian, R.~Sassot and M.~Stratmann, 
 Phys. Rev. D {\bf 75} (2007) 114010. 

\bibitem{Gelis:2009wh} 
 F.~Gelis, T.~Lappi and L.~McLerran,
 Nucl.\ Phys.\ A {\bf 828} (2009) 149.

\bibitem{Schenke:2012wb} 
 B.~Schenke, P.~Tribedy and R.~Venugopalan,
 Phys.\ Rev.\ Lett.\  {\bf 108} (2012) 252301.

\bibitem{Schenke:2013dpa}
  B.~Schenke, P.~Tribedy and R.~Venugopalan,
  Phys.\ Rev.\ C {\bf 89} (2014)  024901.

\bibitem{McLerran:2015qxa} 
 L.~McLerran and P.~Tribedy,
 Nucl.\ Phys.\ A {\bf 945} (2016) 216.
  
\bibitem{Kowalski:2003hm} 
 H.~Kowalski and D.~Teaney,
 Phys.\ Rev.\ D {\bf 68} (2003) 114005.
 
\bibitem{Rezaeian:2012ji}
  A.~H.~Rezaeian, M.~Siddikov, M.~Van de Klundert and R.~Venugopalan,
  Phys.\ Rev.\ D {\bf 87} (2013) 034002.

\bibitem{Schenke:2015aqa}
  B.~Schenke, S.~Schlichting and R.~Venugopalan,
  Phys.\ Lett.\ B {\bf 747} (2015) 76.

\bibitem{Schenke:2016lrs}
  B.~Schenke, S.~Schlichting, P.~Tribedy and R.~Venugopalan,
  Phys.\ Rev.\ Lett.\  {\bf 117} (2016) 162301.


\bibitem{HIJING} 
  X.~N.~Wang and M.~Gyulassy,
 Phys.\ Rev. D {\bf 44} (1991) 3501.

\bibitem{HP2017}
  G.~G.~Barnaf\"oldi {\it et al.},
  arXiv:1701.08496 [hep-ph].

\bibitem{PYTHIAv5} 
  T.~Sjostrand,
 Comput.\ Phys.\ Commun.  {\bf 82} (1994) 74.

\bibitem{FRITIOF} 
B.~Nilsson-Almqvist and E.~Stenlund,
 Comput.\ Phys.\ Commun.  {\bf 43} (1987) 387.

\bibitem{ARIADNE} 
 L.~Lonnblad,
 Comput.\ Phys.\ Commun.  {\bf 71} (1992) 15.

\bibitem{CERNLIB} CERNLib:  https://cernlib.web.cern.ch/cernlib/ (2017)

\bibitem{HIJING2}
  W.~T.~Deng, X.~N.~Wang and R.~Xu,
  Phys.\ Rev. C {\bf 83} (2011) 014915.

\bibitem{HIJINGsh} 
  X.~N.~Wang,
 Phys.\ Rev. C {\bf 61} (2000) 064910.
  
\bibitem{PYTHIA} 
  T.~Sjöstrand {\it et al.},
 Comput.\ Phys.\ Commun.  {\bf 191} (2015) 159.

\bibitem{LHAPDF6}
  A. Buckley {\it et al.}, Eur. Phys. J. C {\bf 75} (2015) 132;
  http://lhapdf.hepforge.org/  (2017)

\bibitem{Monash2013}
  P.~Skands, S.~Carrazza and J.~Rojo,
 Eur.\ Phys.\ J. C {\bf 74} (2014) 3024.

\bibitem{Perugia}
  P.~Z.~Skands, Phys. Rev. D {\bf 82} (2010) 074018.
  
\bibitem{ampt}
This and previous versions of the $\mathtt{AMPT}$ codes are available at http://myweb.ecu.edu/linz/ampt 

\bibitem{Lin:2004en}
Z.~W.~Lin, C.~M.~Ko, B.~-A.~Li, B.~Zhang and S.~Pal,
 Phys.\ Rev. C {\bf 72} (2005) 064901.

\bibitem{Lin:2014tya} 
  Z.~W.~Lin,
  Phys.\ Rev. C {\bf 90} (2014) 014904.

\bibitem{Ma:2016fve} 
  G.~L.~Ma and Z.~W.~Lin,
  Phys.\ Rev. C {\bf 93} (2016) 054911.

\bibitem{Gyulassy:2003mc} 
  M.~Gyulassy, I.~Vitev, X.~-N.~Wang and B.~-W.~Zhang,
  In *Hwa, R.C. (ed.) {\it et al.}: Quark gluon plasma* 123-191
  [nucl-th/0302077].

\bibitem{Vitev:2002pf} 
  I.~Vitev and M.~Gyulassy,
  Phys.\ Rev.\ Lett.  {\bf 89} (2002) 252301.

\bibitem{Kang:2014hha} 
  Z.~B.~Kang, I.~Vitev, E.~Wang, H.~Xing and C.~Zhang,
  Phys.\ Lett. B {\bf 740} (2015)  23.

\bibitem{Qiu:2003pm} 
  J.~w.~Qiu and I.~Vitev,
 Phys.\ Lett. B {\bf 570} (2003) 161,

\bibitem{Vitev:2004gn} 
  I.~Vitev,
   Phys.\ Lett. B {\bf 606} (2005) 303.

  \bibitem{Vitev:2008vk}
    I. Vitev and B.-W. Zhang,
    Phys. Lett. B {\bf  669} (2008)  337. 

\bibitem{Vitev:2007ve} 
  I.~Vitev,
  Phys.\ Rev. C {\bf 75} (2007) 064906.
  
\bibitem{Neufeld:2010dz} 
  R.~B.~Neufeld, I.~Vitev and B.-W.~Zhang,
 Phys.\ Lett. B {\bf 704} (2011) 590.

\bibitem{Kang:2015mta} 
Z.~B.~Kang, I.~Vitev and H.~Xing,
 Phys.\ Rev. C {\bf 92} (2015) 054911.

\bibitem{Qiu:2004qk} 
  J.~-W.~Qiu and I.~Vitev,
 Phys.\ Lett. B {\bf 587} (2004) 52.

\bibitem{YZ02}
 Y.~Zhang, G.~I.~Fai, G.~Papp, G.~G.~Barnafoldi and P.~Levai,
  Phys.\ Rev. C {\bf 65} (2002) 034903.

\bibitem{pgNLO}
 G.~Papp, G.~G.~Barnafoldi, P.~Levai and G.~I.~Fai,
  hep-ph/0212249.

\bibitem{MRST01} 
 A.~D.~Martin, R.~G.~Roberts, W.~J.~Stirling and R.~S.~Thorne,
  Eur.\ Phys.\ J. C {\bf 28} (2003) 455.

\bibitem{KKP}
S.~Albino, B.~A.~Kniehl, G.~Kramer and W.~Ochs,
 Eur.\ Phys.\ J. C {\bf 36} (2004) 49.

\bibitem{Adeola:2009} 
  A.~Adeluyi, G.~G.~Barnafoldi, G.~Fai and P.~Levai,
 Phys.\ Rev. C {\bf 80} (2009) 014903.

\bibitem{Barnafoldi:2011px} 
  G.~G.~Barnafoldi, J.~Barrette, M.~Gyulassy, P.~Levai and V.~Topor Pop,
 Phys.\ Rev. C {\bf 85} (2012) 024903.
  
\bibitem{Cron75}
     J.~W.~Cronin  {\it et al.} [CP Collaboration], 
Phys. Rev. D {\bf 11} (1975) 3105.

\bibitem{Antr79}
  D.~Antreasyan, J.~W.~Cronin, H.~J.~Frisch, M.~J.~Shochet, L.~Kluberg,
  P.~A.~Piroue and R.~L.~Sumner,
 Phys.\ Rev. D {\bf 19} (1979) 764.

\bibitem{Eskola:2009uj}
  K.~J.~Eskola, H.~Paukkunen and C.~A.~Salgado,
   JHEP {\bf 0904} (2009) 065.  
 
\bibitem{Rezaeian:2011ia}
A. H. Rezaeian, Phys. Rev. D {\bf 85} (2012) 014028.

\bibitem{Albacete:2012xq}
J.~L.~Albacete, A.~Dumitru, H.~Fujii and Y.~Nara,
 	Nucl. Phys. A {\bf 897} (2013) 1.

\bibitem{Rezaeian:2012ye}
A. H. Rezaeian, Phys. Lett. B {\bf 718} (2013) 1058.


\bibitem{Levin:2010dw}
  E. Levin and A. H. Rezaeian,
  Phys. Rev. D {\bf 82} (2010) 014022.

\bibitem{Levin:2010br}
  E. Levin and A. H. Rezaeian, AIP Conf.\ Proc.\  {\bf 1350} (2011) 243
  [arXiv:1011.3591 [hep-ph]].

\bibitem{Levin:2010zy}
  E. Levin and A. H. Rezaeian,
  Phys. Rev. D {\bf 82} (2010) 054003.
  
\bibitem{Levin:2011hr}
  E. Levin and A. H. Rezaeian,
  Phys. Rev. D {\bf 83} (2011) 114001.


\bibitem{Rezaeian:2013woa}  
  A. H. Rezaeian, Phys. Lett. B {\bf 727} (2013) 218.

\bibitem{Rezaeian:2013tka}
  A. H. Rezaeian and I. Schmidt, Phys. Rev. D {\bf 88} (2013) 074016.

\bibitem{Armesto:2014sma}  
  N. Armesto and A.~H.~Rezaeian, Phys. Rev. D {\bf 90} (2014) 054003.

\bibitem{Back:2003hx}
  B. B. Back {\it et al.} [PHOBOS Collaboration],
  Phys. Rev. Lett. {\bf 93} (2004) 082301.

\bibitem{Arsene:2004cn}  
  I. Arsene {\it et al.} [BRAHMS Collaboration],
  Phys. Rev. Lett. {\bf 94} (2005) 032301.


\bibitem{Lappi:2003bi}
  T.~Lappi,
  Phys.\ Rev.\ C {\bf 67} (2003) 054903.

\bibitem{Krasnitz:2001qu}
  A.~Krasnitz, Y.~Nara and R.~Venugopalan,
  Phys.\ Rev.\ Lett.\  {\bf 87} (2001) 192302.

\bibitem{Aamodt:2010pp} 
  K.~Aamodt {\it et al.} [ALICE Collaboration],
  Eur.\ Phys.\ J.\ C {\bf 68} (2010) 345.

\bibitem{CMS:2012qk} 
 S.~Chatrchyan {\it et al.} [CMS Collaboration],
 Phys.\ Lett.\ B {\bf 718} (2013) 705.
        
\bibitem{Aad:2015zza}
  G.~Aad {\it et al.} [ATLAS Collaboration],
  Eur.\ Phys.\ J. C {\bf 76} (2016) 199.

\bibitem{1412.6282v2}
  J. Adam {\it et al.} [ALICE Collaboration],
  Phys. Rev. C {\bf 91} (2015) 064905.


  
\bibitem{Khachatryan:2015xaa}
  V.~Khachatryan {\it et al.}  [CMS Collaboration],
 Eur.\ Phys.\ J. C {\bf 75} (2015) 237.

\bibitem{ATLAS:2014cza}
  The ATLAS collaboration [ATLAS Collaboration],
  ATLAS-CONF-2014-029, ATLAS-COM-CONF-2014-031.

\bibitem{Aad:2016zif}
  G.~Aad {\it et al.} [ATLAS Collaboration],
 Phys.\ Lett. B {\bf 763} (2016) 313.

\bibitem{Abelev:2014dsa}
  B.~B.~Abelev {\it et al.}  [ALICE Collaboration],
 Eur.\ Phys.\ J. C {\bf 74} (2014) 3054.

\bibitem{Khachatryan:2016odn}
  V.~Khachatryan {\it et al.} [CMS Collaboration],
 $\sqrt{s_{\mathrm{NN}}}= $ 5.02 TeV,''
  arXiv:1611.01664 [nucl-ex].

\bibitem{d'Enterria:2013vba}
  D.~d'Enterria, K.~J.~Eskola, I.~Helenius and H.~Paukkunen,
 Nucl.\ Phys. B {\bf 883} (2014) 615.
 
\bibitem{Helenius:2012wd}
  I.~Helenius, K.~J.~Eskola, H.~Honkanen and C.~A.~Salgado,
 JHEP {\bf 1207} (2012) 073

\bibitem{Aurenche:1999nz}
  P.~Aurenche, M.~Fontannaz, J.~P.~Guillet, B.~A.~Kniehl and M.~Werlen,
  Eur.\ Phys.\ J. C {\bf 13} (2000) 347.

\bibitem{Lai:2010vv}
  H.~L.~Lai, M.~Guzzi, J.~Huston, Z.~Li, P.~M.~Nadolsky, J.~Pumplin and
  C.-P.~Yuan,
 Phys.\ Rev. D {\bf 82} (2010) 074024.

\bibitem{Kretzer:2000yf}
  S.~Kretzer,
 Phys.\ Rev. D {\bf 62} (2000) 054001.

\bibitem{Kniehl:2000fe}
  B.~A.~Kniehl, G.~Kramer and B.~Potter,
 Nucl.\ Phys. B {\bf 582} (2000) 514.
  
\bibitem{deFlorian:2007hc}
  D.~de Florian, R.~Sassot and M.~Stratmann,
 Phys.\ Rev. D {\bf 76} (2007) 074033.

\bibitem{Dumitru:2005gt}
  A. Dumitru, A. Hayashigaki and J. Jalilian-Marian,
  Nucl. Phys. A {\bf 765} (2006) 464.

\bibitem{Albacete:2010ad}
J. L. Albacete and A. Dumitru, arXiv:1011.5161. 

\bibitem{JalilianMarian:2011dt}
  J. Jalilian-Marian and  A. H. Rezaeian,
  Phys. Rev. D {\bf 85} (2012) 014017.

\bibitem{dEnterria:2003xac}
  D.~G.~d'Enterria,
  nucl-ex/0302016.

\bibitem{Abelev:2013haa}
  B.~B.~Abelev {\it et al.} [ALICE Collaboration],
  Phys.\ Lett.\ B {\bf 728} (2014) 25.

\bibitem{Chatrchyan:2013eya} 
  S.~Chatrchyan {\it et al.} [CMS Collaboration],
  Eur.\ Phys.\ J.\ C {\bf 74}(2014) 2847.

\bibitem{Schnedermann:1993ws} 
  E.~Schnedermann, J.~Sollfrank and U.~W.~Heinz,
  Phys.\ Rev.\ C {\bf 48} (1993) 2462.

\bibitem{Werner:2013tya} 
  K.~Werner, B.~Guiot, I.~Karpenko and T.~Pierog,
  Phys.\ Rev.\ C {\bf 89} (2014) 064903.
  
  \bibitem{Ducloue:2016pqr}
B.~Duclou\'e, T.~Lappi and H.~M\"antysaari, 
Phys. Rev. D {\bf 94} (2016) 074031.
 
\bibitem{Li:2014zza} 
  W.~Li [CMS Collaboration],
  Ann. Phys.  {\bf 352} (2015) 35.

\bibitem{CMS:2013bza} 
  S.~Chatrchyan {\it et al.}  [CMS Collaboration],
  JHEP {\bf 1402} (2014) 088.

\bibitem{Bzdak:2014dia} 
  A.~Bzdak and G.~L.~Ma,
  Phys.\ Rev.\ Lett.  {\bf 113} (2014) 252301.

\bibitem{Chatrchyan:2013nka}
S.~Chatrchyan {\it et al.} [CMS Collaboration],
{\it Phys.\ Lett.\ B} {\bf 724} (2013) 213.
  
\bibitem{Nelson:2012bc}
R. E. Nelson, R. Vogt and A. D. Frawley,  Phys. Rev. C, 
{\bf 87} (2013) 014908.

\bibitem{Vogt:2015uba}
  R.~Vogt,
   Phys.\ Rev.\ C {\bf 92} (2015) 034909.

\bibitem{Lansberg:2016deg}
  J.~P.~Lansberg and H.~S.~Shao,
 Eur.\ Phys.\ J. C {\bf 77} (2017) 1.

\bibitem{LHAPDF5}
M. R. Whalley, D. Bourilkov, and R. C. Group, arXiv:hep-ph/0508110.


\bibitem{Shao:2015vga}
  H.~S.~Shao,
 Comput.\ Phys.\ Commun.  {\bf 198} (2016) 238.

\bibitem{Kovarik:2015cma}
  K.~Kovarik {\it et al.},
  Phys.\ Rev. D {\bf 93} (2016) 085037.


\bibitem{CT14}
S. Dulat {\it et al.}, Phys. Rev. D {\bf 93} (2016) 033006.



\bibitem{Arleo:2012hn}
F.~Arleo and S.~Peign{\'e}, 
 Phys. Rev. Lett. {\bf 109} (2012) 122301.

\bibitem{Arleo:2012rs}
F. Arleo and S. Peign\'e, 
 JHEP {\bf 1303} (2013) 122.

\bibitem{Peigne:2014uha}
S.~Peign{\'e}, F.~Arleo, and R.~Kolevatov, 
 Phys. Rev. D {\bf 98} (2016) 014006.

  \bibitem{Leitch:1999ea}
M.~J. Leitch {\it et~al.} [NuSea Collaboration],
  Phys. Rev. Lett. {\bf 84} (2000) 3256. 

  \bibitem{Arleo:2010rb}
F.~Arleo, S.~Peign{\'e}, and T.~Sami, 
 Phys. Rev. D {\bf 83} (2011) 114036,

\bibitem{ALICE_psiy5} 
  B. Abelev {\it et al.}  [ALICE Collaboration],
JHEP {\bf 1402} (2014) 073.

\bibitem{LHCb_psiy5} R. Aaij {\it et al.} [LHCb Collaboration], 
JHEP {\bf 1402} (2014) 072.


\bibitem{ALICE_psi_pT}
J.~Adam {\it et al.}  [ALICE Collaboration],
JHEP {\bf 1506} (2015) 055.

\bibitem{Adare:2013ezl} 
  A.~Adare {\it et al.}  [PHENIX Collaboration],
  Phys.\ Rev.\ Lett.  {\bf 111} (2013) 202301.
    
\bibitem{Abelev:2014zpa} 
  B.~B.~Abelev {\it et al.}  [ALICE Collaboration],
 JHEP {\bf 1412} (2014) 073.
    
\bibitem{Arnaldi:2014kta} 
  R.~Arnaldi [ALICE Collaboration],
  Nucl.\ Phys. A {\bf 931} (2014) 628.
 
\bibitem{Aaij:2016eyl} 
  R.~Aaij {\it et al.} [LHCb Collaboration],
 JHEP {\bf 1603} (2016) 133.
 
\bibitem{Ferreiro:2014bia} 
  E.~G.~Ferreiro,
 Phys.\ Lett. B {\bf 749} (2015) 98.
  
\bibitem{Gavin:1996yd} 
  S.~Gavin and R.~Vogt,
  Phys.\ Rev.\ Lett.  {\bf 78} (1997) 1006.
 
 \bibitem{Capella97} A.~Capella, A.~Kaidalov, A.~Kouider Akil and
  C.~Gerschel,  Phys.\ Lett. B {\bf 393} (1997) 431.

\bibitem{Armesto98} N.~Armesto and A.~Capella,
 Phys. Lett. B {\bf 430} (1998) 23.

\bibitem{Armesto99} N.~Armesto, A.~Capella and E. G.~Ferreiro,
 Phys. Rev. C {\bf 59} (1999) 395.

\bibitem{Capella00} A.~Capella, E. G.~Ferreiro and A. B.~Kaidalov,
  Phys. Rev. Lett. {\bf 85} (2000) 2080.
 
\bibitem{Capella05} A.~Capella and E. G.~Ferreiro,
 Eur. Phys. J. C {\bf 42} (2005) 419.
    
  \bibitem{Capella:2006mb}
  A.~Capella and E.~G.~Ferreiro,
  Phys.\ Rev. C {\bf 76} (2007) 064906.
 
\bibitem{Capella:2007jv} 
  A.~Capella, L.~Bravina, E.~G.~Ferreiro, A.~B.~Kaidalov, K.~Tywoniuk and
  E.~Zabrodin,
  Eur.\ Phys.\ J. C {\bf 58} (2008) 437.
  
\bibitem{Ferreiro:2012rq} 
  E.~G.~Ferreiro,
  Phys.\ Lett. B {\bf 731} (2014) 57.
 
\bibitem{Ferreiro:2012mm} 
  E.~G.~Ferreiro, F.~Fleuret, J.~P.~Lansberg and A.~Rakotozafindrabe,
   J.\ Phys.\ Conf.\ Ser.  {\bf 422} (2013) 012018.
 
\bibitem{Ferreiro:2013pua} 
  E.~G.~Ferreiro, F.~Fleuret, J.~P.~Lansberg and A.~Rakotozafindrabe,
  Phys.\ Rev. C {\bf 88} (2013) 047901.

  

\bibitem{Ducloue:2015gfa}
B.~Duclou\'e, T.~Lappi and H.~M\"antysaari, 
Phys. Rev. D {\bf 91} (2015) 114005. 

\bibitem{McLerran:1993ni}
L.~D. McLerran and R.~Venugopalan
Phys. Rev. D {\bf  49} (1994) 2233.

\bibitem{McLerran:1993ka}
L.~D. McLerran and R.~Venugopalan
Phys. Rev. D {\bf  49} (1994) 3352.

\bibitem{Iancu:2003xm}
E.~Iancu and R.~Venugopalan
arXiv:hep-ph/0303204  [hep-ph].

\bibitem{Gelis:2010nm}
F.~Gelis, E.~Iancu, J.~Jalilian-Marian, and R.~Venugopalan,
 Ann. Rev. Nucl. Part. Sci. {\bf 60} (2010) 463.

\bibitem{Bodwin:1994jh}
G.~T. Bodwin, E.~Braaten, and G.~P. Lepage,
Phys. Rev. D {\bf 51} (1995) 1125.

\bibitem{Kang:2013hta}
Z.-B. Kang, Y.-Q. Ma, and R.~Venugopalan,
JHEP {\bf 1401}
  (2014) 056.

\bibitem{Ma:2015sia}
  Y.-Q. Ma, R. Venugopalan and H.-F. Zhang, Phys. Rev. D {\bf 92} (2015)
  071901.

\bibitem{Ma:2014mri}
Y.-Q. Ma and R.~Venugopalan
Phys. Rev. Lett.
  {\bf 113} (2014) 192301.

\bibitem{Blaizot:2004wv}
J.~P. Blaizot, F.~Gelis, and R.~Venugopalan,
Nucl. Phys. A
  {\bf 743} (2004) 57.

\bibitem{Chao:2012iv}
K.-T. Chao, Y.-Q. Ma, H.-S. Shao, K.~Wang, and Y.-J. Zhang,
Phys. Rev. Lett.
  {\bf 108} (2012) 242004.

\bibitem{Balitsky:1995ub}
I.~Balitsky
Nucl. Phys. B
  {\bf 463} (1996) 99.

\bibitem{Kovchegov:1999yj}
Y.~V. Kovchegov,
Phys. Rev. D {\bf
  60} (1999) 034008.


\bibitem{Kang:2013ufa}
  Z.~B.~Kang, I.~Vitev and H.~Xing,
  Phys.\ Rev. D {\bf 88} (2013) 054010.

\bibitem{Luo:1994np} 
  M.~Luo, J.~w.~Qiu and G.~F.~Sterman,
   Phys.\ Rev. D {\bf 50} (1994) 1951.

\bibitem{Qiu:2003vd} 
  J.~w.~Qiu and I.~Vitev,
 Phys.\ Rev.\ Lett.  {\bf 93} (2004) 262301.

\bibitem{LHCbD0}
  R.~Aaij {\it et al.} [LHCb Collaboration],
  arXiv:1707.02750 [hep-ex].
 
\bibitem{Aaij:2013noa}
  R.~Aaij {\it et al.} [LHCb Collaboration],
  JHEP {\bf 1308} (2013) 117.


\bibitem{Abelev:2013yxa}
B. B. Abelev {\it et al.}  [ALICE Collaboration], 
 JHEP {\bf 1402} (2014) 073.

\bibitem{Aaij:2013zxa}
R. Aaij {\it et al.} [LHCb Collaboration], 
 JHEP {\bf 1402} (2014) 072.

\bibitem{Arleo:2015qiv}
F.~Arleo and S.~Peign{\'e}, Phys. Rev. D {\bf 95} (2017) 011502.

\bibitem{Catani:2007vq}
  S.~Catani and M.~Grazzini,
 Phys.\ Rev.\ Lett. {\bf 98} (2007) 222002.

\bibitem{Catani:2009sm}
  S.~Catani, L.~Cieri, G.~Ferrera, D.~de Florian and M.~Grazzini,
 Phys.\ Rev.\ Lett.  {\bf 103} (2009) 082001

\bibitem{Martin:2009iq}
  A.~D.~Martin, W.~J.~Stirling, R.~S.~Thorne and G.~Watt,
  Eur.\ Phys.\ J.\ C {\bf 63} (2009) 189.
%
  
\bibitem{Kutak:2014wga}
  K.~Kutak,
 Phys.\ Rev. D {\bf 91} (2015) 034021.

\bibitem{vanHameren:2014ala}
  A.~van Hameren, P.~Kotko, K.~Kutak and S.~Sapeta,
 Phys.\ Lett. B {\bf 737} (2014) 335.

\bibitem{Mueller:2013wwa}
  A.~H.~Mueller, B.~W.~Xiao and F.~Yuan,
 Phys.\ Rev. D {\bf 88} (2013) 114010.

\bibitem{Catani:1990eg}
  S.~Catani, M.~Ciafaloni, F.~Hautmann,
 Nucl.\ Phys. B  {\bf 366 } (1991)  135.

\bibitem{Kotko:2015ura}
  P.~Kotko, K.~Kutak, C.~Marquet, E.~Petreska, S.~Sapeta and A.~van Hameren,
JHEP {\bf 1509} (2015) 106.

\bibitem{vanHameren:2016ftb}
  A.~van Hameren, P.~Kotko, K.~Kutak, C.~Marquet, E.~Petreska and S.~Sapeta,
 JHEP {\bf 1612} (2016) 034.

\bibitem{Deak:2010gk}
  M.~Deak, F.~Hautmann, H.~Jung and K.~Kutak,
  arXiv:1012.6037 [hep-ph].


\bibitem{Kutak:2003bd}
  K.~Kutak and J.~Kwiecinski,
  Eur.\ Phys.\ J. C {\bf 29} (2003) 521.

\bibitem{Kutak:2004ym}
  K.~Kutak and A.~M.~Stasto,
 Eur.\ Phys.\ J. C {\bf 41} (2005) 343.

\bibitem{Kutak:2012rf}
  K.~Kutak and S.~Sapeta,
 Phys.\ Rev. D {\bf 86} (2012) 094043.


\bibitem{Schienbein:2009kk}
  I.~Schienbein, J.~Y.~Yu, K.~Kovarik, C.~Keppel, J.~G.~Morfin, F.~Olness and J.
~F.~Owens,
 Phys.\ Rev. D {\bf 80} (2009) 094004.

\bibitem{Kovarik:2010uv}
  K.~Kovarik, I.~Schienbein, F.~I.~Olness, J.~Y.~Yu, C.~Keppel, J.~G.~Morfin, J.
~F.~Owens and T.~Stavreva,
 Phys.\ Rev.\ Lett. {\bf 106} (2011) 122301.

\bibitem{Kulagin:2004ie}
  S.~A.~Kulagin and R.~Petti,
 Nucl.\ Phys. A {\bf 765} (2006) 126.

\bibitem{Kulagin:2014vsa}
  S.~A.~Kulagin and R.~Petti,
 Phys.\ Rev. C {\bf 90} (2014) 045204.

\bibitem{Alekhin:2015cza}
  S.~Alekhin, J.~Bluemlein, S.~Moch and R.~Placakyte,
  Phys. Rev. D {\bf 94} (2016) 114038.

\bibitem{Ru:2014yma}
  P.~Ru, B.~W.~Zhang, L.~Cheng, E.~Wang and W.~N.~Zhang,
 J.\ Phys. G {\bf 42} (2015) 085104.

\bibitem{Ru:2016wfx}
  P.~Ru, S.~A.~Kulagin, R.~Petti and B.~W.~Zhang,
  Phys. Rev. D {\bf 94} (2016) 113013.

\bibitem{CMS:2014sca}
  A. J. Zsigmond {\it et al.} [CMS Collaboration],
Nucl. Phys. A {\bf 931} (2014) 718.

\bibitem{Kang:2012am}
  Z.~B.~Kang and J.~W.~Qiu,
 Phys.\ Lett. B {\bf 721} (2013) 277.

\bibitem{Ru:2015pfa}
  P.~Ru, B.~W.~Zhang, E.~Wang and W.~N.~Zhang,
 Eur.\ Phys.\ J. C {\bf 75} (2015) 426.

\bibitem{Aaij:2012dz}
  R.~Aaij {\it et al.} [LHCb Collaboration],
  JHEP {\bf 1206} (2012) 141.
   Addendum: [JHEP {\bf 1403} (2014) 108].
 

\bibitem{Czakon:2013goa}M.~Czakon, P.~Fiedler and A.~Mitov,  Phys.\ Rev.\ Lett.\
  {\bf 110} (2013) 252004.

\bibitem{mcfm1} J.~M.~Campbell and R.~K.~Ellis,
  Nucl.\ Phys.\ Proc.\ Suppl.\ {\bf 205-206} (2010) 10.
 
\bibitem{mcfm2}  J.~M.~Campbell and R.~K.~Ellis,
  J.\ Phys.\ G {\bf 42} (2015) 015005.  

\bibitem{d'Enterria:2015mgr} D.~d'Enterria, K.~Krajcz\'ar and H.~Paukkunen,
  Phys.\ Lett.\ B {\bf 746} (2015) 64.

\bibitem{deFlorian:2011fp}
  D.~de Florian, R.~Sassot, P.~Zurita and M.~Stratmann,
  Phys.\ Rev. D {\bf 85} (2012) 074028.

\bibitem{Frankfurt:2011cs} L.~Frankfurt, V.~Guzey and M.~Strikman,
  Phys.\ Rept.\  {\bf 512} (2012) 255.

\bibitem{Alice:2016wka}J.~Adam {\it et al.} [ALICE Collaboration], JHEP
  {\bf 1702} (2017) 077.

\bibitem{Dainese:2016gch}
  A.~Dainese, U. A. Wiedemann, N. Armesto, D. d'Enterria, J. M. Jowett
  {\it et al.}, CERN Yellow Report (2017) 635; [arXiv:1605.01389 [hep-ph]].

\bibitem{Apolinario:2017}%
  L.~Apolin\'ario, N.~Armesto, G.~Milhano, G.~Salam and C.A.~Salgado,
  in progress.

\bibitem{Olive:2016xmw}
  C.~Patrignani {\it et al.} [Particle Data Group], Chin.\ Phys.\ C {\bf 40}
  (2016) 100001.

\bibitem{Gao:2013xoa}
J.~Gao {\it et al.}, Phys.\ Rev.\ D {\bf 89} (2014)  033009.


\bibitem{Chatrchyan:2011ew}S.~Chatrchyan {\it et al.} [CMS Collaboration],
  Eur.\ Phys.\ J.\ C {\bf 71} (2011) 1721.

\bibitem{Chatrchyan:2011nb} S.~Chatrchyan {\it et al.} [CMS Collaboration],
  JHEP {\bf 1107} (2011) 049.

\bibitem{ATLAS:2012aa} G.~Aad {\it et al.} [ATLAS Collaboration],
  JHEP {\bf 1205} (2012) 059.

\bibitem{Chatrchyan:2013kff}
  S.~Chatrchyan {\it et al.} [CMS Collaboration],  Eur.\ Phys.\ J.\ C {\bf 73}
  (2013) 2386.

\bibitem{Aad:2015pga}  G.~Aad {\it et al.} [ATLAS Collaboration],
  Phys.\ Rev.\ D {\bf 91} (2015) 112013.

\bibitem{Khachatryan:2015uqb}
  V.~Khachatryan {\it et al.} [CMS Collaboration],  Phys.\ Rev.\ Lett.\
  {\bf 116} (2016)  052002.
 
\end{thebibliography}
\end{document}